\documentclass[final,3p,times,authoryear]{elsarticle}
\usepackage{graphicx}
\usepackage[T1]{fontenc}
\usepackage{endnotes}

\usepackage{hyperref}
\usepackage{ifthen}

\usepackage{amssymb}

\journal{Physics Reports}

\def\gsim{~\rlap{$>$}{\lower 1.0ex\hbox{$\sim$}}}
\def\lsim{~\rlap{$<$}{\lower 1.0ex\hbox{$\sim$}}}
\def\G{{\rm G}}
\def\clight{{\rm c}}
\def\d{{\rm d}}

\newcommand{\pcite}[1]{(\citealt{#1})}
\newcommand{\sref}[1]{\raisebox{0.8ex}{\scriptsize\ref{#1}}}
\newcounter{AGNDone}
\def\AGN{\ifthenelse{\equal{\arabic{AGNDone}}{0}}{active galactic nuclei (AGN)\setcounter{AGNDone}{1}}{AGN}}
\newcounter{CDMDone}
\def\CDM{\ifthenelse{\equal{\arabic{CDMDone}}{0}}{cold dark matter (CDM)\setcounter{CDMDone}{1}}{CDM}}
\newcounter{CMBDone}
\def\CMB{\ifthenelse{\equal{\arabic{CMBDone}}{0}}{cosmic microwave background (CMB)\setcounter{CMBDone}{1}}{CMB}}
\newcounter{ICMDone}
\def\ICM{\ifthenelse{\equal{\arabic{ICMDone}}{0}}{intracluster medium (ICM)\setcounter{ICMDone}{1}}{ICM}}
\newcounter{ISMDone}
\def\ISM{\ifthenelse{\equal{\arabic{ISMDone}}{0}}{interstellar medium (ISM)\setcounter{ISMDone}{1}}{ISM}}
\newcounter{NFWDone}
\def\NFW{\ifthenelse{\equal{\arabic{NFWDone}}{0}}{Navarro-Frenk-White (NFW)\setcounter{NFWDone}{1}}{NFW}}
\newcounter{SMBHDone}
\def\SMBH{\ifthenelse{\equal{\arabic{SMBHDone}}{0}}{supermassive black hole (SMBH)\setcounter{SMBHDone}{1}}{SMBH}}
\def\SMBHs{\ifthenelse{\equal{\arabic{SMBHDone}}{0}}{supermassive black holes (SMBH)\setcounter{SMBHDone}{1}}{SMBHs}}

\begin{document}

\begin{frontmatter}



\title{Galaxy Formation Theory}


\author{Andrew J. Benson}

\address{California Institute of Technology, MC350-17, 1200 E. California Blvd., Pasadena, CA 91125, U.S.A. (E-Mail: {\tt abenson@caltech.edu})}

\begin{abstract}
We review the current theory of how galaxies form within the cosmological framework provided by the cold dark matter paradigm for structure formation. Beginning with the pre-galactic evolution of baryonic material we describe the analytical and numerical understanding of how baryons condense into galaxies, what determines the structure of those galaxies and how internal and external processes (including star formation, merging, active galactic nuclei, etc.) determine their gross properties and evolution. Throughout, we highlight successes and failures of current galaxy formation theory. We include a review of computational implementations of galaxy formation theory and assess their ability to provide reliable modeling of this complex phenomenon. We finish with a discussion of several ``hot topics'' in contemporary galaxy formation theory and assess future directions for this field.
\end{abstract}


\end{frontmatter}

\tableofcontents

\section{Introduction}

While the concept of galaxies as ``island universes'' can be traced back to \cite{wright_original_1750} and \cite{kant_allgemeine_1755} the study of the formation of galaxies did not begin until after their extra-Galactic status was confirmed by \cite{hubble_spiral_1929}. In fact, much of the early work on galaxy evolution and formation was driven by the necessity of understanding galaxies in order to answer questions of cosmology (such as whether or not the Universe began with a Big Bang). While an understanding of galaxies remains necessary for such reasons even today, the field has since become an important one in its own right.

Modern galaxy formation theory therefore grew out of early studies of cosmology and structure formation and is set within the cold dark matter cosmological model and so proceeds via a fundamentally hierarchical paradigm. Observational evidence and theoretical expectations indicate that galaxy formation is an ongoing process which has been occurring over the vast majority of the Universe's history. The goal of galaxy formation theory then is to describe how simple physics gives rise to the complicated set of phenomena which galaxies encompass.

Galaxy formation is very much an observationally driven field in the sense that we are still decidedly in the stage of making new experimental discoveries rather than performing precision tests of well-specified theoretical models. While this situation shows signs of a gradual shift to the ``precision tests'' phase it seems unlikely that the transition will be completed any time soon. In addition, astronomy is perhaps uniquely hindered by experimental biases, since we are not able to design the experiment, merely to observe what the Universe has decided to put on show. The complicated nature of the resulting selection effects result in a secondary, but very important, role for theoretical models, namely in quantifying these biases and interpreting the data. While this secondary role is well established it needs to become more so, in particular it should become an integral part of any observational campaign and will require direct and simple access to modeling capabilities for all.

\section{Background Material}

Modern galaxy formation theory is set within the larger scale cold dark matter cosmological model \pcite{blumenthal_formation_1984}. The success of that model in explaining the \CMB\ \citep{komatsu_five-year_2009} and large scale structure \citep{seljak_cosmological_2005,percival_shape_2007,ferramacho_constraintscdm_2008,sanchez_cosmological_2009} of the Universe makes it the \emph{de facto} standard. However, it is important to recognize that the scales on which the simplest \CDM\ model has been most precisely tested are much larger than those that matter for galaxy formation\footnote{Although measurements of the Lyman-$\alpha$ forest \protect\citep{slosar_cosmological_2007,viel_cold_2008} and weak-lensing \protect\citep{mandelbaum_density_2006} give interesting constraints on the distribution of dark matter on small scales they typically require modeling of either the nonlinear evolution of dark matter or the behavior of baryons (or both) which complicate any interpretation. They are therefore not as ``clean'' as \CMB\ and large scale structure constraints.}. As such, we cannot fully rule out that dark matter is warm or self-interacting, although good constraints exist on both of these properties \citep{markevitch_constraintsdark_2004,boehm_constraintsdark_2005,ahn_formation_2005,miranda_constraining_2007,randall_constraintsself-interaction_2008,yuksel_strong_2008,boyarsky_lyman-alpha_2008}.

More extensive reviews of the cold dark matter cosmological model can be found in, for example, \cite{narlikar_standard_2001}, \cite{frenk_simulatingformation_2002} and \cite{bertone_particle_2005}.

\subsection{Background Cosmology}

A combination of experimental measures, including studies of the \CMB\ \pcite{dunkley_five-year_2009}, large scale structure \pcite{tegmark_three-dimensional_2004,cole_2df_2005,tegmark_cosmological_2006,percival_shape_2007,percival_measuringmatter_2007}, the Type Ia supernovae magnitude--redshift relation \pcite{kowalski_improved_2008} and galaxy clusters \citep{mantz_observed_2009,vikhlinin_chandra_2009,rozo_cosmological_2010}, have now placed strong constraints on the parameters of the cold dark matter cosmogony. The picture that emerges \citep{komatsu_seven-year_2010} is one in which the energy density of the Universe is shared between dark energy ($\Omega_\Lambda=0.728^{+0.015}_{-0.016}$), dark matter ($\Omega_{\rm c}=0.227\pm0.014$) and baryonic matter ($\Omega_{\rm b}=0.0456\pm0.0016$), with a Hubble parameter of $70.4^{+1.3}_{-1.4}$~km/s/Mpc. Perturbations on the uniform model seem to be well described by a scale-free primordial power spectrum with power-law index $n_{\rm s}=0.963\pm0.012$ and amplitude $\sigma_8=0.809\pm0.024$.

Given such a cosmological model, the Universe is $13.75\pm0.11$~Gyr old. Galaxies probably began forming at $z\sim 20-50$ when the first sufficiently deep dark matter potential wells formed to allow gas to cool and condense to form galaxies \pcite{tegmark_small_1997,gao_first_2007}.

\subsection{Structure Formation}

The formation of structure in the Universe is seeded by minute perturbations in matter density expanded to cosmological scales by inflation. The dark matter component, having no pressure, must undergo gravitational collapse and, as such, these perturbations will grow. The linear theory of cosmological perturbations is well understood and provides an accurate description of the early evolution of these perturbations. Once the perturbations become nonlinear, their evolution is significantly more complicated, but simple arguments (e.g. spherical top-hat collapse and developments thereof; \citealt{gunn_massive_1977,shaw_improved_2008}) provide insight into the basic behavior. Empirical methods to determine the statistical distribution of matter in the nonlinear regime exist \pcite{hamilton_reconstructingprimordial_1991,peacock_non-linear_1996,smith_stable_2003,heitmann_coyote_2009}. These, together with N-body simulations (e.g. \citealt{klypin_three-dimensional_1983,springel_simulations_2005,heitmann_coyote_2008}) show that a network of halos strung along walls and filaments forms, creating a cosmic web. This web is consistent with measurements of galaxy and quasar clustering on a wide range of scales.

\subsection{Halo Formation}

The final result of the nonlinear evolution of a dark matter density perturbation is the formation of a dark matter halo: an approximately stable, near-equilibrium state supported against its own self-gravity by the random motions of its constituent particles. In a hierarchical universe the first halos to form will do so from fluctuations on the smallest scales. Later generations of halos can be thought of as forming from the merging of these earlier generations of halos. For the purposes of galaxy formation, two fundamental properties of the dark matter halos are of primary concern: (i) the distribution of their masses at any given redshift and (ii) the distribution of their formation histories (i.e. the statistical properties of the halos from which they formed).

\subsubsection{Halo Mass Distribution}

The insight of \cite{press_formation_1974} was that halos could be associated with peaks in the Gaussian random density field of dark matter in the early universe. Using the relatively simple statistics of Gaussian random fields they were able to derive the following form for the distribution of dark matter halo masses such that the number of halos per unit volume in the mass range $M$ to $M+\delta M$ is $\delta M (\d n / \d M)$ where\footnote{The original derivation by \protect\cite{press_formation_1974} differed by a factor of 2, resulting in only half of the mass of the Universe being locked up in halos. Later derivations placed the method on a firmer mathematical basis and resolved this problem, a symptom of the ``cloud-in-cloud'' problem \protect\pcite{bond_excursion_1991,bower_evolution_1991,lacey_merger_1993}.}:
\begin{eqnarray}
{\d n\over \d M}(M,t) &=& \left({2\over \pi}\right)^{1/2} {\rho_0 \over M^2} {\delta_{\rm c}(t) \over \sigma(M)} \left| {\d \ln \sigma \over \d \ln M} \right| \nonumber \\
 & & \times \exp \left[ -{\delta_{\rm c}^2(t)\over 2 \sigma^2(M)} \right],
\end{eqnarray}
where $\rho_0$ is the mean density of the Universe, $\sigma(M)$ is the fractional root variance in the density field smoothed using a top-hat filter that contains, on average, a mass $M$, and $\delta_{\rm c}(t)$ is the critical overdensity for spherical top-hat collapse at time $t$ \pcite{eke_cluster_1996}.

While the \cite{press_formation_1974} expression is remarkably accurate given its simplicity, it does not provide a sufficiently accurate description of modern N-body measures of the halo mass function. Several attempts have been made to ``fix'' the \cite{press_formation_1974} theory by appealing to different filters and barriers (e.g. \citealt{sheth_ellipsoidal_2001}) although to date none are able to accurately predict the measured form without adjusting tunable parameters. The most accurate fitting formulae currently available are those of \citeauthor{tinker_towardhalo_2008}~(\citeyear{tinker_towardhalo_2008}; see also \citealt{reed_halo_2007,robertson_collapse_2009}). Specifically, the mass function is given by
\begin{equation}
{\d n \over\d M} = f(\sigma) {\bar{\rho} \over M} {\d \ln \sigma^{-1} \over \d M},
\end{equation}
where
\begin{equation}
 f(\sigma) = A \left[ \left( {\sigma \over b}\right)^{-a} + 1 \right] \exp\left(-{c \over \sigma^2}\right),
\end{equation}
and where $A$, $a$, $b$ and $c$ are parameters determined by fitting to the results of N-body simulations. The mass variance $\sigma^2(M)$ is determined from the power spectrum of density fluctuations
\begin{equation}
 \sigma^2(M) = {1 \over 2 \pi^2} \int_0^\infty P(k) T^2(k) \hat{W}_M^2(k) k^2 \d k
\end{equation}
where $P_(k)$ is the primordial power spectrum (usually taken to be a scale-free power spectrum with index $n_{\rm s}$), $T(k)$ is the cold dark matter transfer function \pcite{eisenstein_power_1999} and $\hat{W}_M(k)$ is the Fourier transform of the real-space top-hat window function.  \cite{tinker_towardhalo_2008} give values for these parameters as a function of the overdensity, $\Delta$, used to define what we consider to be a halo. Additionally, they find that the best fit values are functions of redshift. Earlier studies had hoped to find a ``universal'' form for the mass function (such that the functional form was always the same when expressed in terms of $\nu=\sigma/\delta_{\rm c}$). While this is approximately true, the work of \cite{tinker_towardhalo_2008} demonstrates that universality does not hold when high precision results are considered.

\subsubsection{Halo Formation Distribution}

A statistical description of the formation of halos, specifically the sequence of merging events and the masses of halos involved in those events, can be extracted using similar arguments as the original \cite{press_formation_1974} approach \pcite{lacey_merger_1993}. These show that the distribution of halo progenitor masses, $M_1$m at redshift $z_1$ for a halo of mass $M_2$ at later redshift $z_2$ is given by
\begin{eqnarray}
 {\d N \over \d M_1} &=& \left({2\over \pi}\right)^{1/2} {\d \ln \sigma \over \d \ln M}_1 M_2 {\sigma_1^2 \over M_1^2} {\delta_{\rm c 1}-\delta_{\rm c 2} \over (\sigma_1^2-\sigma_2^2)^{3/2}} \nonumber \\ 
 & & \times \exp\left[-{(\delta_{\rm c 1}-\delta_{\rm c 2})^2\over (\sigma_1^2-\sigma_2^2)}\right],
\end{eqnarray}
where $\sigma_1=\sigma(M_1)$, $\sigma_2=\sigma(M_2)$, $\delta_{\rm c 1}=\delta_{\rm c}(z_1)$, $\delta_{\rm c 2}=\delta_{\rm c}(z_2)$. With a zero time-lag (i.e. as $z_1 \rightarrow z_2$ and therefore $\delta_{\rm c 1}\rightarrow\delta_{\rm c 2}$) this can be interpreted as a merging rate (although see \cite{neistein_merger_2008} for a counter argument). Repeated application of this merging rate can be used to build a merger tree. Finding a suitable algorithm is non-trivial and many attempts have been made \pcite{kauffmann_merging_1993,somerville_semi-analytic_1999,cole_hierarchical_2000,parkinson_generating_2008}. A recent examination of alternative algorithms is given by \cite{zhang_to_2008}. Current implementations of merger tree algorithms are highly accurate and can reproduce the progenitor halo mass distribution over large spans of redshift \pcite{parkinson_generating_2008}.

A fundamental limitation of any \cite{press_formation_1974} based approach is that the merger rates are not symmetric, in the sense that switching the masses $M_1$ and $M_2$ results in two different predictions for the rate of mergers between halos of mass $M_1$ and $M_2$. \cite{benson_self-consistent_2005} and \cite{benson_constraining_2008} showed that a symmetrized form of the \cite{parkinson_generating_2008} merger rate function could be made to approximately solve Smoluchowski's coagulation equation \pcite{smoluchowksi__1916} and thereby provide a solution free from ambiguities. Other empirical determinations of merger rates have been made \pcite{cole_statistical_2008,fakhouri_nearly_2008,fakhouri_environmental_2008}.

In addition to these purely analytic approaches, numerical studies utilizing N-body simulations have lead to the development of an empirical understanding of halo formation histories \citep{wechsler_concentrations_2002,van_den_bosch_universal_2002} and halo--halo merger rates \citep{fakhouri_nearly_2008,fakhouri_environmental_2009,stewart_galaxy_2009,fakhouri_merger_2010,hopkins_mergers_2010}. Merger trees can also be extracted directly from N-body simulations (e.g. \citealt{helly_galaxy_2003,springel_simulations_2005}) which sidesteps these problems but incorporates any limitations of the simulation (spatial, mass and time resolution), and additionally provides information on the spatial distribution of halos. Such N-body merger trees also serve to highlight some, perhaps fundamental, limitations of the \cite{press_formation_1974} type approach. For example, halos in N-body simulations undergo periods of mass loss, which is not expected in pure coagulation scenarios. The existence of systems of substructures within halos (see Section~ref{sec:galaxyOrbits}) can even lead to three-body encounters which cause subhalos to be ejected \pcite{sales_cosmic_2007}.

\subsection{Halo Structure}

Dark matter halos are characterized by their large overdensity with respect to the background density. Spherical top-hat collapse models (e.g. \citealt{eke_cluster_1996}) show that the overdensity, $\Delta$, of a just collapsed halo is approximately 200, with some dependence on cosmological parameters. This overdensity corresponds, approximately, to the virialized region of a halo, and allows us to define a virial radius as
\begin{equation}
 r_{\rm v} = \left({3 M \over 4 \pi \rho_0 \Delta}\right)^{1/3}.
\end{equation}
Studies utilizing N-body simulations show that halos approximately obey the virial theorem within this radius and that $r_{\rm v}$ is a characteristic radius for the transition from ordered inflow of matter (on larger scales) to virialized random motions (on smaller scales) \pcite{cole_structure_1996}. Dark matter halos have triaxial shapes and the distribution of axial ratios has been well characterized from N-body simulations \pcite{frenk_formation_1988,jing_triaxial_2002,bett_spin_2007}.

Recent N-body studies \pcite{navarro_inner_2004,merritt_universal_2005,prada_far_2006} indicate that density profiles of dark matter halos have an approximately universal form in \CDM\ cosmologies that is better described by the Einasto profile \pcite{einasto__1965} than the \NFW\ profile \pcite{navarro_structure_1996,navarro_universal_1997}. The Einasto density profile is given by
\begin{equation}
\rho(r) = \rho_{-2} \exp\left( -{2\over \alpha} \left[ \left({r\over r_{-2}}\right)^\alpha- 1 \right] \right),
\end{equation}
where $r_{-2}$ is a characteristic radius at which the logarithmic slope of the density profile equals $-2$ and $\alpha$ is a parameter which controls how rapidly the logarithmic slope varies with radius. The parameter $\alpha$ seems to correlate well with the height of the peak from which a halo formed, $\nu = \delta_{\rm c}(z)/\sigma(M)$ as has been shown by \cite{gao_redshift_2008} who provide a fitting formula,
\begin{equation}
\alpha = \left\{ \begin{array}{ll} 0.155 + 0.0095\nu^2 & \hbox{if } \nu < 3.907 \\ 0.3 & \hbox{if } \nu \ge 3.907,\end{array} \right.
\end{equation}
which is a good match to halos in the Millennium Simulation\footnote{We have truncated this fit so that $\alpha$ never exceeds $0.3$. \protect\cite{gao_redshift_2008} were not able to probe the behavior of $\alpha$ in the very high $\nu$ regime. Extrapolating their formula to $\nu > 4$ is not justified and we instead choose to truncate it at a maximum of $\alpha=0.3$.} \pcite{springel_simulations_2005}. The value of $r_{-2}$ for each halo is determined from the known virial radius, $r_{\rm v}$, and the concentration, $c_{-2}\equiv r_{\rm v}/r_{-2}$. The NFW profile has a significantly simpler form and is good to within 10--20\% making it still useful. It is given by
\begin{equation}
 \rho(r) = 4 {\rho_{\rm s} \over (r/r_{\rm s}) [ 1 + r/r_{\rm s} ]^2 },
\end{equation}
where $r_{\rm s}$ is a characteristic scale radius and $\rho_{\rm s}$ is the density at $r=r_{\rm s}$. For NFW halos, the concentration is defined as $c_{\rm NFW}=r_{\rm v}/r_{\rm s}$.

Concentrations are found to depend weakly on halo mass and on redshift and can be predicted from the formation history of a halo \pcite{wechsler_concentrations_2002}. Simple algorithms to approximately determine concentrations have been proposed by \cite{navarro_universal_1997}, \cite{eke_power_2001} and \cite{bullock_profiles_2001}. More accurate power-law fits have also been determined from N-body simulations \pcite{gao_redshift_2008,zhao_accurate_2009}.

Integrals over the density and mass distribution are needed to compute the enclosed mass, velocity dispersion, gravitational energy and so on for the halo density profile. For NFW halos the integrals are mostly straightforward, although some require numerical calculation. For the Einasto profile some of these may be expressed analytically in terms of incomplete gamma functions \pcite{cardone_spherical_2005}. Specifically, expressions for the mass and gravitational potential are provided by \cite{cardone_spherical_2005}, other integrals (e.g. gravitational energy) must be computed numerically.

\section{Pre-Galactic Evolution of Baryons}

Baryons are initially distributed near uniformly---they are expected to trace the dark matter distribution on scales above the Jeans length \pcite{arons_jeans_1968,gnedin_probinguniverse_1998}. To form galaxies they must first be concentrated by the forces of gravity which are dominated by the distribution of dark matter. In particular, we expect that baryons will concentrate towards the deep potential wells of dark matter halos. These should therefore be the sites of galaxy formation.

\subsection{Cold or Hot Accretion (Shocks)}

Baryonic material will be dragged along by the gravitationally dominant dark matter such that dark matter halos are expected to accrete baryonic material. How much baryonic material they accrete depends upon the depth of their potential well and the pressure of the baryons. 

According to \citet{okamoto_mass_2008}, the mass of baryons which accrete from the intergalactic medium into a galaxy halo during some interval $\delta t$ after the Universe has been reionized (i.e. the hydrogen content of the Universe has been almost fully ionized as a result of emission from stars and \AGN) is given by
\begin{equation}
 M_{\rm b} = M_{\rm b}^\prime+M_{\rm acc},
\end{equation}
where $M_{\rm acc}$ is given below,
\begin{equation}
 M_{\rm b}^\prime = \sum_{\rm prog} \exp\left(-{\delta t \over t_{\rm evp}}\right) M_{\rm b},
\end{equation}
and where the sum is taken over all progenitors of the current halo, and $t_{\rm evp}$ is the timescale for gas to evaporate from the progenitor halo and is given by
\begin{equation}
 t_{\rm evp}=\left\{ \begin{array}{ll}
                      r_{\rm v}/c_{\rm s}(\Delta_{\rm evp}) & \hbox{if } T_{\rm v} < T_{\rm evp}, \\
                      \infty & \hbox{if } T_{\rm v} > T_{\rm evp},
                     \end{array} \right.
\end{equation}
where $T_{\rm evp}$ is the temperature below which gas will be heated and evaporated from the halo, $c_{\rm s}$ is the sound speed in the halo gas. \cite{okamoto_mass_2008} compute $T_{\rm evp}$ by finding the equilibrium temperature of gas at an overdensity of $\Delta_{\rm evp}=10^6$. The accreted mass, $M_{\rm acc}$, is given by
\begin{equation}
 M_{\rm acc} = \left\{ \begin{array}{ll}
                {\Omega_{\rm b}\over \Omega_0} M_{\rm H} - M_{\rm b}^\prime & \hbox{if } T_{\rm vir} > T_{\rm acc}, \\
                0 & \hbox{if } T_{\rm vir} < T_{\rm acc}.
               \end{array} \right.
\end{equation}
Here, $T_{\rm acc}$ is the larger of $T_{\rm eq}$ and the temperature of intergalactic medium gas adiabatically compressed to the density of accreting gas where $T_{\rm eq}$ is the equilibrium temperature at which radiative cooling balances photoheating for gas at the density expected at the virial radius, for which \cite{okamoto_mass_2008} use one third of the halo overdensity.

An accretion shock is a generic expectation whenever the gas accretes supersonically as it will do if the halo virial temperature exceeds the temperature of the accreting gas \pcite{binney_physics_1977}. Models of accretion shocks have been presented by several authors \pcite{bertschinger_self-similar_1985,tozzi_evolution_2001,voit_origin_2003,book_role_2010} with the general conclusion that the shock occurs at a radius comparable to (or perhaps slightly larger than) the virial radius when cooling times are long compared to dynamical times. In the other limit of short cooling times, it has long been understood that the shock must instead form at much smaller radii, close to the forming galaxy \citep{rees_cooling_1977,white_galaxy_1991}. For example, according to \cite{rees_cooling_1977}: ``Unless pre-galactic
clouds collapse in an exceedingly homogeneous fashion, their kinetic
energy of infall will be thermalized by shocks before collapse has
proceeded by more than a factor $\sim2$. What happens next depends on the
relative value of the cooling and collapse timescales. Masses in the
range $10^{10}$--$10^{12} M_\odot$ cool so efficiently that they always
collapse at the free-fall rate, and probably quickly fragment into
stars. Larger masses may, however, experience a quasistatic
contraction phase\ldots''. Thus, \cite{rees_cooling_1977} clearly understood the difference between the rapid inflow and hydrostatic cooling regimes, and correctly identified the transition mass, suggesting that this be identified this with the characteristic stellar mass of galaxies. Accretion in these two regimes may be expected to result in very different spatial and spectral distributions of cooling radiation, leading to the possibility of observationally distinguishing the two types of accretion \citep{fardal_cooling_2001}.

The distinction between these two regimes has always been an integral part of analytic models of galaxy formation, beginning with \cite{rees_cooling_1977}. For example, \cite{white_galaxy_1991} introduced a transition between rapid and slow cooling regimes at the point where cooling and virial radii become equal, or, equivalently, the point at which cooling and dynamical times at the halo virial radius become equal. In the rapid cooling regime, the accretion rate of gas into the central galaxy was then determined by the cosmological infall rate, while in the slow cooling regime the accretion rate was determined by the cooling time in the gas. Their Figure~2 illustrates that the rapid cooling regime will occur in low mass halos and at high redshifts. All subsequent semi-analytic models of galaxy formation (e.g. \citealt{kauffmann_formation_1993,cole_recipe_1994}) have adopted this prescription, or some variant of it, and it has also been validated by 1D hydrodynamical simulations (e.g \citealt{forcada-miro_radiative_1997}). The validity of this prescription has been confirmed by studies which compared its predictions for the condensed masses of galaxies with those from smoothed particle hydrodynamics simulations across the boundary of the rapid to slow cooling transition \citep{benson_comparison_2001,yoshida_gas_2002,helly_comparison_2003}, although it should be noted that the accuracy of these comparisons is less than that at which semi-analytic models are now being used.

Recent work has once again focused on the formation of accretion shocks. Recent 3D hydrodynamical simulations (\citealt{fardal_cooling_2001}; see also \citealt{kerevs_do_2005,ocvirk_bimodal_2008,kerevs_galaxies_2009}), have suggested that a significant fraction of gas in low mass galaxies has never been shock heated (at least within regions adequately resolved by the simulations; for example numerical simulations may not adequately resolve shocks in the radiative regime due to artificial viscosity, numerical diffusion and other numerical artifacts; \citealt{agertz_fundamental_2007}). Motivated by these results \cite{birnboim_virial_2003} developed an analytic treatment of accretion shock stability\footnote{Unpublished work by \protect\cite{forcada-miro_radiative_1997}, also utilizing a 1D hydrodynamics code, reached similar conclusions.}. The accretion shock relies on the presence of a stable atmosphere of post-shock gas to support itself. If cooling times in the post-shock gas are sufficiently short, this atmosphere cools and collapses and can no longer support the shock. The shock therefore shrinks to smaller radii, where it can be stable. \cite{birnboim_virial_2003} find that
\begin{equation}
 \rho r \Lambda(T_1)/u^3 \mu^2 m_{\rm H}^2 < 0.0126,
\end{equation}
where $\rho$ is gas density, $\Lambda(T)$ is the cooling function for gas at temperature $T$, $T_1=(3/16)\mu u_0^2 /k_{\rm B}N_a$ is the post-shock temperature and $u$ is the infall velocity, all evaluated for the pre-shock gas at radius $r$, is required for the shock to be stable\footnote{The left-hand side of this expression is equivalent, to order of magnitude, to $t_{\rm sc}/t_{\rm cool}$ where $t_{\rm sc}$ is the sound-crossing time in the halo and $t_{\rm cool}$ is the cooling time in the post-shock gas. This provides some physical insight into this condition: if the post-shock gas can cool too quickly sound waves cannot communicate across the halo and thereby form a hydrostatic atmosphere which can support a shock front).}.  For cosmological halos this implies that shocks can only form close to the virial radius in halos with mass greater than $10^{11}M_\odot$ for primordial gas (or around $10^{12}M_\odot$ for gas of Solar metallicity). These values are found to depend only weakly on redshift and are in good agreement with the results of hydrodynamical simulations. It is crucial to note that these new criteria are equivalent to that of \cite{white_galaxy_1991} up to factors of order unity.

As a result, in low mass halos gas tends to accrete into halos ``cold\footnote{Or, more likely, warm \protect\citep{bland-hawthorn_warm_2009}.}''---never being shock heated to the virial temperature and proceeding to flow along filaments towards the center of the halo where it will eventually shock\footnote{While this picture seems reasonable on theoretical grounds, it as yet has little direct observational support \protect\citep{steidel_structure_2010}.}. Halos which do support shocks close to the virial radius are expected to contain a quasi-hydrostatic atmosphere of hot gas. The structure of this atmosphere is determined by the entropy that the gas gains at the accretion shock and that may be later modified by radiative cooling and feedback \pcite{voit_origin_2003,mccarthy_modelling_2007}. In practice, the transition from rapid to slow cooling regimes is not sharp---halos able to support a shock at their virial radius still contain some unshocked gas; because the halos retain a memory of past accretion and because cold filaments may penetrate through the hot halo. At high redshifts in particular, the ``cold'' accretion mode may be active even in halos whose accretion of gas is primarily via an accretion shock close to the virial radius.

The consequences of rapid vs. slow cooling regimes for the properties of the galaxy forming warrant further study. As \cite{croton_many_2006}
have stressed, the absence of a more detailed treatment of the rapid cooling regime may not be important since, by definition the
gas accretion rate in small halos is limited by the growth of the halo rather than by the
system's cooling time. 
In contrast, \cite{brooks_role_2009} demonstrate in hydrodynamical simulations that in the rapid cooling regime accreted gas can reach the galaxy more rapidly, by virtue of the fact that it does not have to cool but instead merely has to free-fall to the center of the halo (starting with a velocity comparable to the virial velocity). This results in earlier star formation than if all gas were assumed to be initially shock heated to the virial temperature close to the virial radius of the halo. It is also clear that the situation needs to be carefully
reassessed in the presence of effective feedback schemes that prevent excessive star formation,
particularly in the high redshift universe. 

While hot atmospheres of gas are clearly present in massive systems such as groups and clusters of galaxies (where the hot gas is easily detected by virtue of its X-ray emission), observational evidence for hot atmospheres of gas arising from cosmological infall surrounding lower mass systems, such as massive but isolated galaxies is sparse. Interesting upper limits have been placed on the X-ray emission from massive, nearby spiral galaxies \pcite{benson_diffuse_2000,rasmussen_hot_2009}, and the hot component of the Milky Way's halo is constrained by the ram pressure that it exerts on the Magellanic Stream \pcite{moore_origin_1994,mastropietro_gravitational_2005}. Additionally, ultraviolet line detections of the so-called warm-hot intergalactic medium (intergalactic gas at temperatures of $10^5$--$10^6$K) show that some of this material must lie close to the Milky Way \pcite{wang_warm-hot_2005,williams_chandra_2006}. What is known is that the Milky Way's halo contains a significant mass of cold, neutral gas in the form of high velocity clouds \pcite{putman_h_2003}. This may indicate, as expected on theoretical grounds \citep{crain_galaxies-intergalactic_2009}, that the Milky Way is in the transition mass range between purely cold and purely hot accretion.

\subsection{Cooling}\label{sec:Cooling}

Gas which does experience a strong virial shock will have its kinetic infall energy thermalized and therefore be heated to of order the virial temperature
\begin{equation}
 T_{\rm v} = {2 \over 3} {\G M_{\rm v} \over r_{\rm v}} {\mu m_{\rm H} \over k_{\rm B}}.
\end{equation}
This gas will proceed to form a hydrostatically supported atmosphere obeying the usual hydrostatic equilibrium equation:
\begin{equation}
 {\d P \over \d r} = - {\G M(r) \over r^2} \rho(r),
\end{equation}
where $P$ is the gas pressure, $\rho(r)$ the gas density and $M(r)$ the total (i.e. dark matter plus baryonic) mass within radius $r$. This distribution may be modified, particularly on small scales, by other contributors to supporting the halo against gravity, such as turbulence \pcite{frenk_santa_1999}, cosmic ray pressure \pcite{guo_feedback_2008} and magnetic pressure \pcite{gonccalves_magnetic_1999}. In any case, the resulting hot atmosphere will fill the dark matter halo with a density still several orders of magnitude lower than typical galactic densities. In the absence of any dissipitative process, the gas would remain in this state indefinitely. Fortunately, however, gas is able to cool radiatively and so will eventually lose energy and, consequently, pressure support, at which point it must fall towards the center of the gravitational potential of the dark matter halo thereby increasing its density. Gas which does not experience a shock close to the virial radius can fall almost ballistically towards the halo center. It must still lose its infall energy at some point, however,  shocking closer to the halo center near the forming galaxy. As such, it too will eventually be heated and must cool down (although it will presumably do so much more rapidly due to its higher density).

In the remainder of this subsection we will review the various mechanisms by which such gas cools.

\subsubsection{Atomic}\label{sec:CoolingAtomic}

\begin{figure}
 \begin{center}
  \includegraphics[width=80mm]{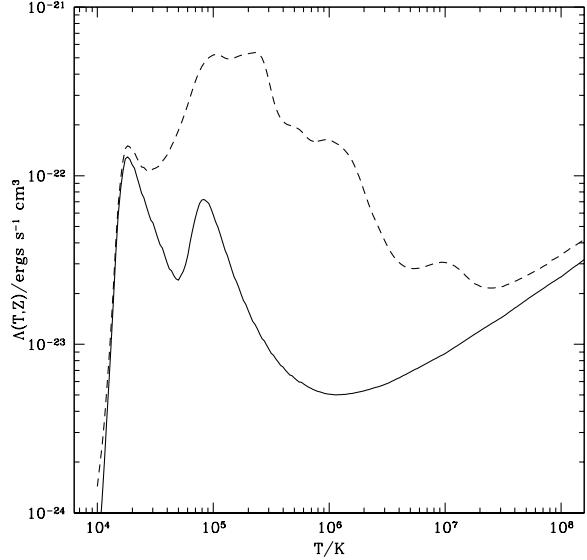}
 \end{center}
 \caption{Cooling functions for gas in collisional ionization equilibrium are shown as a function of temperature. The solid line corresponds to gas of primordial composition, while the dashed line corresponds to gas of Solar composition.}
 \label{fig:CIE}
\end{figure}

For metal rich gas, or gas hot enough to begin to collisionally ionize hydrogen (i.e. $T\gsim 10^4$K), the primary cooling mechanisms at low redshifts are a combination of various atomic processes including recombination radiation, collisional excitation and subsequent decay and Bremmsstahlung. In the absence of any external radiation field (see Section~ref{sec:PhotoHeat}), these are all two-body processes and so the cooling rate is expected to scale as the square of the density for gas of fixed chemical composition and temperature. It is usual, therefore, to write the cooling rate per unit volume of gas as
\begin{equation}
 {\mathcal L} = n_{\rm H}^2 \Lambda(T,\hbox{\boldmath $Z$}),
\end{equation}
where $n_{\rm H}$ is the number density of hydrogen (both neutral and ionized) and $\Lambda(T,\hbox{\boldmath $Z$})$ is the ``cooling function'' and depends on temperature and chemical composition. Typically, the chemical composition is described by a single number, the metallicity $Z$ (defined as the mass fraction of elements heavier than helium), and an assumed set of abundance ratios (e.g. primordial or Solar). The cooling function can then be found by first solving for the ionization state of the gas assuming collisional ionization equilibrium and then summing the cooling rates from the various above mentioned cooling mechanisms. Such calculations can be carried out by using, for example, {\sc Cloudy} \pcite{ferland_cloudy_1998}. Examples of cooling functions computed in this way are shown in Fig.~\ref{fig:CIE}.

In detail, the cooling function depends not just on the overall metallicity, but on the detailed chemical composition of the gas. If the abundances of individual elements are known, the corresponding cooling function is easily calculated. In theoretical calculations of galaxy formation it is often computationally impractical to follow the abundance of numerous chemical species in a large number of galaxies and dark matter halos. Fortunately, \citet{martinez-serrano_chemical_2008} have demonstrated that an optimal linear combination of abundances, which minimizes the variance between cooling/heating rates computed using that linear combination as a parameter and a full calculation using all abundances, provides very accurate estimates of cooling rates. The best linear combination turns out to be a function of temperature. This reduces the problem to tracking the optimal combination of elements for a small number of temperatures. This number can then be used in place of metallicity when computing cooling functions.

\begin{figure}
 \begin{center}
  \includegraphics[width=80mm]{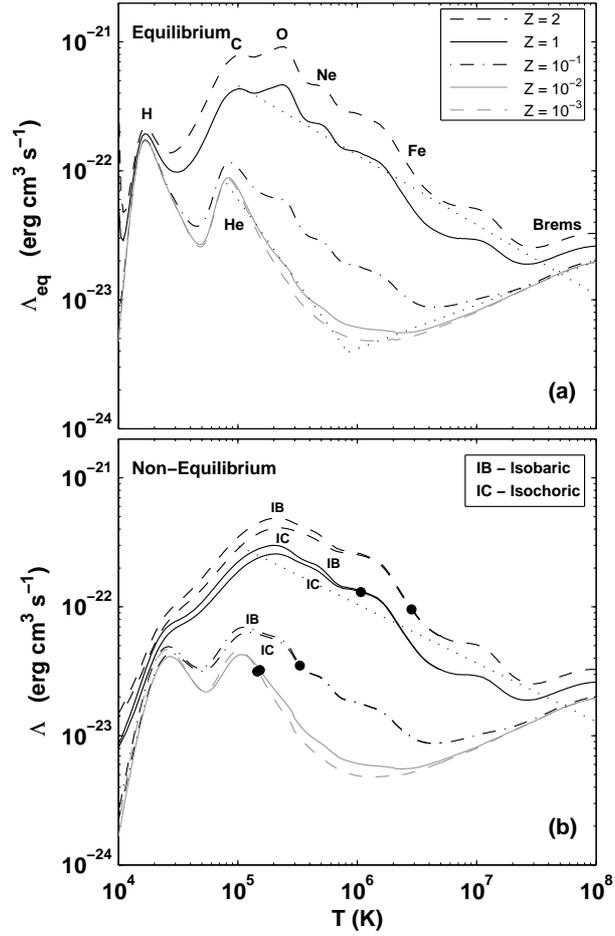}
 \end{center}
 \caption{A comparison of cooling functions for gas in collisional ionization equilibrium (upper panel) with effective cooling functions for gas in which the time-dependent ionization state is computed throughout the cooling (lower panel). Line types indicate different metallicities (shown in the upper panel, values in units of Solar metallicity), while IB and IC labels indicate whether the gas was assumed to be cooling isobarically or isochorically. The primary coolant at each temperature is indicated in the upper panel.\newline
\emph{Source:} Reproduced with permission from \protect\citealt{gnat_time-dependent_2007}.}
 \label{fig:NECool}
\end{figure}

Most calculations of cooling rates in cosmological halos assume that the gas is in collisional or photoionization equilibrium. Even if the gas begins in such a state, as it cools it can drift away from equilibrium as, particularly at low temperatures, the ion--electron recombination timescales can significantly exceed the cooling timescales---as such, the ionization state always lags behind the equilibrium state due to the rapidly changing gas temperature. Figure~\ref{fig:NECool} shows calculations of effective cooling functions for gas initially in collisional ionization equilibrium (and with no external radiation field) with a fully time-dependent calculation of cooling and ionization state \pcite{gnat_time-dependent_2007}. Significant differences can be seen, resulting in cooling timescales being a factor of 2--3 longer when non-equilibrium effects are taken into account.

\subsubsection{Compton Cooling}

At high redshifts, the density of cosmic microwave background photons becomes sufficiently high that the frequent Compton scattering of these photons from electrons in the ionized plasma inside dark matter halos results in significant cooling of that plasma (assuming that its temperature exceeds that of the cosmic microwave background). The Compton cooling timescale is given by \pcite{peebles_recombination_1968}
\begin{equation}
\tau_{\rm Compton} = {3 m_{\rm e}\clight (1+1/x_{\rm e}) \over 8\sigma_{\rm T}aT^4_{\rm CMB}(1-T_{\rm CMB}/T_{\rm e})},
\end{equation}
where $x_{\rm e}=n_{\rm e}/n_{\rm t}$, $n_{\rm e}$ is the electron number density, $n_{\rm t}$ is the number density of all atoms and ions, $T_{\rm CMB}$ is the \CMB\ temperature and $T_{\rm e}$ is the electron temperature of the gas. Unlike the various atomic cooling processes described in Section~ref{sec:CoolingAtomic}, the Compton cooling rate per unit volume does not scale as the square of the particle density, since it involves an interaction between a particle and a cosmic microwave background photon. Since the \CMB\ photon density is the same everywhere, independent of the local gas density, the Compton cooling rate scales linearly with density. As a result, the Compton cooling timescale is independent of gas density. Thus, if gas in a dark matter halo is able to cool via Compton cooling, it can do so at all radii (assuming that it is isothermal and has the same electron fraction everywhere).

In a Universe with WMAP 5~yr cosmological parameters, the Compton cooling time is less than the age of the Universe above $z\approx 6$.

\subsubsection{Molecular Hydrogen Cooling}

Somewhat ironically, the gas in dark matter halos at the highest redshifts is too cold to cool any further. In halos with virial temperatures below $10^4$K even shock heated gas (if shocks can occur) will be mostly neutral and therefore unable to cool via the usual atomic processes described in Section~ref{sec:CoolingAtomic}. Studies show that the dominant coolant in such cases becomes the small fraction of hydrogen in the form of molecular hydrogen \pcite{abel_modeling_1997}. Cooling via molecular hydrogen is crucial for the formation of the first stars \pcite{abel_formation_2002} and galaxies \pcite{bromm_formation_2009} and, therefore, for the sources which cause the reionization of the Universe \pcite{benson_epoch_2006,wise_very_2008}. The details of molecular hydrogen cooling are more complicated than those of atomic cooling: in addition to uncertainties in the molecular chemistry \pcite{glover_uncertainties_2008}, in many cases equilibrium is not reached and the photon background can lead to both negative \pcite{wise_suppression_2007} and positive \pcite{ricotti_feedbackgalaxy_2001} feedbacks. 

Cosmologically, molecular hydrogen forms via the gas-phase reactions \pcite{mcdowell_formation_1961}:
\begin{eqnarray}
 {\rm H} + {\rm e}^- &\rightarrow& {\rm H}^-, + \gamma \nonumber \\
 {\rm H}^- + {\rm H} & \rightarrow& {\rm H}_2 + {\rm e}^-,
\end{eqnarray}
and
\begin{eqnarray}
 {\rm H}^+ + {\rm H} &\rightarrow& {\rm H}^+_2 + \gamma, \nonumber \\
 {\rm H}^+_2 + {\rm H} & \rightarrow& {\rm H}_2 + {\rm H}^+.
\end{eqnarray}
Our discussion of molecular hydrogen cooling will mostly follow \cite{yoshida_formation_2006}.

The molecular hydrogen cooling timescale is found by first estimating the abundance, $f_{{H_2},c}$, of molecular hydrogen that would be present if there is no background of H$_2$-dissociating radiation from stars. For gas with hydrogen number density $n_{\rm H}$ and temperature $T_{\rm V}$ the fraction is \pcite{tegmark_small_1997}:
\begin{eqnarray}
f_{{\rm H_2},c} &=& 3.5 \times 10^{-4}T_3^{1.52} [1+(7.4\times 10^8 (1+z)^{2.13} \nonumber \\
 & & \times \exp\left\{-3173/(1+z)\right\}/n_{\rm H 1})]^{-1},
\end{eqnarray}
where $T_3$ is the temperature $T_{\rm V}$ in units of 1000K and $n_{\rm H 1}$ is the hydrogen density in units of cm$^{-3}$. Using this initial abundance we calculate the final H$_2$ abundance, still in the absence of a photodissociating background, as
\begin{equation}
f_{\rm H_2} = f_{{\rm H_2},c}\exp\left({-T_{\rm V} \over 51920K}\right)
\end{equation}
where the exponential cut-off is included to account for collisional dissociation of H$_2$, as in \citet{benson_epoch_2006}. 

Finally, the cooling timescale due to molecular hydrogen can be computed using \pcite{galli_chemistry_1998}:
\begin{equation}
 \tau_{{\rm H}_2} = 6.56419 \times 10^{-33} T_{\rm V} f^{-1}_{{\rm H}_2} n^{-1}_{{\rm H} 1} \Lambda^{-1}_{{\rm H}_2},
\end{equation}
where
\begin{equation}
 \Lambda_{\rm H_2} = {\Lambda_{\rm LTE} \over 1+n_{\rm cr}/n_{\rm H}},
\end{equation}
\begin{equation}
 {n_{\rm cr}\over n_{\rm H}} = {\Lambda_{\rm H_2}({\rm LTE}) \over \Lambda_{\rm H_2}[n_{\rm H}\rightarrow0]}
\end{equation}
and
\begin{eqnarray}
\log_{10}\Lambda_{\rm H_2}[n_{\rm H}\rightarrow0] &=& -103+97.59 \ln(T) \nonumber \\
 & &-48.05 \ln(T)^2 +10.8 \ln(T)^3 \nonumber \\
 & & -0.9032 \ln(T)^4
\end{eqnarray}
is the cooling function in the low density limit (independent of hydrogen density) and we have used the fit given by \citet{galli_chemistry_1998}, and
\begin{equation}
 \Lambda_{\rm LTE} = \Lambda_r+\Lambda_v
\end{equation}
is the cooling function in local thermodynamic equilibrium and
\begin{eqnarray}
 \Lambda_r &=& {1\over n_{\rm H_1}}\left\{{9.5\times10^{-22} T_3^{3.76}\over1+0.12 T_3^{2.1}} \exp\left(-\left[{0.13\over T_3}\right]^3\right) \right. \nonumber \\
 & & \left. +3\times10^{-24} \exp\left(-{0.51\over T_3}\right) \right\} \hbox{ergs cm}^3\hbox{ s}^{-1} \\
 \Lambda_v &=& {1\over n_{\rm H_1}}\left\{ 6.7\times 10^{-19} \exp\left(-{5.86\over T_3}\right) \right. \nonumber \\
 & & \left. +1.6\times 10^{-18} \exp\left(-{11.7\over T_3}\right)\right\} \hbox{ergs cm}^3\hbox{ s}^{-1}
\end{eqnarray}
are the cooling functions for rotational and vibrational transitions in H$_2$ \pcite{hollenbach_molecule_1979}.

It is also possible to estimate the rate of molecular hydrogen formation on dust grains using the approach of \citet{cazaux_molecular_2004}, who find that the rate of H$_2$ formation via this route can be important in the high redshift Universe. In this case we have to modify equation (13) of \cite{tegmark_small_1997}, which gives the rate of change of the H$_2$ fraction, to account for the dust grain growth path. The molecular hydrogen fraction growth rate becomes:
\begin{equation}
 \dot{f} = k_{\rm d} f (1-x-2f) + k_{\rm m} n (1-x-2f) x,
\end{equation}
where $f$ is the fraction of H$_2$ by number, $x$ is the ionization fraction of H which has total number density $n$, 
\begin{equation}
k_{\rm d}=3.025\times 10^{-17}{\xi_{\rm d}\over0.01} S_{\rm H}(T) \sqrt{{T_{\rm g}\over100\hbox{K}}} \hbox{cm}^3 \hbox{s}^{-1}
\end{equation}
is the dust formation rate coefficient (\citealt{cazaux_molecular_2004}; eqn.~4), and $k_{\rm m}$ is the effective rate coefficient for H$_2$ formation (\citealt{tegmark_small_1997}; eqn.~14). Adopting the expression given by \citeauthor{cazaux_molecular_2004}~(\citeyear{cazaux_molecular_2004}; eqn. 3) for the H sticking coefficient, $S_{\rm H}(T)$ and $\xi_{\rm d}=0.53 Z$ for the dust-to-gas mass ratio as suggested by \cite{cazaux_molecular_2004}, results in $\xi_{\rm d}\approx 0.01$ for Solar metallicity. This must be solved simultaneously with the recombination equation governing the ionized fraction $x$. The solution, assuming $x(t)=x_0/(1+x_0nk_1t)$ and $1-x-2f\approx 1$ as do \cite{tegmark_small_1997}, is
\begin{equation}
f(t) = f_0 {k_{\rm m} \over k_1} \exp\left[ {\tau_{\rm r} +t\over \tau_{\rm d}} \right] \left\{ {\rm E}_{\rm i}\left( {\tau_{\rm r} \over \tau_{\rm d}} \right) - {\rm E}_{\rm i}\left( {\tau_{\rm r} +t \over \tau_{\rm d}} \right) \right\} 
\end{equation}
where $\tau_{\rm r}=1/x_0/n_{\rm H}/k_1$, $\tau_{\rm d}=1/n_{\rm H}/k_{\rm d}$, $k_1$ is the hydrogen recombination coefficient and ${\rm E}_{\rm i}$ is the exponential integral.

\subsection{Heating}

While gas must cool in order to collapse and form a galaxy, there are several physical processes which instead heat the gas. In this subsection we review the nature and effects of those mechanisms.

\subsubsection{Photoheating}\label{sec:PhotoHeat}

Immediately after the epoch of cosmological recombination, the cosmological background light consists of just the blackbody radiation of the \CMB. Once stars (and perhaps AGN) begin to form they emit photons over a range of energies, including some at energies greater than the ionization edges of important coolants such as hydrogen, helium and heavier elements. Such photons can, in principle, photoionize atoms and ions in dark matter halos. This changes the ionization balance in the halo and heats the gas (via the excess energy of the photon above the ionization edge), thereby altering the rate at which this gas can cool to form a galaxy.

The ionizing background at some wavelength, $\lambda$, and redshift, $z$, is given by
\begin{equation}
 F_\lambda = \int_z^\infty {\mathcal E}\left({\lambda\over1+z^\prime},z^\prime\right) {\exp[-\tau(z,z^\prime,\lambda)] \over (1+z^\prime)^4} {\d t \over \d z^\prime} \d z,
\end{equation}
where ${\mathcal E}(\lambda,z)$ is the proper volume-averaged emissivity at redshift $z$ and wavelength $\lambda$ and $\tau(z_1,z_2,\lambda)$ is the optical depth between $z_1$ and $z_2$ for a photon with wavelength $\lambda$ at $z_1$. Calculation of the background requires a knowledge of the emissivity history of the Universe and the optical depth (itself a function of the ionization state and density distribution in the IGM) as a function of redshift.

Detailed theoretical models of the ionizing background, using observational constraints on the emissivity history, together with models of the distribution of neutral gas and calculations of radiative transfer, have been developed by \citeauthor{haardt_radiative_1996}~(\citeyear{haardt_radiative_1996}; see also \citealt{madau_he_2009}) and allow the background to be computed for any redshift.

Once the background is known, its effects on gas in dark matter halos can be determined. Such calculations require the use of photoionization codes, such as {\sc Cloudy} \pcite{ferland_cloudy_1998}, to solve the complex set of coupled equations that describe the photoionization equilibrium and to determine the resulting net cooling or heating rates.

\begin{figure}
 \begin{center}
  \includegraphics[width=80mm]{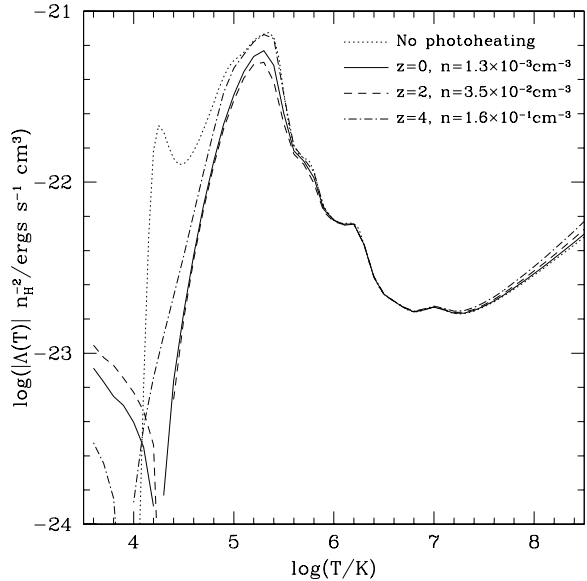}
 \end{center}
 \caption{Cooling functions for gas at the mean density of cosmological halos at three different redshifts (as indicated in the figure). The dotted line shows the cooling function in the absence of any photoionizing background. Other lines show the net cooling/heating function in the presence of the photoionizing background self-consistently computed by \protect\cite{benson_effects_2002}. All calculations assume a metallicity of $0.3Z_\odot$. The discontinuity at low temperatures shows the transition from net heating to net cooling.}
 \label{fig:PIE}
\end{figure}

Figure~\ref{fig:PIE} shows examples of net heating/cooling functions in the presence of a photoionizing background. These functions were calculated using the {\sc MappingsIII} code of \cite{allen_mappings_2008} with a photoionizing background computed self-consistently from the galaxy formation model of \cite{benson_effects_2002} and an assumed metallicity of $Z=0.3Z_\odot$. Unlike collisional ionization equilibrium cooling functions, which are density independent\footnote{Since the simple $n^2$ scaling is factored out of the cooling function by definition.}, photoionization equilibrium cooling curves depend on the density of the gas. The curves shown are therefore computed for gas at densities typical of gas in dark matter halos at each redshift indicated. It can be seen that heating becomes important for temperatures $T\lsim 3 \times 10^4$K.

\subsubsection{Heating from Feedback}\label{sec:HeatFB}

The overcooling problem has been a longstanding issue for galaxy formation theory. Simply put, in massive dark matter halos simple estimates suggest that gas can cool at a sufficiently large rate that, by the present day, galaxies much more massive than any observed will have formed \pcite{white_core_1978,white_galaxy_1991,katz_dissipational_1992,benson_what_2003}. An obvious way to counteract this problem is to heat the cooling gas, thereby offsetting the effects of cooling. Some sort of feedback loop is attractive here in order to couple the heating rate to the cooling rate and thereby balance the two.

In the massive halos where overcooling is a problem energy input from supernovae (see Section~ref{sec:SNeFeedback}) is insufficient to offset cooling\footnote{The total binding energy of gas in a dark matter halo scales as $M^{5/3}$ while the energy available from supernovae, assuming 100\% efficient conversion of gas into stars, scales as $M$. Consequently, in sufficiently massive halos, there will always be insufficient energy from supernovae to heat the entire halo.}. Consequently, much interest has been recently given to the idea that heating caused by \AGN\ may be responsible for solving the overcooling problem. The amount of energy available from \AGN\ is up to a factor 20--50 higher than from supernovae \pcite{benson_what_2003}. Semi-analytic models have demonstrated that such a scenario can work (see \S\ref{sec:AGNFeedback}) --- producing a significant reduction in the mass of gas able to cool in massive halos while simultaneously producing black holes with properties consistent with those observed --- under the assumption that energy output from \AGN\ can efficiently couple to the cooling atmosphere in the surrounding dark matter halo \pcite{croton_many_2006,bower_breakinghierarchy_2006,somerville_semi-analytic_2008}.

The mechanism by which energy output from the \AGN\ is coupled to the surrounding atmosphere remains unclear. Several solutions have been proposed, from effervescent heating (in which \AGN\ jets inflate bubbles which heat the \ICM\ via $P\d V$ work; \citealt{roychowdhury_entropy_2004,dalla_vecchia_quenching_2004,ruszkowski_cosmic_2008}, but see \citealt{vernaleo_agn_2006}) to heating by outflows driven by super-Eddington accretion \pcite{king_heating_2009} and viscous dissipation of sound waves \pcite{ruszkowski_cluster_2004} with turbulence playing an important role \pcite{bruggen_self-regulation_2009}.

\subsubsection{Preheating}

Heating that occurs prior to the collapse into a dark matter halo can also significantly affect the later evolution of baryons. Prior to virial collapse and any associated shock heating the gas evolves approximately adiabatically, maintaining a constant entropy. In cosmological studies the entropy is usually written as
\begin{equation}
 K = k_{\rm B} T/\mu m_{\rm H} \rho_{\rm B}^{\gamma-1},
\end{equation}
where $\gamma$ is a adiabatic index of the gas. An early period of heating (perhaps from the first generation of galaxies) can increase the entropy of the gas by raising its temperature. Since entropy is conserved this preheating will be ``remembered'' by the gas. We can consider what happens to such gas when it accretes into a halo, assuming for now that a shock forms close to the virial radius. A shock randomizes the ordered infall motion of the gas and therefore increases its entropy. At early times, when the shock is relatively weak, this entropy gain will be small compared to the preheated entropy of the gas. At late times, the shock entropy will dominate. The result is that the entropy distribution of preheated gas in a halo looks similar to that of non-preheated gas, except that there is a floor of minimum entropy.

The equation of hydrostatic equilibrium can be re-written in terms of entropy using the fact that $\rho = (P/K)^{1/\gamma}$ giving
\begin{equation}
 {\d P \over \d r} = -{ \G M(<r) \over r^2} \left({P \over K}\right)^{1/\gamma},
\end{equation}
and shows that the gas will arrange itself in the halo with the lowest entropy material in the center. The presence of an entropy floor leads to a density core in the halo center. This increases cooling times in the halo core and may therefore help prevent the formation of supermassive galaxies via the overcooling problem (see \S\ref{sec:HeatFB}; \citealt{borgani_preheatingintracluster_2001,voit_origin_2003,younger_cosmological_2007}). Additionally, X-ray observations of galaxy clusters seem to suggest the presence of entropy floors.

While preheating is an attractive way of explaining entropy floors in clusters and reducing the overcooling problem it is not clear whether it can consistently explain the abundance of galaxies in lower mass halos. If preheating occurred uniformly, it would drastically reduce the number of lower mass galaxies forming \pcite{benson_early_2003}. Preheating would need to occur preferentially in the sites of proto-clusters to avoid this problem---not inconceivable but not demonstrated either.

\subsubsection{Thermal Conduction}

Shock heated gas in dark matter halos is an ionized plasma and is therefore expected to have a thermal conductivity given by \pcite{spitzer_physics_1962}:
\begin{equation}
\kappa_{\rm S} = 32 \left({2\over\pi}\right)^{1/2} {n_{\rm e} \over \nu_0 m^{5/2}} T^{5/2},
\end{equation}
where $n_{\rm e}$ is the electron density and 
\begin{equation}
\nu_0 = {2\pi n_{\rm e} Z e^4 \ln \Lambda \over m^2},
\end{equation}
where $Z$ is the atomic number of the ions. If cooling in the inner regions of the hot gas halo leads to a temperature gradient in the atmosphere then conduction will cause energy to be deposited at radius $r$ at a rate per unit volume $\Sigma$ given by:
\begin{equation}
\Sigma = 4 \pi {\d \over \d r}\left( \kappa_{\rm S} r^2 {\d T \over \d r}\right),
\end{equation}
where $T$ is the temperature in the atmosphere. Thus, thermal conduction can, in principle, act as a heat source for the inner regions of the hot atmosphere---as they cool, heat is conducted inwards from the outer, hotter regions of the atmosphere. The actual conductivity may be substantially reduced below the Spitzer value if the hot atmosphere is threaded by tangled magnetic fields (such that electrons cannot directly transport heat but must effectively diffuse along field lines).

While thermal conduction is an attractive mechanism by which to solve the overcooling problem (since the conductivity is higher in more massive clusters due to the strong temperature dependence of $\kappa_{\rm S}$) it has been demonstrated that it cannot sufficiently offset cooling rates even with conductivities close to the Spizter value \pcite{benson_what_2003,dolag_thermal_2004,pope_effects_2005}, nor can it maintain a stable hot gas atmosphere over cosmological timescales \citep{conroy_thermal_2008,parrish_anisotropic_2009}.

\subsection{Fate of Cooling Gas}

In the simplest picture, gas which cools sufficiently below the virial temperature loses pressure support and flows smoothly towards the minimum of the gravitational potential well, settling there to form a galaxy. Such a picture is likely oversimplified, however.

\cite{maller_multiphase_2004} consider the consequences of the thermal instability in cooling atmospheres. They find that cooling gas fragments into two phases: cold ($T\approx 10^4$), dense clouds in pressure equilibrium with a hot (approximately virial temperature), diffuse component which can persist for cosmological periods of time due to a long cooling time. The masses of the clouds are determined from the thermal conduction limit, known as the Field length\footnote{Above this length scale thermal conduction can damp temperature perturbations in the intracluster medium and prevent cloud formation.} \pcite{field_thermal_1965}, and processes such as Kelvin-Helmholtz instabilities and conductive evaporation which act to destroy clouds. This significantly alters the manner in which fueling of galaxies occurs. The rate of gas supply to a forming galaxy now depends on the rate at which the dense gas clouds can infall, due to processes such a hydrodynamical drag and cloud-cloud collisions. The timescales for these two processes are given by
\begin{equation}
\tau_{\rm ram} = {2 m_{\rm cl} \over \pi C_{\rm d} r_{\rm cl}^2 \bar{\rho}_{h} v_{\rm cl}} \approx 2.6 \hbox{Gyr}
\end{equation}
and
\begin{equation}
\tau_{\rm cc} = {4 m_{\rm cl} R_{\rm c}^3 \over 3 M_{\rm cl} v_{\rm cl} r_{\rm cl}^2} \approx 2.4 \hbox{Gyr}
\end{equation}
respectively, where $m_{\rm cl}$, $r_{\rm cl}$ and $v_{\rm cl}$ are the characteristic mass, radius and velocity of cold clouds respectively, $M_{\rm cl}$ is the total mass in clouds, $C_{\rm d}$ is a drag coefficient, $\bar{\rho}_{\rm h}$ is the mean density of hot gas in the halo and $R_{\rm c}$ is the radius within which gas is sufficiently dense that it has been able to radiate away all of its thermal energy (i.e. the ``cooling radius''; \citealt{maller_multiphase_2004}). The numerical values given in each equation are computed for typical cloud properties taken from \cite{maller_multiphase_2004}.

These timescales can significantly exceed the cooling time for gas in galactic scale dark matter halos. Consequently, \cite{maller_multiphase_2004} find that considering the formation and infall of such clouds can significantly reduce the rate of gas supply to a forming galaxy (by factors of two or so), particularly in more massive halos. This picture has recently been confirmed in numerical experiments by \cite{kaufmann_redistributing_2009}.

\section{Galaxy Interactions}

The original concept of galaxies as ``island universes'' is, of course, the truth, but not the whole truth. While many galaxies do seem to be evolving in isolation there are clear indications that some galaxies are interacting strongly with other galaxies or their larger scale environment. We describe these interactions in the remainder of this section.

\subsection{Galaxy Orbits}\label{sec:galaxyOrbits}

A consequence of the hierarchical nature of structure formation in a cold dark matter universe is that dark matter halos are built through the merging together of earlier generations of less massive halos. While for a long time numerical simulations indicated that all trace of earlier generations of halos was erased during the merging process \pcite{katz_hierarchical_1993,summers_galaxy_1995} it was understood on analytical grounds that this was likely a numerical artifact rather than a physical result \pcite{moore_destruction_1996}. Beginning in the late 1990's, N-body simulations clearly demonstrated that this was indeed the case \pcite{tormen_survival_1998,moore_dark_1999,klypin_galaxies_1999}. Unlike earlier generations of simulations, they found that halos can persist as \emph{subhalos} within larger halos into which they merge. The current highest resolution simulations of individual halos \pcite{kuhlen_via_2008,springel_aquarius_2008} show almost 300,000 subhalos\footnote{This is a lower limit due to the limited resolution of the simulations. The earliest generations of cold dark matter halos may have masses as low as $10^{-12}M_\odot$ (depending on the particle nature of the dark matter) while state of the art simulations resolve only halos with masses greater than around $10^5M_\odot$. The ability of even lower mass halos to survive is a subject of much debate \protect\pcite{berezinsky_destruction_2006,zhao_tidal_2007,goerdt_survival_2007,angus_cold_2007,elahi_subhaloes_2009}.} and even show multiple levels of subclustering (i.e. subhalos within subhalos within subhalos\ldots). Each subhalo may, in principle, have acted as a site of galaxy formation and so may contain a galaxy which becomes a satellite in the host potential.

These subhalos are gravitationally bound to their host halo and, as such, will orbit within it. A subhalo's orbit can take it into regions where interactions affect the properties of any galaxy that it may contain. We begin, therefore, by considering the orbits of subhalos.

\begin{figure*}
 \begin{center}
 \begin{tabular}{cc}
     \includegraphics[width=80mm,bbllx=110mm,bblly=58mm,bburx=200mm,bbury=150mm,clip=]{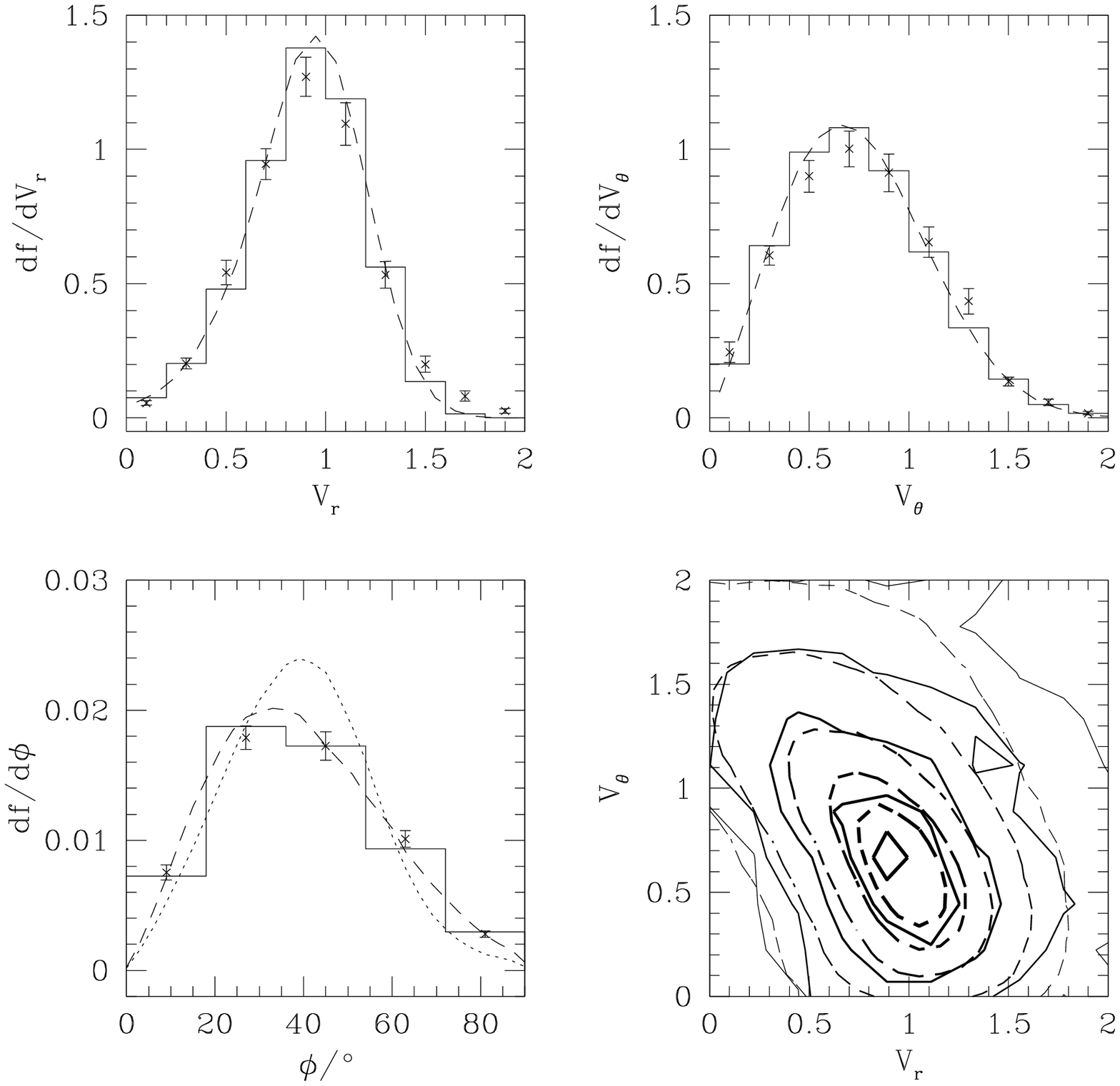} & \includegraphics[width=80mm]{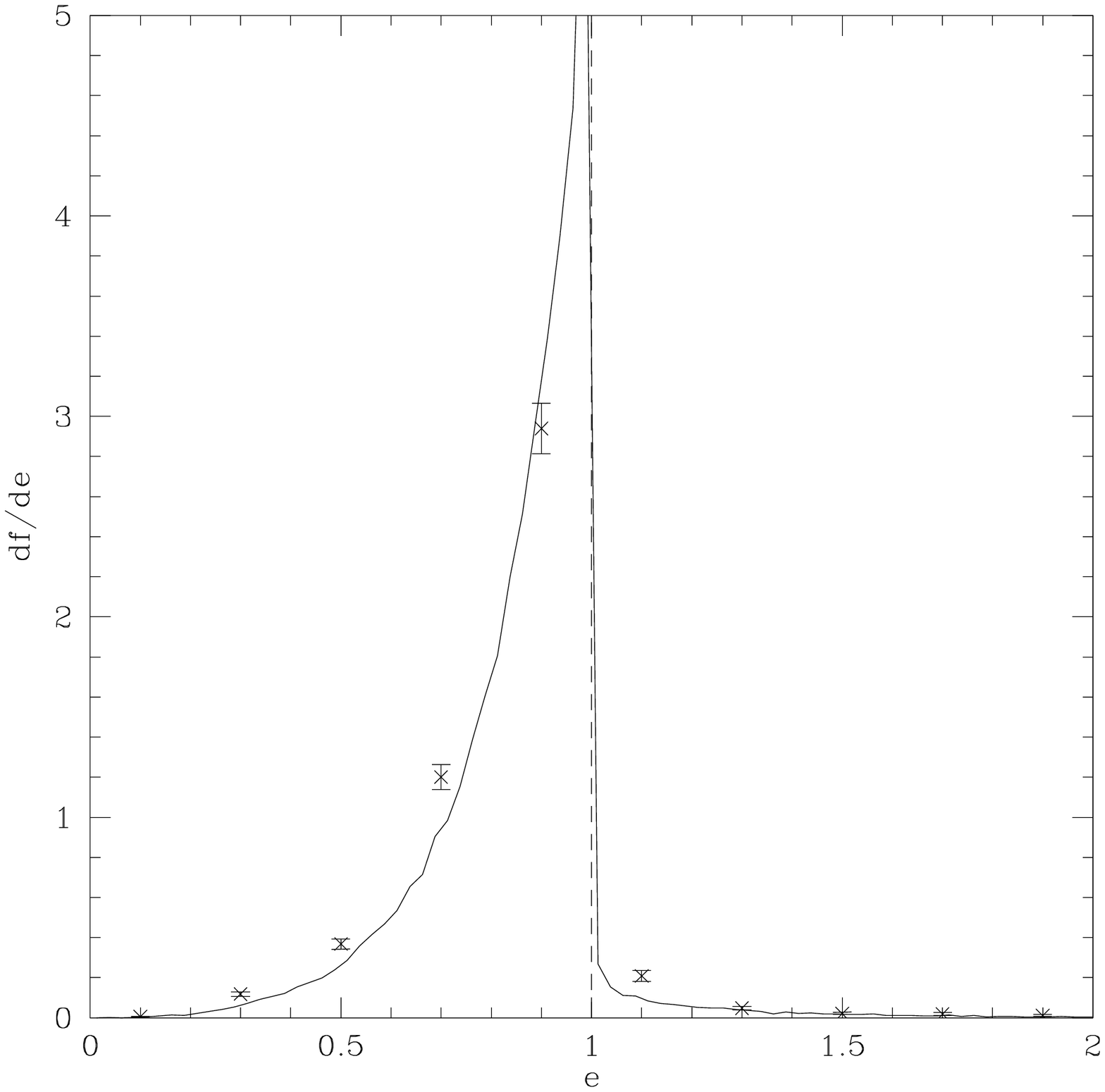}
 \end{tabular}
 \end{center}
 \caption{\emph{Left-hand panel:} The joint distribution of radial and tangential velocities of infalling subhalos as they cross the virial radius of their host halo. Velocities are expressed in units of the virial velocity of the host halo. Solid contours indicate measurements from a compilation of N-body simulations while dashed lines indicate the fitting formula of \protect\cite{benson_orbital_2005}. Contours are drawn at values of $\d^2 f / \d V_\theta \d V_r$ (the normalized distribution function) of 0.01, 0.1, 0.5, 1.0 and 1.4. \emph{Right-hand panel:} The corresponding distribution of orbital eccentricities. The distribution peaks close to $e=1$ (parabolic orbits) and it is apparent that some subhalos are on unbound ($e>1$) orbits. Points indicate the distribution measured from a compilation of N-body simulations, while the solid line indicates the distribution found from the same fitting formula as used in the left-hand panel. Reproduced, with permission, from \protect\cite{benson_orbital_2005}.}
 \label{fig:OrbitalDistribution}
\end{figure*}

At the point of merging, which we will define as the time at which the center of mass of a subhalo-to-be first crosses the virial radius of its future host halo, we expect the orbital parameters (velocities, energy etc.) to be of order unity when expressed in units of the characteristic scales of the host halo. For example, \cite{benson_orbital_2005} shows that the radial and tangential velocities of merging subhalos are distributed close to unity when expressed in units of the virial velocity of the host halo (Fig.~\ref{fig:OrbitalDistribution}). This distribution of velocities reflects the influence of the host halo (infall in its potential well) but also of the surrounding large scale structure which may have torqued the infalling subhalo.

\begin{figure}
 \begin{center}
     \includegraphics[width=80mm]{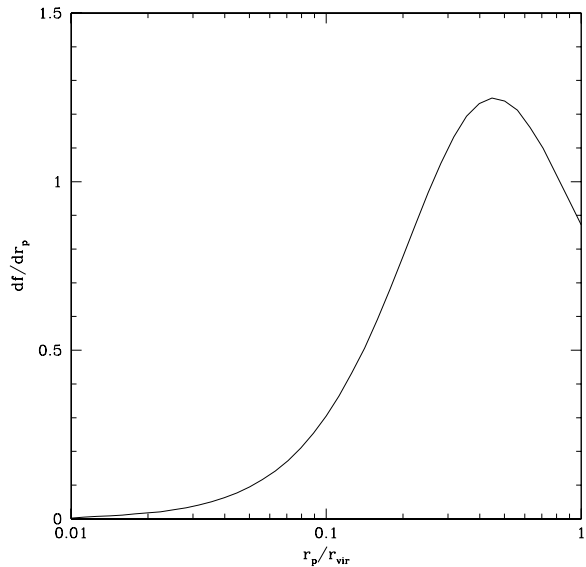}
 \end{center}
 \caption{The distribution of orbital pericenters (in units of the virial radius) in an NFW halo having concentration parameter of $10$, assuming the orbital parameter distribution of \protect\cite{benson_orbital_2005}.}
 \label{fig:PericenterDistribution}
\end{figure}

Such orbits will typically carry subhalos into the inner regions of halos. Figure~\ref{fig:PericenterDistribution} shows the distribution of orbital pericenters assuming an NFW halo with concentration $10$ and the orbital parameter distribution of \cite{benson_orbital_2005}. Most orbits initially reach to 40\% of the virial radius, but a significant tail have orbits which carry them into the inner 10\% of the halo.

\subsection{Gravitational Interactions}

\subsubsection{Mergers}\label{sec:Mergers}

Orbiting subhalos are gravitationally bound to their host halos and, as such, rarely encounter other subhalos at velocities resulting in a bound interaction \pcite{angulo_fate_2008,somerville_semi-analytic_2008,wetzel_clustering_2008}.

To cause gravitationally bound interactions between subhalos and their galaxies typically requires a dissipative process to reduce their orbital energies. Dynamical friction fulfills this role and tends to drag subhalos down towards the center of their host halo, where they may merge with any other galaxy which finds itself there. The classic derivation of dynamical friction acceleration from \cite{chandrasekhar_dynamical_1943} has been used extensively to estimate dynamical friction timescales within dark matter halos. For example, \cite{lacey_merger_1993} applied this formula to estimate merging timescales for subhalos in isothermal dark matter halos, finding:
\begin{equation}
 T_{\rm df} = {f(\epsilon) \tau_{\rm dyn}\over 2 B(1) \ln \Lambda} \left({r_{c} \over R_{\rm v}}\right)^2 \left({M_{\rm v} \over {m_{\rm v}}}\right)
 \label{eq:DynFricTime}
\end{equation}
where $\tau_{\rm dyn}=R_{\rm v} / V_{\rm v}$ is the dynamical time of the halo, $m_{\rm v}$ the virial mass of the orbiting satellite,
\begin{equation}
B(x) = \hbox{erf}(x)-{2x\over\sqrt{\pi}}\exp(-x^2), 
\end{equation}
$\ln\Lambda \approx \ln( r_{\rm v} V_{\rm v}^2 / \G m_{\rm v})\equiv \ln(M_{\rm v}/m_{\rm v})$ is the Coulomb logarithm (treating the satellite as a point mass), $f(\epsilon)$ encapsulates the dependence on the orbital parameters through the quantity $\epsilon = J/J_{\rm c}(E)$ where $J$ is the angular momentum of the satellite and $J_{\rm c}(E)$ is the angular momentum of a circular orbit with the same energy, $E$, as the actual orbit and $r_{\rm c}$ is the radius of that circular orbit. \cite{lacey_merger_1993} found that $f(\epsilon)=\epsilon^{0.78}$ was a good fit to numerical integrations of orbits experiencing dynamical friction.

Besides the fact the dark matter halos are not isothermal, there are a number of other reasons why this simple approach is inaccurate:
\begin{enumerate}
 \item Chandrasekhar's derivation assumes an infinite, uniform medium, not a dark matter halo with a radially varying density profile;
 \item There is the usual uncertainty in how to define the Coulomb logarithm (e.g. what is the maximum impact parameter for interactions between the satellite and dark matter particles);
 \item Subhalos will experience mass loss as they orbit, changing the timescale;
 \item The host halo itself is constantly evolving;
 \item The host halo is non-spherical and has a non-isotropic velocity dispersion.
\end{enumerate}

\begin{figure}
 \begin{center}
     \includegraphics[width=80mm]{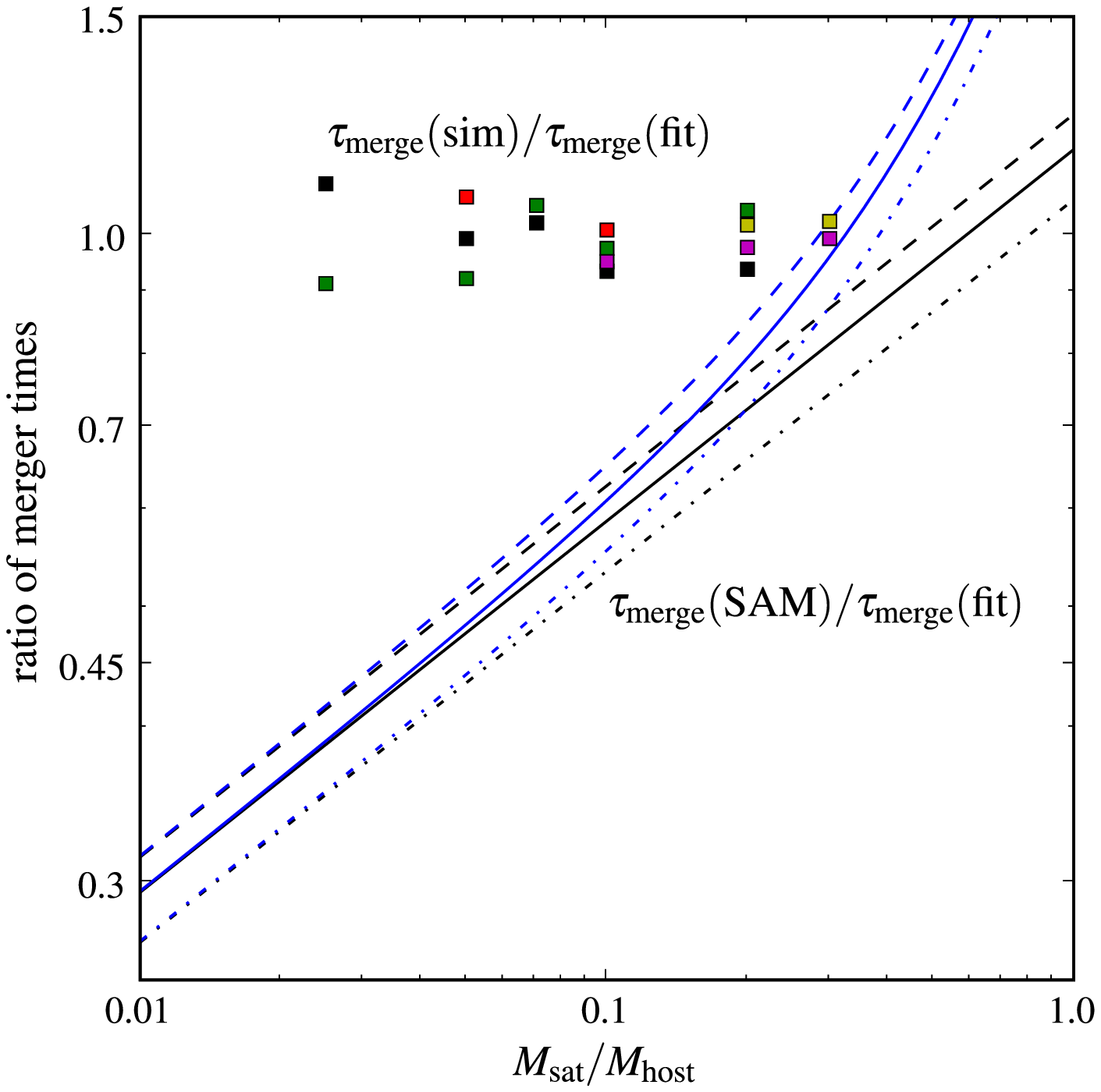}
 \end{center}
 \caption{A comparison of dynamical friction timescales measured from N-body simulations, $\tau_{\rm merge}$(sim),  with the fitting formula of \protect\cite{boylan-kolchin_dynamical_2008}, $\tau_{\rm merge}$(fit), is shown by the colored points, with colors coding for orbital circularity: $\epsilon = 0.33$ (red), 0.46 (green), 0.65 (black), 0.78 (magenta) and 1.0 (yellow). Lines compare $\tau_{\rm merge}$(fit) with the expectation from eqn.~(\protect\ref{eq:DynFricTime}), labeled $\tau_{\rm merge}$(SAM). The different color curves correspond to different choices of Coulomb logarithms in eqn.~(\protect\ref{eq:DynFricTime}): $\ln (1 + M_{\rm V}/m_{\rm V})$
(black curves) and $\frac{1}{2} \ln (1 + M_{\rm V}^2/m_{\rm V}^2)$ (blue curves). Reproduced, with permission, from \protect\cite{boylan-kolchin_dynamical_2008}.}
 \label{fig:BKTauDF}
\end{figure}

While some of these limitations can be overcome (e.g. mass loss can be modeled; \citep{benson_effects_2002,taylor_evolution_2004}; the Coulomb logarithm can be treated as a parameter to be fit to numerical results; results exists for dynamical friction in anisotropic velocity distributions; \citep{binney_dynamical_1977,benson_heating_2004}) others are more problematic. Recently, attempts have been made to find empirical formulae which describe the merging timescale. These usually begin with an expression similar to the one in eqn.~(\ref{eq:DynFricTime}) but add empirical dependencies on subhalo mass and orbital parameters which are constrained to match results from N-body simulations. Results from such studies \pcite{jiang_fitting_2008,boylan-kolchin_dynamical_2008} show that the simple formula in eqn.~(\ref{eq:DynFricTime}) tends to underestimate the timescale for low mass satellites (probably because it ignores mass loss from such systems) and overestimates the timescale for massive satellites (probably due to a failure of several of the assumptions made in this limit). Alternative fitting formulae have been derived from these studies. For example, \cite{boylan-kolchin_dynamical_2008} find
\begin{equation}
T_{\rm df} = \tau_{\rm dyn} A {(M_{\rm V}/m_{\rm V})^b \over \ln(1+M_{\rm V}/m_{\rm V})} \exp\left[c \epsilon \right] \left[{r_{\rm c}(E) \over r_{\rm V}}\right]^d,
\end{equation}
with $A=0.216$, $b=1.3$, $c=1.9$ and $d=1.0$ while \cite{jiang_fitting_2008} finds
\begin{equation}
T_{\rm df} = \tau_{\rm dyn} {0.94\epsilon^{0.60}+0.60\over 2C} {M_{\rm V} \over m_{\rm V}} {1 \over \ln(1+M_{\rm V} / m_{\rm V})},
\end{equation}
with $C=0.43$. A comparison of the \cite{boylan-kolchin_dynamical_2008} fit, eqn.~(\ref{eq:DynFricTime}) and measurements from numerical simulations is shown in Fig.~\ref{fig:BKTauDF}.

The consequences of merging for the galaxies involved are discussed in \S\ref{sec:MajorMergers}.

\subsubsection{Tidal Destruction}

An orbiting subhalo and its galaxy will experience tidal forces which may strip away the outer regions or, in extreme cases, entirely disrupt the galaxy resulting in a stellar stream (as seems to be happening with the Sagittarius dwarf galaxy in orbit around the Milky Way; \citealt{belokurov_field_2006}).

In a rotating frame in which an orbiting satellite instantaneously has zero tangential velocity, the effective tidal field felt by the satellite is
\begin{equation}
 {\mathcal D} = -{\d \over \d r} \left({\G M_{\rm h}(r)\over r^2}\right) + \omega^2 r,
\end{equation}
where $M_{\rm h}(r)$ is the mass enclosed within radius $r$ in the host halo and $\omega$ is the instantaneous angular velocity of the satellite. An estimate of the radius, $r_{\rm t}$, in the satellite subhalo/galaxy system beyond which tidal forces become important can be made by equating the tidal force to the self-gravity of the subhalo
\begin{equation}
 {\mathcal D} r_{\rm t} = {\G M_{\rm s}(r_{\rm t})\over r_{\rm t}^2}.
\end{equation}
Beyond this \emph{tidal radius} material becomes unbound from the satellite, forming a stream of dark matter and, potentially, stars which continue to orbit in the host potential.

This simple estimate ignores the fact that particles currently residing in the inner regions of a subhalo may have orbits which carry them out to larger radii where they may be more easily stripped. As such, the degree of tidal mass loss should depend not only on the density profile of the satellite but also on the velocity distribution of the constituent particles. Attempts to account for this find that particles in an orbiting satellite that are on prograde orbits are more easily stripped than those on radial orbits which are in turn more easily stripped than those on retrograde orbits \pcite{read_tidal_2006}. Additionally, some material will be stripped from within the classical tidal radius, as particles which contribute to the density inside that radius may be on orbits which carry them beyond it. This can lead to more extensive and continuous mass loss as the reduction in the inner potential of the satellite due to this mass loss makes it more susceptible to further tidal stripping. \pcite{kampakoglou_tidal_2007}.

\subsubsection{Harassment}

A less extreme form of tidal interaction arises when tidal forces are not strong enough to actually strip material from a galaxy. The tidal forces can, nevertheless, transfer energy from the orbit to internal motions of stars in the galaxy, effectively heating the galaxy. The generic results of such heating are to cause the galaxy to expand and to destroy cold, ordered structures such as disks \pcite{moore_galaxy_1996,moore_morphological_1998,mayer_tidal_2001,mayer_metamorphosis_2001,gnedin_tidal_2003,mastropietro_morphological_2005,mastropietro_gravitational_2005}. The harassment process works via \emph{tidal shocking} in which the stars in a galaxy experience a rapidly changing tidal field along its orbit and gain energy in the form of random motions, leading to the system expanding and becoming dynamically hotter. During such tidal shocks, the energy per unit mass of the galaxy changes by \pcite{gnedin_tidal_2003}
\begin{equation}
\langle \Delta E \rangle = {1\over 6} I_{\rm tid} \langle r^2\rangle,
\end{equation}
where $\langle r^2\rangle$ is the mean squared radius of the galaxy and
\begin{equation}
I_{\rm tid} = \sum_n \sum_{\alpha,\beta} \left(\int F_{\alpha\beta} \d t \right)^2_n \left(1+{\tau_n^2 \over t^2_{\rm dyn}}\right)^{-3/2},
\end{equation}
where the sums extend over all $n$ peaks in the density field (i.e. the host halo and any other subhalos that it may contain) and over all components of the tidal tensor
\begin{equation}
F_{\alpha\beta} = - {\d^2\Phi \over \d r_\alpha \d r_\beta},
\end{equation}
where $\Phi$ is the gravitational potential. Here, $\tau_n$ is the effective duration of the encounter with peak $n$ and $t_{\rm dyn}$ is the dynamical time at the half-mass radius of the galaxy. The $(1+\tau_n^2 /t^2_{\rm dyn})^{-3/2}$ term describes the transition from the impulsive to adiabatic shock regimes (\citealt{gnedin_self-consistent_1999}; see also \citealt{murali_effect_1997,murali_evolution_1997,murali_globular_1997}).

\subsection{Hydrodynamical Interactions}

While the collisionless dark matter is affected only by gravity the baryonic content of galaxies (and their surrounding atmospheres of gas) can be strongly affected by hydrodyamical forces.

\subsubsection{Ram pressure}

The orbital motion of a subhalo through the hot atmosphere of a host halo leads to a large ram pressure. The characteristic magnitude of that pressure
\begin{equation}
 P_{\rm ram} = \rho_{\rm host} V_{\rm orbit}^2
\end{equation}
can greatly exceed the binding energy per unit volume of both hot gas in subhalos and \ISM\ gas in their galaxies. As such, ram pressure forces may be expected to quite efficiently remove the hot atmospheres of satellite galaxies, a process with several grim aliases including \emph{strangulation} and \emph{starvation}, and the \ISM\ of the galaxy.

The first quantification of this process was made by \cite{gunn_infall_1972} who showed that the ram pressure force could remove material from a galactic disk if it exceeded the gravitational restoring force per unit area which itself cannot exceed
\begin{equation}
 {\mathcal F} = 2 \pi \G \Sigma_\star \Sigma_{\rm gas}.
\end{equation}
For a disk of mass $M_{\rm d}$ with gas mass $M_{\rm g}$ having an exponential surface density profile for both gas and stars with scale length $r_{\rm d}$, the gravitational restoring force per unit area is given by \pcite{abadi_ram_1999}:
\begin{equation}
{\mathcal F} = {\G M_{\rm d} M_{\rm g}\over 4\pi r_{\rm d}^4} x {\rm e}^{-x} \left[I_0\left({x\over 2}\right)K_1\left({x\over 2}\right)-I_1\left({x\over 2}\right)K_0\left({x\over 2}\right)\right],
\end{equation}
where $x=r/r_{\rm d}$ and $I_0$, $I_1$, $K_0$ and $K_1$ are Bessel functions.

The mass loss caused by this ram pressure can, in many cases, be further enhanced by related effects, such as turbulent viscous stripping \pcite{nulsen_transport_1982}. This initial estimate has been revised and calibrated more accurately using numerical simulations \pcite{abadi_ram_1999,mccarthy_ram_2008}. The process of ram pressure stripping has been incorporated into some semi-analytic models of galaxy formation \pcite{lanzoni_galics-_2005,font_colours_2008} where it plays an important role in mediating the transition of cluster galaxies from the blue cloud of star-forming galaxies to the red sequence of passively evolving galaxies \pcite{font_colours_2008}.

\section{Galactic Structure}

The preceding sections describe the path by which gas achieves sufficiently high density to allow a galaxy to form. The next significant question is what structure that galaxy will have. In the following section we will review the process of disk and spheroid formation and make quantitative estimates of the sizes of these structures.

\subsection{Disk Formation}

Disks are a common astrophysical phenomenon and galactic disks owe their origins to the same fundamental process as other astrophysical disks: conservation of angular momentum in a system collapsing under gravity eventually leading to arrest of the collapse by rotational support. Understanding the physical properties of galactic disks therefore requires knowledge of their angular momentum content and the gravitational potential in which they form.

The angular momentum of the gas which will eventually form a galactic disk arises in the same way as that of dark matter halos, namely from tidal torques from surrounding large scale structure \cite{hoyle_origin_1949}\footnote{An alternative view of this process, although fundamentally based upon the same physics, involves considering the angular momentum supplied to a halo by accreting satellites and has been developed by \protect\cite{vitvitska_origin_2002}, \protect\cite{maller_modelling_2002} and \protect\cite{benson_galaxy_2010}.}. The magnitude of the angular momentum content of the baryonic component of a halo is close to being a scaled version of that of the dark matter---for example, \cite{bosch_angular_2002} find that the distributions of spin parameters and angular momentum distributions within individual halos for dark matter and non-radiative gas are very similar. They find that there is, however, a significant (typically $30^\circ$) misalignment between the dark matter and gas angular momenta vectors\footnote{These authors note that a significant fraction (5--50\%) of the gas in a halo can have a negative angular momentum relative to the total angular momentum of the halo, and suggest that this could lead to the formation of low angular momentum spheroids.}. Numerical hydrodynamical studies show that the angular momenta vectors of galactic disks \emph{are} well aligned with the angular momenta of the inner regions of dark matter halos \citep{bett_angular_2009,hahn_large-scale_2010}.

The distribution of the angular momentum is less well studied but recent investigations \pcite{sharma_angular_2005} using non-radiative hydrodynamical simulations have shown that the differential distribution of specific angular momenta, $j$, is given by
\begin{equation}
 {1\over M}{\d M \over \d j} = {1 \over j_{\rm d}^\alpha\Gamma(\alpha)}j^{\alpha-1}{\rm e}^{-j/j_{\rm d}},
\end{equation}
where $\Gamma$ is the gamma function, $M$ is the total mass of gas, $j_{\rm d}=j_{\rm tot}/\alpha$ and $j_{\rm tot}$ is the mean specific angular momentum of the gas. \cite{sharma_angular_2005} find that the simulated halos have a median value of $\alpha=0.89$. The fraction of mass with specific angular momentum less than $j$ is then given by
\begin{equation}
 f(<j) = \gamma\left(\alpha,{j\over j_{\rm d}}\right),
\end{equation}
where $\gamma$ is the incomplete gamma function.

We have assumed so far that the angular momentum of the pre-galactic gas is conserved during collapse. This may not be precisely true and indeed did not seem to be in earlier generations of hydrodynamical simulations which typically found that disk galaxies lost significant fractions of their angular momentum and, as a result, were too small \pcite{navarro_dynamics_1991,navarro_simulations_1994,navarro_assembly_1995}. However, more recent simulations \pcite{thacker_star_2001,steinmetz_hierarchical_2002} do show approximate conservation of angular momentum (most likely due to the inclusion of effective feedback in these later generations of simulations) and, furthermore, conservation of angular momentum leads to disks with sizes comparable to those observed---certainly the gas cannot lose a significant fraction of its angular momentum if it is to form a disk of comparable size to observed galactic disks. Confirmation of these ideas has recently been made by \cite{zavala_bulges_2008} who convincingly show in N-body+hydrodynamical simulations that the particular feedback prescription used can lead to the formation of a disk dominated or spheroid dominated galaxy in the same dark matter halo, with stronger feedback leading to a disk galaxy. The angular momentum of the mass in the disk galaxy tracks that of the dark matter halo as a whole, growing as expected during the linear regime and remaining nearly constant after halo collapse. With weaker feedback a spheroid forms instead. Its angular momentum also grows during the linear regime but then declines rapidly (as does that of the inner regions of the dark matter halo) due to angular momentum transfer from dense, progenitor blobs to the surrounding, diffuse dark matter. Feedback processes are clearly of crucial importance here---for example, \cite{wise_resolvingformation_2008} show that feedback during the formation of protogalaxies at early times can effectively boost the angular momentum content of such systems by factors of three to five. This is achieved via forces associated with HII region and SNe blast waves which produce torques much stronger than the usual cosmological torques.

\subsubsection{Sizes}

The physics which determines the sizes of galaxy disks was originally described by \cite{fall_formation_1980} and later expanded upon by many authors (see, in particular, \citealt{mo_formation_1998,mao_evolution_1998}).

Once the angular momentum distribution of that part of the baryonic component of the halo which cools to form the galaxy is known, finding the structure of the resulting rotationally supported disk is reduced to solving the following equation
\begin{equation}
 {j^2(M) \over R^3(M)} =  {\partial \over \partial R} \Phi(R),
\end{equation}
where $j(M)$ is the specific angular momentum enclosing mass $M$ and this equation is solved for $R(M)$. The potential is a sum of the self-gravity of the disk and that of any external potential (dark matter halo and bulge for example) which may have responded adiabatically to the formation of the disk. For a thin disk, the potential is generically given by \pcite{binney_galactic_2008}
\begin{equation}
 \Phi(R,0) = -4 \G \int_0^R {\d a \over \sqrt{R^2-a^2}}\int_a^\infty \d R^\prime {R^\prime \Sigma(R^\prime) \over \sqrt{R^{\prime 2}-a^2}},
\end{equation}
which, for an exponential surface density profile\footnote{These equations assume a razor thin disk. For a thickened disk a density distribution $\rho(R,z)=\rho_0 \exp(-R/R_{\rm d})\hbox{sech}^2(2/2z_0)$, with $z_0$ being the scale height of the disk, is often assumed as it is the self-consistent distribution for an isothermal population of stars \protect\pcite{spitzer_dynamics_1942}. The resulting gravitational potential can be found (e.g. \protect\citealt{kuijken_mass_1989}) although the calculation is significantly more involved.}
\begin{equation}
\Sigma(R) = \Sigma_0 \exp\left(-{R \over R_{\rm d}}\right),
\end{equation}
simplifies to
\begin{equation}
 \Phi(R,0) = \pi \G \Sigma_0 R [I_0(y)K_1(y)-I_1(y)K_0(y)],
\end{equation}
where $y=R/2R_{\rm d}$ and $I_0$, $I_1$, $K_0$ and $K_1$ are Bessel functions. Knowing the functional form of $j(M)$ it is possible to solve for $R(M)$ and therefore the density profile of the disk. Often, for simplicity, a particular form for the density (e.g. exponential) is assumed which leaves a single free parameter (the scale length) to be solved for.

A significant complication to this picture arises from the fact that the external potential in which the disk forms is likely to change in response to the formation of the disk. If the disk forms slowly, such that the timescale for changes in the potential greatly exceeds the dynamical time of the dark matter halo, then we can use adiabatically invariant quantities \pcite{binney_galactic_2008} to estimate the response of the halo to the forming disk. The original formulation of this argument for galaxies was given by \cite{blumenthal_contraction_1986} and has been used extensively ever since. In this simplified picture, the dark matter particles are considered to be on circular orbits in a spherical potential. In this case, the only adiabatic invariant whose corresponding angle variable has a non-infinite period is the magnitude of the angular momentum, $L_\phi$. Prior to the formation of any galaxy, this angular momentum is given by
\begin{equation}
 L^2_\phi = (1-f_{\rm b})^{-1} \G M_{\rm DM}(r_0) r_0,
\end{equation}
where $M_{\rm DM}(r)$ is the mass of dark matter within radius $r$, $r_0$ is the initial radius of the circular orbit and we have assumed that baryons are present in the halo at the universal fraction, $f_{\rm b}(\equiv \Omega_{\rm b}/[\Omega_{\rm b}+\Omega_{\rm CDM}])$ and distributed as the dark matter. If $L_\phi$ is conserved during the formation of a galaxy, then, at any point after galaxy formation has begun we have
\begin{equation}
 L^2_\phi = (1-f_{\rm b}^\prime)^{-1} \G [M^\prime_{\rm DM}(r_{\rm f}) + M_{\rm gal}(r_{\rm f})] r_{\rm f},
\end{equation}
where $M^\prime(r)$ is the mass of dark matter enclosed within radius $r$ after responding to the forming galaxy, $M_{\rm gal}(r)$ is the mass of the galaxy within that same radius\footnote{Note that we are approximating the galaxy as a spherical mass distribution here.}, $f_{\rm b}^\prime$ is the baryon fraction remaining uncondensed into the galactic phase and $r_{\rm f}$ is the radius of the new circular orbit for this dark matter particle. If we assume no shell crossing, then $M^\prime_{\rm DM}(r_{\rm f}) = M_{\rm DM}(r_0)$. Given a knowledge of the mass and density distribution of the galaxy this allows us to solve for $r_{\rm f}$ as a function of the initial radius $r_0$.

This approach makes several simplifying assumptions, including adiabatic growth, a spherical halo and circular orbits for dark matter particles. Many of these assumptions have yet to be adequately tested. However, recent work has attempted to address the last of these. With a more realistic distribution function for the dark matter, the particles will occupy a range of orbital eccentricities. In a spherical potential, the angular momentum is joined by a second adiabatic invariant, the radial action:
\begin{equation}
 L_r = {1 \over \pi} \int_{r_{\rm min}}^{r_{\rm max}} v_r(r) \d r,
\end{equation}
where the integral is taken along the orbit of a particle from its pericenter to apocenter. The effects of taking into account this second invariant and a physically reasonable orbital distribution have been explored by \cite{gnedin_response_2004}. \cite{gnedin_response_2004} find that the simple model described above systematically overpredicts the degree of contraction of the halo. However, they find that a simple modification in which the combination $M(r)r$ in the above is replaced with $M(\overline{r})r$, where $\overline{r}$ is the orbit-averaged radius, approximately accounts for the orbital eccentricities and gives good agreement with numerical simulations. \cite{gnedin_response_2004} further find that the orbit-averaged radius in cosmological halos can be well fit by the relation $\overline{r}/r_{\rm vir} = A (r/r_{\rm vir})^w$ with $A=0.85\pm 0.05$ and $w = 0.8 \pm 0.02$. More recent simulations \pcite{gustafsson_baryonic_2006} support this picture, but show that the optimal values of $A$ and $w$ vary from galaxy to galaxy and seem to depend on the formation history (i.e. on the adopted rules for star formation and feedback; see also \citealt{abadi_galaxy-induced_2009}). This suggests that the process of halo response to galaxy growth is more complicated than is captured by these simple models, perhaps because the assumed invariants are not precisely invariant in non-spherical potentials or perhaps because galaxy growth is not sufficiently slow to be truly adiabatic.

Galaxy disks are, of course, not razor thin. The origins of the vertical extent of galaxy disks remain a topic of active research but various possibilities are currently considered seriously:
\begin{enumerate}
 \item External origins (accretion): 
 \begin{enumerate}
  \item The hierarchical nature of galaxy formation implies that a galactic disk can expect to accrete pre-existing, smaller stellar systems frequently during its life. The stars from these smaller galaxies are often found in numerical simulations to form a thickened disk structure, in the same plane as the pre-existing disk of the galaxy \pcite{abadi_simulations_2003,villalobos_simulations_2008}. This is to be expected as dynamical friction will tend to drag orbiting satellites into the plane of the disk where tidal forces will proceed to shred the satellite, leaving its stars orbiting in the plane of the disk.
  \item Early, chaotic gas accretion: In a slightly different scenario, \cite{brook_emergence_2004} find, in numerical simulations of a forming galaxy, that many of the thick disk stars form from accreted gaseous systems during an early, chaotic period of merging.
 \end{enumerate}
 \item Internal origins (dynamical heating):
 \begin{enumerate}
  \item Dark matter substructure: Building upon earlier works \pcite{toth_galactic_1992,quinn_heating_1993,sellwood_resonant_1998,velazquez_sinking_1999}, in recent years there have been numerous studies focused on the question of whether relatively thin galactic disks can survive in the rather violent environment of a hierarchically formed dark matter halo \pcite{font_halo_2001,benson_heating_2004,kazantzidis_cold_2008,read_thin_2008,kazantzidis_cold_2009,purcell_destruction_2009}. A careful treatment of dark matter substructure orbital evolution and the build-up of the substructure population over time is required to address this problem. The consensus conclusion from these studies is that galactic disks can survive in the currently accepted cold dark matter cosmogony, but that interactions with orbiting dark matter substructure must be a significant contribution to the thickening of galaxy disks. The heating is dominated by the most massive dark matter substructures and so is a rather stochastic process, depending on the details of the merger history of each galaxy's dark matter halo and the orbital properties of those massive substructures.
  \item Molecular clouds: Massive molecular clouds can gravitationally scatter stars that happen to pass by them, effectively transforming some of their orbital energy in the plane of the galaxy disk into motion perpendicular to that plane, thereby effectively thickening the disk \pcite{spitzer_possible_1953,lacey_influence_1984}.

  \item Spiral arms: Spiral density waves can also act as scatterers of stars, but predominantly increase the stellar velocity dispersion in the plane of the galaxy, resulting in little thickening \pcite{jenkins_spiral_1990,minchev_radial_2006}.
 \end{enumerate}
\end{enumerate}
Most likely, a combination of these processes is at work. A recent review of these mechanisms is given by \cite{wen_thick_2004}.

\subsubsection{Stability}

Spiral arms and other non-axisymmetric features such as bars in galactic disks are a visually impressive reminder that these systems possess interesting dynamics. The basic theory of spiral arms, that they are density waves, was first proposed by \citeauthor{lin_spiral_1964}~(\citeyear{lin_spiral_1964}; see also \citealt{marochnik_milky_1996,binney_galactic_2008}) and has come to be widely accepted. From our current standpoint, the question in which we are interested in is whether these perturbations to an otherwise smooth disk are stable or unstable and, if unstable, how they affect the evolution of the galaxy as a whole.

\begin{figure}
\begin{center}
  \includegraphics[height=80mm,angle=270]{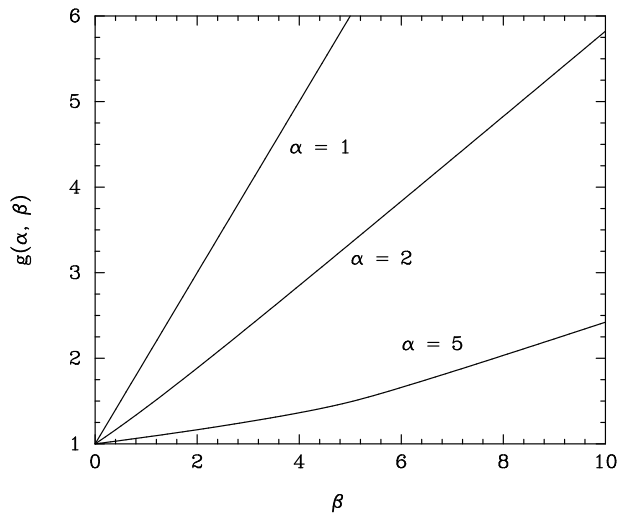}
\end{center}
\caption{The function $g(\alpha,\beta)$ appearing in eqn.~(\protect\ref{eq:QJoint}). $g(\alpha,\beta)$ is shown as a function of $\beta$ (the ratio of stellar to gas velocity dispersions) for various values of $\alpha$ (the ratio of stellar to gas surface density). Reproduced, with permission, from \protect\cite{efstathiou_model_2000}.}
\label{fig:gAlphaBeta}
\end{figure}

\citeauthor{toomre_gravitational_1964}~(\citeyear{toomre_gravitational_1964}; see also \citealt{goldreich_ii._1965}) derived an expression for the local stability of thin disks to axisymmetric modes in the tight-winding approximation which turns out to be extremely useful (and often approximately correct even in regimes where its assumptions fail). Disks will be unstable to axisymmetric modes if $Q < 1$ where
\begin{equation}
 Q = {\kappa \sigma_{\rm gas} \over \pi \G \Sigma_{\rm gas}},
\end{equation}
where
\begin{equation}
\kappa = \left( R {\d \Omega^2 \over \d R} + 4 \Omega^2 \right)^{1/2}
\end{equation}
is the epicyclic frequency and $\Omega$ is the angular frequency of the disk, for a gaseous disk of surface density $\Sigma_{\rm gas}$ and velocity dispersion $\sigma_{\rm gas}$, and
\begin{equation}
 Q = {\kappa \sigma_\star \over 3.36 \G \Sigma_\star}
\end{equation}
for a stellar disk of surface density $\Sigma_\star$ and velocity dispersion $\sigma_\star$. For a disk consisting of two components, gas and stars, a joint stability analysis was carried out by \cite{jog_two-fluid_1984}. \citeauthor{efstathiou_model_2000}~(\citeyear{efstathiou_model_2000}; see also \citealt{bertin_global_1988}) solves the resulting cubic equation for the most unstable mode and finds a criterion
\begin{equation}
 Q = {\kappa \sigma_{\rm gas} \over \pi \G \Sigma_{\rm gas} g(\alpha,\beta)},
 \label{eq:QJoint}
\end{equation}
where $\alpha$ and $\beta$ are the ratio of stellar to gas velocity dispersions and surface densities respectively and the function $g(\alpha,\beta)$ is as given by \cite{efstathiou_model_2000} and is shown in Fig.~\ref{fig:gAlphaBeta}. \cite{wang_gravitational_1994} also show that the stability of a two-component disk can be approximated by
\begin{equation}
 Q = (Q_\star^{-1}+Q_{\rm gas}^{-1})^{-1},
\end{equation}
while a more general criterion taking into account the thickness of the disk was found by \cite{romeo_stability_1992} and \cite{romeo_faithful_1994}.

Toomre's criterion applies to local perturbations. Perturbations on the scale of the disk can occur also. Study of these global instabilities is less amenable to analytic treatment (since one can no longer ignore contributions from distant parts of the disk). Significant work on this subject began in the 1970's. The classic result from that time is due to \cite{ostriker_numerical_1973} who found that rapidly rotating, self-gravitating stellar systems would become violently unstable to non-axisymmetric $m=2$ modes if $T/|W|\gsim 0.14$ (for a stellar system) or $0.27$ (for a fluid system) where $T$ and $W$ are the kinetic and potential energies of the system. Such systems can be unstable to global perturbations even if they satisfy Toomre's stability criterion. These values were approximately confirmed by numerous numerical studies, but also challenged by others (see \cite{christodoulou_new_1995} for a discussion of this). Later work \pcite{christodoulou_phase-transition_1995} has demonstrated that $T/|W|$ is not a sufficiently general parameter to determine stability in all systems. More recent studies have proposed improved stability criteria. For example \cite{christodoulou_new_1995} find that
\begin{equation}
 \alpha = {T/|W|\over \Omega/\Omega_{\rm J}},
\end{equation}
where $\Omega_{\rm J}$ is the Jeans frequency in the radial direction and $\Omega$ is the mean angular velocity of the system, is a much better indicator of stability. This can be approximated by
\begin{equation}
 \alpha = T_J/|W|,
\end{equation}
where $T_J=L\Omega_J/2$ and $L$ is the total angular momentum of the system. \cite{christodoulou_new_1995} found that $\alpha \le 0.254$--$0.258$ is required for stability in stellar systems while $\alpha \le 0.341$--$0.354$ is required for gaseous systems. \cite{christodoulou_new_1995} also demonstrate that this criterion is approximately equivalent to the alternative form found by \cite{efstathiou_stability_1982} on the basis of numerical simulations of exponential stellar disks:
\begin{equation}
 \epsilon_m \equiv {v_m \over (\G M_{\rm D}/R_{\rm S})^{1/2}} \gsim 1.1,
\end{equation}
and show that an equivalent result for gaseous disks would replace the $1.1$ with $0.9$. A much more extensive review of disk stability is given by \cite{sellwood_dynamics_2010}.

Global instabilities most likely lead to the formation of a very strong bar which effectively disrupts the disk leaving, after a few dynamical times, a boxy/peanut bulge or a disky bulge \pcite{athanassoula_disc_2008}. This may therefore be a possible formation scenario for pseudo-bulges \pcite{kormendy_secular_2004}, i.e. bulges formed through secular processes in the disk (see \S\ref{sec:Secular}) rather than as the result of a merger event (see \S\ref{sec:MajorMergers}).

\subsubsection{Bars/Spiral Arms}

\begin{figure}
\begin{center}
  \includegraphics[width=80mm]{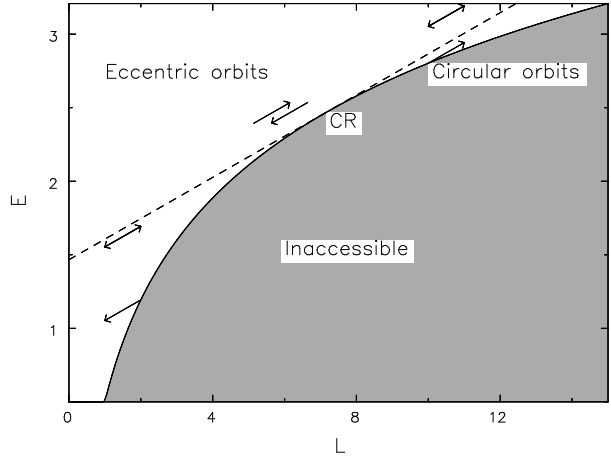}
\end{center}
\caption{The Lindblad diagram showing allowed regions of the energy--angular momentum plane (white, with forbidden regions shaded gray) bounded by the curve corresponding to circular orbits. Arrows indicate how scattering from spiral density waves moves stars in this diagram. At the corotation resonance (CR) this motion parallels the circular orbit curve, implying that scattering at corotation of stars initially on circular orbits leaves those stars on approximately circular orbits. Reproduced, with permission, from \protect\cite{sellwood_radial_2002}.}
\label{fig:SellwoodLindblad}
\end{figure}

\begin{figure}
\begin{center}
  \includegraphics[width=80mm,bbllx=0mm,bblly=0mm,bburx=90mm,bbury=90mm,clip=]{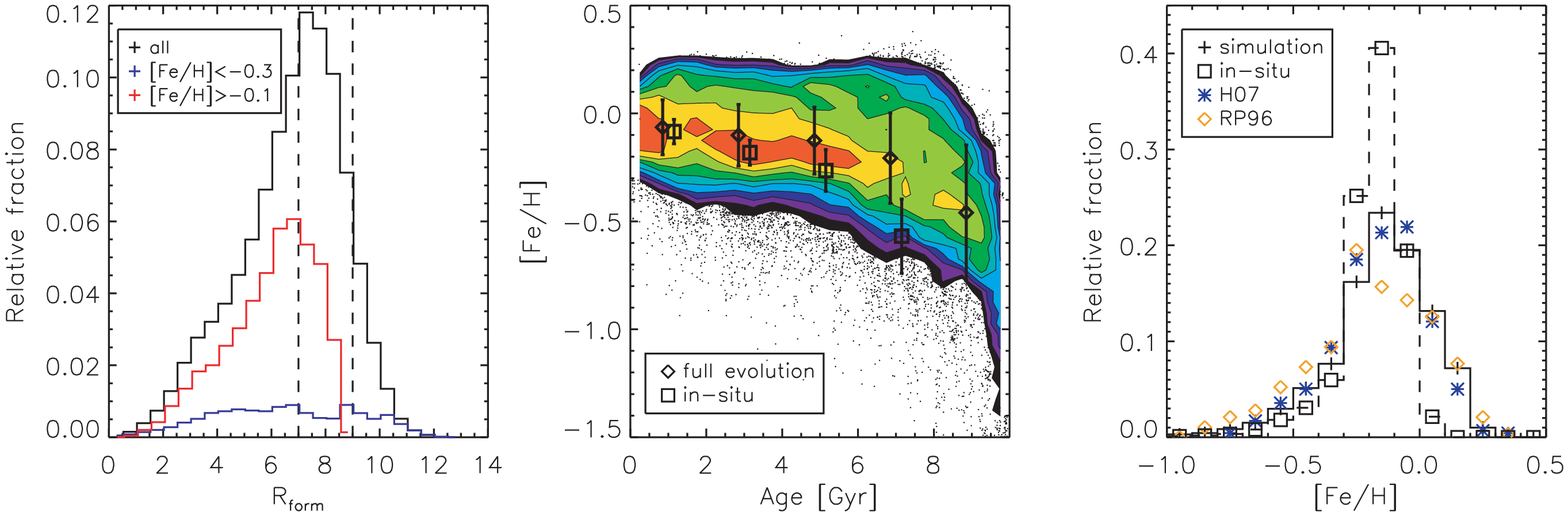}
\end{center}
\caption{The distribution of formation radii, $R_{\rm form}$ (in kpc), for stars in the Solar neighborhood. The black histogram shows the distribution from all stars while the red and blue histograms show the contributions from high and low metallicity stars respectively. Reproduced, with permission, from \protect\cite{roskar_beyond_2008}.}
\label{fig:RoskarMigration}
\end{figure}

The presence of perturbations in the disk can also affect its structure in less dramatic but still significant ways. (Once again, the work by \cite{sellwood_dynamics_2010} gives a more in depth review of these subjects and is highly recommended.) For example, numerical simulations by \citeauthor{roskar_beyond_2008}~(\citeyear{roskar_beyond_2008}; see also \citealt{roskar_ridingspiral_2008}) show that a significant fraction of stars in galaxy disks undergo large migrations in radius (as shown in Fig.~\ref{fig:RoskarMigration}) due to resonant scattering from spiral arms. The resonant scattering of a star initially on an approximately circular orbit tends to change its energy and angular momentum in such a way as to keep it in a circular orbit, but to change the radius of that orbit. In a frame corotating with the spiral perturbation there is an energy-like invariant quantity know as Jacobi's integral, $E_J$, defined as \pcite{binney_galactic_2008}
\begin{equation}
 E_J = E - \Omega_{\rm p} L,
\end{equation}
where $E$ and $L$ are the energy and angular momentum of an orbit respectively and $\Omega_{\rm p}$ is the pattern speed of the perturbation. This implies that changes in energy and angular momentum are related by
\begin{equation}
 \Delta E = \Omega_{\rm p} \Delta L.
\end{equation}
As shown in Fig.~\ref{fig:SellwoodLindblad} the slope of the $E$--$L$ relation for circular orbits at the corotation resonance is $\Omega_{\rm p}$. As such, stars initially on circular orbits near the corotation resonance will tend to be scattered into approximately circular orbits. \cite{roskar_beyond_2008} show that this phenomenon can lead to the formation of an outer stellar disk with a steeper exponential decline than the inner disk and formed entirely of stars scattered from the inner regions of the galaxy. This process also smooths out age and metallicity gradients in disks that otherwise form inside out and so tend to be older and less metal rich in the center.

\subsection{Spheroid Formation}

The formation of galactic spheroids (which we take here to mean both the bulges of disk galaxies and isolated elliptical galaxies) can proceed via two distinct routes. The first, through the destruction of pre-existing stellar systems in violent mergers, is a natural consequence of hierarchical galaxy formation. The second, secular evolution of galactic disks, is a natural consequence of the dynamics of self-gravitating disk systems. It is worth noting that, observationally, it has been proposed that bulges be divided into two broad classes \pcite{kormendy_secular_2004}: ``classical'' bulges (those which look like ellipticals in terms of their light distribution and kinematics but happen to live inside a disk) and ``pseudo-bulges'' (those which do not look like ellipticals\footnote{Typically, pseudo-bulges are better fit by a \cite{sersic_atlas_1968} profile with index $n\approx 1$ (as opposed to the $n=4$ profiles of classical bulges), have significant rotation and may show signs of disk phenomena such as bars or spiral features.}). There is good evidence (see \S\ref{sec:Mergers}) that ellipticals (and therefore presumably classical bulges also) form through major mergers, and suggestions that pseudo-bulges are the result of secular evolution of galactic disks.

\subsubsection{Major Mergers}\label{sec:MajorMergers}

As was discussed in \S\ref{sec:Mergers}, dissipative processes can lead to gravitationally bound interactions between galaxies. When the masses of the galaxies in question are comparable we may expect significant changes in their structure and the formation of a merger remnant which is very different from either of the merging galaxies. Such \emph{major mergers} are thought to be responsible for the formation of spheroidal (i.e. elliptical) galaxies from pre-existing galaxies (which could be, in principle, disk galaxies, other spheroidals or something intermediate).

The process of violent relaxation (\citealt{lynden-bell_statistical_1967}; see also \citealt{tremaine_h-functions_1986}), in which the energy of orbits undergoes order unity changes due to signficant time-variable fluctuations in the gravitational potential, leads to a randomization of the orbits leading to a Maxwellian distribution of energies but with temperature proportional to stellar mass. This can turn the ordered motions of disks into the random motions seen in spheroids. This process is, however, rather poorly understood---it seeks an equilibrium state which maximizes the entropy of the system, but the usual entropy is unbounded in gravitating systems implying no equilibrium state exists. \cite{arad_inconsistency_2005} demonstrate this problem by showing hysteresis effects in violently relaxed systems (i.e. the final state depends on how the system goes from the initial to final states).

The remnants of major mergers of purely stellar disk systems, while spheroidal, do not look like elliptical galaxies. As shown by \cite{hernquist_structure_1993} their phase space densities are too low in the central regions compared to observed ellipticals. This implies that mergers between reasonably gas rich (gas fractions of around 25\%) galaxies are required---the presence of gas allows for dissipation and the formation of higher phase-space density cores. \cite{robertson_fundamental_2006} find that similar gas fractions and the subsequent dissipation are required to produce the observed tilt in the fundamental plane of elliptical galaxies---mergers of purely stellar systems instead follow the expected virial scalings.

Mergers are often separated into \emph{major} (mergers between galaxies of comparable mass) and \emph{minor} (mergers in which one galaxy is significantly less massive than the other). Numerical simulations (e.g. \citealt{bournaud_galaxy_2005}) show that mergers with a mass ratio $\mu\equiv M_2/M_1 \gsim 0.25$ are able to destroy any disks in the ingoing galaxies and leave a spheroidal remnant, while mergers with lower mass ratio tend to leave disks in place (although somewhat thickened). As this transition is unlikely to be a sharp one, \cite{somerville_semi-analytic_2008} propose a more gradual transition\footnote{\protect\cite{somerville_semi-analytic_2008} give no justification for this particular functional form, it is merely intended to give a smooth but rapid transition.}, with a fraction
\begin{equation}
 f_{\rm sph} = 1 - \left[1+\left({\mu \over f_{\rm ellip}}\right)^8\right]^{-1},
\end{equation}
where $f_{\rm ellip}\approx 0.25$--$0.30$, of the disk stellar mass being put into the spheroid component of the merger remnant.

Mergers are expected to trigger a, possibly very large, enhancement in the star formation rate in the merging system. The strengths of such \emph{bursts} of star formation depend upon the properties of the merging galaxies. \cite{cox_effect_2008} have calibrated this enhancement against a large suite of N-body+hydrodynamics merger simulations. They define the ``burst efficiency'', $e$, to be the fraction of gas consumed during the interaction above that which would have been consumed by the constituent galaxies in isolation during the same time period. They find that the burst efficiency is well fit by\footnote{The equation in \protect\cite{cox_effect_2008} is missing the coefficient of $e_{1:1}$ (Cox, private communication).}
\begin{equation}
 e = \left\{ \begin{array}{ll}
              e_{1:1}\left({M_{\rm sat}\over M_{\rm primary}}-e_0\right)^\gamma & \hbox{if } M_{\rm sat}/M_{\rm primary} > e_0 \\
               0                   & \hbox{if } M_{\rm sat}/M_{\rm primary} \le e_0,
             \end{array} \right.
\end{equation}
where $e_{1:1}=0.56$, $\gamma=0.50$ and $e_0=0.09$.

It has recently become clear that not all major mergers lead to the formation of a spheroid. Under certain conditions, major mergers of very gas rich systems can lead to the reformation of a disk after the merger is over \pcite{barnes_formation_2002,springel_formation_2005,robertson_merger-driven_2006}. This requires a high gas fraction (greater than about 50\%) just prior to the final coalescence of the merging galaxies and therefore may preferentially occur under conditions which prevent the rapid depletion of gas after the first passage of the galaxies.

\subsubsection{Secular Evolution}\label{sec:Secular}

Major mergers are not the only way to form a spheroid. Internal, secular processes\footnote{Generically, any internal dynamical process operating on a timescale significantly longer than the dynamical time.} in galaxies can also disrupt the cold and relatively fragile disks \pcite{kormendy_secular_2004}. In particular, bars (a disk phenomenon) can efficiently redistribute mass and angular momentum and lead to the build-up of dense central mass concentrations, reminiscent in many ways of bulges formed through mergers. To distinguish such secularly formed bulges from their merger-formed (or ``classical'') counterparts, they are referred to as ``pseudo-bulges'' \pcite{kormendy_secular_2004}.

Such secular processes are the result of quite generic dynamical considerations\footnote{As pointed out by \protect\cite{kormendy_secular_2004} disks are fundamentally prone to spreading in the presence of any dissipative process, where mass is transported inwards and angular momentum outwards, because this lowers the energy of the system while conserving angular momentum \protect\pcite{lynden-bell_generating_1972}. This result can be traced back to the negative specific heat of gravitating systems, and is analogous to the process of core collapse in three dimensional systems \protect\pcite{lynden-bell_generating_1972,binney_galactic_2008}.} and so most likely operate in all galaxies. Whether or not they are important depends upon their timescale. For example, relaxation due to star-star encounters in a galaxy operates on a timescale many orders of magnitude longer than the age of the Universe and so can be safely neglected. Instead, most relevant secular processes involve the interaction of stars (or gas elements) with collective phenomena such as bars or spiral arms.

A general picture of how secular evolution leads to the formation of pseudo-bulges has emerged. As a bar spontaneously begins to form\footnote{The bar instability in galactic disks involves some fascinating dynamics. \protect\cite{binney_galactic_2008} give a clear explanation of the physics involved. Briefly, the bar instability involves the joint actions of the swing amplifier and a feedback mechanism. A randomly occurring leading spiral density wave in a disk will unwind and, as it does so, will rotate faster. As it swings from leading to trailing it reaches a maximum rotation speed which is close to the average orbital speed of stars in the disk. This leads to a resonance condition, in which the wave can strongly perturb the orbits of those stars, the self-gravity of which enhances the bar further, leading to an amplification of the wave. If there is some mechanism to convert the amplified trailing wave that results into a leading wave once more (e.g. if the wave can pass through the center of the galaxy and emerge as a leading wave, or if nonlinear couplings of waves can generate leading waves) the whole process can repeat and the wave will grow stronger and stronger.} it transfers angular momentum to the outer disk and increases in amplitude. The response of gas to this bar is crucial---gas accelerates as it enters the bar potential and decelerates as it leaves. This leads to shocks forming in the gas which lie approximately along the ridge line of the bar. These shocks lead to dissipation of orbital energy and, consequently, inflow of the gas. The enhancement in the gas density as it is concentrated towards the galactic center inevitably leads to star formation and the build-up of a pseudo-bulge. Bars eventually destroy themselves in this way---the increase in the central mass of the galaxy effectively prevents the bar instability from working. 

While this general picture seems to be incontrovertible, many of the details (e.g. the redshift evolution of bars, the importance of the dark matter halo and the cosmological evolution of the disk) remain poorly understood, but actively studied \pcite{friedli_secular_1993,shlosman_effects_1993,el-zant_dark_2002,athanassoula_bar-halo_2002,bournaud_gas_2002,shen_destruction_2004}.

\subsubsection{Sizes}

The sizes of spheroidal galaxies formed through major mergers can be determined given the properties of the progenitor galaxies and their orbit and some knowledge of the extent to which mass and energy is conserved through the merging process. If the galaxies are dissipationless, purely stellar systems (a so-called \emph{dry merger}) then energy is at least approximately conserved\footnote{In principle, energy may be lost from the stellar system by being transferred to the dark matter, or high energy stars may be flung out beyond the escape velocity, removing both mass and energy from the system.} In a gas rich (or \emph{wet}) merger the gaseous component is likely to radiate significant amounts of energy prior to forming stars.

A simple model to compute the size of a galaxy formed via a major merger was described by \cite{cole_hierarchical_2000}, who assumed perfect conservation of mass and energy. The size of the merger remnant, $r_{\rm f}$, is then given by
\begin{eqnarray}
 {c_f \G (M_1+M_2)^2 \over r_{\rm f}} &=& {c_1 \G M_1^2 \over r_1} + {c_2 \G M_2^2 \over r_2} \nonumber \\
 & & + {f_{\rm orbit} \G M_1 M_2 \over r_1+r_2}.
 \label{eq:Cole2000SpheroidSize}
\end{eqnarray}
The left-hand side represents the gravitational binding energy of the final system while the first two terms on the right represent the gravitational binding energy of the two merging galaxies, which have masses $M_1$ and $M_2$ respectively and half-mass radii $r_1$ and $r_2$ respectively. The ``$c$'' coefficients relate the actual binding energy to the characteristic value of $\G M^2/r$ and depend on the density distribution of the galaxy\footnote{In principle, the binding energy terms should also account for the stellar--dark matter binding energy. \protect\cite{cole_hierarchical_2000} ignore the dark matter contribution of the binding energy, effectively assuming that its contribution to each term in eqn.~(\ref{eq:Cole2000SpheroidSize}) scales in proportion to the stellar masses of the galaxies.}. \cite{cole_hierarchical_2000} find $c=0.45$ for a de Vaucouler's spheroid and $c=0.49$ for an exponential disk and so adopt $c=0.5$ for all galaxies (which may be composite systems consisting of both a disk and a spheroid) for simplicity. The final term in eqn.~(\ref{eq:Cole2000SpheroidSize}) represents the orbital energy of the two merging galaxies just prior to merging and is parameterized in terms of their gravitational binding energy when separated by the sum of the half-mass radii. \cite{cole_hierarchical_2000} adopt $f_{\rm orbit}=1$ so that this orbital term corresponds to a circular orbit. According to this approach, a merger of two identical galaxies of radii $r_1$ results in a remnant of size $r_{\rm f}=(4/3)r_1$.

More recently, \cite{covington_predictingproperties_2008} performed numerical simulations of merging galaxies including gaseous components, and found significant deviations from the simple model of \cite{cole_hierarchical_2000} as a result of energy loss due to radiative processes in merging galaxies containing gas. \cite{covington_predictingproperties_2008} propose an improved model which accounts for this energy loss and results in a merger remnant size given by
\begin{equation}
 E_{\rm int,f} = E_{\rm int,i} + E_{\rm rad} + E_{\rm orb},
\end{equation}
where
\begin{equation}
 E_{\rm int,f} = -{c \G M_{\rm f}^2 \over r_{\rm f}}
\end{equation}
is the internal binding energy of the final galaxy,
\begin{equation}
  E_{\rm int,i} = -{c \G M_1^2 \over r_1}-{c \G M_2^2 \over r_2}
\end{equation}
is the internal binding energy of the initial galaxies,
\begin{equation}
 E_{\rm orb} = -{\G M_1 M_2 \over r_{\rm sep}} + {1\over 2}M_1V_1^2 + {1\over 2}M_2v_2^2
\end{equation}
is the orbital energy of the galaxies just prior to the merger when their separation is $r_{\rm sep}$ and their center of mass velocities are $V_1$ and $V_2$, and
\begin{equation}
 E_{\rm rad} = -c_{\rm rad} \sum_{i=1}^2 K_i f_{{\rm g},i} f_{{\rm k},i} (1+f_{{\rm k},i})
\end{equation}
is the radiative energy loss where $K_i$ and $f_{{\rm g},i}$ are the initial internal kinetic energy and gas fraction for progenitor $i$. The impulse, $f_{{\rm k},i}$ is defined to be $f_{\rm k}=\Delta E/K_{\rm tot}$ where $K_{\rm tot}$ is the total initial internal kinetic energy of the galaxy and
\begin{equation}
 \Delta E = {A \G^2 M_{\rm 1,tot}^2 M_{\rm 2,tot} \over V^2_{\rm peri}(r_{\rm peri}^2 + B r_{\rm 1,tot}r_{\rm peri}+C r_{\rm 1,tot}^2)},
\end{equation}
where $M_{\rm 1,tot}$ and $M_{\rm 2,tot}$ are the total masses of the galaxies (baryonic plus dark matter), $r_{\rm 1,tot}$ is the total half-mass radius of the primary galaxy and $r_{\rm peri}$ is the pericentric distance of the first passage of the two galaxies, treating them as point masses from their initial orbit. \cite{covington_predictingproperties_2008} find that values $A=1.6$, $B=1.0$, $C=0.006$ and $c_{\rm rad}=1$ best fit their simulation results.

Similar arguments have been applied to the formation of spheroids through secular processes (e.g. \citealt{cole_hierarchical_2000}), but the assumptions of this approach are far less well tested in such cases.

\section{Star Formation, AGN and Feedback}

So far, we have not considered how a galaxy converts its available gas into stars and what effect those stars, and associated supernovae explosions, may have on the evolution of the galaxy. Additionally, observational evidence suggests that all galaxies contain a supermassive black hole at their center, which may play a key role in limiting the process of galaxy formation. We will discuss these aspects of galaxy formation theory below. We do not explore in detail the role of the associated galactic winds in galaxy evolution, but refer the reader to \cite{veilleux_galactic_2005} for a thorough review.

\subsection{Star Formation}

Star formation theory warrants an entire review to itself (for a recent example see \citealt{mckee_theory_2007}) and so we will summarize only those aspects most pertinent to galaxy formation theory. The past decade has seen a greatly improved understanding of how the key processes of turbulence, magnetic fields and self-gravity interact to form molecular clouds and stars. Nevertheless, there remain numerous unsolved problems in star formation theory. These problems propagate into galaxy formation theory if we wish to understand the rate at which stars form in a galaxy and any consequences that may have for further galactic evolution. This problem is somewhat mitigated by the fact that, for galaxy formation theory, we do not necessarily care about the details of how stars form. Instead, we would simply like to know, given the large scale properties of a galaxy (e.g. mass, size, density, dynamical time, gas fraction, chemical composition), what is the resulting rate of star formation.

Traditionally, this question has been answered by appealing to empirical rules or dimensional analysis. For example, much use has been made of the empirically derived scaling relations of \cite{schmidt_rate_1959} and \cite{kennicutt_star_1989,kennicutt_global_1998}. The Schmidt-Kennicutt law states that the rate of star formation per unit surface area, $\dot{\Sigma}_\star$, depends on the surface density of gas, $\Sigma_{\rm gas}$ as
\begin{equation}
 \dot{\Sigma}_\star \propto \Sigma_{\rm gas}^n,
\end{equation}
where $n$ is measured to be approximately $1.4$. While this relation is practically useful for galaxy formation theory, insofar as it allows one to bypass the question of star formation and move directly to a star formation rate, it suffers from the same issue as all empirical relations used in theoretical models: there is no way to be sure that it is valid beyond the regimes where it was originally measured.

Recently, new observations have suggested a relatively simple model for star formation \pcite{krumholz_slow_2007}. Stars form in molecular clouds and so it is natural that it would be the density of molecular gas (rather than total) gas which matters. Furthermore, it is observed that a constant fraction (around 2\%) of molecular gas turns into stars per free-fall timescale. This implies $\dot{\rho}_\star = \epsilon_{\rm ff} f_{\rm H_2} \rho_{\rm gas} / t_{\rm ff}$ where $\epsilon_{\rm ff}$ is the fraction of molecular gas turned into stars per free-fall time, $f_{\rm H_2}$ is the molecular fraction and $t_{\rm ff}$ is the local free-fall timescale.

In the model of \citeauthor{robertson_molecular_2008}~(\citeyear{robertson_molecular_2008}, see also \citealt{krumholz_star_2009}) the star formation rate is given by
\begin{equation}
 \dot{\Sigma}_\star \propto f_{\rm H_2} h_{\rm SFR} h_{\rm gas}^{-1.5} \Sigma_{\rm gas}^{1.5},
\end{equation}
where $f_{\rm H_2}$ is the fraction of hydrogen in molecular form, $h_{\rm SFR}$ is the scale height of the star-forming gas and $h_{\rm gas}$ is the scale height of the interstellar medium. \cite{robertson_molecular_2008} compute $f_{\rm H_2}$ and the scale height self-consistently in the presence of the interstellar radiation field produced by earlier generations of stars. This results in a net scaling with total gas surface density of
\begin{equation}
 \dot{\Sigma}_\star \propto \Sigma_{\rm gas}^{n_{\rm tot}},
\end{equation}
where $n_{\rm tot}\approx 2$ for massive galaxies and $n_{\rm tot}\gsim 4$ for dwarf galaxies. When expressed in terms of the molecular hydrogen gas surface density a scaling
\begin{equation}
 \dot{\Sigma}_\star \propto \Sigma_{\rm H_2}^{n_{\rm mol}},
\end{equation}
with $n_{\rm mol}\approx 1.3$ is found for all galaxies. These scalings are consistent with a broad range of observations of star formation in different galaxies.

Similarly, \cite{krumholz_star_2009} find that the following expression encapsulates the complex physics of star formation:
\begin{eqnarray}
\dot{\Sigma}_\star &=& f_{\rm H_2}(\Sigma_{\rm g},c,Z^\prime){\Sigma_{\rm g} \over 2.6\hbox{ Gyr}} \nonumber \\
 & & \times \left\{ \begin{array}{ll} 
\left({\Sigma_{\rm g}\over 85 M_\odot \hbox{pc}^{-2}}\right)^{-0.33}, & \hbox{if } {\Sigma_{\rm g}\over 85 M_\odot \hbox{pc}^{-2}} < 1 \\
\left({\Sigma_{\rm g}\over 85 M_\odot \hbox{pc}^{-2}}\right)^{0.33}, & \hbox{if } {\Sigma_{\rm g}\over 85 M_\odot \hbox{pc}^{-2}} > 1,
 \end{array} \right.
\end{eqnarray}
where the molecular hydrogen fraction is
\begin{equation}
f_{\rm H_2}(\Sigma_{\rm g},c,Z^\prime) \approx 1 - \left[ 1 + \left({3\over 4} {s\over 1+\delta} \right)^{-5} \right]^{-1/5},
\end{equation}
and $s=\ln(1+0.6\chi)/(0.04\Sigma_{\rm comp,0}Z^\prime)$, $\chi=0.77(1+3.1Z^{\prime0.365})$, $\delta=0.0712(0.1s^{-1}+0.675)^{-2.8}$, $\Sigma_{\rm comp,0}=\Sigma_{\rm comp}/(1M_\odot\hbox{pc}^{-2})$ and $Z^\prime$ is the metallicity normalized to the Solar value. The transition at gas surface densities of $85 M_\odot \hbox{pc}^{-2}$ corresponds to the point at which the ambient pressure becomes comparable to the internal pressure in molecular clouds (and therefore influences the properties of those molecular clouds). Here, $\Sigma_{\rm comp}$ is the surface density of a $\sim 100$ pc-sized atomic--molecular complex. Since simulations and semi-analytic models typically only predict the gas distribution on scales significantly larger than this we can write $\Sigma_{\rm comp}=c\Sigma_{\rm g}$ where $c$ is a clumping factor that accounts for structures which are unresolved in the simulation. This clumping factor should approach unity as the resolution approaches 100 pc at which point molecular cloud complexes should be adequately resolved.

\subsection{Black Hole Formation}

Over the past ten years it has become possible to measure the masses of supermassive black holes residing at the centers of galaxies for relatively large samples. The existence of strong correlations between the masses of these black holes and the properties of their host galaxy --- such as spheroid mass \pcite{magorrian_demography_1998}, velocity dispersion \pcite{ferrarese_fundamental_2000,gebhardt_relationship_2000,gebhardt_black_2000}, number of globular clusters \pcite{burkert_correlation_2010} or even host dark matter halo \pcite{ferrarese_beyondbulge:fundamental_2002} --- is suggestive of some interaction between forming galaxies and supermassive black holes. Of course, correlation does not imply causation \pcite{munroe_correlation_2009} and \cite{jahnke_non-causal_2010} show that a black hole--host galaxy mass relation can arise from uncorrelated initial conditions via simple merging, but the theoretical need for large amounts of energy to inhibit galaxy formation in massive halos naturally leads to the idea that \SMBHs\ and galaxy formation are connected \pcite{benson_what_2003}. In addition, understanding the formation of these most massive of black holes is interesting in its own right and has important observational consequences for both studies of active galactic nuclei and gravitational wave detection experiments such as the \emph{Laser Interferometer Space Antenna}\footnote{\href{http://lisa.nasa.gov/}{\tt http://lisa.nasa.gov/}}.

In light of these reasons, several studies have attempted to follow the process of \SMBH\ formation within forming galaxies \pcite{silk_quasars_1998,monaco_joint_2000,kauffmann_unified_2000,king_black_2003,wyithe_self-regulated_2003,matteo_black_2003,volonteri_assembly_2003,king_agn-starburst_2005,begelman_self-regulated_2005,malbon_black_2007,matteo_direct_2008,volonteri_evolution_2008,sijacki_growingfirst_2009,volonteri_journey_2009}. We will not review all details of this process here. Instead, we will focus on the basic formation mechanisms and, in \S\ref{sec:AGNFeedback}, on the interaction between \SMBHs\ and galaxy.

Before supermassive black holes can grow via accretion or merging, there must be some pre-existing (probably not supermassive) \emph{seed} black holes. Most plausibly, these seeds form at high redshifts as the remnants of the earliest generation of Population III stars which have reached the end of their stellar lifetimes. Details of the formation of these first stars remain incompletely understood, but hydrodynamical simulations suggest that they have masses in the range of a few hundred Solar masses \pcite{abel_formation_2002,bromm_formation_2002}, leaving intermediate mass black hole remnants.

To determine the rate at which gas accretes onto a black hole (or, more precisely, the black hole and any associated accretion disk system) we must consider how the black hole affects the gas through which it is moving. The gravity of a fast moving black hole will deflect gas that passes by it, focusing it into a wake behind the black hole which will then accrete onto the black hole. This problem, and the equivalent for a slowly (subsonically) moving black hole, was first studied by \cite{hoyle_effect_1939} and \cite{bondi_mechanism_1944}. This leads to an accretion rate of
\begin{equation}
 \dot{M}_{\rm BHL} = {4 \pi \G^2 M_\bullet^2 \rho \over (c_{\rm s}^2+v^2)^{3/2}},
\end{equation}
where $M_\bullet$ is the mass of the black hole, $c_{\rm s}$ is the sound speed and $v$ the relative velocity of black hole and gas. The accretion occurs from a characteristic radius of
\begin{equation}
 r_{\rm BHL} = {\G M_\bullet \over c_{\rm s}^2}.
\end{equation}
An in-depth review of Bondi-Hoyle-Lyttleton accretion is given by \cite{edgar_review_2004}. The growth of black holes will be enhanced by any process which increases the density in the central regions of the galaxy in which they reside. At early times, this may occur due to gravitationally unstable disks forming bars and driving gas towards the center \pcite{volonteri_evolution_2008}, while at later times galaxy-galaxy mergers can results in dissipation and gas flows to the center \pcite{matteo_energy_2005}. Other mechanisms for delivering gas to the centers of galaxies are also possible, for example \cite{mckernan_new_2010} show that warm clouds from the surrounding halo occasionally impact galactic centers, potentially delivering $10^4$--$10^6M_\odot$ of gas.

Bondi-Hoyle-Lyttleton accretion causes gas to flow towards the black hole. At some point, the angular momentum of the gas will become important and the accreting gas must form a disk. The final stage of accretion is then governed by this accretion disk, which may be a geometrically thin, radiative \pcite{shakura_black_1973} or a geometrically thick, radiatively inefficient (e.g. ADAF) flow \pcite{narayan_advection-dominated_1994}. The details of the flow may be important for determining the spin of the black hole (see below) and the effects of feedback from any nuclear activity (see \S\ref{sec:AGNFeedback}).

Galaxy-galaxy mergers can also lead to galaxies containing two (or potentially more) \SMBHs, resulting in the potential for black hole mergers. The process of bringing two \SMBHs\ together begins by dynamical friction against the background of dark matter (the same process which is causing the black hole host galaxies to merge). The subsequent merging process was originally outlined by \cite{begelman_massive_1980} and assumes that the two black holes form a binary system at an initial separation of
\begin{equation}
 a_{\rm b} = {\G (M_1+M_2) \over 2 \sigma^2},
\end{equation}
where $M_1>M_2$ are the masses of the black holes and $\sigma$ is the velocity dispersion of the host galaxy, such that the binary orbit contains a mass in stars and dark matter comparable to the sum of the black holes masses. Around this binary, the stellar distribution is expected to form a cusp with density profile $\rho_\star \propto r^{-7/4}$ \pcite{bahcall_star_1976}. Initially, the binary hardens due to dynamical friction against the stellar background acting on each black hole individually. As the binary hardens this process becomes less effective as perturbations from distant stars tend to perturb the center of mass of the binary without changing its semi-major axis. However, once the binary becomes sufficiently hard, at a separation of \pcite{quinlan_dynamical_1996}
\begin{equation}
 a_{\rm h} = {\G M_2 \over 4 \sigma^2},
\end{equation}
it can harden further by three-body interactions in which a passing star is captured and then ejected at high velocity. The timescale, $a/\dot{a}$, for hardening in this regime is \pcite{quinlan_dynamical_1996}
\begin{equation}
 t_{\rm h} = {\sigma \over \G \rho_\star a H},
\end{equation}
where the dimensionless hardening rate $H\approx 15$. If hardening continues long enough, the binary eventually becomes sufficiently hard that gravitational radiation dominates the evolution of the system which then coalesces on a timescale of \pcite{peters_gravitational_1964}
\begin{equation}
 t_{\rm GR} = {5 \clight^4 a^4 \over 256 \G^3 M_1M_2(M_1+M_2)}.
\end{equation}
The fly in the ointment of this neat picture is that the above estimates for the rate of hardening by stellar encounters assumes a fixed stellar background. In reality, as stars are ejected in three-body encounters the parts of phase-space containing stars that can be captured becomes depleted (so-called ``loss cone depletion''). This inevitably slows the hardening process. The past ten years have seen numerous studies of this process and examination of various mechanisms by which the loss cone may be refilled. For example, \cite{yu_evolution_2002} finds that in triaxial potentials scattering of stars can efficiently refill the loss cone while \cite{gould_binary_2000} suggest that the presence of a gaseous disk surrounding the black holes can help harden the binary. 
Numerical and analytical works have also indicated that the random walking of the binary center of mass may help mitigate loss cone depletion \pcite{quinlan_dynamical_1997,milosavljevic_formation_2001}, while the effective refilling of the loss cone by ejected stars returning on eccentric orbits \pcite{milosavljevic_long-term_2003,sesana_interaction_2007} and interactions with stars bound to the binary \pcite{sesana_interaction_2008} may enhance the rate of hardening. While the details remain uncertain it seems that this basic process can lead to black holes merging in less than 10~Gyr.

In addition to their mass, cosmological black holes are characterized by one other parameter, their angular momentum\footnote{They will not possess any significant charge, as this would be quickly neutralized by accretion of oppositely charged material.}. The spin of a black hole can have a strong influence on the radiative efficiency and jet power of black holes (see \S\ref{sec:AGNFeedback}) and so the cosmological evolution of this quantity is important to understand. There are fundamentally two mechanisms which change the spin of a black hole: merging with another hole and accretion of material.

The outcomes of binary black hole mergers have proven very difficult to simulate numerically. However, recent advances in numerical techniques have allowed for successful simulation of the entire merging process (e.g. \citealt{tichy_binary_2007}) and, therefore, determination of the final spin of the merger product. While the number of simulations carried out to date is small, \cite{boyle_binary_2008} exploit symmetries of the problem to construct simple fitting formula which accurately predict the spin of the final black hole as a function of the incoming black hole masses, spins and orbital properties.

As first considered by \cite{bardeen_kerr_1970}, material accreted from an accretion disk carries with it some angular momentum (approximately equal to its angular momentum at the latest stable circular orbit, before it began its plunge into the black hole). Defining a dimensionless spin parameter for a black hole through
\begin{equation}
 j=J\clight/\G M_\bullet^2,
\end{equation}
where $J$ is the angular momentum of the hole of mass $M_\bullet$, such that $0\le j\le 1$, then \cite{shapiro_spin_2005} defines a dimensionless spin-up function $s(j)$ by
\begin{equation}
s(j) = {\d j\over \d t} {M_\bullet \over \dot{M}_{\bullet,0}},
\label{eqn:spinup}
\end{equation}
where $\dot{M}_{\bullet,0}$ is the rate of rest mass accretion. For a standard, relativistic, Keplerian thin disk accretion flow with no magnetic fields, the expected spin-up function due to accretion is given by \pcite{shapiro_spin_2005}
\begin{equation}
s(j) = {\mathcal L}_{\rm ISCO}-2 j E_{\rm ISCO},
\end{equation}
where
\begin{equation}
{\mathcal L}_{\rm ISCO}(j) = {\sqrt{r_{\rm ISCO}} (r_{\rm ISCO}^2-2j\sqrt{r_{\rm ISCO}}+j^2) \over r_{\rm ISCO} \sqrt{r_{\rm ISCO}^2-3r_{\rm ISCO}+2j\sqrt{r_{\rm ISCO}}}},
\end{equation}
and
\begin{equation}
E_{\rm ISCO}(j) = {r_{\rm ISCO}^2-2r_{\rm ISCO}+j\sqrt{r_{\rm ISCO}} \over r_{\rm ISCO} \sqrt{r_{\rm ISCO}^2-3r_{\rm ISCO}+2j\sqrt{r_{\rm ISCO}}}},
\end{equation}
are the (dimensionless) specific angular momentum and specific energy of the innermost stable circular orbit (ISCO) of the black hole respectively. 
The radius of the ISCO orbit (in units of the gravitational radius $\G M_\bullet/\clight^2$) is
\begin{eqnarray}
r_{\rm ISCO} &=& -\sqrt{[3-A_1(j)][3+A_1(j)+2A_2(j)]}\nonumber \\
 & & +3+A_2(j)
\label{eq:isco}
\end{eqnarray}
where
\begin{eqnarray}
A_1(j) & = & 1+[(1-j^2)^{1/3}] \nonumber \\
 & & \times [(1+j)^{1/3}+(1-j)^{1/3}], \\
A_2(j) & = & \sqrt{3j^2+A_1(j)^2}.
\end{eqnarray}
\cite{thorne_disk-accretion_1974} found a small correction to this formula due to the fact that the hole preferentially swallows negative angular momentum photons resulting in a spin-up function for the standard thin disk that is positive for all $j<0.998$, and therefore lets the black hole spin up to $j\approx 0.998$ in finite time as noted by \cite{shapiro_spin_2005}. For a thick accretion flow the result is somewhat different (since the flow is no longer supported against gravity by its rotation alone as it has significant thermal and magnetic pressure).  \cite{benson_maximum_2009} compute the spin-up function for ADAF models. Additionally, \cite{benson_maximum_2009} compute how the magnetic torques which allow black holes to drive jets (via the Blandford-Znajek and/or Blandford-Payne mechanisms; \citealt{blandford_electromagnetic_1977,blandford_hydromagnetic_1982}) result in a braking torque on the black hole, spinning it down and resulting in an equilibrium spin of $j\approx 0.93$ for a hole accreting from an ADAF.

The relative importance of mergers and accretion for determining the spins of cosmological black holes depends upon the rate of galaxy mergers, the supply of gas to the black hole and additional factors such as the alignment of accretion disks and black hole spins and merging black hole spins and orbits. Many of these factors are not too well understood. However, cosmological calculations (\citealt{berti_cosmological_2008}; see also \citealt{volonteri_distribution_2005}) suggest that accretion dominates over mergers in terms of determining the spins of supermassive black holes, with the consequence that most such holes are predicted to be rapidly spinning.

One additional consequence of black hole mergers is that the gravitational waves emitted during the final inspiral carry away linear momentum resulting in the black hole recoiling in the opposite direction with a velocity potentially large enough to unbind it from the galaxy \pcite{fitchett_influence_1983,favata_black_2004,merritt_consequences_2004,blanchet_gravitational_2005}. Studies of this effect within a cosmological framework suggest that such ejected black holes (wandering through intergalactic space) can make up 2--3\% of the total mass density of supermassive black holes, while in individual cases they can account for up to half of the total black hole mass associated with a galaxy \pcite{libeskind_effect_2006}.

\subsection{Feedback}

As early as 1974 (\citealt{larson_effects_1974}; see also \citealt{white_core_1978}; \citealt{dekel_origin_1986}) it was realized that star formation could not proceed with 100\% efficiency in all dark matter halos. Evidence for this comes from multiple observed facts, but the two crucial ones are:
\begin{enumerate}
 \item The total mass density in stars is $\Omega_\star = (2.3 \pm 0.34)\times 10^{-3}$ \pcite{cole_2df_2001}, much less than the total baryonic mass density of the Universe $\Omega_{\rm b}=0.0462\pm 0.0015$ \pcite{dunkley_five-year_2009}. Therefore, only a small fraction of all baryons have been able to turn into stars.
 \item The distribution of galaxy luminosities (as described by the luminosity function) is very different from the distribution of dark matter halo masses. In particular there are many fewer faint galaxies relative to bright galaxies than there are low mass to high mass dark matter halos. If each halo contained baryons at the universal mix and turned all of them into stars we would expect these two ratios to be equal \pcite{benson_what_2003}.
\end{enumerate}
This second point can be made even stronger: if all dark matter halos turned a constant fraction of their mass into stars we would still have too many faint relative to bright galaxies. There are a wealth of other observational constraints which indicate that the efficiency of galaxy formation must depend strongly upon halo mass---a useful summary of these constraints is given by \cite{behroozi_comprehensive_2010}.

Clearly what is needed is some process which preferentially suppresses the formation of stars in lower mass dark matter halos. The usual suspect for this process is energy/momentum input from supernovae explosions, perhaps augmented by stellar winds.

A similar problem occurs in the most massive dark matter halos. Although cooling is relatively inefficient in such halos they can nevertheless cool significant mass of gas over cosmic time. Unchecked, this leads to the formation of galaxies significantly more luminous than any that are observed. The energetic requirements of this problem suggest that \AGN\ are a possible solution.

\subsubsection{Supernovae/Stellar Winds}\label{sec:SNeFeedback}

\begin{figure*}
 \begin{center}
 \begin{tabular}{cc}
   \includegraphics[width=65mm]{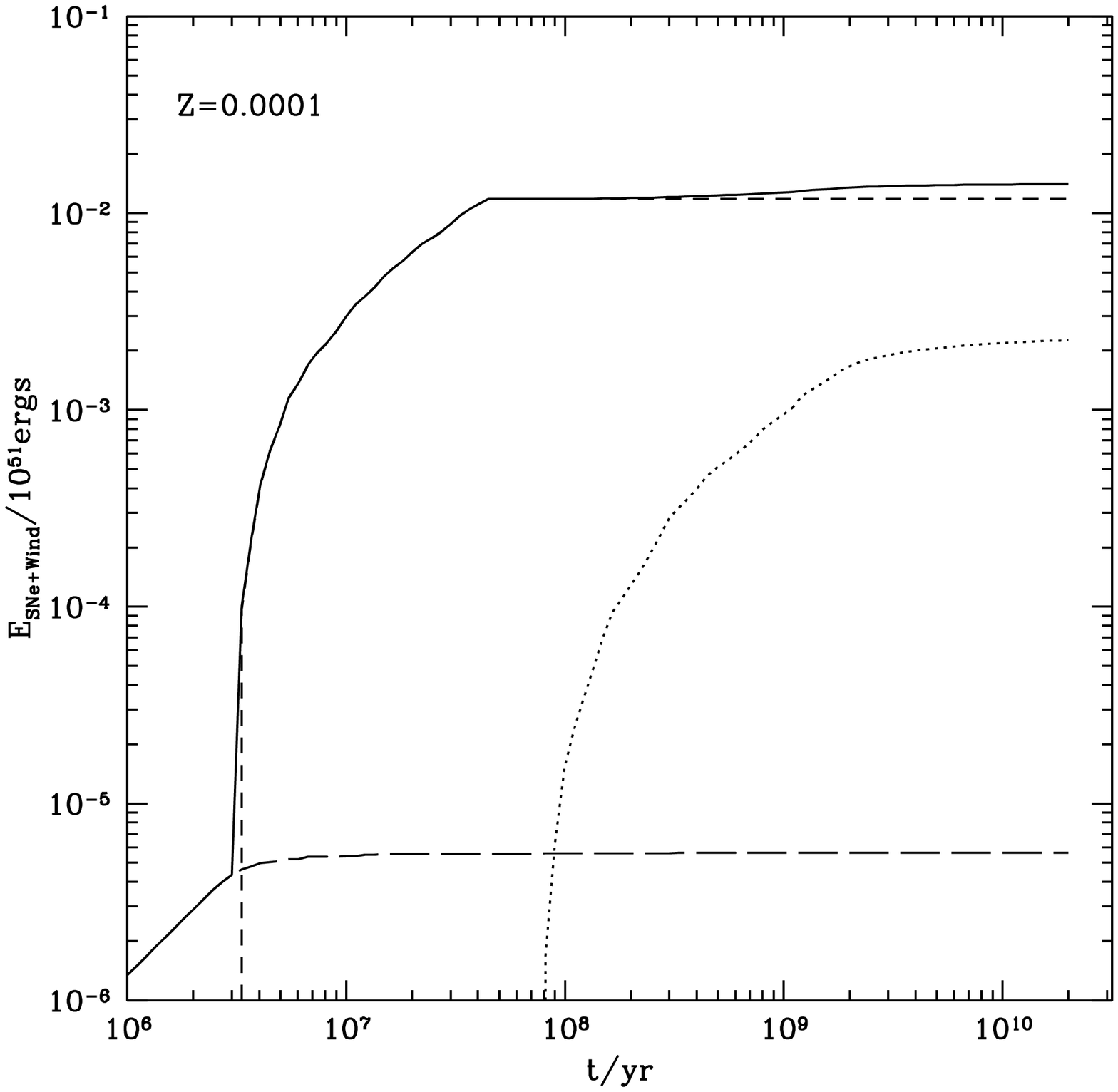} & \includegraphics[width=65mm]{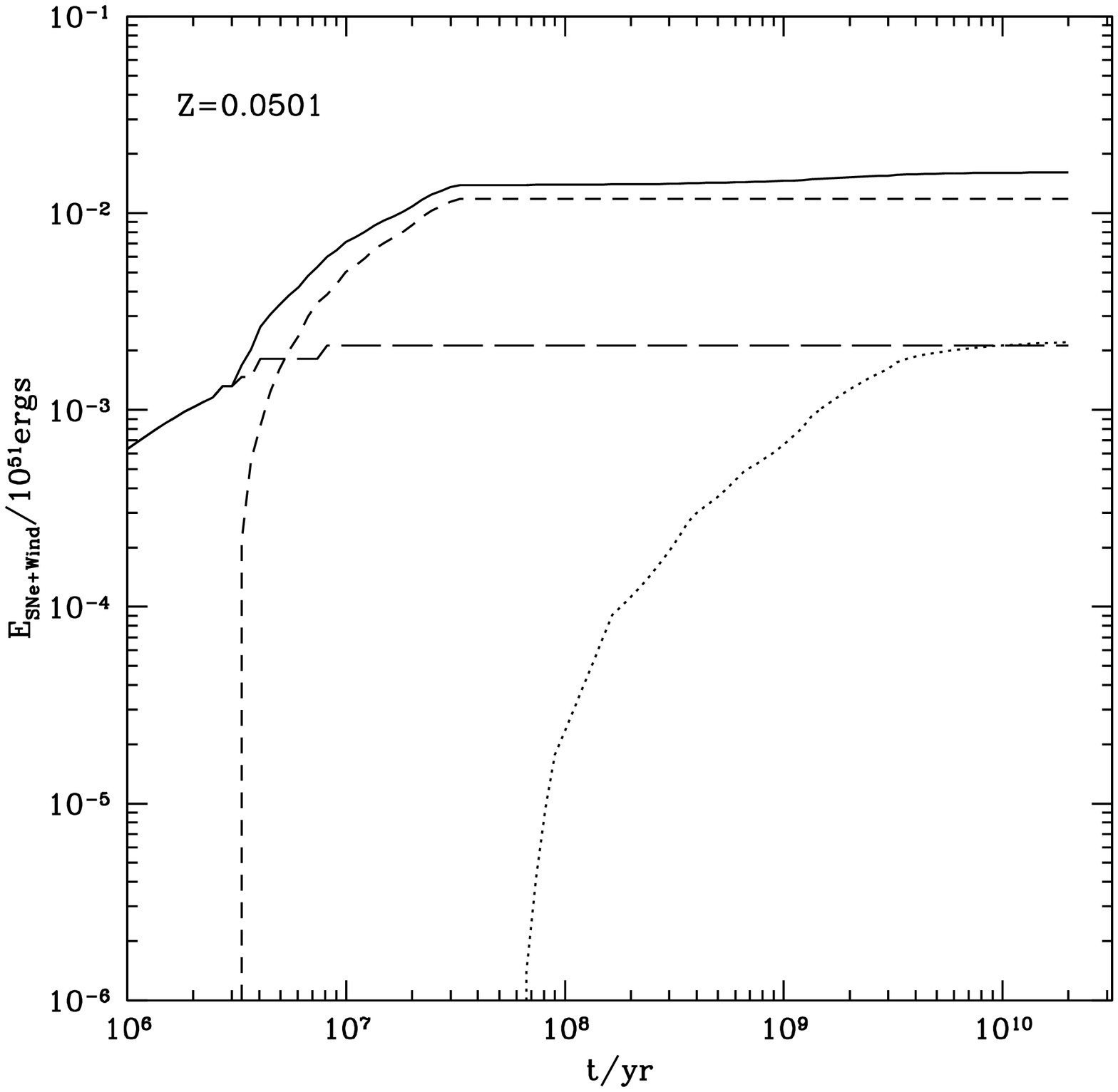}
 \end{tabular}
 \end{center}
 \caption{Cumulative energy input into the interstellar medium, expressed as the number of equivalent supernovae, per unit mass of stars formed as a function of time for a Chabrier initial mass function. The two panels show results for two different metallicities as indicated in the panels. The long dashed line indicates the contribution from stellar winds, the dotted line the contribution from Type~Ia supernovae and the short dashed line the contribution from Type~II supernovae. The solid line shows the sum total of all contributions.}
 \label{fig:SNeInput}
\end{figure*}

Energy input from SNe due to a single stellar population are shown in Fig.~\ref{fig:SNeInput}. If we assume that a fraction $\epsilon$ of this energy is coupled into an outflow of \ISM\ gas which leaves the galaxy at the escape velocity (so that we do not waste energy by giving the gas a velocity at infinity) this results in an outflow rate $\dot{M}_{\rm out}$ given by
\begin{eqnarray}
 {1 \over 2} \dot{M}_{\rm out} V_{\rm esc}^2 &=& \epsilon \int_0^t \dot{M}_\star(t^\prime) \dot{E}_{\rm SNe+winds}(t-t^\prime)\d t^\prime \nonumber \\
&\approx& \dot{M}_\star(t) E_{\rm SN+winds},
\end{eqnarray}
where in the last step we have approximated energy input from SNe and winds as occurring instantaneously after star formation. This implies an outflow rate of
\begin{equation}
 \dot{M}_{\rm out} = \dot{M}_\star(t) {2 \epsilon E_{\rm SN+winds} \over V_{\rm esc}^2}.
\end{equation}
This has the required features: given some initial mass of cold gas, a fraction $1/(1+\beta)$ where $\beta\equiv2 \epsilon E_{\rm SN+winds} / V_{\rm esc}^2$ will be turned into stars, resulting in much lower star formation efficiencies in low mass galaxies.

Figure~\ref{fig:SNeInput} shows the cumulative energy input\footnote{We assume that a Type~II supernovae releases $10^{51}$ ergs of usable energy. We express the energy input in units of $10^{51}$ergs so that it can be interpreted as the equivalent number of supernovae.} into the interstellar medium as a function of time from a $1M_\odot$ burst of star formation with a \cite{chabrier_galactic_2003} initial mass function.

Similar arguments can be made using the momentum of supernovae explosions to drive the wind, which may be more relevant if energy is efficiently radiated from expanding SNe bubbles\footnote{The momentum, of course, cannot be radiated away.} \pcite{murray_maximum_2005}. The rate of momentum injection from supernovae is given by \pcite{murray_maximum_2005}
\begin{equation}
 \dot{P}_{\rm SN} \approx 2\times 10^{28} \left( {\dot{M}_\star \over 1 M_\odot \hbox{ yr}^{-1}} \right) \hbox{ kg m s}^{-2}.
\end{equation}
Since the momentum in any wind driven by this momentum input must satisfy $\dot{P}_{\rm W} = \dot{M}_{\rm out} V_\infty$, where $V_\infty$ is the wind velocity at infinity we have that
\begin{equation}
 \dot{M}_{\rm out} \approx {\dot{P}_{\rm SN} \over V_{\rm esc}},
\end{equation}
where we have assumed that $V_\infty \approx V_{\rm esc}$ \pcite{murray_maximum_2005}. This gives a different scaling with the galaxy escape velocity compared to energy driven winds.

The details of how supernovae feedback actually operates remain somewhat unclear, and is complicated by the necessity to understand how an expanding supernovae remnant interacts with a complex, multiphase interstellar medium. Our discussion below will mostly follow \citeauthor{efstathiou_model_2000}~(\citeyear{efstathiou_model_2000}; see also \citealt{mckee_theory_1977}).

The basic picture considered involves a two phase interstellar medium consisting of hot, diffuse material (which fills most of the volume) and cold, dense clouds (which contain most of the mass). Evaporation of cold clouds by supernovae moves gas from the cold phase into the hot phase. Outflows, and suppression of star formation, from the hot phase will result if the hot phase:
\begin{enumerate}
 \item fills most of the volume;
 \item is sufficiently low density that radiative cooling times are long, and;
 \item its temperature exceeds the virial temperature of galaxy (so that the gas may climb out of the potential).
\end{enumerate}

The effectiveness of feedback in such a model can be examined by considering the mass of cold clouds swept up and evaporated by an expanding supernovae remnant, $M_{\rm ev}$. \cite{efstathiou_model_2000} finds
\begin{equation}
 M_{\rm ev} \approx 1390 E_{51}^{6/5} f_\Sigma^{-3/5} \phi_\kappa^{3/5} n_{\rm h -2}^{-4/5} M_\odot,
\end{equation}
where $E_{51}$ is the energy released by a supernovae in units of $10^{51}$ ergs, $n_{\rm h -2}$ is the number density of hydrogen in units of $10^{-2}$~cm$^{-3}$,  and the evaporation parameter as defined by \cite{mckee_theory_1977} is
\begin{equation}
 \Sigma = {\gamma \over 4 \pi a_1 N_{\rm cl} \phi_\kappa} \equiv f_\Sigma \Sigma_\odot,
\end{equation}
where $\gamma = v_{\rm h}/c_{\rm h}\approx 2.5$ relates the blast wave velocity, $v_{\rm h}$, to the isothermal sound speed, $c_{\rm h}$, $a_1$ is a typical cloud radius, $N_{\rm cl}$ is the number of clouds per unit volume, $\phi_\kappa$ relates the effective conductivity, $\kappa_{\rm eff}$, to the classical thermal conductivity, $\kappa$, via $\kappa_{\rm eff}=\phi_\kappa \kappa$ and $\Sigma_\odot\approx 95$ pc$^2$ is the value of the evaporation parameter in the Solar neighborhood. (Units are chosen so that all variables have values of order unity in a typical galaxy.)

Since $M_{\rm ev}$ is significantly larger than the mass of stars formed per supernovae then we may expect that if a significant fraction of this evaporated gas can escape as a wind then star formation may be efficiently suppressed.

Gas in the hot phase will escape the galaxy if its specific enthalpy
\begin{equation}
 {1 \over 2} v^2 + {5 \over 2} {p \over \rho}
\end{equation}
exceeds its specific gravitational binding energy. This therefore requires that we determine the thermodynamic properties of the hot phase. \cite{efstathiou_model_2000} does this by assuming that supernovae remnants have a porosity of unity (i.e. the total volume of remnants equals the volume of the galaxy, thereby creating an overlapping network of remnants). Given a model for the expanding supernovae remnant this allows the thermodynamic properties of the hot phase to be determined.

\cite{efstathiou_model_2000} shows that such a model of feedback can indeed produce self-regulated star formation in a galaxy disk that is quiescently forming stars (bursts are not required) and that both negative and positive (due to pressure-enhanced star formation) feedback can work simultaneously.

\subsubsection{AGN}\label{sec:AGNFeedback}

Very simple energetic arguments suggest that the energy output involved in building a black hole in a galaxy can have significant effects on the formation of the galaxy itself \pcite{benson_what_2003}. To order of magnitude, let us assume that each galaxy contains a supermassive black hole with a mass equal to 0.1\% of its stellar mass. The energy released by the formation of the black hole per unit mass of stars formed is then
\begin{equation}
 {\mathcal E}_\bullet = 10^{-3} {\epsilon \over 1-\epsilon} \clight^2,
\end{equation}
where $\epsilon$ is the efficiency of conversion of rest mass into energy output (either in radiation or mechanical outflow). For typical values of $\epsilon \sim 0.1$ this implies ${\mathcal E}=1.8\times 10^{50}\hbox{ ergs }M_\odot^{-1}$, which is an order of magnitude greater than that released by supernova explosions and stellar winds and is comparable to the energy released by cooling gas in forming some of the most massive galaxies.

Feedback from \AGN\ has the potential to directly link the properties of supermassive black holes and their host galaxies. As such, it may be a natural explanation for the observed correlations between supermassive black hole mass and galaxy mass or velocity dispersion. The first study of how \AGN\ feedback can lead to such correlations\footnote{Once again, assuming that any causative process is needed at all \protect\pcite{jahnke_non-causal_2010}.} was described by \cite{silk_quasars_1998}. They showed that, on quite general grounds, coupling of the energy released by the formation of the supermassive black hole to the surrounding forming galaxy would lead to a relationship close to the observed $M_\bullet$--$\sigma$ relation. The generality of these arguments imply that this result may be reasonably independent of the details of any more specific feedback model.

The physical mechanism through which \AGN\ feedback operates remains somewhat unclear. \AGN\ likely radiatively heat cooling gas in the atmosphere surrounding galaxies, reducing the rate at which that gas can cool. Additionally, radiatively driven winds originating from the broad line region surrounding the black hole may result in mechanical feedback on the galaxy itself. Finally, in low accretion states, \AGN\ may drive highly collimated and powerful jets which can reach out well into the surrounding halo. Plausibly all of these mechanisms could be active during galaxy formation. 

\cite{ciotti_feedbackcentral_2009} examined the roles of radiative and mechanical feedback modes. They find that radiative feedback alone, while able to forestall the ``cooling catastrophe'' (see \S\ref{sec:Overcooling}), is unable to sufficiently limit the growth of black holes, resulting in them being too massive for a given galaxy (by a factor of around four). Mechanical feedback is found to be successful in limiting the growth of black holes. This is in agreement with other numerical simulations, such as \citealt{springel_modelling_2005,matteo_energy_2005,sijacki_unified_2007,sijacki_growingfirst_2009,booth_cosmological_2009}, which incorporate energy injection from quasars (typically triggered by major mergers) and show that \AGN\ activity can effectively expel gas from a galaxy and can establish the observed correlations between supermassive black hole and galaxy properties. However, \cite{ciotti_feedbackcentral_2009} find that mechanical feedback of high efficiency (as is required to match the scaling relations) is too effective, depleting galaxies of gas to a degree greater than is observed, while low efficiency mechanical feedback allows too much star formation to occur at late times, resulting in blue cores in elliptical galaxies that are inconsistent with observations. Precisely how efficiently jets from an \AGN\ can couple their energy to the surrounding hot gas remains a topic of intense study. The general picture that is emerging is that the jets inflate bubbles or cavities in the hot atmosphere. For example, \cite{omma_heating_2004} use numerical simulations to show that jets inflate cavities in the hot gas that can excite g modes in the cluster atmosphere and heat the gas, while \cite{roychowdhury_entropy_2004} explore a model in which the bubbles are buoyant and heat the atmosphere by doing $p\d V$ work as they rise.

Recent years have seen numerous implementations of detailed models of galaxy formation incorporating \AGN\ feedback \pcite{scannapieco_quasar_2004,croton_many_2006,bower_breakinghierarchy_2006,cattaneo_modellinggalaxy_2006,somerville_semi-analytic_2008}. These have shown that \AGN\ feedback can explain the long standing issue of the exponential break in the galaxy luminosity function (without \AGN\ feedback such models tend to produce far too many bright galaxies; \citealt{benson_what_2003}) and helps produce a bimodal distribution of galaxy colors (as \AGN\ feedback makes massive galaxies ``red and dead''). These models typically assume a ``radio mode'' feedback in which the \AGN\ drives jets out of the galaxy while in a radiatively quiet state. \cite{benson_maximum_2009} give an expression for the jet power produced by a spinning black hole accreting from a geometrically thick accretion flow\footnote{The jet power from a hole accreting from a thin accretion disk is much lower, due to the much weaker magnetic field in such a disk.} relevant to this situation. As a result this type of \AGN\ feedback is effective only when there is a quasi-hydrostatic atmosphere of hot gas in a dark matter halo for it to couple to. Therefore, \AGN\ feedback only works in the virial shocking regime, resulting in it being important above a critical halo mass of order $10^{12}M_\odot$ (with a weak redshift dependence).

\subsection{Chemical Enrichment}\label{sec:ChemicalEvol}

The first generation of stars (known as Population~III) and galaxies must have formed from primordial gas which is (almost) metal free. Stellar nucleosynthesis and subsequent pollution of the \ISM\ and IGM (through stellar winds and supernovae explosions) with heavy elements has a significant impact on later generations of galaxies. In particular, the presence of heavy elements significantly alters the rate at which gas can cool (see \S\ref{sec:Cooling}) and leads to the formation of dust which both attenuates optical and UV light from galaxies and re-emits that light at longer wavelengths. To accurately model the properties of galaxies therefore requires a treatment of chemical enrichment.

\begin{figure}
 \begin{center}
  \includegraphics[width=80mm]{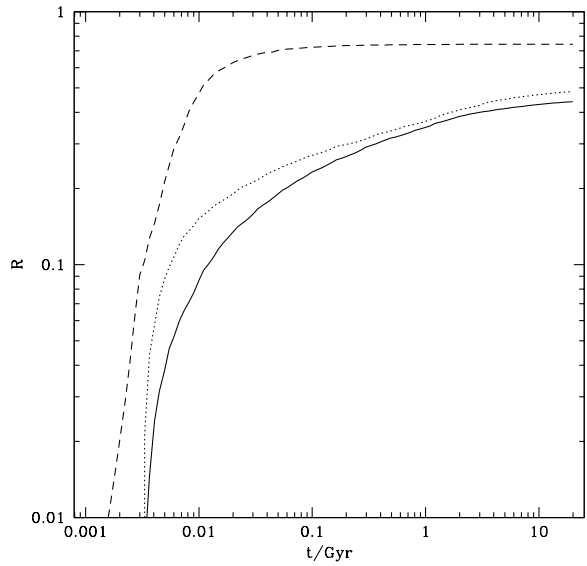}
 \end{center}
 \caption{The fraction of mass from a single stellar population born at time, $t=0$, recycled to the interstellar medium after time $t$. Solid and dotted lines show results for a Chabrier initial mass function with zero and Solar metallicity respectively, while the dashed line shows results for zero metallicity (Pop.~III) stars with a lognormal IMF (case A from \protect\citealt{tumlinson_chemical_2006}).}
 \label{fig:Recycling}
\end{figure}

The fraction of material returned to the \ISM\ by a stellar population as a function of time is given by
\begin{equation}
 R(t) = \int_{M(t;Z)}^{M_{\rm u}} [M-M_{\rm r}(M;Z)]\phi(M) {\d M \over M}
\end{equation}
where $\phi(M)$ is the initial mass function normalized to unit stellar mass and $M_{\rm r}(M)$ is the remnant mass of a star of initial mass $M$. Here, $M(t)$ is the mass of a star with lifetime $t$. Figure~\ref{fig:Recycling} shows the fraction of material recycled to the interstellar medium from a single stellar population as a function of time, $t$, since the birth of that population. Typical Pop.~II initial mass functions (e.g. \citealt{salpeter_luminosity_1955,chabrier_galactic_2003}) lead to around 40\% of mass being recycled after 10~Gyr, dropping to around 30\% at 1~Gyr. A Pop.~III initial mass function recycles much more mass (since it consists of much more massive stars), recycling 70\% of the mass within a few tens of millions of years.

\begin{figure}
 \begin{center}
  \includegraphics[width=80mm]{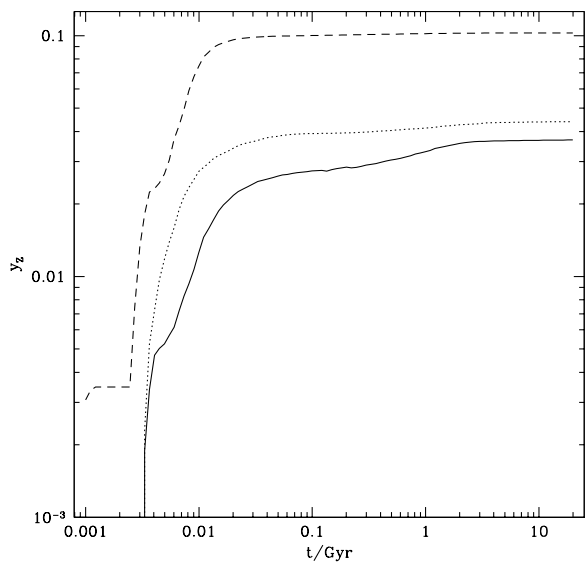}
 \end{center}
 \caption{The total metal yield from a single stellar population born at time, $t=0$, after time $t$. Solid and dotted lines show results for a Chabrier initial mass function with zero and Solar metallicity respectively, while the dashed line shows results for zero metallicity (Pop.~III) stars with a lognormal IMF (case A from \protect\citealt{tumlinson_chemical_2006}).}
 \label{fig:Yields}
\end{figure}

Similarly, the yield of element $i$ is given by
\begin{equation}
 y_i(t) = \int_{M(t;Z)}^{M_{\rm u}} M_i(M_0;Z)\phi(M_0) {\d M_0\over M_0}
\end{equation}
where $M_i(M_0;Z)$ is the mass of metals produced by stars of initial mass $M_0$. For a specified IMF we can compute $y_i(t;Z)$ for all times and elements of interest. Figure~\ref{fig:Yields} shows examples of the total metal yield for Pop.~II and Pop.~III initial mass functions. Typical total metal yields for Pop.~II are around $0.04$ increasing to $0.1$ for the Pop.~III initial mass function shown here. Stellar data are taken from \citet{portinari_galactic_1998} for low and intermediate mass stars and \citet{marigo_chemical_2001} for high mass stars.

In many cases, the process of chemical enrichment has been simplified by adopting the instantaneous recycling approximation, in which mass and metals are assumed to be returned to the interstellar medium instantaneously after the formation of a population of stars (essentially assuming that stellar evolution happens instantaneously). Typically, the recycled fractions and yields are evaluated at some time, $t$, of order the age of the Universe. This approximation is reasonable at late times, when the ages of typical stellar populations in galaxies is several Gyr. For example, Fig.~\ref{fig:Yields} shows that the metal yield is approximately constant after about $0.1$~Gyr, although the recycled fraction (see Fig.~\ref{fig:Recycling}) does not converge so rapidly. The advantage of the instantaneous recycling approximation is that it greatly simplifies the equations governing chemical enrichment such that we have
\begin{equation}
 R(t) = \int_{M(t_{\rm ira};Z)}^{M_{\rm u}} [M-M_{\rm r}(M;Z)]\phi(M) {\d M \over M} \equiv R
\end{equation}
with a similar result for yields. The normal convolution integral to determine the rate of return of mass to the interstellar medium,
\begin{equation}
 \dot{M}_{\rm R}(t) = \int_0^t \dot{R}(t-t^\prime) \dot{M}_\star(t^\prime) \d t^\prime,
\end{equation}
then simplifies to
\begin{equation}
 \dot{M}_{\rm R}(t) = R \dot{M}_\star(t).
\end{equation}
This removes the need for knowledge of the entire star formation history of a galaxy to determine the rate of recycling (which can greatly reduce computational demand in numerical models of galaxy formation).

While the instantaneous recycling approximation has its advantages, it is, of course, an approximation and one which breaks down in regimes of rapid star formation and at high redshifts where stars are much less than 10~Gyr old (since the Universe itself is very young). Detailed models of chemical evolution are well established within the monolithic collapse scenario \pcite{matteucci_relative_1986,matteucci_chemical_1995,timmes_galactic_1995,franccois_evolution_2004,pipino_photochemical_2004,pipino_photochemical_2006,romano_quantifyinguncertainties_2005}, but have only received limited study within the physically motivated context of hierarchical galaxy formation. Numerous hydrodynamical simulation codes now contain implementations of chemical enrichment \pcite{scannapieco_feedback_2005,kobayashi_simulations_2007,pontzen_damped_2008,gnedin_modeling_2009}. Semi-analytic models of galaxy formation are beginning\footnote{Implementation of chemical enrichment models of this type is, in many ways, significantly easier in N-body simulations than in semi-analytic models. N-body simulations that include star formation typically spawn new star particles during each timestep to represent the stars formed at that time. Each particle can therefore be trivially tagged with a formation time and metallicity. On subsequent timesteps the rate of mass and metal return from that particle is easily computed and applied to surrounding gas particles for example. Semi-analytic models on the other hand have traditionally not recorded the full star formation history of each galaxy (due to computational resource requirements) as is needed to compute the chemical enrichment at each timestep. This can, of course, be done, as we are now seeing.} to implement detailed chemical enrichment models also \pcite{nagashima_metal_2005,nagashima_metal_2005-1,pipino_galics_2008,arrigoni_galactic_2009}. Within the context of hierarchical structure formation models such chemical enrichment modeling has been shown to be in agreement with measurements of the intracluster medium metallicities of individual elements \pcite{nagashima_metal_2005}, to reproduce the observed trend of metallicity and [$\alpha$/Fe] with stellar mass in elliptical galaxies \pcite{arrigoni_galactic_2009} and to reproduce the measured distribution of metallicities in damped Lyman-$\alpha$ systems \pcite{pontzen_damped_2008}.

\subsection{Stellar Populations}

Observational studies of galaxies make use of radiation emitted by them (or, sometimes, the lack of radiation due to absorption by galaxies) to infer their physical properties. As such, it is often crucial to be able to compute the emergent spectrum from each galaxy in a theoretical model. This is two-step process, involving first computing the spectrum of light emitted by all of the stars (and possibly the \AGN) in the galaxy and, secondly, computing how this light is reprocessed by the gas and dust in the galaxy and along the line of sight from the galaxy to the observer.

\subsubsection{Steller Population Synthesis}\label{sec:SPS}

In the absence of absorbing dust or gas, the luminosity as a function of frequency for a galaxy, the spectral energy distribution (SED), is simply a sum over the SEDs of its constituent stars. We can imagine a galaxy as consisting of numerous ``simple stellar populations''---populations of stars of the same age, metallicity and initial mass function. The luminosity, $L_\nu^{\rm (galaxy)}$, of a galaxy at frequency $\nu$ is then simply given by the following convolution integral
\begin{eqnarray}
 L_\nu^{\rm (galaxy)} &=& \int_0^t \d t^\prime \int_0^\infty \d Z^\prime \dot{M}_\star(t^\prime,Z^\prime) \nonumber \\
 & & \times L_\nu^{\rm (SSP)}(t-t^\prime,Z^\prime,\phi)
\end{eqnarray}
where $\dot{M}_\star(t,Z)$ is the rate of formation of stars of metallicity $Z$ in the galaxy at time $t$ and $L_\nu^{\rm (SSP)}(t,Z,\phi)$ is the luminosity of a simple stellar population of age $t$, metallicity $Z$ and with initial mass function $\phi(M)$. Given a model which predicts the rate of star formation in a galaxy as a function of time and metallicity (which is the galaxy formation aspect of this calculation), the problem is reduced to determining suitable $L_\nu^{\rm (SSP)}$ (the stellar astrophysics aspect of the calculation). This, in turn depends upon the spectra of individual stars, $L_\nu^{\rm (star)}(t,Z)$, and the stellar initial mass function, $\phi(M)$, such that
\begin{equation}
 L_\nu^{\rm (SSP)}(t,Z,\phi) = \int_{M_{\rm min}}^{M_{\rm max}} \phi(M^\prime) L_\nu^{\rm (star)}(t,Z) \d M^\prime,
\end{equation}
where $M_{\rm min}$ and $M_{\rm max}$ are the minimum and maximum masses for stars respectively.

Several libraries exist which provide $L_\nu^{\rm (SSP)}(t,Z,\phi)$ for several different ages, metallicities and initial mass functions \pcite{bruzual_stellar_2003,maraston_evolutionary_2005,dotter_stellar_2007,conroy_propagation_2008,lee_stellar_2009}. These are constructed using a combination of theoretical stellar evolution models, observations of stars of known age and metallicity and theoretical models of stellar atmospheres where no good observations exist. While enormous progress has been made in understanding the spectra of stars significant uncertainties remain. For example, \cite{conroy_propagation_2008}, who performed a study of how uncertainties in such models propagate into constraints derived from galaxy observations, find that current models do not fully characterize the metallicity dependence of the thermally pulsating asymptotic giant branch phase and that uncertainties in the slope of the initial mass function\footnote{See \protect\cite{kroupa_variation_2001} for a recent discussion of the difficulties associated with determining the local initial mass function. \protect\cite{kroupa_variation_2001}, along with \protect\cite{chary_stellar_2008} and \protect\cite{dokkum_evidence_2008}, also discuss evidence for the non-universality of the initial mass function, although no definitive evidence for such yet exists.} lead to an uncertainty in the evolution of the K-band magnitude of a stellar population of around 0.4 magnitudes per unit redshift (which leads to significant uncertainty when trying to interpret or predict the evolution of galaxy populations).

\subsubsection{Dust Absorption and Re-emission}

The presence of dust in galaxies has been known for a long time, and the effects of this dust on the observed properties of galaxies have been extensively studied. Simply put, dust absorbs light emitted by stars (and \AGN), particularly at short wavelengths, is heated by this light and therefore re-emits it at longer wavelengths (typically in the infrared and sub-mm). The presence of dust in galaxies can therefore significantly affect their observed luminosities at optical and UV wavelengths. Models of galaxy formation must therefore take into account the effects of dust before comparing their predictions to observational data. The simplest such approach, adopted by many early models, is a ``obscuring screen'' or ``slab'' geometry, in which a plane of dust is placed in front of the model galaxy, given an extinction curve measured from the Milky Way (for example) and normalized to have an optical depth at optical wavelengths based on properties of the model galaxy (gas content and metallicity). This provides a simple estimate of the amount of extinction, but clearly does not reflect the true geometry of the dust (which is distributed throughout the galaxy) or the fact that stars may preferentially form in dense, dust regions.

\citet{cole_hierarchical_2000} introduced a model for dust extinction in galaxies which significantly improved upon earlier ``slab'' models. In \citet{cole_hierarchical_2000} the mass of dust is assumed to be proportional to the mass and metallicity of the \ISM\ and to be mixed homogeneously with the \ISM\ (possibly with a different scale height from the stars) and to have properties consistent with the extinction law observed in the
Milky Way. To compute the extinction of any galaxy, a random inclination angle is selected and the extinction computed using the results of radiative transfer calculations carried out by \citet{ferrara_atlas_1999}.

\begin{figure*}
 \begin{center}
 \begin{tabular}{cc}
 \includegraphics[width=80mm]{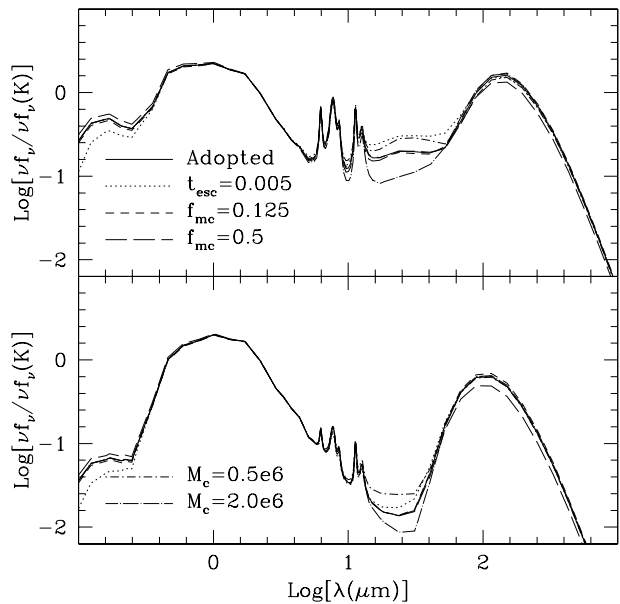} &
 \includegraphics[width=80mm]{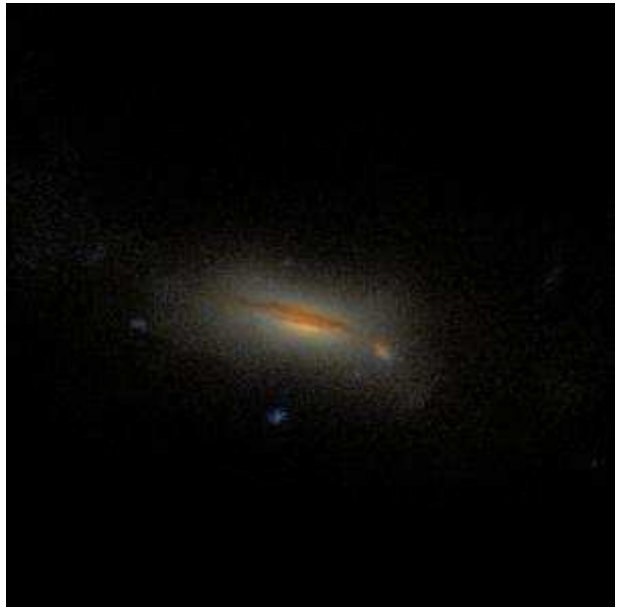}
 \end{tabular}
 \end{center}
 \caption{\emph{Left-hand panel:} An example of the spectral energy distributions from two face-on spiral galaxies as computed by {\sc Grasil} \protect\pcite{granato_infrared_2000}. The SED shows direct emission from stars at short wavelengths, with reprocessed light emitted from dust at long wavelengths. In the intermediate region lines associated with PAHs are visible. Reproduced, with permission, from \protect\cite{granato_infrared_2000}. \emph{Right-hand panel:} An image of a post-merger galaxy from a simulation by \protect\cite{cox_effect_2008} as computed by {\sc Sunrise} \protect\pcite{jonsson_sunrise:_2006}. A complex dust morphology, including a strong dust lane, is clearly visible. Reproduced, with permission, from \protect\cite{cox_effect_2008}.}
 \label{fig:Dust}
\end{figure*}

Beyond these relatively simple models of dust extinction it has recently become possible to employ much more realistic ray tracing techniques to compute the effects of dust on galaxy spectra. For example, the {\sc Grasil} software \pcite{silva_modelingeffects_1998} is designed to compute the radiative transfer of star light through an idealized galactic geometry consisting of a disk and a bulge each of which may contain both diffuse and clumpy gas and dust. {\sc Grasil} takes an input galactic SED together with physical parameters of a galaxy (size of each component, metallicity and mass of gas present) and computes the resulting SED including absorption, scattering and remission from the dust, taking into account a realistic distribution of grains and polycyclic aromatic hydrocarbons (PAHs) the temperature distribution of which are computed self-consistently, and assuming that stars are born in the dense molecular clouds and escape from these on some timescale (leading to enhanced absorption in the UV which is produced primarily by young stars) (Fig.~\ref{fig:Dust}). This code is therefore ideal for (semi-)analytic studies of galaxies and has been employed to examine the expected properties of submillimeter and infrared galaxies in hierarchical cosmologies \pcite{granato_infrared_2000,baugh_predictions_2004,baugh_canfaint_2005,lacey_galaxy_2008,swinbank_properties_2008}.

Similarly, the {\sc Sunrise} code of \cite{jonsson_sunrise:_2006} and {\sc Radishe} by \cite{chakrabarti_panchromatic_2009} solve essentially the same problem (radiative transfer through a dusty medium) but work for arbitrary geometry using Monte Carlo, polychromatic algorithms and so are particularly well suited to hydrodynamical simulations of galaxy formation, allowing realistic images of simulated galaxies to be made at any wavelength. {\sc Sunrise} has been applied to studies of luminous infrared galaxies \pcite{younger_merger-driven_2009}, quantitative morphology of merger remnants \pcite{lotz_galaxy_2008}, physical models to infer star formation rates from molecular indicators \pcite{narayanan_formation_2010} and tests of our ability to recover physical parameters of galaxies from their broadband SEDs \pcite{wuyts_color_2009}.

Since these ray tracing methods are computationally expensive they cannot yet be applied to large samples of model galaxies. Therefore, attempts have been made to construct simpler algorithms which capture most of their results. For example, \citet{gonzalez-perez_massive_2008} extended the model of \cite{cole_hierarchical_2000} by assuming that some fraction, $f_{\rm cloud}$, of the dust is in the form of dense molecular clouds where the stars form (see \citealt{baugh_canfaint_2005}). Stars were assumed to form in these clouds and to escape on a timescale of $\tau_{\rm quies}$ (for quiescent star formation in disks) or $\tau_{\rm burst}$ (for star formation in bursts), which is a parameter of the dust model \pcite{granato_infrared_2000}. Since massive, short-lived stars dominate the UV emission of a galaxy this enhances the extinction at short wavelengths, so these stars spend a significant fraction of their lifetime inside the clouds. 

To compute emission from dust they assumed a far infrared opacity of
\begin{equation}
   \kappa = \left\{ \begin{array}{ll}
          \kappa_1 (\lambda/\lambda_1)^{-\beta_1} & \hbox{for } \lambda<\lambda_{\rm break} \\
        \kappa_1 (\lambda_{\rm break}/\lambda_1)^{-\beta_1} (\lambda/\lambda_{\rm break})^{-\beta_2} & \hbox{for } \lambda>\lambda_{\rm\
 break},
\end{array} \right.
\end{equation}
where the opacity normalization at $\lambda_1=30~\mu$m is chosen to be $\kappa_1=140$cm$^2$/g  to reproduce the dust opacity model used in 
{\sc Grasil}, as described in \cite{silva_modelingeffects_1998}. The dust grain 
model in {\sc Grasil} is a slightly modified version of that proposed by \cite{draine_optical_1984}. Both the \cite{draine_optical_1984} and {\sc Grasil} dust models have been adjusted to fit data on dust extinction and emission in the local \ISM\ (with much more extensive \ISM\ dust emission data being used by \citealt{silva_modelingeffects_1998}).
The normalization is set at 30$\mu$m because the dust opacity in the \cite{draine_optical_1984} and {\sc Grasil} 
models is well fit by a power-law longwards of that wavelength, but not shortwards. The dust luminosity is then assumed to be
\begin{equation}
 L_\nu = 4\pi\kappa(\nu)B_\nu(T) M_{\rm Z,gas},
 \end{equation}
where $B_\nu(T) = [2{\rm h}\nu^3/{\rm c}^2]/[\exp({\rm h}\nu/{\rm k}T)-1]$ is the Planck blackbody spectrum and $M_{\rm Z,gas}$ is the mass of metals in gas. The dust temperature, $T$, is chosen such that the bolometric dust luminosity equals the luminosity absorbed by dust.

This extended dust model, including diffuse and molecular cloud dust components, provides a better match to the detailed results of {\sc Grasil} while being orders of magnitude faster, although it does not capture details such as PAH features. 

\cite{fontanot_evaluating_2009} have explored similar models which aim to reproduce the results of {\sc Grasil} using simple, analytic prescriptions. They found that by fitting the results from {\sc Grasil} they were able to obtain a better match to the extinction in galaxies than previous, simplistic models of dust extinction had been able to attain. 

\subsection{Absorption by the Intergalactic Medium}

In addition to having to pass through the internal gas and dust in each galaxy, light emitted from a galaxy must pass through the entire intervening intergalactic medium between it and ourselves before we can observe it. The intergalactic medium contains significant amounts of neutral hydrogen (even at relatively low redshifts, long after reionization) which is known to be clumped into clouds by virtue of observations of the Lyman-$\alpha$ forest of absorption systems in quasar spectra. Light emitted at redshift $z_{\rm em}$ at some wavelength $\lambda_{\rm em}$, shortwards of the Lyman-$\alpha$ wavelength of 1216\AA, will eventually be redshifted into the Lyman-$\alpha$ line and so will be absorbed by any clouds of neutral hydrogen at redshift $1 + z_{\rm abs} = (1 + z_{\rm em}) [\lambda_{\rm em}/1216\hbox{\AA}]$. Light emitted at even shorter wavelengths may be additionally absorbed by higher lines in the Lyman series, or by the Lyman continuum shortwards of $912$\AA.

Models of the effective optical depth due to this absorption (based upon the observed distribution of Lyman-$\alpha$ forest absorber properties) have been described by \cite{madau_radiative_1995} and \cite{meiksin_colour_2006}. The resulting absorption of starlight shortwards of Lyman-$\alpha$ is so severe that galaxies appear essentially dark at wavelengths shorter than $\lambda_{\rm obs} = 1216\hbox{\AA} (1+z_{\rm em})$. This forms the basis of the ``dropout'' techniques for identifying high redshift galaxies and quasars on the basis of their broadband colors \pcite{madau_high-redshift_1996,fan_high-redshift_2001}.

\section{Computational Techniques}

The process of galaxy formation involves nonlinear physics and a wide variety of physical processes. As such, it is impossible to treat in full detail using analytic techniques. There are two major approaches that have been developed to circumvent this problem. The first, numerical N-body simulation, attempts to directly and numerically solve the fully nonlinear equations governing the physical processes inherent to galaxy formation. The second, semi-analytic modeling, attempts to construct a coherent set of analytic approximations which describe these same physics. Each has its strengths and weaknesses, as will be discussed below. As a result of this and of our ignorance of how some key processes work, both approaches contain a little of each other (rather like yin yang). A different, more empirical approach, utilizing so-called ``halo occupation distributions'' has become widely used in the past ten years and will be discussed briefly below.

\subsection{N-body/Hydro}

The most accurate computational method for solving the physics of galaxy formation is via direct simulation, in which the fundamental equations of gravitation, hydrodynamics and perhaps radiative cooling and transfer are solved far a large number of points (arranged either on a grid or following the trajectories of the fluid flow). I will not attempt to review the numerical methods utilized in this approach in any detail here (recent treatments of this can be found in \citealt{bertschinger_simulations_1998,agertz_fundamental_2007,rosswog_astrophysical_2009}), but will instead merely highlight some of the approaches used.

Collisionless dark matter is (relatively) simple to model in this way, since it responds only to the gravitational force. For the velocities and gravitational fields occurring during structure and galaxy formation non-relativistic Newtonian dynamics is more than adequate and so solving the evolution of some initial distribution of dark matter (usually a Gaussian random field of density perturbations consistent with the power spectrum of the \CMB) reduces to summing large numbers of $1/r^2$ forces between pairs of particles. In practice, clever numerical techniques (such as particle--mesh, tree algorithms etc.) are usually used to reduce this $N^2$ problem into something more manageable \pcite{kravtsov_adaptive_1997,springel_cosmological_2005}. The largest pure dark matter simulations carried out to date contain around 10 billion particles \pcite{springel_simulations_2005}. Dark matter only simulations of this type (carried out primarily for the cold dark matter scenario, but see \citealt{white_size_1984,klypin_structure_1993,bode_halo_2001,dave_halo_2001,colin_structure_2002,ahn_formation_2005}) have been highly successful in determining the large scale structure of the Universe, as embodied in the so-called ``cosmic web''. As a result, the spatial and velocity correlation properties of dark matter and dark matter halos \pcite{davis_evolution_1985,white_galaxy_1987,white_clusters_1987,efstathiou_gravitational_1988,eke_cluster_1996,jenkins_evolution_1998,padilla_cluster_2002,bahcall_evolution_2004,kravtsov_dark_2004,reed_clustering_2009}, together with the density profiles \pcite{navarro_universal_1997,bullock_profiles_2001,navarro_inner_2004,merritt_universal_2005,prada_far_2006}, angular momenta \pcite{barnes_angular_1987,efstathiou_gravitational_1988,warren_dark_1992,cole_structure_1996,lemson_environmental_1999,bullock_universal_2001,bosch_angular_2002,bett_spin_2007,gao_redshift_2008} and internal structure \pcite{moore_dark_1999,klypin_galaxies_1999,kuhlen_via_2008,springel_aquarius_2008} of dark matter halos are known to very high accuracy.

Of course, to study galaxy formation dark matter alone is insufficient, and baryonic material must be added in to the mix. This makes the problem much more difficult since, at the very least, pressure forces must be computed and the internal energy of the baryonic fluid tracked. Particle-based methods (most prominently smoothed particle hydrodynamics; \citealt{springel_cosmological_2005}) have been successful in this area, as have Eulerian grid methods \pcite{ricker_cosmos:hybrid_2000,fryxell_flash:adaptive_2000,plewa_amra:adaptive_2001,quilis_new_2004}. The addition of radiative cooling is relatively straightforward (at least while the gas remains optically thin to its own radiation) by simply tabulating the rate at which gas of given density and temperature radiates energy.

Going beyond this level of detail becomes extremely challenging. Numerous simulation codes are now able to include star formation and feedback from supernovae explosions, while some even attempt to follow the formation of supermassive black holes in galactic centers. It should be kept in mind though that for galaxy scale simulations the real physics of these processes is happening on scales well below the resolution of the simulation and so the treatment of the physics is often at the ``subgrid'' level, which essentially means that it is put in by hand using a semi-analytic approach \pcite{thacker_implementing_2000,kay_including_2002,marri_smoothed_2003,white_feedback_2004,stinson_star_2006,cox_feedback_2006,scannapieco_feedback_2006,tasker_simulating_2006,stinson_supernova_2007,vecchia_simulating_2008,oppenheimer_mass_2008,okamoto_impact_2008,booth_cosmological_2009,ciotti_agn_2009}.

Beyond this, problems such as the inclusion of radiative transfer or magnetic fields complicate the problem further by introducing new sets of equations to be solved and the requirement to follow additional fields. For example, in radiative transfer one must follow photons (or photon packets) which have a position, direction of travel and wavelength, and determine the absorption of these photons as they traverse the baryonic material of the simulation, while simultaneously accounting for the re-emission of absorbed photons at other wavelengths. Despite these complexities, progress has been made on these issues using a variety of ingenious numerical techniques \pcite{abel_photon-conserving_1999,ciardi_cosmological_2001,gnedin_multi-dimensional_2001,ricotti_fate_2002,petkova_implementation_2009,li_cosmomhd:cosmological_2008,aubert_radiative_2008,collins_cosmological_2009,dolag_mhd_2008,finlator_new_2009,laursen_ly_2009}. Providing Moore's Law continues to hold true, the ability of simulations to provide ever more accurate pictures of galaxy formation should hold strong. A brief survey of current cosmological hydrodynamical codes and their functionalities is given in Table~\ref{tb:Hydros}.

\begin{table*}
 \begin{minipage}{\textwidth}
 \caption{A survey of physical processes included in several of the major hydrodynamical codes. The primary reference is indicated next to the name of the code. Where implementations of major physical processes are described elsewhere the reference is given next to the entry in the relevant row.}
 \label{tb:Hydros}
 \begin{center}
 \begin{tabular}{lccccc}
  {\bf Feature} & {\sc Gadget-3}\endnote{``GAlaxies with Dark matter and Gas intEracT'' \protect\citep{springel_cosmological_2005};~} & {\sc Gasoline}\endnote{\protect\cite{wadsley_gasoline:flexible_2004};~} & {\sc HART}\endnote{Hydrodynamic Adaptive Refinement Tree \protect\citep{kravtsov_constrained_2002};~} & {\sc Enzo(Zeus)}\endnote{\protect\cite{oshea_introducing_2004};~} & {\sc Flash}\endnote{{\tt http://flash.uchicago.edu} \protect\citep{fryxell_flash:adaptive_2000};~} \\
 \hline
  Gravity & Tree & Tree & AMR\endnote{Adaptive Mesh Refinement;\label{enote:AMR}} PM\endnote{Particle--mesh; \label{enote:PM}} & AMR\sref{enote:AMR} PM\sref{enote:PM} & Multi-grid \\
  Hydrodynamics & SPH\endnote{Smoothed Particle Hydrodynamics; \label{enote:SPH}} & SPH\sref{enote:SPH} & AMR\sref{enote:AMR} & AMR\sref{enote:AMR} & AMR\sref{enote:AMR} \\
  \hspace{5mm}$\rightarrow$ Multiphase subgrid model\endnote{Applicable only to SPH codes---used correctly, AMR codes naturally resolve multiphase media;} & $\checkmark$\endnote{\protect\cite{scannapieco_feedback_2006};~} & $\times$ & N/A & N/A & N/A \\
  Radiative Cooling & $\checkmark$ & $\checkmark$ & $\checkmark$ & $\checkmark$ & $\checkmark$\endnote{\protect\cite{banerjee_supersonic_2006};\label{enote:flashRadCool}~} \\
  \hspace{5mm}$\rightarrow$ Metal dependent & $\checkmark$\endnote{\protect\cite{scannapieco_feedback_2005};~} & $\times$ & $\checkmark$\endnote{\protect\cite{tassis_scaling_2008};\label{enote:harttassis}~} & $\checkmark$\endnote{\protect\cite{smith_three_2009};~} & $\checkmark$\sref{enote:flashRadCool} \\
  \hspace{5mm}$\rightarrow$ Molecular chemistry & $\checkmark$\endnote{\protect\cite{yoshida_simulations_2003};~} & $\times$ & $\checkmark$\sref{enote:harttassis}$^,$\endnote{Equilibrium only;~} & $\checkmark$\endnote{\protect\cite{turk_formation_2009};~} & $\times$ \\
  Thermal Conduction & $\checkmark$\endnote{\protect\cite{jubelgas_thermal_2004};~} & $\times$ & $\times$& $\times$ & $\checkmark$ \\
  Star formation & $\checkmark$\endnote{\protect\cite{scannapieco_feedback_2005};\label{enote:gadgetSF}~} & $\checkmark$\endnote{\protect\cite{governato_forming_2007};\label{enote:gasolineSF}~} & $\checkmark$\sref{enote:harttassis} & $\checkmark$\endnote{\protect\cite{tasker_effect_2008};\label{enote:enzoSF}~} & $\times$ \\
  \hspace{5mm}$\rightarrow$ SNe feedback & $\checkmark$\sref{enote:gadgetSF} & $\checkmark$\sref{enote:gasolineSF} & $\checkmark$\sref{enote:harttassis} & $\checkmark$\sref{enote:enzoSF} & $\times$ \\
  \hspace{5mm}$\rightarrow$ Chemical enrichment & $\checkmark$\sref{enote:gadgetSF} & $\checkmark$\sref{enote:gasolineSF} & $\checkmark$\sref{enote:harttassis} & $\checkmark$\sref{enote:enzoSF} & $\times$ \\
  Black hole formation  & $\checkmark$\endnote{\protect\cite{matteo_energy_2005};\label{enote:gadgetBH}~} & $\times$ & $\times$ & $\times$ & $\checkmark$\endnote{\protect\cite{federrath_modeling_2010};~} \\
  \hspace{5mm}$\rightarrow$ \AGN\ feedback & $\checkmark$\sref{enote:gadgetBH} & $\times$ & $\times$ & $\times$ & $\times$\\
  Radiative transfer & OTVET\endnote{Optically Thin Variable Eddington Tensor;\label{enote:otvet}~}$^,$\endnote{\protect\cite{petkova_implementation_2009};~} & $\times$ & OTVET\sref{enote:otvet} & $\checkmark$\endnote{Flux-limited diffusion approximation (\protect\citealt{norman_cosmological_2009}; see also \citealt{wise_resolvingformation_2008});~} & $\checkmark$\endnote{\protect\cite{rijkhorst_hybrid_2006,peters_limiting_2010};~} \\
  Magnetic fields & $\checkmark$\endnote{\protect\cite{dolag_mhd_2008};~} & $\times$ & $\times$ & $\checkmark$\endnote{\protect\citeauthor{collins_cosmological_2009}~(\protect\citeyear{collins_cosmological_2009}; see also \citealt{wang_magnetohydrodynamic_2009});~} & $\checkmark$\endnote{\protect\cite{robinson_morphology_2004};~} \\
 \end{tabular}
 \end{center}
 \theendnotes
 \setcounter{endnote}{0}
 \end{minipage}
\end{table*}

\subsection{Semi-Analytic}

The technique of ``semi-analytic modeling'' or, perhaps, ``phenomenological galaxy formation modeling'' takes the approach of treating the various physical processes associated with galaxy formation using approximate, analytic techniques. As with N-body/hydro simulations, the degree of approximation varies considerably with the complexity of the physics being treated, ranging from precision-calibrated estimates of dark matter merger rates to empirically motivated scaling functions with large parameter uncertainty (e.g. in the case of star formation and feedback---just as in N-body/hydro simulations).

The primary advantage of the semi-analytic approach is that it is computationally inexpensive compared to N-body/hydro simulations. This facilitates the construction of samples of galaxies orders of magnitude larger than possible with N-body techniques and for the rapid exploration of parameter space \pcite{henriques_monte_2009} and model space (i.e. adding in new physics and assessing the effects). The primary disadvantage is that they involve a larger degree of approximation. The extent to which this actually matters has not yet been well assessed. Comparison studies of semi-analytic vs. N-body/hydro calculations have shown overall quite good agreement (at least on mass scales well above the resolution limit of the simulation) but have been limited to either simplified physics (e.g. hydrodynamics and cooling only; \citealt{benson_comparison_2001,yoshida_gas_2002,helly_comparison_2003}) or to simulations of individual galaxies \pcite{stringer_analytic_2010}.

Some of the earliest attempts to construct a self-consistent semi-analytic model of galaxy formation began with \cite{white_galaxy_1991}, \cite{cole_modeling_1991} and \cite{lacey_tidally_1991}, drawing on earlier work by \cite{rees_cooling_1977} and \cite{white_core_1978}. Since then numerous studies \pcite{kauffmann_formation_1993,baugh_semianalytic_1999,baugh_epoch_1998,somerville_semi-analytic_1999,cole_hierarchical_2000,benson_effects_2002,hatton_galics-_2003,monaco_morgana_2007} have extended and improved this original framework. A recent review of semi-analytic techniques is given by \cite{baugh_primerhierarchical_2006}. Semi-analytic models have been used to investigate many aspects of galaxy formation including:
\begin{itemize}
\item galaxy counts \pcite{kauffmann_faint_1994,devriendt_galaxy_2000};

\item galaxy clustering \pcite{diaferio_clustering_1999,kauffmann_clustering_1999,kauffmann_clustering_1999-1,baugh_modellingevolution_1999,benson_nature_2000,benson_dependence_2000,wechsler_galaxy_2001,blaizot_galics_2006};

\item galaxy colors and metallicities \pcite{kauffmann_chemical_1998,springel_populatingcluster_2001,lanzoni_galics-_2005,nagashima_metal_2005-1,font_colours_2008};

\item sub-mm and infrared galaxies \pcite{guiderdoni_semi-analytic_1998,granato_infrared_2000,baugh_canfaint_2005,lacey_galaxy_2008};

\item abundance and properties of Local Group galaxies \pcite{benson_effects_2002-1,somerville_can_2002,li_nature_2010};

\item the reionization of the Universe \pcite{devriendt_contribution_1998,benson_non-uniform_2001,somerville_star_2003,benson_epoch_2006};

\item the heating of galactic disks \pcite{benson_heating_2004};

\item the properties of Lyman-break galaxies \pcite{governato_seeds_1998,blaizot_predicting_2003,blaizot_galics-_2004,lacey_evolution_2010};

\item supermassive black hole formation and \AGN\ feedback \pcite{kauffmann_unified_2000,croton_many_2006,bower_breakinghierarchy_2006,malbon_black_2007,somerville_semi-analytic_2008,fontanot_many_2009};

\item damped Lyman-$\alpha$ systems \pcite{maller_damped_2001,maller_damped_2003};

\item the X-ray properties of galaxy clusters \pcite{bower_impact_2001,bower_flip_2008};

\item chemical enrichment of the \ICM\ and IGM \pcite{lucia_chemical_2004,nagashima_metal_2005};

\item the formation histories and morphological evolution of galaxies \pcite{kauffmann_age_1996,lucia_formation_2006,fontanot_reproducingassembly_2007,somerville_explanation_2008}.
\end{itemize}

The term ``semi-analytic'' model has become somewhat insufficient, since this name now encompasses such a diverse range of models that the name alone does not convey enough information. Semi-analytic models in the literature contain a widely disparate range of physical phenomena and implementations. Table~\ref{tb:SAMs} is an attempt to assess in detail the physics included in currently implemented semi-analytic models\footnote{One component missing from Table~\protect\ref{tb:SAMs} is the specific implementation of stellar population synthesis used by each code. We have chosen to not include this because all the models listed treat this calculation in precisely the same way (see \S\protect\ref{sec:SPS}), differing only in which compilation of stellar spectra they choose to employ. Since it is trivial to replace one compilation of stellar spectra with another we do not consider this a fundamental difference between the models. Nevertheless, the choice of which stellar spectra to use can have important consequences for the predicted properties of galaxies \protect\citep{conroy_propagation_2010} and so should not be overlooked.}.

\begin{table*}
\begin{minipage}{\textwidth}
 \caption{A survey of physical processes included in major semi-analytic models of galaxy formation. In each case we indicate how this process is implemented and give references where relevant. In many cases a single model has implemented a given physical process at different levels of complexity/realism. In such cases, we list the most ``advanced'' implementation that the model is capable of.}
 \label{tb:SAMs}
 \begin{center}
 \begin{tabular}{lccccc}
  & \multicolumn{5}{c}{Model} \\
  \cline{2-6}
  {\bf Feature} & {\sc Durham}\endnote{\protect\cite{cole_hierarchical_2000};~} & {\sc Munich}\endnote{\protect\cite{croton_many_2006};~} & {\sc Santa-Cruz}\endnote{\protect\cite{somerville_semi-analytic_2008};~} & {\sc Morgana}\endnote{\protect\cite{monaco_morgana_2007};~} & {\sc Galics}\endnote{\protect\cite{hatton_galics-_2003};~}  \\
  Merger Trees \\
  \hspace{5mm}$\rightarrow$ Analytic & Modified ePS\endnote{\protect\cite{parkinson_generating_2008};~} & ePS\endnote{\protect\cite{kauffmann_merging_1993};~} & ePS & {\sc Pinocchio}\endnote{\protect\cite{monaco_pinocchio_2002};~} & $\times$ \\
  \hspace{5mm}$\rightarrow$ N-body & $\checkmark$\endnote{\protect\cite{helly_galaxy_2003};~} & $\checkmark$ & $\checkmark$ & $\times$ & $\checkmark$ \\
  Halo Profiles & Einasto\endnote{\protect\cite{benson_galaxy_2010};\label{enote:galformBB09}~} & Isothermal & NFW & NFW & Empirical\endnote{A ``dark matter'' core is included in calculations of disk sizes with an empirically selected dark matter fraction;~} \\
  Cooling Model &  \\
  \hspace{5mm}$\rightarrow$ Metal-dependent & $\checkmark$ & $\checkmark$ & $\checkmark$ & $\checkmark$ & $\checkmark$ \\
  Star Formation & $\checkmark$ & $\checkmark$ & $\checkmark$ & $\checkmark$ & $\checkmark$ \\
  Feedbacks \\
  \hspace{5mm}$\rightarrow$ SNe & $\checkmark$ & $\checkmark$ & $\checkmark$ & $\checkmark$ & $\checkmark$ \\
  \hspace{5mm}$\rightarrow$ \AGN\ & $\checkmark$\endnote{\protect\cite{bower_breakinghierarchy_2006};~} & $\checkmark$ & $\checkmark$ & $\checkmark$ & $\checkmark$\endnote{\protect\cite{cattaneo_modellinggalaxy_2006};~} \\
  \hspace{5mm}$\rightarrow$ Reionization & $\checkmark$\sref{enote:galformBB09} & $\times$ & $\checkmark$ & $\checkmark$\endnote{\protect\cite{maccio_origin_2009};~} & $\checkmark$\endnote{\protect\cite{lanzoni_galics-_2005};~} \\
  Merging \\
  \hspace{5mm}$\rightarrow$ Substructure\endnote{How does the model track substructures within halos?;~} & N-body\endnote{Substructure orbits and merging times are determined from N-body simulations;~\label{enote:subNbody}} & N-body\sref{enote:subNbody} & DF\endnote{Dynamical Friction: substructure merging times are computed from analytic estimates of dynamical friction timescales;~\label{enote:subDF}} & DF\sref{enote:subDF} & N-body\sref{enote:subNbody} \\
  \hspace{5mm}$\rightarrow$ Substructure--Substructure\endnote{Does the model allow merging between pairs of subhalos orbiting in the same host halo?;~} & $\checkmark$\endnote{Using hierarchically nested substructures;~}$^,$\sref{enote:galformBB09} & $\times$ & $\checkmark$\endnote{Using random collisions of subhalos;\label{enote:satsatran}~}$^,$\endnote{\protect\cite{somerville_semi-analytic_1999};~} & $\times$ & $\checkmark$\sref{enote:satsatran} \\
  Environments \\
  \hspace{5mm}$\rightarrow$ Ram Pressure Stripping & $\checkmark$\endnote{\protect\cite{font_colours_2008};~} & $\checkmark$\endnote{\protect\cite{bruggen_ram-pressure_2008};~} & $\times$ & $\times$ & $\checkmark$\endnote{\protect\cite{lanzoni_galics-_2005};~} \\
  \hspace{5mm}$\rightarrow$ Tidal Stripping & $\checkmark$\sref{enote:galformBB09} & $\times$ & $\checkmark$ & $\checkmark$ & $\checkmark$ \\
  \hspace{5mm}$\rightarrow$ Harassment & $\times$ & $\times$ & $\times$ & $\times$ & $\times$ \\
  Disks \\
  \hspace{5mm}$\rightarrow$ Disk Stability & $\checkmark$ & $\checkmark$ & $\checkmark$\endnote{\protect\cite{somerville_explanation_2008};~} & $\checkmark$ & $\checkmark$ \\
  \hspace{5mm}$\rightarrow$ Dynamical Friction\endnote{Does the model include dynamical friction forces exerted by a galaxy disk on orbiting satellites?;~} & $\checkmark$\endnote{\protect\cite{benson_heating_2004};\label{enote:galformheat}~} & $\times$ & $\times$ & $\times$ & $\times$ \\
  \hspace{5mm}$\rightarrow$ Thickness & $\checkmark$\sref{enote:galformheat} & $\times$ & $\times$ & $\times$ & $\times$ \\
  Sizes \\
  \hspace{5mm}$\rightarrow$ Adiabatic contraction & $\checkmark$ & $\times$ & $\checkmark$ & $\checkmark$ & $\times$ \\
  Chemical Enrichment & $\checkmark$ [delayed\sref{enote:galformBB09}] & $\checkmark$ [instant\endnote{\protect\cite{lucia_chemical_2004};~}] & $\checkmark$ [delayed\endnote{\protect\cite{arrigoni_galactic_2009};~}] & $\checkmark$ [instant] & $\checkmark$ [delayed\endnote{\protect\cite{pipino_galics_2008};~}] \\
  Dust & {\sc Grasil}\endnote{\protect\cite{silva_modelingeffects_1998};\label{enote:grasil}~} & Screen\endnote{A plane-parallel screen of dust placed in front of the galaxy;~} & Slab\endnote{Empirically calibrated model utilizing a slab of dust mixed uniformly with the stars;\label{enote:slab}~} & {\sc Grasil}\sref{enote:grasil}$^,$\endnote{\protect\cite{fontanot_evaluating_2009};~} & Slab\sref{enote:slab} \\
 \end{tabular}
 \end{center}
 \theendnotes
 \setcounter{endnote}{0}
\end{minipage}
\end{table*}

\subsection{Halo Occupation Distributions}

Given the complexity of galaxy formation it is sometimes desirable to take a more empirical approach to the problem. This can be advantageous both to relate observations to the assumed underlying physical structure of the Universe (e.g. \citealt{zheng_galaxy_2007,tinker_interpretingclustering_2010}), and to make predictions based on extrapolations from current data and which have a cosmological underpinning. ``Halo occupation distributions'', first described by \cite{neyman_theory_1952} and used to study galaxy clustering (see, for example, \citealt{benson_nature_2000,peacock_halo_2000}) in their simplest form specify the probability of finding $N$ galaxies of some prescribed type in a dark matter halo of mass $M$, $P(N|M)$. Given this probability distribution, knowledge of the distribution of dark matter halo masses and their spatial distribution plus some assumptions about the locations of galaxies within halos, one can construct (statistically or within an N-body simulation) the distribution of galaxies. This allows, for example, the abundance and clustering properties of those galaxies to be inferred. A detailed discussion of the use of halo occupation distributions in studying galaxy clustering is given by \cite{cooray_halo_2002}.

This approach has recently been extended by incorporating simple prescriptions relating, for example, star formation rates or quasar activity to halo mass and redshift. Several authors have demonstrated that such empirical models, while simple, can fit a wide variety of galaxy data and can be used to gain insight into phenomena such as ``downsizing'' \citep{yang_constraining_2003,conroy_connecting_2009,croton_simple_2009}. At a fundamental level, the success of simple, empirical models such as these suggests that, despite the complexity of galaxy formation physics, its outcome is relatively simple.

\section{Topics of Current Interest}

Having reviewed the major inputs to our current theory of galaxy formation and explored some of the tools employed to solve them, in this section we briefly explore five topics of particular interest in contemporary galaxy formation theory.

\subsection{The First Galaxies}

After the initial excitement of the Big Bang, inflation, nucleosynthesis and the recombination the Universe enters a protracted period of relative quiescence, known as the ``Dark Ages'', during which there are no luminous sources and structure grows primarily only in the dark sector. The end of the Dark Ages is thought to occur when the first star forms. The formation of these first stars is an unusual problem in astrophysics in that it represents a clean, well-defined problem that can be tackled numerically---the initial conditions are simply those of an expanding Universe containing small perturbations in the dark matter and baryonic components which is a simple mixture of hydrogen and helium. As such, this problem has been examined and explored in significant detail \pcite{abel_formation_2002,yoshida_formation_2006,gao_first_2007}. The consensus picture that has emerged from this work is that the first star will form due to gas collapse and cooling (via the gas-phase formation of molecular hydrogen) in a halo of mass around $10^6M_\odot$ at a redshift of $z\approx 50$. The first star is likely to be very massive (current simulations cannot follow the evolution of the proto-star beyond the point at which it becomes optically thick in various lines of molecular hydrogen due to a lack of treatment of the necessary radiative transfer, but at this stage the collapsing core typically has an accretion rate that is expected to lead to the formation of a star with a mass of several tens of Solar masses) and, therefore, short lived. What happens after this first star forms rapidly becomes a much more complicated problem, involving the poorly understood physics of supernovae explosions and the numerically challenging problem of radiative transfer.

Needless to say then, studying the formation of the first galaxies is more complicated still, although no more complicated than modeling the formation of later generations. In the next decade new facilities such as JWST and thirty meter class telescopes will begin to open a window on the earliest generations of galaxies and, as such, there is a need for theoretical understanding and predictions for this epoch of cosmic history. In particular, this represents a real opportunity to make testable predictions from galaxy formation theory which can be confronted with data in the near future.

In considering the earliest generations of galaxies there are various physical processes which, while often neglected at lower redshifts, are of potentially crucial importance. For example, understanding the properties of Population III stars, their various feedback processes (both positive and negative) and initial mass function \pcite{glover_uncertainties_2008,trenti_formation_2009,ohkubo_evolution_2009,machida_magnetohydrodynamics_2008,lawlor_single_2008,oshea_population_2007,oshea_population_2008,harris_non-grey_2007,stacy_impact_2007} may be crucial for understanding how this population of galaxies grows \pcite{tegmark_small_1997,ricotti_feedbackgalaxy_2001,wise_suppression_2007,wise_resolvingformation_2007,greif_first_2008,wise_resolvingformation_2008,wise_resolvingformation_2008-1,bromm_formation_2009}. Additionally, connecting these galaxies to later generations involves understanding how they evolve through (and cause) the epoch of cosmic reionization \pcite{benson_effects_2002,alvarez_connecting_2008}, modeling non-local feedbacks (i.e. interactions between neighboring galaxies or the galaxy population as a while) and feedbacks in general \pcite{busha_impact_2009}.

Various interesting questions await definitive answers. For example, how long is the period of Population III star formation? This seemingly simple question depends on how we define Population III---theoretical reasoning suggests that we may expect a significant shift in the initial mass function once gas is raised above a critical metallicity of $10^{-4}Z_\odot$ \pcite{santoro_critical_2006} and the ability of Population III stars to enrich their surroundings (and, therefore, future generations of star formation). Since Population III stars are likely to produce many more hydrogen ionizing photons per unit mass than their Population II counterparts this question may have important consequences for the reionization history of the Universe.

Reionization itself remains an interesting problem. While it has been convincingly demonstrated theoretically \pcite{chiu_semianalytic_2000,ciardi_early_2003,somerville_epoch_2003,benson_epoch_2006} that the universe could be reionized by galaxies sufficiently early to match constraints on the optical depth to the \CMB\ \pcite{dunkley_five-year_2009} these calculations are currently forced to make assumptions about the distribution of mass and ionized regions in the IGM and about the ionizing photon escape probability from high redshift galaxies \pcite{dove_escape_2000,ricotti_feedbackgalaxy_2000,razoumov_escape_2006,gnedin_escape_2008,yajima_escape_2009,wise_ionizing_2009} which are still poorly understood.

In addition to observing high redshift galaxies directly, it may be possible to learn much about them from the study of their present day remnants. ``Galactic archeology'' of this sort has been investigated quite extensively \pcite{freeman_new_2002,scannapieco_spatial_2006,brook_spatial_2007}.

The conclusions are that Population III stars (usually taken as a convenient proxy for ``first galaxies'') that formed in progenitor galaxies of the Milky Way likely formed over a fairly broad range of redshifts, with the formation rate peaking at $z\approx 10$ but continuing until $z\approx4$--5 (in chemically isolated halos which have not been contaminated by metals produced by star formation in earlier forming halos). As such, any remnants of these stars are probably spread over most of the nearby Galactic halo. The lack of detection of such metal-free stars therefore places constraints on their initial mass function---they must have been sufficiently massive to have evolved off of the main sequence by the present day. Currently, this suggests that such stars must have been more massive then about $0.8M_\odot$ \pcite{brook_spatial_2007}. If we examine truly old stars, however, the situation is rather different. Simulations suggest that these formed in halos close to the high density peak that eventually formed the final system, and therefore are preferentially located in the central regions (typically the bulge) of the Milky Way.

\subsection{The Formation and Sizes of Galaxy Disks}

Galactic disks are, perhaps, the most prominent feature of galaxies in the local Universe. Additionally, recent observations \pcite{law_kiloparsec-scale_2009,schreiber_sins_2009} have demonstrated that massive disks are already in place as long ago as $z=2$. Understanding the details of how disks form and grow is therefore of crucial importance for galaxy formation theory \pcite{silk_formation_2001}.

The basic picture has been in place for a long time \pcite{fall_formation_1980}. Collapsing systems of dark matter and gas are imparted some small amount of angular momentum as a result of tidal torques from inhomogeneities in the surrounding matter distribution \pcite{hoyle_origin_1949,barnes_angular_1987}. While gas can radiate energy and thereby collapse to form a galaxy it cannot radiate angular momentum. As such, angular momentum is conserved\footnote{It is conserved if we ignore the possibility of torques between baryons and dark matter which could transfer angular momentum from the gas to the dark halo. We will review this assumption later.} during the collapse and is eventually responsible for halting that collapse when the proto-galaxy reaches a radius at which its angular momentum is sufficient to provide support against the gravitational potential in which it resides. Back of the envelope type estimates of the resulting sizes of disks show that this picture produces disks with sizes comparable to those that are observed \pcite{fall_formation_1980,mo_formation_1998}. However, this basic picture makes several simplifying assumptions, some of them difficult to justify. A careful analysis of these assumptions is therefore warranted to understand the process of disk formation in greater detail.

It has been convincingly established that the distribution of angular momentum of gas in halos prior to any cooling and galaxy formation is similar to that of the dark matter \pcite{bosch_angular_2002,bosch_angular_2003} in so far as the distribution of spin parameters is very similar (the directions of the angular momenta vectors of dark matter and gas are often misaligned; \citealt{bosch_angular_2002}). The distribution of the angular momentum of gas within an individual halo is less well studied, but recent hydrodynamical simulations \pcite{sharma_angular_2005} have demonstrated that the distribution of the specific angular momentum, $j$, can be well described by a simple form:
\begin{equation}
 {1\over M}{\d M \over \d j} = {1 \over j_{\rm d}^{\alpha_j}\Gamma(\alpha_j)}j^{\alpha_j-1}{\rm e}^{-j/j_{\rm d}},
\end{equation}
where $\Gamma$ is the gamma function, $M$ is the total mass of gas, $j_{\rm d}=j_{\rm tot}/\alpha$ and $j_{\rm tot}$ is the mean specific angular momentum of the gas. The parameter $\alpha_j$ takes on different values in different halos, with a median value of $0.89$. The fraction of mass with specific angular momentum less than $j$ is then given by
\begin{equation}
 f(<j) = \gamma\left(\alpha_j,{j\over j_{\rm d}}\right),
\end{equation}
where $\gamma$ is the incomplete gamma function. Despite this knowledge of the initial distribution of angular momentum in the baryonic component, the question of how well angular momentum is conserved during the process of cooling and collapse is less clear. However, recent simulation studies have begun to shed some light on this issue. \cite{zavala_bulges_2008} describe high resolution SPH simulations in which the baryonic physics (particularly the star formation and feedback from supernovae) is varied such that one simulation results in the formation of a disk galaxy and another in the formation of a spheroidal galaxy. They show that, in the case of the disk galaxy the angular momentum of the baryonic material that will eventually form the galaxy tracks the behavior of the dark matter angular momentum: growing as predicted by linear theory at early times and then remaining constant. In the case of the spheroidal galaxy, the angular momentum still grows initially as expected from linear theory, but then 90\% is rapidly lost as pre-galactic fragments undergo mergers which transfer their angular momentum to the dark matter. In the disk galaxy case, strong feedback prevents gas from cooling into these small halos at early times, and so avoids angular momentum loss to the dark halo.

In addition to macroscopic quantities such as the size of a disk, the picture described above can predict the radial density profiles of disks. This is interesting since observed disks typically follow an exponential radial profile (at least out to some radius, beyond which the stellar light profile typically truncates---gaseous disks typically extend out further beyond this stellar truncation radius). In a hierarchical Universe, in which dark matter halos and their gas content grow in mass and size with time, the angular momentum of material accreting onto a galaxy will, on average, increase with time also. As a result, disk formation is an ``inside out'' process, with early infall leading to the formation of the inner regions of the disk and later infall adding primarily to the outer regions. Knowing the distribution of specific angular momenta of the gas from which the disk formed it is possible to determine the distribution of radii at which that gas will settle and, consequently, the radial density profile. Of course, this assumes that the angular momenta of gas elements are unchanged after they arrive in the disk. In reality this may not be the case---viscosity in a gaseous disk may redistribute angular momentum and even in a purely stellar disk stars can scatter from perturbations such as spiral density waves and thereby change their angular momentum and may help to establish the exponential radial profile if the viscous timescale is of order the star formation timescale \pcite{silk_dissipational_1981,lin_formation_1987,clarke_chemical_1989,clarke_solar_1991,firmani_viscous_1996}. Current results suggest that exponential disks can plausibly arise given our present understanding of the initial distribution of angular momentum, star formation efficiency and supernovae feedback without the need for viscous redistribution of angular momentum \pcite{dutton_origin_2009,stringer_analytic_2010}, although the detailed properties (e.g. scale length distributions) of the disks do not necessarily agree perfectly with what is observed \pcite{dutton_revised_2007}. Of course, comparison with measurements of stellar disks requires determination of not just the mass density profile but the stellar density profile. Star formation will proceed most rapidly where the gas is densest (i.e. in the inner regions of the disk) and may be inhibited altogether at larger radii where the background radiation can prevent the formation of molecular hydrogen and where the disk may become stable to perturbations and so unable to form molecular cloud complexes \pcite{martin_star_2001,schaye_star_2004}.

The assumption that the angular momentum distribution of the disk remains constant after formation is not precisely correct even in the absence of hydrodynamical effects such as viscosity. Stars in a disk can interact with perturbations such as spiral density waves, exchanging energy and angular momentum with them. \cite{sellwood_radial_2002} show that interactions between stars and spiral waves at the corotation resonance lead to an exchange of energy and angular momentum that changes the radius of a star's orbit while keeping it circular. A star can, in principle, interact with multiple spiral waves and thereby migrate over a significant radial range. The question of how much radial mixing occurs due to this effect has been examined recently by \cite{roskar_ridingspiral_2008} and \cite{roskar_beyond_2008}. They find that this is a significant process which can lead to the formation of a population of stars beyond the star formation truncation radius in a disk, and which also flattens metallicity gradients (particularly in older stellar populations).

Disks are dynamically rather fragile systems and so not only must they form but they must survive across cosmological spans of time. This requires that the disk be stable to the growth of large scale perturbations which could destroy it. Additionally, in a hierarchical universe, survival is a non-trivial feat as the disk must survive frequent merging with other galaxies and constant bombardment by dark matter substructures orbiting within its own dark matter halo. Significant work has gone into studying the effects of dark matter substructures on galactic disks\footnote{Primarily because this is potentially a means by which to rule our the cold dark matter scenario: if the predicted abundance of substructure was sufficient to destroy galactic disks, the model could be ruled out.} \pcite{toth_galactic_1992,velazquez_sinking_1999,font_halo_2001,benson_heating_2004,kazantzidis_cold_2008,read_thin_2008,kazantzidis_cold_2009,purcell_destruction_2009}. The current consensus seems to be that the cold dark matter model is consistent with the presence of thin galactic disks, but only just so, and some fraction of disks have probably been significantly thickened or destroyed by interactions with dark matter substructures.

\subsection{The Overcooling Problem}\label{sec:Overcooling}

\cite{white_core_1978} proposed that galaxies would form as gas cools inside of dark matter halos and demonstrated that this provided a reasonable estimate of the typical mass scale of galaxies. This simple picture has a long standing problem however. The mass function of dark matter halos rises steeply (approximately as $M^{-1.9}$; \citealt{reed_halo_2007}) at low masses. Since cooling is very efficient in these low mass halos we might expect the galaxy mass and/or luminosity function to show a similar slope at the low mass/luminosity end. In fact, measured slopes are much shallower (typically around $-1$; e.g. \citealt{cole_2df_2001}). Rectifying this discrepancy is usually achieved by postulating some form of feedback, typically from supernovae, which can inhibit star formation in these low mass systems (e.g. by driving a wind out of the galaxy). However, as shown by \cite{benson_what_2003} this causes another problem---too much gas is now left over to accrete into massive halos (group and cluster mass objects) at late time wherein it cools and forms over-massive galaxies (much more massive than any galaxy observed).

This ``overcooling'' problem is not easy to solve, for the simple reason that the energy scales involved are much larger than for lower mass systems (the characteristic potential well depth of a dark matter halos scales as $M^{2/3}$ for halos of the same mean density). Several possible solutions have been proposed however, ranging from thermal conduction \pcite{benson_what_2003,dolag_thermal_2004,pope_effects_2005}, massive outflows \pcite{benson_what_2003}, multiphase cooling (in which gas cools but is locked up into clouds which are inefficiently transferred to the galaxy; \citealt{maller_multiphase_2004,kaufmann_redistributing_2009}) and feedback from active galactic nuclei.

The \AGN\ feedback scenario has gained considerable favor in the past few years for a variety of reasons. Firstly, observations have indicated that all galaxies seem to contain a central supermassive black hole with a mass that scales roughly in proportion to the mass of the galaxy \pcite{magorrian_demography_1998,gebhardt_relationship_2000,ferrarese_fundamental_2000,gebhardt_black_2000,tremaine_slope_2002}. The formation of these black holes must have involved the liberation of large amounts of energy and material sinks deep into the potential well of the black hole in some form of accretion flow \pcite{benson_what_2003}. If this energy can be successfully utilized to counteract the overly rapid cooling of gas onto the galaxy it would provide a natural source of sufficient energy available in every galaxy. Additionally, some feedback loop connection between supermassive black hole and galaxy formation of this type is attractive as it provides a means to explain the correlation between galaxy and black hole properties.

Feedback from \AGN\ can be divided into two categories: quasar mode and radio mode. The quasar mode is triggered when large amounts of gas are funneled into the circumnuclear disk from much larger scales by, for example, a merger between two galaxies in which torques act to transfer angular momentum from the gas, causing it to flow inwards. This is likely the dominant mechanism for black hole mass growth and results in a high (relative to Eddington) accretion rate --- most likely through a thin accretion disk --- and significant optical luminosity. In contrast, radio mode feedback occurs when the black hole is accreting at a more modest rate due to Bondi accretion from a diffuse hot atmosphere of gas and is in an optically dim but radio loud phase. In this phase, the black hole is thought to drive powerful jets \pcite{meier_magnetically_1999,meier_association_2001,villiers_magnetically_2005,nemmen_models_2007,komissarov_magnetic_2007,benson_maximum_2009} which can reach to large distances and have been seen to have a significant impact on their surroundings \pcite{birzan_systematic_2004}. The mechanism via which energy from the jets is efficiently coupled to the hot atmosphere of gas remains poorly understood: a combination of observational evidence (e.g. \citealt{owen_m87_2000}) and theoretical insights suggest that jets inflate bubbles or cavities in the hot atmosphere (see \S\ref{sec:AGNFeedback}). Despite these uncertainties, semi-analytic treatments which simply assume an efficient coupling have demonstrated that this can effectively shut down cooling in massive halos, resulting in a reduction in the masses of the largest galaxies and good agreement with luminosity functions and the bimodal distribution of 
galaxy colors \pcite{croton_many_2006,bower_breakinghierarchy_2006,somerville_semi-analytic_2008}. Observational evidence in support of \AGN\ feedback is beginning to emerge \pcite{schawinski_observational_2007} and seems to favor a radio mode scenario \pcite{schawinski_role_2009}.

\subsection{Local Group Dwarf Satellites}

As with any theory, galaxy formation theory is often best tested by exploring its predictions in extremes of physical conditions. For example, while our current theory may do well at explaining the properties of galaxies of average luminosity, it may perform less well in explaining the properties of the lowest mass galaxies that can form. Unfortunately, low mass galaxies are also low luminosity and so are difficult to observe. Fortunately, our own privileged position within the Milky Way and the Local Group means that there exists a population of faint, dwarf galaxies much closer by than we might otherwise expect. Study of the population of Local Group galaxies therefore has the potential to teach us much about galaxy formation at the lowest mass scales (and therefore, in a hierarchical Universe, at the earliest times). We must be cautious, however, of over-interpreting conclusions drawn from the Local Group which is, after all, just one small patch of the Universe.

Until the advent of the Sloan Digital Sky Survey around 11 satellite galaxies were known to exist within the virial radius of the Milky Way's halo\footnote{We are following the definition of virial radius given by \protect\cite{benson_effects_2002} here. Under this definition, 9 Local Group satellites are within the virial radius of M31, while a further 19 lie outside of the virial radii of the halos of both the Milky Way and M31.} \pcite{mateo_dwarf_1998}. The deep and uniform photometry of the SDSS has allowed many new such galaxies to be found by searching for overdensities in the stellar distribution in both position and color \pcite{willman_new_2005}. The current roster of dwarf satellites within the Milky Way's halo amounts to 24 \pcite{tollerud_hundreds_2008}. However, the SDSS covers only around one fifth of the sky and detection algorithms are imperfect. Recent estimates of the total number of dwarf satellites in the Milky Way's halo alone are of the order a few hundred \pcite{tollerud_hundreds_2008}. The faintest galaxy currently known, Segue 1, has a luminosity of only 340 times that of the Sun \pcite{belokurov_cats_2007}.

These dwarf galaxies are interesting from a theoretical perspective for several reasons. Firstly, their low mass makes them highly sensitive to feedback effects, both internal (from supernovae for example; \citealt{font_modelingmilky_2010}) and external (photoheating of the IGM by the entire population of galaxies can, in principle, inhibit the formation of these satellites; \citealt{gnedin_effect_2000,okamoto_mass_2008}). Furthermore, they may have gained at least some of their mass as a result of molecular hydrogen cooling, and have very low metallicities which may provide insight into the process of chemical enrichment. Beyond these galaxy-related aspects, the dwarf satellites appear to be highly dark matter dominated \pcite{strigari_common_2008}, making them excellent systems in which to explore the properties of dark matter and, therefore, to place interesting constraints on the behavior of dark matter on small scales.

Consequently, these galaxies have attracted significant theoretical attention over the past few years. In particular, when simulations demonstrated that dark matter substructure could survive within a host halo for significant periods of time, it became clear that the Milky Way's halo should contain thousands of small dark matter halos that could potentially have formed a dwarf galaxy within them \pcite{moore_dark_1999,kuhlen_via_2008,springel_aquarius_2008}. This was in stark contrast to the 10 or so satellites known at that time. This ``missing satellites'' problem appears to have been largely solved. Theoretical models have demonstrated that a combination of supernovae feedback, suppression of galaxy formation by a photoionizing background and tidal plus ram pressure stripping \pcite{bullock_reionization_2000,somerville_can_2002,benson_effects_2002,busha_impact_2009,maccio_origin_2009,wadepuhl_satellite_2010,font_modelingmilky_2010} can reduce the number of satellites significantly (leaving a population of truly dark subhalos; \citealt{kormendy_scaling_2004}), while the discovery of new satellites (and accounting for ones as yet undetected) has raised the observational target closer to the theoretical predictions.

Beyond this first order question though, there are many other interesting aspects to the Local Group satellites. For example, while they all contain populations of old stars, many of them show evidence of periodic star formation at much more recent times \pcite{tolstoy_star_2009}. What triggers this star formation? Given that feedback can strongly affect these systems, it is interesting to ask if we can gain insight into the feedback mechanism by studying these galaxies (e.g. their metallicities are highly sensitive to the strength of feedback so may place useful constraints on any feedback model; \citealt{font_modelingmilky_2010}). Finally, some of these galaxies must have been torn apart by the tidal field of the Milky Way (the Sagittarius dwarf is currently suffering this fate)---the dwarf satellite population may therefore have contributed to the build-up of the Milky Way's stellar halo \pcite{cooper_stellar_2009}, and there may be remnant tidal streams which could provide information about the hierarchical growth of the Milky Way's halo \pcite{bullock_tracing_2005,robertson__2005,font_chemical_2006,font_phase-space_2006,johnston_tracing_2008}.

\subsection{The Origins of the Hubble Sequence}

The variety of morphologies of galaxies is, perhaps, the most obvious observed characteristic of galaxies. Traditionally, morphology has been measured ``by eye'' by a trained observer who classifies each galaxy into a different morphological class based upon (amongst other things) the prominence of any central bulge, how concentrated the light distribution is and the presence or otherwise of dust lanes. \citeauthor{hubble_realm_1936}~(\citeyear{hubble_realm_1936}; see also \citealt{vaucouleurs_classification_1959}) placed galaxies into a morphological classification scheme using such an approach, and this basic morphological classification has persisted to the present \pcite{vaucouleurs_third_1991}. Applying this type of morphological classification to today's large datasets is difficult, but has been achieved by utilizing ``crowdsourcing'' techniques \pcite{lintott_galaxy_2008}. A key observational goal is to assess how the Hubble sequence evolves over time, as this should place strong constraints on theoretical models. This is observationally challenging, but progress is being made (e.g. \citealt{kriek_hubble_2009}).

Theoretically, morphological evolution is inevitable in a hierarchical universe in which galaxies interact with each other. Early work  demonstrated that major mergers between galaxies could transform disks into spheroids leading to morphological evolution \pcite{barnes_dynamics_1992}, but it is clear that the picture is more complicated, with secular evolution playing an important role in transforming disks into spheroids \pcite{kormendy_secular_2004} and major mergers of very gas rich systems can lead to the reformation of a disk after the merger is over \pcite{barnes_formation_2002,springel_formation_2005,robertson_merger-driven_2006}.

A fundamental difficulty in assessing the ability of any given theoretical model to explain the morphological properties of galaxies is that the definition of morphology itself is very complicated, and somewhat nebulously defined. This problem is beginning to be circumvented, both by the ability of numerical simulations of galaxy formation to produce realistic ``mock images'' of galaxies \pcite{jonsson_sunrise:_2006} which can, in principle, be classified by eye just as a real galaxy, and by the use of more quantitative morphological measures such as bulge--disk decomposition \citep{schade_evolution_1996,ratnatunga_disk_1999,simard_deep_2002,benson_luminosity_2007}, Gini-M20 \citep{lotz_new_2004}, concentration-asymmetry-smoothness \citep{watanabe_digital_1985, abraham_galaxy_1996, bell_optical_2003} and so on.

Understanding the build-up and evolution of the Hubble sequence of morphologies is, nevertheless, an important task for galaxy formation theory. The morphological structure of a galaxy clearly tells us something about its formation history and therefore captures information which its stellar population, for example, does not. Several studies have explored this aspect of galaxy formation theory \pcite{baugh_faint_1996,baugh_evolution_1996,kauffmann_age_1996,governato_origin_1999,firmani_physical_2003,lucia_formation_2006,lucia_hierarchical_2007,governato_forming_2007,parry_galaxy_2009,benson_origin_2009} with the general conclusion that hierarchical cosmologies can plausibly give rise to the observed mix of morphological types, although with significant uncertainties remaining in both the modeling of morphological transformation and in the comparison with observed morphologies. The consensus opinion is that massive elliptical galaxies form most of their stars in smaller progenitor galaxies and only assemble them into a single galaxy significantly later, while spiral galaxies are dominated by in situ star formation. Secular evolution of disks into spheroids also seems to be a crucial ingredient, particularly in the production of lower mass spheroidal systems \pcite{parry_galaxy_2009,kormendy_structure_2009}.

\section{Future Directions}

The fundamental goal of galaxy formation studies is to comprehend how the laws of nature turned a Gaussian random distribution of density fluctuations laid down by inflation into a complex population of galaxies seen at the present day. At this time, this author does not see any convincing evidence that any new physics is needed to explain the phenomena of galaxies\footnote{By ``new physics'' here I mean modifications to established physical laws, new forces or fields etc. Of course, dark matter and dark energy probably require ``new physics'' of one type or another, but I will leave those as a problem for cosmology\ldots}. The problem is more one of complexity: can we tease out the underlying mechanisms that drive different aspects of galaxy formation and evolution. The key here then is ``understanding''. One can easily comprehend how a $1/r^2$ force works and can, by extrapolation, understand how this force applies to the billions of particles of dark matter in an N-body simulation. However, it is not directly obvious (at least not to this author) how a $1/r^2$ force leads to the formation of complex filamentary structures and collapsed virialized objects. Instead, we have developed simplified analytic models (e.g. the Zel'dovich approximation, spherical top-hat collapse models etc.) which explain these phenomena in terms more accessible to the human intellect. It seems that this is what we must strive for in galaxy formation theory---a set of analytic models that we can comprehend and which allow us to understand the physics and a complementary set of precision numerical tools to allow us to determine the quantitative outcomes of that physics (in order to make precision tests of our understanding).

The division of galaxy formation models into N-body/hydro and semi-analytic is rather idealized. In reality there is significant overlap between the two---many semi-analytic models make use of dark matter halo merger trees drawn from N-body simulations while many hydrodynamics simulations include recipes for star formation and feedback which are semi-analytic in nature. It seems most likely that these two techniques will continue to develop and may, in fact, grow less distinguishable, incorporating aspects of each other into each other (once again, yin yang).

Galaxy formation benefits from a wealth of observational data, often to the degree that it is an observationally lead field in which theory plays the role of trying to explain observed phenomena. This deluge of data is unlikely to cease any time soon --- we can expect more and higher quality data on local galaxies and also the arrival of usefully sized datasets of galaxies at the highest redshifts. Galaxy formation theory should continue to attempt to develop our comprehension of these observed phenomena but should also strive to move into the regime of making true predictions for as-yet-unobserved regions of parameter space (e.g. the high redshift, $z\gsim 6$, Universe; e.g. \citealt{finlator_smoothly-rising_2010,lacey_evolution_2010}). Only in this way can we grow our confidence that we have truly understood the physics of galaxy formation.

\section{Summary}

Our theory of galaxy formation is gradually becoming more and more complete, but it is clear that large gaps in our understanding remain. This is, perhaps, not suprising --- galaxy formation incorporates a wide array of physical processes, many of which we can currently observe the consequences of only indirectly.  Most of the physics of galaxy formation is inherently nonlinear, making it difficult to obtain accurate solutions. Finally, what we observe from galaxies is usually several steps removed from the underlying physical properties (mass, density etc.) that we would ideally like to know about. Despite all of these difficulties, rapid progress is being made. The next decade should see this trend continuing via a combination of ever better observational datasets (both low and high redshift) and the continued development of novel theoretical tools.

\section*{Acknowledgments}

The author acknowledges the support of the Gordon \& Betty Moore Foundation. This work has benefited greatly from discussions with the Durham galaxy formation Group (Carlos Frenk, Shaun Cole, Richard Bower, Carlton Baugh, Cedric Lacey, Martin Stringer, Andreea Font and John Helly) and with Juna Kollmeier, Alyson Brooks, Annika Peter, Brant Robertson, TJ Cox, Dan Grin, Laura Book and Elisabeth Krause. The author would particularly like to thank Carlos Frenk for carefully reading through this manuscript and the anonymous referee for numerous valuable suggestions.

\bibliographystyle{model2-names}
\bibliography{BensonGalaxyFormationTheory}

\begin{thebibliography}{627}
\expandafter\ifx\csname natexlab\endcsname\relax\def\natexlab#1{#1}\fi
\expandafter\ifx\csname url\endcsname\relax
  \def\url#1{\texttt{#1}}\fi
\expandafter\ifx\csname urlprefix\endcsname\relax\def\urlprefix{URL }\fi
\providecommand{\eprint}[2][]{\url{#2}}
\providecommand{\bibinfo}[2]{#2}
\ifx\xfnm\relax \def\xfnm[#1]{\unskip,\space#1}\fi
\bibitem[{Abadi et~al.(1999)Abadi, Moore and Bower}]{abadi_ram_1999}
\bibinfo{author}{Abadi, M.G.}, \bibinfo{author}{Moore, B.},
  \bibinfo{author}{Bower, R.G.}, \bibinfo{year}{1999}.
\newblock \bibinfo{journal}{{MNRAS}} \bibinfo{volume}{308},
  \bibinfo{pages}{947--954}.
\bibitem[{Abadi et~al.(2009)Abadi, Navarro, Fardal, Babul and
  Steinmetz}]{abadi_galaxy-induced_2009}
\bibinfo{author}{Abadi, M.G.}, \bibinfo{author}{Navarro, J.F.},
  \bibinfo{author}{Fardal, M.}, \bibinfo{author}{Babul, A.},
  \bibinfo{author}{Steinmetz, M.}, \bibinfo{year}{2009}.
\bibitem[{Abadi et~al.(2003)Abadi, Navarro, Steinmetz and
  Eke}]{abadi_simulations_2003}
\bibinfo{author}{Abadi, M.G.}, \bibinfo{author}{Navarro, J.F.},
  \bibinfo{author}{Steinmetz, M.}, \bibinfo{author}{Eke, V.R.},
  \bibinfo{year}{2003}.
\newblock \bibinfo{journal}{Astrophysical Journal} \bibinfo{volume}{597},
  \bibinfo{pages}{21--34}.
\bibitem[{Abel et~al.(1997)Abel, Anninos, Zhang and
  Norman}]{abel_modeling_1997}
\bibinfo{author}{Abel, T.}, \bibinfo{author}{Anninos, P.},
  \bibinfo{author}{Zhang, Y.}, \bibinfo{author}{Norman, M.L.},
  \bibinfo{year}{1997}.
\newblock \bibinfo{journal}{New Astronomy} \bibinfo{volume}{2},
  \bibinfo{pages}{181--207}.
\bibitem[{Abel et~al.(2002)Abel, Bryan and Norman}]{abel_formation_2002}
\bibinfo{author}{Abel, T.}, \bibinfo{author}{Bryan, G.L.},
  \bibinfo{author}{Norman, M.L.}, \bibinfo{year}{2002}.
\newblock \bibinfo{journal}{Science} \bibinfo{volume}{295},
  \bibinfo{pages}{93--98}.
\bibitem[{Abel et~al.(1999)Abel, Norman and
  Madau}]{abel_photon-conserving_1999}
\bibinfo{author}{Abel, T.}, \bibinfo{author}{Norman, M.L.},
  \bibinfo{author}{Madau, P.}, \bibinfo{year}{1999}.
\newblock \bibinfo{journal}{Astrophysical Journal} \bibinfo{volume}{523},
  \bibinfo{pages}{66--71}.
\bibitem[{Abraham et~al.(1996)Abraham, Tanvir, Santiago, Ellis, Glazebrook and
  van~den Bergh}]{abraham_galaxy_1996}
\bibinfo{author}{Abraham, R.G.}, \bibinfo{author}{Tanvir, N.R.},
  \bibinfo{author}{Santiago, B.X.}, \bibinfo{author}{Ellis, R.S.},
  \bibinfo{author}{Glazebrook, K.}, \bibinfo{author}{van~den Bergh, S.},
  \bibinfo{year}{1996}.
\newblock \bibinfo{journal}{Monthly Notices of the Royal Astronomical Society}
  \bibinfo{volume}{279}, \bibinfo{pages}{L47}.
\bibitem[{Agertz et~al.(2007)Agertz, Moore, Stadel, Potter, Miniati, Read,
  Mayer, Gawryszczak, Kravtsov, Nordlund, Pearce, Quilis, Rudd, Springel,
  Stone, Tasker, Teyssier, Wadsley and Walder}]{agertz_fundamental_2007}
\bibinfo{author}{Agertz, O.}, \bibinfo{author}{Moore, B.},
  \bibinfo{author}{Stadel, J.}, \bibinfo{author}{Potter, D.},
  \bibinfo{author}{Miniati, F.}, \bibinfo{author}{Read, J.},
  \bibinfo{author}{Mayer, L.}, \bibinfo{author}{Gawryszczak, A.},
  \bibinfo{author}{Kravtsov, A.}, \bibinfo{author}{Nordlund, {\r{A}ke}.},
  \bibinfo{author}{Pearce, F.}, \bibinfo{author}{Quilis, V.},
  \bibinfo{author}{Rudd, D.}, \bibinfo{author}{Springel, V.},
  \bibinfo{author}{Stone, J.}, \bibinfo{author}{Tasker, E.},
  \bibinfo{author}{Teyssier, R.}, \bibinfo{author}{Wadsley, J.},
  \bibinfo{author}{Walder, R.}, \bibinfo{year}{2007}.
\newblock \bibinfo{journal}{Monthly Notices of the Royal Astronomical Society}
  \bibinfo{volume}{380}, \bibinfo{pages}{963--978}.
\bibitem[{Ahn and Shapiro(2005)}]{ahn_formation_2005}
\bibinfo{author}{Ahn, K.}, \bibinfo{author}{Shapiro, P.R.},
  \bibinfo{year}{2005}.
\newblock \bibinfo{journal}{Monthly Notices of the Royal Astronomical Society}
  \bibinfo{volume}{363}, \bibinfo{pages}{1092--1110}.
\bibitem[{Allen et~al.(2008)Allen, Groves, Dopita, Sutherland and
  Kewley}]{allen_mappings_2008}
\bibinfo{author}{Allen, M.G.}, \bibinfo{author}{Groves, B.A.},
  \bibinfo{author}{Dopita, M.A.}, \bibinfo{author}{Sutherland, R.S.},
  \bibinfo{author}{Kewley, L.J.}, \bibinfo{year}{2008}.
\newblock \bibinfo{journal}{Astrophysical Journal Supplement Series}
  \bibinfo{volume}{178}, \bibinfo{pages}{20--55}.
\bibitem[{Alvarez et~al.(2008)Alvarez, Busha, Abel and
  Wechsler}]{alvarez_connecting_2008}
\bibinfo{author}{Alvarez, M.A.}, \bibinfo{author}{Busha, M.T.},
  \bibinfo{author}{Abel, T.}, \bibinfo{author}{Wechsler, R.H.},
  \bibinfo{year}{2008}.
\bibitem[{Angulo et~al.(2008)Angulo, Lacey, Baugh and Frenk}]{angulo_fate_2008}
\bibinfo{author}{Angulo, R.E.}, \bibinfo{author}{Lacey, C.G.},
  \bibinfo{author}{Baugh, C.M.}, \bibinfo{author}{Frenk, C.S.},
  \bibinfo{year}{2008}.
\bibitem[{Angus and Zhao(2007)}]{angus_cold_2007}
\bibinfo{author}{Angus, G.W.}, \bibinfo{author}{Zhao, H.},
  \bibinfo{year}{2007}.
\newblock \bibinfo{journal}{Monthly Notices of the Royal Astronomical Society}
  \bibinfo{volume}{375}, \bibinfo{pages}{1146--1156}.
\bibitem[{Arad and {Lynden-Bell}(2005)}]{arad_inconsistency_2005}
\bibinfo{author}{Arad, I.}, \bibinfo{author}{{Lynden-Bell}, D.},
  \bibinfo{year}{2005}.
\newblock \bibinfo{journal}{Monthly Notices of the Royal Astronomical Society}
  \bibinfo{volume}{361}, \bibinfo{pages}{385--395}.
\bibitem[{Arons and Silk(1968)}]{arons_jeans_1968}
\bibinfo{author}{Arons, J.}, \bibinfo{author}{Silk, J.}, \bibinfo{year}{1968}.
\newblock \bibinfo{journal}{Monthly Notices of the Royal Astronomical Society}
  \bibinfo{volume}{140}, \bibinfo{pages}{331}.
\bibitem[{Arrigoni et~al.(2009)Arrigoni, Trager, Somerville and
  Gibson}]{arrigoni_galactic_2009}
\bibinfo{author}{Arrigoni, M.}, \bibinfo{author}{Trager, S.C.},
  \bibinfo{author}{Somerville, R.S.}, \bibinfo{author}{Gibson, B.K.},
  \bibinfo{year}{2009}.
\bibitem[{Athanassoula(2002)}]{athanassoula_bar-halo_2002}
\bibinfo{author}{Athanassoula, E.}, \bibinfo{year}{2002}.
\newblock \bibinfo{journal}{Astrophysical Journal} \bibinfo{volume}{569},
  \bibinfo{pages}{L83--L86}.
\bibitem[{Athanassoula(2008)}]{athanassoula_disc_2008}
\bibinfo{author}{Athanassoula, E.}, \bibinfo{year}{2008}.
\newblock \bibinfo{journal}{Monthly Notices of the Royal Astronomical Society}
  \bibinfo{volume}{390}, \bibinfo{pages}{L69--L72}.
\bibitem[{Aubert and Teyssier(2008)}]{aubert_radiative_2008}
\bibinfo{author}{Aubert, D.}, \bibinfo{author}{Teyssier, R.},
  \bibinfo{year}{2008}.
\newblock \bibinfo{journal}{Monthly Notices of the Royal Astronomical Society}
  \bibinfo{volume}{387}, \bibinfo{pages}{295--307}.
\bibitem[{Bahcall and Wolf(1976)}]{bahcall_star_1976}
\bibinfo{author}{Bahcall, J.N.}, \bibinfo{author}{Wolf, R.A.},
  \bibinfo{year}{1976}.
\newblock \bibinfo{journal}{Astrophysical Journal} \bibinfo{volume}{209},
  \bibinfo{pages}{214--232}.
\bibitem[{Bahcall et~al.(2004)Bahcall, Hao, Bode and
  Dong}]{bahcall_evolution_2004}
\bibinfo{author}{Bahcall, N.A.}, \bibinfo{author}{Hao, L.},
  \bibinfo{author}{Bode, P.}, \bibinfo{author}{Dong, F.}, \bibinfo{year}{2004}.
\newblock \bibinfo{journal}{Astrophysical Journal} \bibinfo{volume}{603},
  \bibinfo{pages}{1--6}.
\bibitem[{Banerjee et~al.(2006)Banerjee, Pudritz and
  Anderson}]{banerjee_supersonic_2006}
\bibinfo{author}{Banerjee, R.}, \bibinfo{author}{Pudritz, R.E.},
  \bibinfo{author}{Anderson, D.W.}, \bibinfo{year}{2006}.
\newblock \bibinfo{journal}{Monthly Notices of the Royal Astronomical Society}
  \bibinfo{volume}{373}, \bibinfo{pages}{1091--1106}.
\bibitem[{Bardeen(1970)}]{bardeen_kerr_1970}
\bibinfo{author}{Bardeen, J.M.}, \bibinfo{year}{1970}.
\newblock \bibinfo{journal}{Nature} \bibinfo{volume}{226}, \bibinfo{pages}{64}.
\bibitem[{Barnes and Efstathiou(1987)}]{barnes_angular_1987}
\bibinfo{author}{Barnes, J.}, \bibinfo{author}{Efstathiou, G.},
  \bibinfo{year}{1987}.
\newblock \bibinfo{journal}{{ApJ}} \bibinfo{volume}{319},
  \bibinfo{pages}{575--600}.
\bibitem[{Barnes(2002)}]{barnes_formation_2002}
\bibinfo{author}{Barnes, J.E.}, \bibinfo{year}{2002}.
\newblock \bibinfo{journal}{Monthly Notices of the Royal Astronomical Society}
  \bibinfo{volume}{333}, \bibinfo{pages}{481--494}.
\bibitem[{Barnes and Hernquist(1992)}]{barnes_dynamics_1992}
\bibinfo{author}{Barnes, J.E.}, \bibinfo{author}{Hernquist, L.},
  \bibinfo{year}{1992}.
\newblock \bibinfo{journal}{Annual Review of Astronomy and Astrophysics}
  \bibinfo{volume}{30}, \bibinfo{pages}{705--742}.
\bibitem[{Baugh(2006)}]{baugh_primerhierarchical_2006}
\bibinfo{author}{Baugh, C.M.}, \bibinfo{year}{2006}.
\newblock \bibinfo{journal}{Reports on Progress in Physics}
  \bibinfo{volume}{69}, \bibinfo{pages}{3101--3156}.
\bibitem[{Baugh et~al.(1999a)Baugh, Benson, Cole, Frenk and
  Lacey}]{baugh_modellingevolution_1999}
\bibinfo{author}{Baugh, C.M.}, \bibinfo{author}{Benson, A.J.},
  \bibinfo{author}{Cole, S.}, \bibinfo{author}{Frenk, C.S.},
  \bibinfo{author}{Lacey, C.G.}, \bibinfo{year}{1999}a.
\newblock \bibinfo{journal}{{MNRAS}} \bibinfo{volume}{305},
  \bibinfo{pages}{L21--L25}.
\bibitem[{Baugh et~al.(1996a)Baugh, Cole and Frenk}]{baugh_faint_1996}
\bibinfo{author}{Baugh, C.M.}, \bibinfo{author}{Cole, S.},
  \bibinfo{author}{Frenk, C.S.}, \bibinfo{year}{1996}a.
\newblock \bibinfo{journal}{Monthly Notices of the Royal Astronomical Society}
  \bibinfo{volume}{282}, \bibinfo{pages}{L27--L32}.
\bibitem[{Baugh et~al.(1996b)Baugh, Cole and Frenk}]{baugh_evolution_1996}
\bibinfo{author}{Baugh, C.M.}, \bibinfo{author}{Cole, S.},
  \bibinfo{author}{Frenk, C.S.}, \bibinfo{year}{1996}b.
\newblock \bibinfo{journal}{Monthly Notices of the Royal Astronomical Society}
  \bibinfo{volume}{283}, \bibinfo{pages}{1361--1378}.
\bibitem[{Baugh et~al.(1998)Baugh, Cole, Frenk and Lacey}]{baugh_epoch_1998}
\bibinfo{author}{Baugh, C.M.}, \bibinfo{author}{Cole, S.},
  \bibinfo{author}{Frenk, C.S.}, \bibinfo{author}{Lacey, C.G.},
  \bibinfo{year}{1998}.
\newblock \bibinfo{journal}{{ApJ}} \bibinfo{volume}{498}, \bibinfo{pages}{504}.
\bibitem[{Baugh et~al.(1999b)Baugh, Lacey, Cole and
  Frenk}]{baugh_semianalytic_1999}
\bibinfo{author}{Baugh, C.M.}, \bibinfo{author}{Lacey, C.G.},
  \bibinfo{author}{Cole, S.}, \bibinfo{author}{Frenk, C.S.},
  \bibinfo{year}{1999}b.
\newblock in: \bibinfo{booktitle}{The Most Distant Radio Galaxies},
  \bibinfo{publisher}{Royal Netherlands Academy of Arts and Science},
  \bibinfo{address}{Amsterdam}. p. \bibinfo{pages}{265}.
\bibitem[{Baugh et~al.(2004)Baugh, Lacey, Frenk, Benson, Cole, Granato, Silva
  and Bressan}]{baugh_predictions_2004}
\bibinfo{author}{Baugh, C.M.}, \bibinfo{author}{Lacey, C.G.},
  \bibinfo{author}{Frenk, C.S.}, \bibinfo{author}{Benson, A.J.},
  \bibinfo{author}{Cole, S.}, \bibinfo{author}{Granato, G.L.},
  \bibinfo{author}{Silva, L.}, \bibinfo{author}{Bressan, A.},
  \bibinfo{year}{2004}.
\newblock \bibinfo{journal}{New Astronomy Review} \bibinfo{volume}{48},
  \bibinfo{pages}{1239--1246}.
\bibitem[{Baugh et~al.(2005)Baugh, Lacey, Frenk, Granato, Silva, Bressan,
  Benson and Cole}]{baugh_canfaint_2005}
\bibinfo{author}{Baugh, C.M.}, \bibinfo{author}{Lacey, C.G.},
  \bibinfo{author}{Frenk, C.S.}, \bibinfo{author}{Granato, G.L.},
  \bibinfo{author}{Silva, L.}, \bibinfo{author}{Bressan, A.},
  \bibinfo{author}{Benson, A.J.}, \bibinfo{author}{Cole, S.},
  \bibinfo{year}{2005}.
\newblock \bibinfo{journal}{{MNRAS}} \bibinfo{volume}{356},
  \bibinfo{pages}{1191--1200}.
\bibitem[{Begelman et~al.(1980)Begelman, Blandford and
  Rees}]{begelman_massive_1980}
\bibinfo{author}{Begelman, M.C.}, \bibinfo{author}{Blandford, R.D.},
  \bibinfo{author}{Rees, M.J.}, \bibinfo{year}{1980}.
\newblock \bibinfo{journal}{Nature} \bibinfo{volume}{287},
  \bibinfo{pages}{307--309}.
\bibitem[{Begelman and Nath(2005)}]{begelman_self-regulated_2005}
\bibinfo{author}{Begelman, M.C.}, \bibinfo{author}{Nath, B.B.},
  \bibinfo{year}{2005}.
\newblock \bibinfo{journal}{Monthly Notices of the Royal Astronomical Society}
  \bibinfo{volume}{361}, \bibinfo{pages}{1387--1392}.
\bibitem[{Behroozi et~al.(2010)Behroozi, Conroy and
  Wechsler}]{behroozi_comprehensive_2010}
\bibinfo{author}{Behroozi, P.S.}, \bibinfo{author}{Conroy, C.},
  \bibinfo{author}{Wechsler, R.H.}, \bibinfo{year}{2010}.
\newblock
  \bibinfo{howpublished}{{http://adsabs.harvard.edu/abs/2010arXiv1001.0015B}}.
\bibitem[{Bell et~al.(2003)Bell, {McIntosh}, Katz and
  Weinberg}]{bell_optical_2003}
\bibinfo{author}{Bell, E.F.}, \bibinfo{author}{{McIntosh}, D.H.},
  \bibinfo{author}{Katz, N.}, \bibinfo{author}{Weinberg, M.D.},
  \bibinfo{year}{2003}.
\newblock \bibinfo{journal}{Astrophysical Journal Supplement Series}
  \bibinfo{volume}{149}, \bibinfo{pages}{289--312}.
\bibitem[{Belokurov et~al.(2006)Belokurov, Zucker, Evans, Gilmore, Vidrih,
  Bramich, Newberg, Wyse, Irwin, Fellhauer, Hewett, Walton, Wilkinson, Cole,
  Yanny, Rockosi, Beers, Bell, Brinkmann, Ivezi\'c and
  Lupton}]{belokurov_field_2006}
\bibinfo{author}{Belokurov, V.}, \bibinfo{author}{Zucker, D.B.},
  \bibinfo{author}{Evans, N.W.}, \bibinfo{author}{Gilmore, G.},
  \bibinfo{author}{Vidrih, S.}, \bibinfo{author}{Bramich, D.M.},
  \bibinfo{author}{Newberg, H.J.}, \bibinfo{author}{Wyse, R.F.G.},
  \bibinfo{author}{Irwin, M.J.}, \bibinfo{author}{Fellhauer, M.},
  \bibinfo{author}{Hewett, P.C.}, \bibinfo{author}{Walton, N.A.},
  \bibinfo{author}{Wilkinson, M.I.}, \bibinfo{author}{Cole, N.},
  \bibinfo{author}{Yanny, B.}, \bibinfo{author}{Rockosi, C.M.},
  \bibinfo{author}{Beers, T.C.}, \bibinfo{author}{Bell, E.F.},
  \bibinfo{author}{Brinkmann, J.}, \bibinfo{author}{Ivezi\'c, Å.},
  \bibinfo{author}{Lupton, R.}, \bibinfo{year}{2006}.
\newblock \bibinfo{journal}{The Astrophysical Journal} \bibinfo{volume}{642},
  \bibinfo{pages}{L137--L140}.
\bibitem[{Belokurov et~al.(2007)Belokurov, Zucker, Evans, Kleyna, Koposov,
  Hodgkin, Irwin, Gilmore, Wilkinson, Fellhauer, Bramich, Hewett, Vidrih, Jong,
  Smith, Rix, Bell, Wyse, Newberg, Mayeur, Yanny, Rockosi, Gnedin, Schneider,
  Beers, Barentine, Brewington, Brinkmann, Harvanek, Kleinman, Krzesinski,
  Long, Nitta and Snedden}]{belokurov_cats_2007}
\bibinfo{author}{Belokurov, V.}, \bibinfo{author}{Zucker, D.B.},
  \bibinfo{author}{Evans, N.W.}, \bibinfo{author}{Kleyna, J.T.},
  \bibinfo{author}{Koposov, S.}, \bibinfo{author}{Hodgkin, S.T.},
  \bibinfo{author}{Irwin, M.J.}, \bibinfo{author}{Gilmore, G.},
  \bibinfo{author}{Wilkinson, M.I.}, \bibinfo{author}{Fellhauer, M.},
  \bibinfo{author}{Bramich, D.M.}, \bibinfo{author}{Hewett, P.C.},
  \bibinfo{author}{Vidrih, S.}, \bibinfo{author}{Jong, J.T.A.D.},
  \bibinfo{author}{Smith, J.A.}, \bibinfo{author}{Rix, H.},
  \bibinfo{author}{Bell, E.F.}, \bibinfo{author}{Wyse, R.F.G.},
  \bibinfo{author}{Newberg, H.J.}, \bibinfo{author}{Mayeur, P.A.},
  \bibinfo{author}{Yanny, B.}, \bibinfo{author}{Rockosi, C.M.},
  \bibinfo{author}{Gnedin, O.Y.}, \bibinfo{author}{Schneider, D.P.},
  \bibinfo{author}{Beers, T.C.}, \bibinfo{author}{Barentine, J.C.},
  \bibinfo{author}{Brewington, H.}, \bibinfo{author}{Brinkmann, J.},
  \bibinfo{author}{Harvanek, M.}, \bibinfo{author}{Kleinman, S.J.},
  \bibinfo{author}{Krzesinski, J.}, \bibinfo{author}{Long, D.},
  \bibinfo{author}{Nitta, A.}, \bibinfo{author}{Snedden, S.A.},
  \bibinfo{year}{2007}.
\newblock \bibinfo{journal}{Astrophysical Journal} \bibinfo{volume}{654},
  \bibinfo{pages}{897--906}.
\bibitem[{Benson and Devereux(2009)}]{benson_origin_2009}
\bibinfo{author}{Benson, A.}, \bibinfo{author}{Devereux, N.A.},
  \bibinfo{year}{2009}.
\newblock \bibinfo{journal}{Monthly Notices of the Royal Astronomical Society}
  \bibinfo{volume}{402}, \bibinfo{pages}{2321--2334}.
\bibitem[{Benson(2005)}]{benson_orbital_2005}
\bibinfo{author}{Benson, A.J.}, \bibinfo{year}{2005}.
\newblock \bibinfo{journal}{{MNRAS}} \bibinfo{volume}{358},
  \bibinfo{pages}{551--562}.
\bibitem[{Benson(2008)}]{benson_constraining_2008}
\bibinfo{author}{Benson, A.J.}, \bibinfo{year}{2008}.
\newblock \bibinfo{journal}{Monthly Notices of the Royal Astronomical Society}
  \bibinfo{volume}{388}, \bibinfo{pages}{1361--1371}.
\bibitem[{Benson and Babul(2009)}]{benson_maximum_2009}
\bibinfo{author}{Benson, A.J.}, \bibinfo{author}{Babul, A.},
  \bibinfo{year}{2009}.
\bibitem[{Benson et~al.(2000a)Benson, Baugh, Cole, Frenk and
  Lacey}]{benson_dependence_2000}
\bibinfo{author}{Benson, A.J.}, \bibinfo{author}{Baugh, C.M.},
  \bibinfo{author}{Cole, S.}, \bibinfo{author}{Frenk, C.S.},
  \bibinfo{author}{Lacey, C.G.}, \bibinfo{year}{2000}a.
\newblock \bibinfo{journal}{{MNRAS}} \bibinfo{volume}{316},
  \bibinfo{pages}{107--119}.
\bibitem[{Benson and Bower(2010)}]{benson_galaxy_2010}
\bibinfo{author}{Benson, A.J.}, \bibinfo{author}{Bower, R.},
  \bibinfo{year}{2010}.
\newblock \bibinfo{journal}{Monthly Notices of the Royal Astronomical Society}
  , \bibinfo{pages}{no--no}.
\bibitem[{Benson et~al.(2003)Benson, Bower, Frenk, Lacey, Baugh and
  Cole}]{benson_what_2003}
\bibinfo{author}{Benson, A.J.}, \bibinfo{author}{Bower, R.G.},
  \bibinfo{author}{Frenk, C.S.}, \bibinfo{author}{Lacey, C.G.},
  \bibinfo{author}{Baugh, C.M.}, \bibinfo{author}{Cole, S.},
  \bibinfo{year}{2003}.
\newblock \bibinfo{journal}{Astrophysical Journal} \bibinfo{volume}{599},
  \bibinfo{pages}{38--49}.
\bibitem[{Benson et~al.(2000b)Benson, Bower, Frenk and
  White}]{benson_diffuse_2000}
\bibinfo{author}{Benson, A.J.}, \bibinfo{author}{Bower, R.G.},
  \bibinfo{author}{Frenk, C.S.}, \bibinfo{author}{White, S.D.M.},
  \bibinfo{year}{2000}b.
\newblock \bibinfo{journal}{Monthly Notices of the Royal Astronomical Society}
  \bibinfo{volume}{314}, \bibinfo{pages}{557--565}.
\bibitem[{Benson et~al.(2000c)Benson, Cole, Frenk, Baugh and
  Lacey}]{benson_nature_2000}
\bibinfo{author}{Benson, A.J.}, \bibinfo{author}{Cole, S.},
  \bibinfo{author}{Frenk, C.S.}, \bibinfo{author}{Baugh, C.M.},
  \bibinfo{author}{Lacey, C.G.}, \bibinfo{year}{2000}c.
\newblock \bibinfo{journal}{{MNRAS}} \bibinfo{volume}{311},
  \bibinfo{pages}{793--808}.
\bibitem[{Benson et~al.(2007)Benson, Dzanovic, Frenk and
  Sharples}]{benson_luminosity_2007}
\bibinfo{author}{Benson, A.J.}, \bibinfo{author}{Dzanovic, D.},
  \bibinfo{author}{Frenk, C.S.}, \bibinfo{author}{Sharples, R.},
  \bibinfo{year}{2007}.
\newblock \bibinfo{journal}{Monthly Notices of the Royal Astronomical Society}
  \bibinfo{volume}{379}, \bibinfo{pages}{841--866}.
\bibitem[{Benson et~al.(2005)Benson, Kamionkowski and
  Hassani}]{benson_self-consistent_2005}
\bibinfo{author}{Benson, A.J.}, \bibinfo{author}{Kamionkowski, M.},
  \bibinfo{author}{Hassani, S.H.}, \bibinfo{year}{2005}.
\newblock \bibinfo{journal}{Monthly Notices of the Royal Astronomical Society}
  \bibinfo{volume}{357}, \bibinfo{pages}{847--858}.
\bibitem[{Benson et~al.(2002a)Benson, Lacey, Baugh, Cole and
  Frenk}]{benson_effects_2002}
\bibinfo{author}{Benson, A.J.}, \bibinfo{author}{Lacey, C.G.},
  \bibinfo{author}{Baugh, C.M.}, \bibinfo{author}{Cole, S.},
  \bibinfo{author}{Frenk, C.S.}, \bibinfo{year}{2002}a.
\newblock \bibinfo{journal}{{MNRAS}} \bibinfo{volume}{333},
  \bibinfo{pages}{156--176}.
\bibitem[{Benson et~al.(2002b)Benson, Lacey, Baugh, Cole and
  Frenk}]{benson_effects_2002-1}
\bibinfo{author}{Benson, A.J.}, \bibinfo{author}{Lacey, C.G.},
  \bibinfo{author}{Baugh, C.M.}, \bibinfo{author}{Cole, S.},
  \bibinfo{author}{Frenk, C.S.}, \bibinfo{year}{2002}b.
\newblock \bibinfo{journal}{Monthly Notices of the Royal Astronomical Society}
  \bibinfo{volume}{333}, \bibinfo{pages}{156--176}.
\bibitem[{Benson et~al.(2004)Benson, Lacey, Frenk, Baugh and
  Cole}]{benson_heating_2004}
\bibinfo{author}{Benson, A.J.}, \bibinfo{author}{Lacey, C.G.},
  \bibinfo{author}{Frenk, C.S.}, \bibinfo{author}{Baugh, C.M.},
  \bibinfo{author}{Cole, S.}, \bibinfo{year}{2004}.
\newblock \bibinfo{journal}{{MNRAS}} \bibinfo{volume}{351},
  \bibinfo{pages}{1215--1236}.
\bibitem[{Benson and Madau(2003)}]{benson_early_2003}
\bibinfo{author}{Benson, A.J.}, \bibinfo{author}{Madau, P.},
  \bibinfo{year}{2003}.
\newblock \bibinfo{journal}{Monthly Notices of the Royal Astronomical Society}
  \bibinfo{volume}{344}, \bibinfo{pages}{835--846}.
\bibitem[{Benson et~al.(2001a)Benson, Nusser, Sugiyama and
  Lacey}]{benson_non-uniform_2001}
\bibinfo{author}{Benson, A.J.}, \bibinfo{author}{Nusser, A.},
  \bibinfo{author}{Sugiyama, N.}, \bibinfo{author}{Lacey, C.G.},
  \bibinfo{year}{2001}a.
\newblock \bibinfo{journal}{{MNRAS}} \bibinfo{volume}{320},
  \bibinfo{pages}{153--176}.
\bibitem[{Benson et~al.(2001b)Benson, Pearce, Frenk, Baugh and
  Jenkins}]{benson_comparison_2001}
\bibinfo{author}{Benson, A.J.}, \bibinfo{author}{Pearce, F.R.},
  \bibinfo{author}{Frenk, C.S.}, \bibinfo{author}{Baugh, C.M.},
  \bibinfo{author}{Jenkins, A.}, \bibinfo{year}{2001}b.
\newblock \bibinfo{journal}{Monthly Notices of the Royal Astronomical Society}
  \bibinfo{volume}{320}, \bibinfo{pages}{261--280}.
\bibitem[{Benson et~al.(2006)Benson, Sugiyama, Nusser and
  Lacey}]{benson_epoch_2006}
\bibinfo{author}{Benson, A.J.}, \bibinfo{author}{Sugiyama, N.},
  \bibinfo{author}{Nusser, A.}, \bibinfo{author}{Lacey, C.G.},
  \bibinfo{year}{2006}.
\newblock \bibinfo{journal}{{MNRAS}} \bibinfo{volume}{369},
  \bibinfo{pages}{1055--1080}.
\bibitem[{Berezinsky et~al.(2006)Berezinsky, Dokuchaev and
  Eroshenko}]{berezinsky_destruction_2006}
\bibinfo{author}{Berezinsky, V.}, \bibinfo{author}{Dokuchaev, V.},
  \bibinfo{author}{Eroshenko, Y.}, \bibinfo{year}{2006}.
\newblock \bibinfo{journal}{Physical Review D} \bibinfo{volume}{73},
  \bibinfo{pages}{63504}.
\bibitem[{Berti and Volonteri(2008)}]{berti_cosmological_2008}
\bibinfo{author}{Berti, E.}, \bibinfo{author}{Volonteri, M.},
  \bibinfo{year}{2008}.
\newblock \bibinfo{journal}{Astrophysical Journal} \bibinfo{volume}{684},
  \bibinfo{pages}{822--828}.
\bibitem[{Bertin and Romeo(1988)}]{bertin_global_1988}
\bibinfo{author}{Bertin, G.}, \bibinfo{author}{Romeo, A.B.},
  \bibinfo{year}{1988}.
\newblock \bibinfo{journal}{Astronomy and Astrophysics} \bibinfo{volume}{195},
  \bibinfo{pages}{105--113}.
\bibitem[{Bertone et~al.(2005)Bertone, Hooper and Silk}]{bertone_particle_2005}
\bibinfo{author}{Bertone, G.}, \bibinfo{author}{Hooper, D.},
  \bibinfo{author}{Silk, J.}, \bibinfo{year}{2005}.
\newblock \bibinfo{journal}{Physics Reports} \bibinfo{volume}{405},
  \bibinfo{pages}{279--390}.
\bibitem[{Bertschinger(1985)}]{bertschinger_self-similar_1985}
\bibinfo{author}{Bertschinger, E.}, \bibinfo{year}{1985}.
\newblock \bibinfo{journal}{Astrophysical Journal Supplement Series}
  \bibinfo{volume}{58}, \bibinfo{pages}{39--65}.
\bibitem[{Bertschinger(1998)}]{bertschinger_simulations_1998}
\bibinfo{author}{Bertschinger, E.}, \bibinfo{year}{1998}.
\newblock \bibinfo{journal}{Annual Review of Astronomy and Astrophysics}
  \bibinfo{volume}{36}, \bibinfo{pages}{599--654}.
\bibitem[{Bett et~al.(2007)Bett, Eke, Frenk, Jenkins, Helly and
  Navarro}]{bett_spin_2007}
\bibinfo{author}{Bett, P.}, \bibinfo{author}{Eke, V.}, \bibinfo{author}{Frenk,
  C.S.}, \bibinfo{author}{Jenkins, A.}, \bibinfo{author}{Helly, J.},
  \bibinfo{author}{Navarro, J.}, \bibinfo{year}{2007}.
\newblock \bibinfo{journal}{{MNRAS}} \bibinfo{volume}{376},
  \bibinfo{pages}{215--232}.
\bibitem[{Bett et~al.(2009)Bett, Eke, Frenk, Jenkins and
  Okamoto}]{bett_angular_2009}
\bibinfo{author}{Bett, P.}, \bibinfo{author}{Eke, V.}, \bibinfo{author}{Frenk,
  C.S.}, \bibinfo{author}{Jenkins, A.}, \bibinfo{author}{Okamoto, T.},
  \bibinfo{year}{2009}.
\bibitem[{Binney(1977a)}]{binney_physics_1977}
\bibinfo{author}{Binney, J.}, \bibinfo{year}{1977}a.
\newblock \bibinfo{journal}{Astrophysical Journal} \bibinfo{volume}{215},
  \bibinfo{pages}{483--491}.
\bibitem[{Binney(1977b)}]{binney_dynamical_1977}
\bibinfo{author}{Binney, J.}, \bibinfo{year}{1977}b.
\newblock \bibinfo{journal}{Monthly Notices of the Royal Astronomical Society}
  \bibinfo{volume}{181}, \bibinfo{pages}{735--746}.
\bibitem[{Binney and Tremaine(2008)}]{binney_galactic_2008}
\bibinfo{author}{Binney, J.}, \bibinfo{author}{Tremaine, S.},
  \bibinfo{year}{2008}.
\bibitem[{Birnboim and Dekel(2003)}]{birnboim_virial_2003}
\bibinfo{author}{Birnboim, Y.}, \bibinfo{author}{Dekel, A.},
  \bibinfo{year}{2003}.
\newblock \bibinfo{journal}{Monthly Notices of the Royal Astronomical Society}
  \bibinfo{volume}{345}, \bibinfo{pages}{349--364}.
\bibitem[{Birzan et~al.(2004)Birzan, Rafferty, {McNamara}, Wise and
  Nulsen}]{birzan_systematic_2004}
\bibinfo{author}{Birzan, L.}, \bibinfo{author}{Rafferty, D.A.},
  \bibinfo{author}{{McNamara}, B.R.}, \bibinfo{author}{Wise, M.W.},
  \bibinfo{author}{Nulsen, P.E.J.}, \bibinfo{year}{2004}.
\newblock \bibinfo{journal}{Astrophysical Journal} \bibinfo{volume}{607},
  \bibinfo{pages}{800--809}.
\bibitem[{Blaizot et~al.(2003)Blaizot, Guiderdoni, Devriendt, Bouchet and
  Hatton}]{blaizot_predicting_2003}
\bibinfo{author}{Blaizot, J.}, \bibinfo{author}{Guiderdoni, B.},
  \bibinfo{author}{Devriendt, J.E.G.}, \bibinfo{author}{Bouchet, F.R.},
  \bibinfo{author}{Hatton, S.}, \bibinfo{year}{2003}.
\newblock \bibinfo{journal}{Astrophysics and Space Science}
  \bibinfo{volume}{284}, \bibinfo{pages}{373--376}.
\bibitem[{Blaizot et~al.(2004)Blaizot, Guiderdoni, Devriendt, Bouchet, Hatton
  and Stoehr}]{blaizot_galics-_2004}
\bibinfo{author}{Blaizot, J.}, \bibinfo{author}{Guiderdoni, B.},
  \bibinfo{author}{Devriendt, J.E.G.}, \bibinfo{author}{Bouchet, F.R.},
  \bibinfo{author}{Hatton, S.J.}, \bibinfo{author}{Stoehr, F.},
  \bibinfo{year}{2004}.
\newblock \bibinfo{journal}{{MNRAS}} \bibinfo{volume}{352},
  \bibinfo{pages}{571--588}.
\bibitem[{Blaizot et~al.(2006)Blaizot, Szapudi, Colombi, Budav\`ari, Bouchet,
  Devriendt, Guiderdoni, Pan and Szalay}]{blaizot_galics_2006}
\bibinfo{author}{Blaizot, J.}, \bibinfo{author}{Szapudi, I.},
  \bibinfo{author}{Colombi, S.}, \bibinfo{author}{Budav\`ari, T.},
  \bibinfo{author}{Bouchet, F.R.}, \bibinfo{author}{Devriendt, J.E.G.},
  \bibinfo{author}{Guiderdoni, B.}, \bibinfo{author}{Pan, J.},
  \bibinfo{author}{Szalay, A.}, \bibinfo{year}{2006}.
\newblock \bibinfo{journal}{{MNRAS}} \bibinfo{volume}{369},
  \bibinfo{pages}{1009--1020}.
\bibitem[{Blanchet et~al.(2005)Blanchet, Qusailah and
  Will}]{blanchet_gravitational_2005}
\bibinfo{author}{Blanchet, L.}, \bibinfo{author}{Qusailah, M.S.S.},
  \bibinfo{author}{Will, C.M.}, \bibinfo{year}{2005}.
\newblock \bibinfo{journal}{Astrophysical Journal} \bibinfo{volume}{635},
  \bibinfo{pages}{508--515}.
\bibitem[{{Bland-Hawthorn}(2009)}]{bland-hawthorn_warm_2009}
\bibinfo{author}{{Bland-Hawthorn}, J.}, \bibinfo{year}{2009}.
\newblock in: \bibinfo{booktitle}{The Galaxy Disk in Cosmological Context},
  \bibinfo{publisher}{International Astronomical Union}. pp.
  \bibinfo{pages}{241--254}.
\bibitem[{Blandford and Payne(1982)}]{blandford_hydromagnetic_1982}
\bibinfo{author}{Blandford, R.D.}, \bibinfo{author}{Payne, D.G.},
  \bibinfo{year}{1982}.
\newblock \bibinfo{journal}{Monthly Notices of the Royal Astronomical Society}
  \bibinfo{volume}{199}, \bibinfo{pages}{883--903}.
\bibitem[{Blandford and Znajek(1977)}]{blandford_electromagnetic_1977}
\bibinfo{author}{Blandford, R.D.}, \bibinfo{author}{Znajek, R.L.},
  \bibinfo{year}{1977}.
\newblock \bibinfo{journal}{Monthly Notices of the Royal Astronomical Society}
  \bibinfo{volume}{179}, \bibinfo{pages}{433--456}.
\bibitem[{Blumenthal et~al.(1986)Blumenthal, Faber, Flores and
  Primack}]{blumenthal_contraction_1986}
\bibinfo{author}{Blumenthal, G.R.}, \bibinfo{author}{Faber, S.M.},
  \bibinfo{author}{Flores, R.}, \bibinfo{author}{Primack, J.R.},
  \bibinfo{year}{1986}.
\newblock \bibinfo{journal}{Astrophysical Journal} \bibinfo{volume}{301},
  \bibinfo{pages}{27--34}.
\bibitem[{Blumenthal et~al.(1984)Blumenthal, Faber, Primack and
  Rees}]{blumenthal_formation_1984}
\bibinfo{author}{Blumenthal, G.R.}, \bibinfo{author}{Faber, S.M.},
  \bibinfo{author}{Primack, J.R.}, \bibinfo{author}{Rees, M.J.},
  \bibinfo{year}{1984}.
\newblock \bibinfo{journal}{Nature} \bibinfo{volume}{311},
  \bibinfo{pages}{517--525}.
\bibitem[{Bode et~al.(2001)Bode, Ostriker and Turok}]{bode_halo_2001}
\bibinfo{author}{Bode, P.}, \bibinfo{author}{Ostriker, J.P.},
  \bibinfo{author}{Turok, N.}, \bibinfo{year}{2001}.
\newblock \bibinfo{journal}{Astrophysical Journal} \bibinfo{volume}{556},
  \bibinfo{pages}{93--107}.
\bibitem[{Boehm and Schaeffer(2005)}]{boehm_constraintsdark_2005}
\bibinfo{author}{Boehm, C.}, \bibinfo{author}{Schaeffer, R.},
  \bibinfo{year}{2005}.
\newblock \bibinfo{journal}{Astronomy and Astrophysics} \bibinfo{volume}{438},
  \bibinfo{pages}{419--442}.
\bibitem[{Bond et~al.(1991)Bond, Cole, Efstathiou and
  Kaiser}]{bond_excursion_1991}
\bibinfo{author}{Bond, J.R.}, \bibinfo{author}{Cole, S.},
  \bibinfo{author}{Efstathiou, G.}, \bibinfo{author}{Kaiser, N.},
  \bibinfo{year}{1991}.
\newblock \bibinfo{journal}{Astrophysical Journal} \bibinfo{volume}{379},
  \bibinfo{pages}{440--460}.
\bibitem[{Bondi and Hoyle(1944)}]{bondi_mechanism_1944}
\bibinfo{author}{Bondi, H.}, \bibinfo{author}{Hoyle, F.}, \bibinfo{year}{1944}.
\newblock \bibinfo{journal}{Monthly Notices of the Royal Astronomical Society}
  \bibinfo{volume}{104}, \bibinfo{pages}{273}.
\bibitem[{Book and Benson(2010)}]{book_role_2010}
\bibinfo{author}{Book, L.G.}, \bibinfo{author}{Benson, A.J.},
  \bibinfo{year}{2010}.
\newblock
  \bibinfo{howpublished}{{http://adsabs.harvard.edu/abs/2010arXiv1001.2305B}}.
\bibitem[{Booth and Schaye(2009)}]{booth_cosmological_2009}
\bibinfo{author}{Booth, C.M.}, \bibinfo{author}{Schaye, J.},
  \bibinfo{year}{2009}.
\newblock \bibinfo{journal}{Monthly Notices of the Royal Astronomical Society}
  \bibinfo{volume}{398}, \bibinfo{pages}{53--74}.
\bibitem[{Borgani et~al.(2001)Borgani, Governato, Wadsley, Menci, Tozzi, Lake,
  Quinn and Stadel}]{borgani_preheatingintracluster_2001}
\bibinfo{author}{Borgani, S.}, \bibinfo{author}{Governato, F.},
  \bibinfo{author}{Wadsley, J.}, \bibinfo{author}{Menci, N.},
  \bibinfo{author}{Tozzi, P.}, \bibinfo{author}{Lake, G.},
  \bibinfo{author}{Quinn, T.}, \bibinfo{author}{Stadel, J.},
  \bibinfo{year}{2001}.
\newblock \bibinfo{journal}{Astrophysical Journal} \bibinfo{volume}{559},
  \bibinfo{pages}{L71--L74}.
\bibitem[{van~den Bosch(2002)}]{van_den_bosch_universal_2002}
\bibinfo{author}{van~den Bosch, F.C.}, \bibinfo{year}{2002}.
\newblock \bibinfo{journal}{Monthly Notices of the Royal Astronomical Society}
  \bibinfo{volume}{331}, \bibinfo{pages}{98--110}.
\bibitem[{van~den Bosch et~al.(2002)van~den Bosch, Abel, Croft, Hernquist and
  White}]{bosch_angular_2002}
\bibinfo{author}{van~den Bosch, F.C.}, \bibinfo{author}{Abel, T.},
  \bibinfo{author}{Croft, R.A.C.}, \bibinfo{author}{Hernquist, L.},
  \bibinfo{author}{White, S.D.M.}, \bibinfo{year}{2002}.
\newblock \bibinfo{journal}{Astrophysical Journal} \bibinfo{volume}{576},
  \bibinfo{pages}{21--35}.
\bibitem[{van~den Bosch et~al.(2003)van~den Bosch, Abel and
  Hernquist}]{bosch_angular_2003}
\bibinfo{author}{van~den Bosch, F.C.}, \bibinfo{author}{Abel, T.},
  \bibinfo{author}{Hernquist, L.}, \bibinfo{year}{2003}.
\newblock \bibinfo{journal}{Monthly Notices of the Royal Astronomical Society}
  \bibinfo{volume}{346}, \bibinfo{pages}{177--185}.
\bibitem[{Bournaud and Combes(2002)}]{bournaud_gas_2002}
\bibinfo{author}{Bournaud, F.}, \bibinfo{author}{Combes, F.},
  \bibinfo{year}{2002}.
\newblock \bibinfo{journal}{Astronomy and Astrophysics} \bibinfo{volume}{392},
  \bibinfo{pages}{83--102}.
\bibitem[{Bournaud et~al.(2005)Bournaud, Jog and Combes}]{bournaud_galaxy_2005}
\bibinfo{author}{Bournaud, F.}, \bibinfo{author}{Jog, C.J.},
  \bibinfo{author}{Combes, F.}, \bibinfo{year}{2005}.
\newblock \bibinfo{journal}{Astronomy and Astrophysics} \bibinfo{volume}{437},
  \bibinfo{pages}{69--85}.
\bibitem[{Bower(1991)}]{bower_evolution_1991}
\bibinfo{author}{Bower, R.G.}, \bibinfo{year}{1991}.
\newblock \bibinfo{journal}{Monthly Notices of the Royal Astronomical Society}
  \bibinfo{volume}{248}, \bibinfo{pages}{332--352}.
\bibitem[{Bower et~al.(2001)Bower, Benson, Lacey, Baugh, Cole and
  Frenk}]{bower_impact_2001}
\bibinfo{author}{Bower, R.G.}, \bibinfo{author}{Benson, A.J.},
  \bibinfo{author}{Lacey, C.G.}, \bibinfo{author}{Baugh, C.M.},
  \bibinfo{author}{Cole, S.}, \bibinfo{author}{Frenk, C.S.},
  \bibinfo{year}{2001}.
\newblock \bibinfo{journal}{{MNRAS}} \bibinfo{volume}{325},
  \bibinfo{pages}{497--508}.
\bibitem[{Bower et~al.(2006)Bower, Benson, Malbon, Helly, Frenk, Baugh, Cole
  and Lacey}]{bower_breakinghierarchy_2006}
\bibinfo{author}{Bower, R.G.}, \bibinfo{author}{Benson, A.J.},
  \bibinfo{author}{Malbon, R.}, \bibinfo{author}{Helly, J.C.},
  \bibinfo{author}{Frenk, C.S.}, \bibinfo{author}{Baugh, C.M.},
  \bibinfo{author}{Cole, S.}, \bibinfo{author}{Lacey, C.G.},
  \bibinfo{year}{2006}.
\newblock \bibinfo{journal}{{MNRAS}} \bibinfo{volume}{370},
  \bibinfo{pages}{645--655}.
\bibitem[{Bower et~al.(2008)Bower, {McCarthy} and Benson}]{bower_flip_2008}
\bibinfo{author}{Bower, R.G.}, \bibinfo{author}{{McCarthy}, I.G.},
  \bibinfo{author}{Benson, A.J.}, \bibinfo{year}{2008}.
\newblock \bibinfo{journal}{{MNRAS}} \bibinfo{volume}{390},
  \bibinfo{pages}{1399--1410}.
\bibitem[{Boyarsky et~al.(2008)Boyarsky, Lesgourgues, Ruchayskiy and
  Viel}]{boyarsky_lyman-alpha_2008}
\bibinfo{author}{Boyarsky, A.}, \bibinfo{author}{Lesgourgues, J.},
  \bibinfo{author}{Ruchayskiy, O.}, \bibinfo{author}{Viel, M.},
  \bibinfo{year}{2008}.
\bibitem[{{Boylan-Kolchin} et~al.(2008){Boylan-Kolchin}, Ma and
  Quataert}]{boylan-kolchin_dynamical_2008}
\bibinfo{author}{{Boylan-Kolchin}, M.}, \bibinfo{author}{Ma, C.},
  \bibinfo{author}{Quataert, E.}, \bibinfo{year}{2008}.
\newblock \bibinfo{journal}{{MNRAS}} \bibinfo{volume}{383},
  \bibinfo{pages}{93--101}.
\bibitem[{Boyle et~al.(2008)Boyle, Kesden and Nissanke}]{boyle_binary_2008}
\bibinfo{author}{Boyle, L.}, \bibinfo{author}{Kesden, M.},
  \bibinfo{author}{Nissanke, S.}, \bibinfo{year}{2008}.
\newblock \bibinfo{journal}{Physical Review Letters} \bibinfo{volume}{100},
  \bibinfo{pages}{151101}.
\bibitem[{Bromm et~al.(2002)Bromm, Coppi and Larson}]{bromm_formation_2002}
\bibinfo{author}{Bromm, V.}, \bibinfo{author}{Coppi, P.S.},
  \bibinfo{author}{Larson, R.B.}, \bibinfo{year}{2002}.
\newblock \bibinfo{journal}{Astrophysical Journal} \bibinfo{volume}{564},
  \bibinfo{pages}{23--51}.
\bibitem[{Bromm et~al.(2009)Bromm, Yoshida, Hernquist and
  {McKee}}]{bromm_formation_2009}
\bibinfo{author}{Bromm, V.}, \bibinfo{author}{Yoshida, N.},
  \bibinfo{author}{Hernquist, L.}, \bibinfo{author}{{McKee}, C.F.},
  \bibinfo{year}{2009}.
\newblock \bibinfo{journal}{Nature} \bibinfo{volume}{459},
  \bibinfo{pages}{49--54}.
\bibitem[{Brook et~al.(2004)Brook, Kawata, Gibson and
  Freeman}]{brook_emergence_2004}
\bibinfo{author}{Brook, C.B.}, \bibinfo{author}{Kawata, D.},
  \bibinfo{author}{Gibson, B.K.}, \bibinfo{author}{Freeman, K.C.},
  \bibinfo{year}{2004}.
\newblock \bibinfo{journal}{Astrophysical Journal} \bibinfo{volume}{612},
  \bibinfo{pages}{894--899}.
\bibitem[{Brook et~al.(2007)Brook, Kawata, Scannapieco, Martel and
  Gibson}]{brook_spatial_2007}
\bibinfo{author}{Brook, C.B.}, \bibinfo{author}{Kawata, D.},
  \bibinfo{author}{Scannapieco, E.}, \bibinfo{author}{Martel, H.},
  \bibinfo{author}{Gibson, B.K.}, \bibinfo{year}{2007}.
\newblock \bibinfo{journal}{Astrophysical Journal} \bibinfo{volume}{661},
  \bibinfo{pages}{10--18}.
\bibitem[{Brooks et~al.(2009)Brooks, Governato, Quinn, Brook and
  Wadsley}]{brooks_role_2009}
\bibinfo{author}{Brooks, A.M.}, \bibinfo{author}{Governato, F.},
  \bibinfo{author}{Quinn, T.}, \bibinfo{author}{Brook, C.B.},
  \bibinfo{author}{Wadsley, J.}, \bibinfo{year}{2009}.
\newblock \bibinfo{journal}{Astrophysical Journal} \bibinfo{volume}{694},
  \bibinfo{pages}{396--410}.
\bibitem[{Br\"uggen and Lucia(2008)}]{bruggen_ram-pressure_2008}
\bibinfo{author}{Br\"uggen, M.}, \bibinfo{author}{Lucia, G.D.},
  \bibinfo{year}{2008}.
\newblock \bibinfo{journal}{Monthly Notices of the Royal Astronomical Society}
  \bibinfo{volume}{383}, \bibinfo{pages}{1336--1342}.
\bibitem[{Br\"uggen and Scannapieco(2009)}]{bruggen_self-regulation_2009}
\bibinfo{author}{Br\"uggen, M.}, \bibinfo{author}{Scannapieco, E.},
  \bibinfo{year}{2009}.
\newblock \bibinfo{journal}{Monthly Notices of the Royal Astronomical Society}
  \bibinfo{volume}{398}, \bibinfo{pages}{548--560}.
\bibitem[{Bruzual and Charlot(2003)}]{bruzual_stellar_2003}
\bibinfo{author}{Bruzual, G.}, \bibinfo{author}{Charlot, S.},
  \bibinfo{year}{2003}.
\newblock \bibinfo{journal}{{MNRAS}} \bibinfo{volume}{344},
  \bibinfo{pages}{1000--1028}.
\bibitem[{Bullock et~al.(2001a)Bullock, Dekel, Kolatt, Kravtsov, Klypin,
  Porciani and Primack}]{bullock_universal_2001}
\bibinfo{author}{Bullock, J.S.}, \bibinfo{author}{Dekel, A.},
  \bibinfo{author}{Kolatt, T.S.}, \bibinfo{author}{Kravtsov, A.V.},
  \bibinfo{author}{Klypin, A.A.}, \bibinfo{author}{Porciani, C.},
  \bibinfo{author}{Primack, J.R.}, \bibinfo{year}{2001}a.
\newblock \bibinfo{journal}{The Astrophysical Journal} \bibinfo{volume}{555},
  \bibinfo{pages}{240--257}.
\bibitem[{Bullock and Johnston(2005)}]{bullock_tracing_2005}
\bibinfo{author}{Bullock, J.S.}, \bibinfo{author}{Johnston, K.V.},
  \bibinfo{year}{2005}.
\newblock \bibinfo{journal}{Astrophysical Journal} \bibinfo{volume}{635},
  \bibinfo{pages}{931--949}.
\bibitem[{Bullock et~al.(2001b)Bullock, Kolatt, Sigad, Somerville, Kravtsov,
  Klypin, Primack and Dekel}]{bullock_profiles_2001}
\bibinfo{author}{Bullock, J.S.}, \bibinfo{author}{Kolatt, T.S.},
  \bibinfo{author}{Sigad, Y.}, \bibinfo{author}{Somerville, R.S.},
  \bibinfo{author}{Kravtsov, A.V.}, \bibinfo{author}{Klypin, A.A.},
  \bibinfo{author}{Primack, J.R.}, \bibinfo{author}{Dekel, A.},
  \bibinfo{year}{2001}b.
\newblock \bibinfo{journal}{Monthly Notices of the Royal Astronomical Society}
  \bibinfo{volume}{321}, \bibinfo{pages}{559--575}.
\bibitem[{Bullock et~al.(2000)Bullock, Kravtsov and
  Weinberg}]{bullock_reionization_2000}
\bibinfo{author}{Bullock, J.S.}, \bibinfo{author}{Kravtsov, A.V.},
  \bibinfo{author}{Weinberg, D.H.}, \bibinfo{year}{2000}.
\newblock \bibinfo{journal}{Astrophysical Journal} \bibinfo{volume}{539},
  \bibinfo{pages}{517--521}.
\bibitem[{Burkert and Tremaine(2010)}]{burkert_correlation_2010}
\bibinfo{author}{Burkert, A.}, \bibinfo{author}{Tremaine, S.},
  \bibinfo{year}{2010}.
\newblock
  \bibinfo{howpublished}{{http://adsabs.harvard.edu/abs/2010arXiv1004.0137B}}.
\bibitem[{Busha et~al.(2009)Busha, Alvarez, Wechsler, Abel and
  Strigari}]{busha_impact_2009}
\bibinfo{author}{Busha, M.T.}, \bibinfo{author}{Alvarez, M.A.},
  \bibinfo{author}{Wechsler, R.H.}, \bibinfo{author}{Abel, T.},
  \bibinfo{author}{Strigari, L.E.}, \bibinfo{year}{2009}.
\bibitem[{Cardone et~al.(2005)Cardone, Piedipalumbo and
  Tortora}]{cardone_spherical_2005}
\bibinfo{author}{Cardone, V.F.}, \bibinfo{author}{Piedipalumbo, E.},
  \bibinfo{author}{Tortora, C.}, \bibinfo{year}{2005}.
\newblock \bibinfo{journal}{{MNRAS}} \bibinfo{volume}{358},
  \bibinfo{pages}{1325--1336}.
\bibitem[{Cattaneo et~al.(2006)Cattaneo, Dekel, Devriendt, Guiderdoni and
  Blaizot}]{cattaneo_modellinggalaxy_2006}
\bibinfo{author}{Cattaneo, A.}, \bibinfo{author}{Dekel, A.},
  \bibinfo{author}{Devriendt, J.}, \bibinfo{author}{Guiderdoni, B.},
  \bibinfo{author}{Blaizot, J.}, \bibinfo{year}{2006}.
\newblock \bibinfo{journal}{Monthly Notices of the Royal Astronomical Society}
  \bibinfo{volume}{370}, \bibinfo{pages}{1651--1665}.
\bibitem[{Cazaux and Spaans(2004)}]{cazaux_molecular_2004}
\bibinfo{author}{Cazaux, S.}, \bibinfo{author}{Spaans, M.},
  \bibinfo{year}{2004}.
\newblock \bibinfo{journal}{{ApJ}} \bibinfo{volume}{611},
  \bibinfo{pages}{40--51}.
\bibitem[{Chabrier(2003)}]{chabrier_galactic_2003}
\bibinfo{author}{Chabrier, G.}, \bibinfo{year}{2003}.
\newblock \bibinfo{journal}{Publications of the Astronomical Society of the
  Pacific} \bibinfo{volume}{115}, \bibinfo{pages}{763--795}.
\bibitem[{Chakrabarti and Whitney(2009)}]{chakrabarti_panchromatic_2009}
\bibinfo{author}{Chakrabarti, S.}, \bibinfo{author}{Whitney, B.A.},
  \bibinfo{year}{2009}.
\newblock \bibinfo{journal}{Astrophysical Journal} \bibinfo{volume}{690},
  \bibinfo{pages}{1432--1451}.
\bibitem[{Chandrasekhar(1943)}]{chandrasekhar_dynamical_1943}
\bibinfo{author}{Chandrasekhar, S.}, \bibinfo{year}{1943}.
\newblock \bibinfo{journal}{Astrophysical Journal} \bibinfo{volume}{97},
  \bibinfo{pages}{255}.
\bibitem[{Chary(2008)}]{chary_stellar_2008}
\bibinfo{author}{Chary, R.}, \bibinfo{year}{2008}.
\newblock \bibinfo{journal}{Astrophysical Journal} \bibinfo{volume}{680},
  \bibinfo{pages}{32--40}.
\bibitem[{Chiu and Ostriker(2000)}]{chiu_semianalytic_2000}
\bibinfo{author}{Chiu, W.A.}, \bibinfo{author}{Ostriker, J.P.},
  \bibinfo{year}{2000}.
\newblock \bibinfo{journal}{Astrophysical Journal} \bibinfo{volume}{534},
  \bibinfo{pages}{507--532}.
\bibitem[{Christodoulou et~al.(1995a)Christodoulou, Kazanas, Shlosman and
  Tohline}]{christodoulou_phase-transition_1995}
\bibinfo{author}{Christodoulou, D.M.}, \bibinfo{author}{Kazanas, D.},
  \bibinfo{author}{Shlosman, I.}, \bibinfo{author}{Tohline, J.E.},
  \bibinfo{year}{1995}a.
\newblock \bibinfo{journal}{Astrophysical Journal} \bibinfo{volume}{446},
  \bibinfo{pages}{472}.
\bibitem[{Christodoulou et~al.(1995b)Christodoulou, Shlosman and
  Tohline}]{christodoulou_new_1995}
\bibinfo{author}{Christodoulou, D.M.}, \bibinfo{author}{Shlosman, I.},
  \bibinfo{author}{Tohline, J.E.}, \bibinfo{year}{1995}b.
\newblock \bibinfo{journal}{Astrophysical Journal} \bibinfo{volume}{443},
  \bibinfo{pages}{551--562}.
\bibitem[{Ciardi et~al.(2001)Ciardi, Ferrara, Marri and
  Raimondo}]{ciardi_cosmological_2001}
\bibinfo{author}{Ciardi, B.}, \bibinfo{author}{Ferrara, A.},
  \bibinfo{author}{Marri, S.}, \bibinfo{author}{Raimondo, G.},
  \bibinfo{year}{2001}.
\newblock \bibinfo{journal}{Monthly Notices of the Royal Astronomical Society}
  \bibinfo{volume}{324}, \bibinfo{pages}{381--388}.
\bibitem[{Ciardi et~al.(2003)Ciardi, Ferrara and White}]{ciardi_early_2003}
\bibinfo{author}{Ciardi, B.}, \bibinfo{author}{Ferrara, A.},
  \bibinfo{author}{White, S.D.M.}, \bibinfo{year}{2003}.
\newblock \bibinfo{journal}{Monthly Notices of the Royal Astronomical Society}
  \bibinfo{volume}{344}, \bibinfo{pages}{L7--L11}.
\bibitem[{Ciotti(2009)}]{ciotti_agn_2009}
\bibinfo{author}{Ciotti, L.}, \bibinfo{year}{2009}.
\bibitem[{Ciotti et~al.(2009)Ciotti, Ostriker and
  Proga}]{ciotti_feedbackcentral_2009}
\bibinfo{author}{Ciotti, L.}, \bibinfo{author}{Ostriker, J.P.},
  \bibinfo{author}{Proga, D.}, \bibinfo{year}{2009}.
\newblock \bibinfo{journal}{The Astrophysical Journal} \bibinfo{volume}{699},
  \bibinfo{pages}{89--104}.
\bibitem[{Clarke(1989)}]{clarke_chemical_1989}
\bibinfo{author}{Clarke, C.J.}, \bibinfo{year}{1989}.
\newblock \bibinfo{journal}{Monthly Notices of the Royal Astronomical Society}
  \bibinfo{volume}{238}, \bibinfo{pages}{283--292}.
\bibitem[{Clarke(1991)}]{clarke_solar_1991}
\bibinfo{author}{Clarke, C.J.}, \bibinfo{year}{1991}.
\newblock \bibinfo{journal}{Monthly Notices of the Royal Astronomical Society}
  \bibinfo{volume}{249}, \bibinfo{pages}{704--706}.
\bibitem[{Cole(1991)}]{cole_modeling_1991}
\bibinfo{author}{Cole, S.}, \bibinfo{year}{1991}.
\newblock \bibinfo{journal}{Astrophysical Journal} \bibinfo{volume}{367},
  \bibinfo{pages}{45--53}.
\bibitem[{Cole et~al.(1994)Cole, {Aragon-Salamanca}, Frenk, Navarro and
  Zepf}]{cole_recipe_1994}
\bibinfo{author}{Cole, S.}, \bibinfo{author}{{Aragon-Salamanca}, A.},
  \bibinfo{author}{Frenk, C.S.}, \bibinfo{author}{Navarro, J.F.},
  \bibinfo{author}{Zepf, S.E.}, \bibinfo{year}{1994}.
\newblock \bibinfo{journal}{Monthly Notices of the Royal Astronomical Society}
  \bibinfo{volume}{271}, \bibinfo{pages}{781}.
\bibitem[{Cole et~al.(2008)Cole, Helly, Frenk and
  Parkinson}]{cole_statistical_2008}
\bibinfo{author}{Cole, S.}, \bibinfo{author}{Helly, J.},
  \bibinfo{author}{Frenk, C.S.}, \bibinfo{author}{Parkinson, H.},
  \bibinfo{year}{2008}.
\newblock \bibinfo{journal}{Monthly Notices of the Royal Astronomical Society}
  \bibinfo{volume}{383}, \bibinfo{pages}{546--556}.
\bibitem[{Cole and Lacey(1996)}]{cole_structure_1996}
\bibinfo{author}{Cole, S.}, \bibinfo{author}{Lacey, C.}, \bibinfo{year}{1996}.
\newblock \bibinfo{journal}{{MNRAS}} \bibinfo{volume}{281},
  \bibinfo{pages}{716}.
\bibitem[{Cole et~al.(2000)Cole, Lacey, Baugh and
  Frenk}]{cole_hierarchical_2000}
\bibinfo{author}{Cole, S.}, \bibinfo{author}{Lacey, C.G.},
  \bibinfo{author}{Baugh, C.M.}, \bibinfo{author}{Frenk, C.S.},
  \bibinfo{year}{2000}.
\newblock \bibinfo{journal}{{MNRAS}} \bibinfo{volume}{319},
  \bibinfo{pages}{168--204}.
\bibitem[{Cole et~al.(2001)Cole, Norberg, Baugh, Frenk, {Bland-Hawthorn},
  Bridges, Cannon, Colless, Collins, Couch, Cross, Dalton, Propris, Driver,
  Efstathiou, Ellis, Glazebrook, Jackson, Lahav, Lewis, Lumsden, Maddox,
  Madgwick, Peacock, Peterson, Sutherland and Taylor}]{cole_2df_2001}
\bibinfo{author}{Cole, S.}, \bibinfo{author}{Norberg, P.},
  \bibinfo{author}{Baugh, C.M.}, \bibinfo{author}{Frenk, C.S.},
  \bibinfo{author}{{Bland-Hawthorn}, J.}, \bibinfo{author}{Bridges, T.},
  \bibinfo{author}{Cannon, R.}, \bibinfo{author}{Colless, M.},
  \bibinfo{author}{Collins, C.}, \bibinfo{author}{Couch, W.},
  \bibinfo{author}{Cross, N.}, \bibinfo{author}{Dalton, G.},
  \bibinfo{author}{Propris, R.D.}, \bibinfo{author}{Driver, S.P.},
  \bibinfo{author}{Efstathiou, G.}, \bibinfo{author}{Ellis, R.S.},
  \bibinfo{author}{Glazebrook, K.}, \bibinfo{author}{Jackson, C.},
  \bibinfo{author}{Lahav, O.}, \bibinfo{author}{Lewis, I.},
  \bibinfo{author}{Lumsden, S.}, \bibinfo{author}{Maddox, S.},
  \bibinfo{author}{Madgwick, D.}, \bibinfo{author}{Peacock, J.A.},
  \bibinfo{author}{Peterson, B.A.}, \bibinfo{author}{Sutherland, W.},
  \bibinfo{author}{Taylor, K.}, \bibinfo{year}{2001}.
\newblock \bibinfo{journal}{Monthly Notices of the Royal Astronomical Society}
  \bibinfo{volume}{326}, \bibinfo{pages}{255--273}.
\bibitem[{Cole et~al.(2005)Cole, Percival, Peacock, Norberg, Baugh, Frenk,
  Baldry, {Bland-Hawthorn}, Bridges, Cannon, Colless, Collins, Couch, Cross,
  Dalton, Eke, Propris, Driver, Efstathiou, Ellis, Glazebrook, Jackson,
  Jenkins, Lahav, Lewis, Lumsden, Maddox, Madgwick, Peterson, Sutherland and
  Taylor}]{cole_2df_2005}
\bibinfo{author}{Cole, S.}, \bibinfo{author}{Percival, W.J.},
  \bibinfo{author}{Peacock, J.A.}, \bibinfo{author}{Norberg, P.},
  \bibinfo{author}{Baugh, C.M.}, \bibinfo{author}{Frenk, C.S.},
  \bibinfo{author}{Baldry, I.}, \bibinfo{author}{{Bland-Hawthorn}, J.},
  \bibinfo{author}{Bridges, T.}, \bibinfo{author}{Cannon, R.},
  \bibinfo{author}{Colless, M.}, \bibinfo{author}{Collins, C.},
  \bibinfo{author}{Couch, W.}, \bibinfo{author}{Cross, N.J.G.},
  \bibinfo{author}{Dalton, G.}, \bibinfo{author}{Eke, V.R.},
  \bibinfo{author}{Propris, R.D.}, \bibinfo{author}{Driver, S.P.},
  \bibinfo{author}{Efstathiou, G.}, \bibinfo{author}{Ellis, R.S.},
  \bibinfo{author}{Glazebrook, K.}, \bibinfo{author}{Jackson, C.},
  \bibinfo{author}{Jenkins, A.}, \bibinfo{author}{Lahav, O.},
  \bibinfo{author}{Lewis, I.}, \bibinfo{author}{Lumsden, S.},
  \bibinfo{author}{Maddox, S.}, \bibinfo{author}{Madgwick, D.},
  \bibinfo{author}{Peterson, B.A.}, \bibinfo{author}{Sutherland, W.},
  \bibinfo{author}{Taylor, K.}, \bibinfo{year}{2005}.
\newblock \bibinfo{journal}{Monthly Notices of the Royal Astronomical Society}
  \bibinfo{volume}{362}, \bibinfo{pages}{505--534}.
\bibitem[{Col\'in et~al.(2002)Col\'in, {Avila-Reese}, Valenzuela and
  Firmani}]{colin_structure_2002}
\bibinfo{author}{Col\'in, P.}, \bibinfo{author}{{Avila-Reese}, V.},
  \bibinfo{author}{Valenzuela, O.}, \bibinfo{author}{Firmani, C.},
  \bibinfo{year}{2002}.
\newblock \bibinfo{journal}{Astrophysical Journal} \bibinfo{volume}{581},
  \bibinfo{pages}{777--793}.
\bibitem[{Collins et~al.(2009)Collins, Xu, Norman, Li and
  Li}]{collins_cosmological_2009}
\bibinfo{author}{Collins, D.C.}, \bibinfo{author}{Xu, H.},
  \bibinfo{author}{Norman, M.L.}, \bibinfo{author}{Li, H.},
  \bibinfo{author}{Li, S.}, \bibinfo{year}{2009}.
\bibitem[{Conroy and Gunn(2010)}]{conroy_propagation_2010}
\bibinfo{author}{Conroy, C.}, \bibinfo{author}{Gunn, J.E.},
  \bibinfo{year}{2010}.
\newblock \bibinfo{journal}{The Astrophysical Journal} \bibinfo{volume}{712},
  \bibinfo{pages}{833--857}.
\bibitem[{Conroy et~al.(2008)Conroy, Gunn and White}]{conroy_propagation_2008}
\bibinfo{author}{Conroy, C.}, \bibinfo{author}{Gunn, J.E.},
  \bibinfo{author}{White, M.}, \bibinfo{year}{2008}.
\bibitem[{Conroy and Ostriker(2008)}]{conroy_thermal_2008}
\bibinfo{author}{Conroy, C.}, \bibinfo{author}{Ostriker, J.P.},
  \bibinfo{year}{2008}.
\newblock \bibinfo{journal}{The Astrophysical Journal} \bibinfo{volume}{681},
  \bibinfo{pages}{151--166}.
\bibitem[{Conroy and Wechsler(2009)}]{conroy_connecting_2009}
\bibinfo{author}{Conroy, C.}, \bibinfo{author}{Wechsler, R.H.},
  \bibinfo{year}{2009}.
\newblock \bibinfo{journal}{The Astrophysical Journal} \bibinfo{volume}{696},
  \bibinfo{pages}{620--635}.
\bibitem[{Cooper et~al.(2009)Cooper, Cole, Frenk, White, Helly, Benson, Lucia,
  Helmi, Jenkins, Navarro, Springel and Wang}]{cooper_stellar_2009}
\bibinfo{author}{Cooper, A.P.}, \bibinfo{author}{Cole, S.},
  \bibinfo{author}{Frenk, C.S.}, \bibinfo{author}{White, S.D.M.},
  \bibinfo{author}{Helly, J.C.}, \bibinfo{author}{Benson, A.J.},
  \bibinfo{author}{Lucia, G.D.}, \bibinfo{author}{Helmi, A.},
  \bibinfo{author}{Jenkins, A.}, \bibinfo{author}{Navarro, J.F.},
  \bibinfo{author}{Springel, V.}, \bibinfo{author}{Wang, J.},
  \bibinfo{year}{2009}.
\newblock \bibinfo{journal}{{MNRAS}} \bibinfo{volume}{submitted}.
\bibitem[{Cooray and Sheth(2002)}]{cooray_halo_2002}
\bibinfo{author}{Cooray, A.}, \bibinfo{author}{Sheth, R.},
  \bibinfo{year}{2002}.
\newblock \bibinfo{journal}{Physics Reports} \bibinfo{volume}{372},
  \bibinfo{pages}{1--129}.
\bibitem[{Covington et~al.(2008)Covington, Dekel, Cox, Jonsson and
  Primack}]{covington_predictingproperties_2008}
\bibinfo{author}{Covington, M.}, \bibinfo{author}{Dekel, A.},
  \bibinfo{author}{Cox, T.J.}, \bibinfo{author}{Jonsson, P.},
  \bibinfo{author}{Primack, J.R.}, \bibinfo{year}{2008}.
\newblock \bibinfo{journal}{Monthly Notices of the Royal Astronomical Society}
  \bibinfo{volume}{384}, \bibinfo{pages}{94--106}.
\bibitem[{Cox et~al.(2006)Cox, Jonsson, Primack and
  Somerville}]{cox_feedback_2006}
\bibinfo{author}{Cox, T.J.}, \bibinfo{author}{Jonsson, P.},
  \bibinfo{author}{Primack, J.R.}, \bibinfo{author}{Somerville, R.S.},
  \bibinfo{year}{2006}.
\newblock \bibinfo{journal}{Monthly Notices of the Royal Astronomical Society}
  \bibinfo{volume}{373}, \bibinfo{pages}{1013--1038}.
\bibitem[{Cox et~al.(2008)Cox, Jonsson, Somerville, Primack and
  Dekel}]{cox_effect_2008}
\bibinfo{author}{Cox, T.J.}, \bibinfo{author}{Jonsson, P.},
  \bibinfo{author}{Somerville, R.S.}, \bibinfo{author}{Primack, J.R.},
  \bibinfo{author}{Dekel, A.}, \bibinfo{year}{2008}.
\newblock \bibinfo{journal}{Monthly Notices of the Royal Astronomical Society}
  \bibinfo{volume}{384}, \bibinfo{pages}{386--409}.
\bibitem[{Crain et~al.(2009)Crain, Theuns, Vecchia, Eke, Frenk, Jenkins, Kay,
  Peacock, Pearce, Schaye, Springel, Thomas, White and
  Wiersma}]{crain_galaxies-intergalactic_2009}
\bibinfo{author}{Crain, R.A.}, \bibinfo{author}{Theuns, T.},
  \bibinfo{author}{Vecchia, C.D.}, \bibinfo{author}{Eke, V.R.},
  \bibinfo{author}{Frenk, C.S.}, \bibinfo{author}{Jenkins, A.},
  \bibinfo{author}{Kay, S.T.}, \bibinfo{author}{Peacock, J.A.},
  \bibinfo{author}{Pearce, F.R.}, \bibinfo{author}{Schaye, J.},
  \bibinfo{author}{Springel, V.}, \bibinfo{author}{Thomas, P.A.},
  \bibinfo{author}{White, S.D.M.}, \bibinfo{author}{Wiersma, R.P.C.},
  \bibinfo{year}{2009}.
\newblock \bibinfo{journal}{Monthly Notices of the Royal Astronomical Society}
  \bibinfo{volume}{399}, \bibinfo{pages}{1773--1794}.
\bibitem[{Croton(2009)}]{croton_simple_2009}
\bibinfo{author}{Croton, D.J.}, \bibinfo{year}{2009}.
\newblock \bibinfo{journal}{Monthly Notices of the Royal Astronomical Society}
  \bibinfo{volume}{394}, \bibinfo{pages}{1109--1119}.
\bibitem[{Croton et~al.(2006)Croton, Springel, White, Lucia, Frenk, Gao,
  Jenkins, Kauffmann, Navarro and Yoshida}]{croton_many_2006}
\bibinfo{author}{Croton, D.J.}, \bibinfo{author}{Springel, V.},
  \bibinfo{author}{White, S.D.M.}, \bibinfo{author}{Lucia, G.D.},
  \bibinfo{author}{Frenk, C.S.}, \bibinfo{author}{Gao, L.},
  \bibinfo{author}{Jenkins, A.}, \bibinfo{author}{Kauffmann, G.},
  \bibinfo{author}{Navarro, J.F.}, \bibinfo{author}{Yoshida, N.},
  \bibinfo{year}{2006}.
\newblock \bibinfo{journal}{{MNRAS}} \bibinfo{volume}{365},
  \bibinfo{pages}{11--28}.
\bibitem[{Dav\'e et~al.(2001)Dav\'e, Spergel, Steinhardt and
  Wandelt}]{dave_halo_2001}
\bibinfo{author}{Dav\'e, R.}, \bibinfo{author}{Spergel, D.N.},
  \bibinfo{author}{Steinhardt, P.J.}, \bibinfo{author}{Wandelt, B.D.},
  \bibinfo{year}{2001}.
\newblock \bibinfo{journal}{Astrophysical Journal} \bibinfo{volume}{547},
  \bibinfo{pages}{574--589}.
\bibitem[{Davis et~al.(1985)Davis, Efstathiou, Frenk and
  White}]{davis_evolution_1985}
\bibinfo{author}{Davis, M.}, \bibinfo{author}{Efstathiou, G.},
  \bibinfo{author}{Frenk, C.S.}, \bibinfo{author}{White, S.D.M.},
  \bibinfo{year}{1985}.
\newblock \bibinfo{journal}{Astrophysical Journal} \bibinfo{volume}{292},
  \bibinfo{pages}{371--394}.
\bibitem[{Dekel and Silk(1986)}]{dekel_origin_1986}
\bibinfo{author}{Dekel, A.}, \bibinfo{author}{Silk, J.}, \bibinfo{year}{1986}.
\newblock \bibinfo{journal}{Astrophysical Journal} \bibinfo{volume}{303},
  \bibinfo{pages}{39--55}.
\bibitem[{Devriendt and Guiderdoni(2000)}]{devriendt_galaxy_2000}
\bibinfo{author}{Devriendt, J.E.G.}, \bibinfo{author}{Guiderdoni, B.},
  \bibinfo{year}{2000}.
\newblock \bibinfo{journal}{{A\&A}} \bibinfo{volume}{363},
  \bibinfo{pages}{851--862}.
\bibitem[{Devriendt et~al.(1998)Devriendt, Sethi, Guiderdoni and
  Nath}]{devriendt_contribution_1998}
\bibinfo{author}{Devriendt, J.E.G.}, \bibinfo{author}{Sethi, S.K.},
  \bibinfo{author}{Guiderdoni, B.}, \bibinfo{author}{Nath, B.B.},
  \bibinfo{year}{1998}.
\newblock \bibinfo{journal}{{MNRAS}} \bibinfo{volume}{298},
  \bibinfo{pages}{708--718}.
\bibitem[{Diaferio et~al.(1999)Diaferio, Kauffmann, Colberg and
  White}]{diaferio_clustering_1999}
\bibinfo{author}{Diaferio, A.}, \bibinfo{author}{Kauffmann, G.},
  \bibinfo{author}{Colberg, J.M.}, \bibinfo{author}{White, S.D.M.},
  \bibinfo{year}{1999}.
\newblock \bibinfo{journal}{{MNRAS}} \bibinfo{volume}{307},
  \bibinfo{pages}{537--552}.
\bibitem[{van Dokkum(2008)}]{dokkum_evidence_2008}
\bibinfo{author}{van Dokkum, P.G.}, \bibinfo{year}{2008}.
\newblock \bibinfo{journal}{Astrophysical Journal} \bibinfo{volume}{674},
  \bibinfo{pages}{29--50}.
\bibitem[{Dolag et~al.(2004)Dolag, Jubelgas, Springel, Borgani and
  Rasia}]{dolag_thermal_2004}
\bibinfo{author}{Dolag, K.}, \bibinfo{author}{Jubelgas, M.},
  \bibinfo{author}{Springel, V.}, \bibinfo{author}{Borgani, S.},
  \bibinfo{author}{Rasia, E.}, \bibinfo{year}{2004}.
\newblock \bibinfo{journal}{Astrophysical Journal} \bibinfo{volume}{606},
  \bibinfo{pages}{L97--L100}.
\bibitem[{Dolag and Stasyszyn(2008)}]{dolag_mhd_2008}
\bibinfo{author}{Dolag, K.}, \bibinfo{author}{Stasyszyn, F.A.},
  \bibinfo{year}{2008}.
\newblock \bibinfo{journal}{0807.3553} .
\bibitem[{Dotter et~al.(2007)Dotter, Chaboyer, Ferguson, chul Lee, Worthey,
  Jevremovic and Baron}]{dotter_stellar_2007}
\bibinfo{author}{Dotter, A.}, \bibinfo{author}{Chaboyer, B.},
  \bibinfo{author}{Ferguson, J.W.}, \bibinfo{author}{chul Lee, H.},
  \bibinfo{author}{Worthey, G.}, \bibinfo{author}{Jevremovic, D.},
  \bibinfo{author}{Baron, E.}, \bibinfo{year}{2007}.
\newblock \bibinfo{journal}{Astrophysical Journal} \bibinfo{volume}{666},
  \bibinfo{pages}{403--412}.
\bibitem[{Dove et~al.(2000)Dove, Shull and Ferrara}]{dove_escape_2000}
\bibinfo{author}{Dove, J.B.}, \bibinfo{author}{Shull, J.M.},
  \bibinfo{author}{Ferrara, A.}, \bibinfo{year}{2000}.
\newblock \bibinfo{journal}{Astrophysical Journal} \bibinfo{volume}{531},
  \bibinfo{pages}{846--860}.
\bibitem[{Draine and Lee(1984)}]{draine_optical_1984}
\bibinfo{author}{Draine, B.T.}, \bibinfo{author}{Lee, H.M.},
  \bibinfo{year}{1984}.
\newblock \bibinfo{journal}{Astrophysical Journal} \bibinfo{volume}{285},
  \bibinfo{pages}{89--108}.
\bibitem[{Dunkley et~al.(2009)Dunkley, Komatsu, Nolta, Spergel, Larson,
  Hinshaw, Page, Bennett, Gold, Jarosik, Weiland, Halpern, Hill, Kogut, Limon,
  Meyer, Tucker, Wollack and Wright}]{dunkley_five-year_2009}
\bibinfo{author}{Dunkley, J.}, \bibinfo{author}{Komatsu, E.},
  \bibinfo{author}{Nolta, M.R.}, \bibinfo{author}{Spergel, D.N.},
  \bibinfo{author}{Larson, D.}, \bibinfo{author}{Hinshaw, G.},
  \bibinfo{author}{Page, L.}, \bibinfo{author}{Bennett, C.L.},
  \bibinfo{author}{Gold, B.}, \bibinfo{author}{Jarosik, N.},
  \bibinfo{author}{Weiland, J.L.}, \bibinfo{author}{Halpern, M.},
  \bibinfo{author}{Hill, R.S.}, \bibinfo{author}{Kogut, A.},
  \bibinfo{author}{Limon, M.}, \bibinfo{author}{Meyer, S.S.},
  \bibinfo{author}{Tucker, G.S.}, \bibinfo{author}{Wollack, E.},
  \bibinfo{author}{Wright, E.L.}, \bibinfo{year}{2009}.
\newblock \bibinfo{journal}{Astrophysical Journal Supplement Series}
  \bibinfo{volume}{180}, \bibinfo{pages}{306--329}.
\bibitem[{Dutton(2009)}]{dutton_origin_2009}
\bibinfo{author}{Dutton, A.A.}, \bibinfo{year}{2009}.
\newblock \bibinfo{journal}{Monthly Notices of the Royal Astronomical Society}
  \bibinfo{volume}{396}, \bibinfo{pages}{121--140}.
\bibitem[{Dutton et~al.(2007)Dutton, van~den Bosch, Dekel and
  Courteau}]{dutton_revised_2007}
\bibinfo{author}{Dutton, A.A.}, \bibinfo{author}{van~den Bosch, F.C.},
  \bibinfo{author}{Dekel, A.}, \bibinfo{author}{Courteau, S.},
  \bibinfo{year}{2007}.
\newblock \bibinfo{journal}{Astrophysical Journal} \bibinfo{volume}{654},
  \bibinfo{pages}{27--52}.
\bibitem[{Edgar(2004)}]{edgar_review_2004}
\bibinfo{author}{Edgar, R.}, \bibinfo{year}{2004}.
\newblock \bibinfo{journal}{New Astronomy Reviews} \bibinfo{volume}{48},
  \bibinfo{pages}{843--859}.
\bibitem[{Efstathiou(2000)}]{efstathiou_model_2000}
\bibinfo{author}{Efstathiou, G.}, \bibinfo{year}{2000}.
\newblock \bibinfo{journal}{Monthly Notices of the Royal Astronomical Society}
  \bibinfo{volume}{317}, \bibinfo{pages}{697--719}.
\bibitem[{Efstathiou et~al.(1988)Efstathiou, Frenk, White and
  Davis}]{efstathiou_gravitational_1988}
\bibinfo{author}{Efstathiou, G.}, \bibinfo{author}{Frenk, C.S.},
  \bibinfo{author}{White, S.D.M.}, \bibinfo{author}{Davis, M.},
  \bibinfo{year}{1988}.
\newblock \bibinfo{journal}{{MNRAS}} \bibinfo{volume}{235},
  \bibinfo{pages}{715--748}.
\bibitem[{Efstathiou et~al.(1982)Efstathiou, Lake and
  Negroponte}]{efstathiou_stability_1982}
\bibinfo{author}{Efstathiou, G.}, \bibinfo{author}{Lake, G.},
  \bibinfo{author}{Negroponte, J.}, \bibinfo{year}{1982}.
\newblock \bibinfo{journal}{Monthly Notices of the Royal Astronomical Society}
  \bibinfo{volume}{199}, \bibinfo{pages}{1069--1088}.
\bibitem[{Einasto(1965)}]{einasto__1965}
\bibinfo{author}{Einasto, J.}, \bibinfo{year}{1965}.
\newblock \bibinfo{journal}{Trudy Inst. Astrofiz. {Alma-Ata}}
  \bibinfo{volume}{51}, \bibinfo{pages}{87}.
\bibitem[{Eisenstein and Hu(1999)}]{eisenstein_power_1999}
\bibinfo{author}{Eisenstein, D.J.}, \bibinfo{author}{Hu, W.},
  \bibinfo{year}{1999}.
\newblock \bibinfo{journal}{{ApJ}} \bibinfo{volume}{511},
  \bibinfo{pages}{5--15}.
\bibitem[{Eke et~al.(1996)Eke, Cole, Frenk and Navarro}]{eke_cluster_1996}
\bibinfo{author}{Eke, V.R.}, \bibinfo{author}{Cole, S.},
  \bibinfo{author}{Frenk, C.S.}, \bibinfo{author}{Navarro, J.F.},
  \bibinfo{year}{1996}.
\newblock \bibinfo{journal}{Monthly Notices of the Royal Astronomical Society}
  \bibinfo{volume}{281}, \bibinfo{pages}{703}.
\bibitem[{Eke et~al.(2001)Eke, Navarro and Steinmetz}]{eke_power_2001}
\bibinfo{author}{Eke, V.R.}, \bibinfo{author}{Navarro, J.F.},
  \bibinfo{author}{Steinmetz, M.}, \bibinfo{year}{2001}.
\newblock \bibinfo{journal}{Astrophysical Journal} \bibinfo{volume}{554},
  \bibinfo{pages}{114--125}.
\bibitem[{{El-Zant} and Shlosman(2002)}]{el-zant_dark_2002}
\bibinfo{author}{{El-Zant}, A.}, \bibinfo{author}{Shlosman, I.},
  \bibinfo{year}{2002}.
\newblock \bibinfo{journal}{Astrophysical Journal} \bibinfo{volume}{577},
  \bibinfo{pages}{626--650}.
\bibitem[{Elahi et~al.(2009)Elahi, Thacker, Widrow and
  Scannapieco}]{elahi_subhaloes_2009}
\bibinfo{author}{Elahi, P.J.}, \bibinfo{author}{Thacker, R.J.},
  \bibinfo{author}{Widrow, L.M.}, \bibinfo{author}{Scannapieco, E.},
  \bibinfo{year}{2009}.
\newblock \bibinfo{journal}{Monthly Notices of the Royal Astronomical Society}
  \bibinfo{volume}{395}, \bibinfo{pages}{1950--1962}.
\bibitem[{Fakhouri and Ma(2008a)}]{fakhouri_nearly_2008}
\bibinfo{author}{Fakhouri, O.}, \bibinfo{author}{Ma, C.},
  \bibinfo{year}{2008}a.
\newblock \bibinfo{journal}{Monthly Notices of the Royal Astronomical Society}
  \bibinfo{volume}{386}, \bibinfo{pages}{577--592}.
\bibitem[{Fakhouri and Ma(2008b)}]{fakhouri_environmental_2008}
\bibinfo{author}{Fakhouri, O.}, \bibinfo{author}{Ma, C.},
  \bibinfo{year}{2008}b.
\bibitem[{Fakhouri and Ma(2009)}]{fakhouri_environmental_2009}
\bibinfo{author}{Fakhouri, O.}, \bibinfo{author}{Ma, C.}, \bibinfo{year}{2009}.
\newblock \bibinfo{journal}{Monthly Notices of the Royal Astronomical Society}
  \bibinfo{volume}{394}, \bibinfo{pages}{1825--1840}.
\bibitem[{Fakhouri et~al.(2010)Fakhouri, Ma and
  {Boylan-Kolchin}}]{fakhouri_merger_2010}
\bibinfo{author}{Fakhouri, O.}, \bibinfo{author}{Ma, C.},
  \bibinfo{author}{{Boylan-Kolchin}, M.}, \bibinfo{year}{2010}.
\newblock
  \bibinfo{howpublished}{{http://adsabs.harvard.edu/abs/2010arXiv1001.2304F}}.
\bibitem[{Fall and Efstathiou(1980)}]{fall_formation_1980}
\bibinfo{author}{Fall, S.M.}, \bibinfo{author}{Efstathiou, G.},
  \bibinfo{year}{1980}.
\newblock \bibinfo{journal}{Monthly Notices of the Royal Astronomical Society}
  \bibinfo{volume}{193}, \bibinfo{pages}{189--206}.
\bibitem[{Fan et~al.(2001)Fan, Strauss, Schneider, Gunn, Lupton, Becker, Davis,
  Newman, Richards, White, Anderson, Annis, Bahcall, Brunner, Csabai, Hennessy,
  Hindsley, Fukugita, Kunszt, Ivezic, Knapp, {McKay}, Munn, Pier, Szalay and
  York}]{fan_high-redshift_2001}
\bibinfo{author}{Fan, X.}, \bibinfo{author}{Strauss, M.A.},
  \bibinfo{author}{Schneider, D.P.}, \bibinfo{author}{Gunn, J.E.},
  \bibinfo{author}{Lupton, R.H.}, \bibinfo{author}{Becker, R.H.},
  \bibinfo{author}{Davis, M.}, \bibinfo{author}{Newman, J.A.},
  \bibinfo{author}{Richards, G.T.}, \bibinfo{author}{White, R.L.},
  \bibinfo{author}{Anderson, J.E.}, \bibinfo{author}{Annis, J.},
  \bibinfo{author}{Bahcall, N.A.}, \bibinfo{author}{Brunner, R.J.},
  \bibinfo{author}{Csabai, I.}, \bibinfo{author}{Hennessy, G.S.},
  \bibinfo{author}{Hindsley, R.B.}, \bibinfo{author}{Fukugita, M.},
  \bibinfo{author}{Kunszt, P.Z.}, \bibinfo{author}{Ivezic, Z.},
  \bibinfo{author}{Knapp, G.R.}, \bibinfo{author}{{McKay}, T.A.},
  \bibinfo{author}{Munn, J.A.}, \bibinfo{author}{Pier, J.R.},
  \bibinfo{author}{Szalay, A.S.}, \bibinfo{author}{York, D.G.},
  \bibinfo{year}{2001}.
\newblock \bibinfo{journal}{Astronomical Journal} \bibinfo{volume}{121},
  \bibinfo{pages}{54--65}.
\bibitem[{Fardal et~al.(2001)Fardal, Katz, Gardner, Hernquist, Weinberg and
  Dav\'e}]{fardal_cooling_2001}
\bibinfo{author}{Fardal, M.A.}, \bibinfo{author}{Katz, N.},
  \bibinfo{author}{Gardner, J.P.}, \bibinfo{author}{Hernquist, L.},
  \bibinfo{author}{Weinberg, D.H.}, \bibinfo{author}{Dav\'e, R.},
  \bibinfo{year}{2001}.
\newblock \bibinfo{journal}{The Astrophysical Journal} \bibinfo{volume}{562},
  \bibinfo{pages}{605--617}.
\bibitem[{Favata et~al.(2004)Favata, Hughes and Holz}]{favata_black_2004}
\bibinfo{author}{Favata, M.}, \bibinfo{author}{Hughes, S.A.},
  \bibinfo{author}{Holz, D.E.}, \bibinfo{year}{2004}.
\newblock \bibinfo{journal}{Astrophysical Journal} \bibinfo{volume}{607},
  \bibinfo{pages}{L5--L8}.
\bibitem[{Federrath et~al.(2010)Federrath, Banerjee, Clark and
  Klessen}]{federrath_modeling_2010}
\bibinfo{author}{Federrath, C.}, \bibinfo{author}{Banerjee, R.},
  \bibinfo{author}{Clark, P.C.}, \bibinfo{author}{Klessen, R.S.},
  \bibinfo{year}{2010}.
\newblock \bibinfo{journal}{The Astrophysical Journal} \bibinfo{volume}{713},
  \bibinfo{pages}{269--290}.
\bibitem[{Ferland et~al.(1998)Ferland, Korista, Verner, Ferguson, Kingdon and
  Verner}]{ferland_cloudy_1998}
\bibinfo{author}{Ferland, G.J.}, \bibinfo{author}{Korista, K.T.},
  \bibinfo{author}{Verner, D.A.}, \bibinfo{author}{Ferguson, J.W.},
  \bibinfo{author}{Kingdon, J.B.}, \bibinfo{author}{Verner, E.M.},
  \bibinfo{year}{1998}.
\newblock \bibinfo{journal}{Publications of the Astronomical Society of the
  Pacific} \bibinfo{volume}{110}, \bibinfo{pages}{761--778}.
\bibitem[{Ferramacho et~al.(2008)Ferramacho, Blanchard and
  Zolnierowski}]{ferramacho_constraintscdm_2008}
\bibinfo{author}{Ferramacho, L.D.}, \bibinfo{author}{Blanchard, A.},
  \bibinfo{author}{Zolnierowski, Y.}, \bibinfo{year}{2008}.
\bibitem[{Ferrara et~al.(1999)Ferrara, Bianchi, Cimatti and
  Giovanardi}]{ferrara_atlas_1999}
\bibinfo{author}{Ferrara, A.}, \bibinfo{author}{Bianchi, S.},
  \bibinfo{author}{Cimatti, A.}, \bibinfo{author}{Giovanardi, C.},
  \bibinfo{year}{1999}.
\newblock \bibinfo{journal}{{ApJ} Supplement Series} \bibinfo{volume}{123},
  \bibinfo{pages}{437--445}.
\bibitem[{Ferrarese(2002)}]{ferrarese_beyondbulge:fundamental_2002}
\bibinfo{author}{Ferrarese, L.}, \bibinfo{year}{2002}.
\newblock \bibinfo{journal}{Astrophysical Journal} \bibinfo{volume}{578},
  \bibinfo{pages}{90--97}.
\bibitem[{Ferrarese and Merritt(2000)}]{ferrarese_fundamental_2000}
\bibinfo{author}{Ferrarese, L.}, \bibinfo{author}{Merritt, D.},
  \bibinfo{year}{2000}.
\newblock \bibinfo{journal}{Astrophysical Journal} \bibinfo{volume}{539},
  \bibinfo{pages}{L9--L12}.
\bibitem[{Field(1965)}]{field_thermal_1965}
\bibinfo{author}{Field, G.B.}, \bibinfo{year}{1965}.
\newblock \bibinfo{journal}{Astrophysical Journal} \bibinfo{volume}{142},
  \bibinfo{pages}{531}.
\bibitem[{Finlator et~al.(2010)Finlator, Oppenheimer and
  Dav\'e}]{finlator_smoothly-rising_2010}
\bibinfo{author}{Finlator, K.}, \bibinfo{author}{Oppenheimer, B.D.},
  \bibinfo{author}{Dav\'e, R.}, \bibinfo{year}{2010}.
\newblock
  \bibinfo{howpublished}{{http://adsabs.harvard.edu/abs/2010arXiv1005.4066F}}.
\bibitem[{Finlator et~al.(2009)Finlator, {\"Ozel} and
  Dav\'e}]{finlator_new_2009}
\bibinfo{author}{Finlator, K.}, \bibinfo{author}{{\"Ozel}, F.},
  \bibinfo{author}{Dav\'e, R.}, \bibinfo{year}{2009}.
\newblock \bibinfo{journal}{Monthly Notices of the Royal Astronomical Society}
  \bibinfo{volume}{393}, \bibinfo{pages}{1090--1106}.
\bibitem[{Firmani and {Avila-Reese}(2003)}]{firmani_physical_2003}
\bibinfo{author}{Firmani, C.}, \bibinfo{author}{{Avila-Reese}, V.},
  \bibinfo{year}{2003}.
\newblock pp. \bibinfo{pages}{107--120}.
\bibitem[{Firmani et~al.(1996)Firmani, Hernandez and
  Gallagher}]{firmani_viscous_1996}
\bibinfo{author}{Firmani, C.}, \bibinfo{author}{Hernandez, X.},
  \bibinfo{author}{Gallagher, J.}, \bibinfo{year}{1996}.
\newblock \bibinfo{journal}{Astronomy and Astrophysics} \bibinfo{volume}{308},
  \bibinfo{pages}{403--414}.
\bibitem[{Fitchett(1983)}]{fitchett_influence_1983}
\bibinfo{author}{Fitchett, M.J.}, \bibinfo{year}{1983}.
\newblock \bibinfo{journal}{Monthly Notices of the Royal Astronomical Society}
  \bibinfo{volume}{203}, \bibinfo{pages}{1049--1062}.
\bibitem[{Font et~al.(2010)Font, Benson, Bower and
  Frenk}]{font_modelingmilky_2010}
\bibinfo{author}{Font, A.}, \bibinfo{author}{Benson, A.},
  \bibinfo{author}{Bower, R.}, \bibinfo{author}{Frenk, C.},
  \bibinfo{year}{2010}.
\newblock \bibinfo{journal}{in prep} .
\bibitem[{Font et~al.(2008)Font, Bower, {McCarthy}, Benson, Frenk, Helly,
  Lacey, Baugh and Cole}]{font_colours_2008}
\bibinfo{author}{Font, A.S.}, \bibinfo{author}{Bower, R.G.},
  \bibinfo{author}{{McCarthy}, I.G.}, \bibinfo{author}{Benson, A.J.},
  \bibinfo{author}{Frenk, C.S.}, \bibinfo{author}{Helly, J.C.},
  \bibinfo{author}{Lacey, C.G.}, \bibinfo{author}{Baugh, C.M.},
  \bibinfo{author}{Cole, S.}, \bibinfo{year}{2008}.
\newblock \bibinfo{journal}{{MNRAS}} \bibinfo{volume}{389},
  \bibinfo{pages}{1619--1629}.
\bibitem[{Font et~al.(2006a)Font, Johnston, Bullock and
  Robertson}]{font_chemical_2006}
\bibinfo{author}{Font, A.S.}, \bibinfo{author}{Johnston, K.V.},
  \bibinfo{author}{Bullock, J.S.}, \bibinfo{author}{Robertson, B.E.},
  \bibinfo{year}{2006}a.
\newblock \bibinfo{journal}{Astrophysical Journal} \bibinfo{volume}{638},
  \bibinfo{pages}{585--595}.
\bibitem[{Font et~al.(2006b)Font, Johnston, Bullock and
  Robertson}]{font_phase-space_2006}
\bibinfo{author}{Font, A.S.}, \bibinfo{author}{Johnston, K.V.},
  \bibinfo{author}{Bullock, J.S.}, \bibinfo{author}{Robertson, B.E.},
  \bibinfo{year}{2006}b.
\newblock \bibinfo{journal}{Astrophysical Journal} \bibinfo{volume}{646},
  \bibinfo{pages}{886--898}.
\bibitem[{Font et~al.(2001)Font, Navarro, Stadel and Quinn}]{font_halo_2001}
\bibinfo{author}{Font, A.S.}, \bibinfo{author}{Navarro, J.F.},
  \bibinfo{author}{Stadel, J.}, \bibinfo{author}{Quinn, T.},
  \bibinfo{year}{2001}.
\newblock \bibinfo{journal}{Astrophysical Journal} \bibinfo{volume}{563},
  \bibinfo{pages}{L1--L4}.
\bibitem[{Fontanot et~al.(2009a)Fontanot, Lucia, Monaco, Somerville and
  Santini}]{fontanot_many_2009}
\bibinfo{author}{Fontanot, F.}, \bibinfo{author}{Lucia, G.D.},
  \bibinfo{author}{Monaco, P.}, \bibinfo{author}{Somerville, R.S.},
  \bibinfo{author}{Santini, P.}, \bibinfo{year}{2009}a.
\newblock \bibinfo{journal}{0901.1130} .
\bibitem[{Fontanot et~al.(2007)Fontanot, Monaco, Silva and
  Grazian}]{fontanot_reproducingassembly_2007}
\bibinfo{author}{Fontanot, F.}, \bibinfo{author}{Monaco, P.},
  \bibinfo{author}{Silva, L.}, \bibinfo{author}{Grazian, A.},
  \bibinfo{year}{2007}.
\newblock \bibinfo{journal}{{MNRAS}} \bibinfo{volume}{382},
  \bibinfo{pages}{903--914}.
\bibitem[{Fontanot et~al.(2009b)Fontanot, Somerville, Silva, Monaco and
  Skibba}]{fontanot_evaluating_2009}
\bibinfo{author}{Fontanot, F.}, \bibinfo{author}{Somerville, R.S.},
  \bibinfo{author}{Silva, L.}, \bibinfo{author}{Monaco, P.},
  \bibinfo{author}{Skibba, R.}, \bibinfo{year}{2009}b.
\newblock \bibinfo{journal}{{MNRAS}} \bibinfo{volume}{392},
  \bibinfo{pages}{553--569}.
\bibitem[{{Forcada-Miro} and White(1997)}]{forcada-miro_radiative_1997}
\bibinfo{author}{{Forcada-Miro}, M.I.}, \bibinfo{author}{White, S.D.M.},
  \bibinfo{year}{1997}.
\newblock
  \bibinfo{howpublished}{{http://adsabs.harvard.edu/abs/1997astro.ph.12204F}}.
\bibitem[{Fran\c{c}ois et~al.(2004)Fran\c{c}ois, Matteucci, Cayrel, Spite,
  Spite and Chiappini}]{franccois_evolution_2004}
\bibinfo{author}{Fran\c{c}ois, P.}, \bibinfo{author}{Matteucci, F.},
  \bibinfo{author}{Cayrel, R.}, \bibinfo{author}{Spite, M.},
  \bibinfo{author}{Spite, F.}, \bibinfo{author}{Chiappini, C.},
  \bibinfo{year}{2004}.
\newblock \bibinfo{journal}{Astronomy and Astrophysics} \bibinfo{volume}{421},
  \bibinfo{pages}{613--621}.
\bibitem[{Freeman and {Bland-Hawthorn}(2002)}]{freeman_new_2002}
\bibinfo{author}{Freeman, K.}, \bibinfo{author}{{Bland-Hawthorn}, J.},
  \bibinfo{year}{2002}.
\newblock \bibinfo{journal}{Annual Review of Astronomy and Astrophysics}
  \bibinfo{volume}{40}, \bibinfo{pages}{487--537}.
\bibitem[{Frenk(2002)}]{frenk_simulatingformation_2002}
\bibinfo{author}{Frenk, C.S.}, \bibinfo{year}{2002}.
\newblock \bibinfo{journal}{Royal Society of London Philosophical Transactions
  Series A} \bibinfo{volume}{360}, \bibinfo{pages}{1277}.
\bibitem[{Frenk et~al.(1999)Frenk, White, Bode, Bond, Bryan, Cen, Couchman,
  Evrard, Gnedin, Jenkins, Khokhlov, Klypin, Navarro, Norman, Ostriker, Owen,
  Pearce, Pen, Steinmetz, Thomas, Villumsen, Wadsley, Warren, Xu and
  Yepes}]{frenk_santa_1999}
\bibinfo{author}{Frenk, C.S.}, \bibinfo{author}{White, S.D.M.},
  \bibinfo{author}{Bode, P.}, \bibinfo{author}{Bond, J.R.},
  \bibinfo{author}{Bryan, G.L.}, \bibinfo{author}{Cen, R.},
  \bibinfo{author}{Couchman, H.M.P.}, \bibinfo{author}{Evrard, A.E.},
  \bibinfo{author}{Gnedin, N.}, \bibinfo{author}{Jenkins, A.},
  \bibinfo{author}{Khokhlov, A.M.}, \bibinfo{author}{Klypin, A.},
  \bibinfo{author}{Navarro, J.F.}, \bibinfo{author}{Norman, M.L.},
  \bibinfo{author}{Ostriker, J.P.}, \bibinfo{author}{Owen, J.M.},
  \bibinfo{author}{Pearce, F.R.}, \bibinfo{author}{Pen, U.},
  \bibinfo{author}{Steinmetz, M.}, \bibinfo{author}{Thomas, P.A.},
  \bibinfo{author}{Villumsen, J.V.}, \bibinfo{author}{Wadsley, J.W.},
  \bibinfo{author}{Warren, M.S.}, \bibinfo{author}{Xu, G.},
  \bibinfo{author}{Yepes, G.}, \bibinfo{year}{1999}.
\newblock \bibinfo{journal}{Astrophysical Journal} \bibinfo{volume}{525},
  \bibinfo{pages}{554--582}.
\bibitem[{Frenk et~al.(1988)Frenk, White, Davis and
  Efstathiou}]{frenk_formation_1988}
\bibinfo{author}{Frenk, C.S.}, \bibinfo{author}{White, S.D.M.},
  \bibinfo{author}{Davis, M.}, \bibinfo{author}{Efstathiou, G.},
  \bibinfo{year}{1988}.
\newblock \bibinfo{journal}{Astrophysical Journal} \bibinfo{volume}{327},
  \bibinfo{pages}{507--525}.
\bibitem[{Friedli and Benz(1993)}]{friedli_secular_1993}
\bibinfo{author}{Friedli, D.}, \bibinfo{author}{Benz, W.},
  \bibinfo{year}{1993}.
\newblock \bibinfo{journal}{Astronomy and Astrophysics} \bibinfo{volume}{268},
  \bibinfo{pages}{65--85}.
\bibitem[{Fryxell et~al.(2000)Fryxell, Olson, Ricker, Timmes, Zingale, Lamb,
  {MacNeice}, Rosner, Truran and Tufo}]{fryxell_flash:adaptive_2000}
\bibinfo{author}{Fryxell, B.}, \bibinfo{author}{Olson, K.},
  \bibinfo{author}{Ricker, P.}, \bibinfo{author}{Timmes, F.X.},
  \bibinfo{author}{Zingale, M.}, \bibinfo{author}{Lamb, D.Q.},
  \bibinfo{author}{{MacNeice}, P.}, \bibinfo{author}{Rosner, R.},
  \bibinfo{author}{Truran, J.W.}, \bibinfo{author}{Tufo, H.},
  \bibinfo{year}{2000}.
\newblock \bibinfo{journal}{Astrophysical Journal Supplement Series}
  \bibinfo{volume}{131}, \bibinfo{pages}{273--334}.
\bibitem[{Galli and Palla(1998)}]{galli_chemistry_1998}
\bibinfo{author}{Galli, D.}, \bibinfo{author}{Palla, F.}, \bibinfo{year}{1998}.
\newblock \bibinfo{journal}{{A\&A}} \bibinfo{volume}{335},
  \bibinfo{pages}{403--420}.
\bibitem[{Gao et~al.(2008)Gao, Navarro, Cole, Frenk, White, Springel, Jenkins
  and Neto}]{gao_redshift_2008}
\bibinfo{author}{Gao, L.}, \bibinfo{author}{Navarro, J.F.},
  \bibinfo{author}{Cole, S.}, \bibinfo{author}{Frenk, C.S.},
  \bibinfo{author}{White, S.D.M.}, \bibinfo{author}{Springel, V.},
  \bibinfo{author}{Jenkins, A.}, \bibinfo{author}{Neto, A.F.},
  \bibinfo{year}{2008}.
\newblock \bibinfo{journal}{{MNRAS}} \bibinfo{volume}{387},
  \bibinfo{pages}{536--544}.
\bibitem[{Gao et~al.(2007)Gao, Yoshida, Abel, Frenk, Jenkins and
  Springel}]{gao_first_2007}
\bibinfo{author}{Gao, L.}, \bibinfo{author}{Yoshida, N.},
  \bibinfo{author}{Abel, T.}, \bibinfo{author}{Frenk, C.S.},
  \bibinfo{author}{Jenkins, A.}, \bibinfo{author}{Springel, V.},
  \bibinfo{year}{2007}.
\newblock \bibinfo{journal}{Monthly Notices of the Royal Astronomical Society}
  \bibinfo{volume}{378}, \bibinfo{pages}{449--468}.
\bibitem[{Gebhardt et~al.(2000a)Gebhardt, Bender, Bower, Dressler, Faber,
  Filippenko, Green, Grillmair, Ho, Kormendy, Lauer, Magorrian, Pinkney,
  Richstone and Tremaine}]{gebhardt_relationship_2000}
\bibinfo{author}{Gebhardt, K.}, \bibinfo{author}{Bender, R.},
  \bibinfo{author}{Bower, G.}, \bibinfo{author}{Dressler, A.},
  \bibinfo{author}{Faber, S.M.}, \bibinfo{author}{Filippenko, A.V.},
  \bibinfo{author}{Green, R.}, \bibinfo{author}{Grillmair, C.},
  \bibinfo{author}{Ho, L.C.}, \bibinfo{author}{Kormendy, J.},
  \bibinfo{author}{Lauer, T.R.}, \bibinfo{author}{Magorrian, J.},
  \bibinfo{author}{Pinkney, J.}, \bibinfo{author}{Richstone, D.},
  \bibinfo{author}{Tremaine, S.}, \bibinfo{year}{2000}a.
\newblock \bibinfo{journal}{Astrophysical Journal} \bibinfo{volume}{539},
  \bibinfo{pages}{L13--L16}.
\bibitem[{Gebhardt et~al.(2000b)Gebhardt, Kormendy, Ho, Bender, Bower,
  Dressler, Faber, Filippenko, Green, Grillmair, Lauer, Magorrian, Pinkney,
  Richstone and Tremaine}]{gebhardt_black_2000}
\bibinfo{author}{Gebhardt, K.}, \bibinfo{author}{Kormendy, J.},
  \bibinfo{author}{Ho, L.C.}, \bibinfo{author}{Bender, R.},
  \bibinfo{author}{Bower, G.}, \bibinfo{author}{Dressler, A.},
  \bibinfo{author}{Faber, S.M.}, \bibinfo{author}{Filippenko, A.V.},
  \bibinfo{author}{Green, R.}, \bibinfo{author}{Grillmair, C.},
  \bibinfo{author}{Lauer, T.R.}, \bibinfo{author}{Magorrian, J.},
  \bibinfo{author}{Pinkney, J.}, \bibinfo{author}{Richstone, D.},
  \bibinfo{author}{Tremaine, S.}, \bibinfo{year}{2000}b.
\newblock \bibinfo{journal}{Astrophysical Journal} \bibinfo{volume}{543},
  \bibinfo{pages}{L5--L8}.
\bibitem[{Glover and Abel(2008)}]{glover_uncertainties_2008}
\bibinfo{author}{Glover, S.C.O.}, \bibinfo{author}{Abel, T.},
  \bibinfo{year}{2008}.
\newblock \bibinfo{journal}{Monthly Notices of the Royal Astronomical Society}
  \bibinfo{volume}{388}, \bibinfo{pages}{1627--1651}.
\bibitem[{Gnat and Sternberg(2007)}]{gnat_time-dependent_2007}
\bibinfo{author}{Gnat, O.}, \bibinfo{author}{Sternberg, A.},
  \bibinfo{year}{2007}.
\newblock \bibinfo{journal}{Astrophysical Journal Supplement Series}
  \bibinfo{volume}{168}, \bibinfo{pages}{213--230}.
\bibitem[{Gnedin(2000)}]{gnedin_effect_2000}
\bibinfo{author}{Gnedin, N.Y.}, \bibinfo{year}{2000}.
\newblock \bibinfo{journal}{{ApJ}} \bibinfo{volume}{542},
  \bibinfo{pages}{535--541}.
\bibitem[{Gnedin and Abel(2001)}]{gnedin_multi-dimensional_2001}
\bibinfo{author}{Gnedin, N.Y.}, \bibinfo{author}{Abel, T.},
  \bibinfo{year}{2001}.
\newblock \bibinfo{journal}{New Astronomy} \bibinfo{volume}{6},
  \bibinfo{pages}{437--455}.
\bibitem[{Gnedin and Hui(1998)}]{gnedin_probinguniverse_1998}
\bibinfo{author}{Gnedin, N.Y.}, \bibinfo{author}{Hui, L.},
  \bibinfo{year}{1998}.
\newblock \bibinfo{journal}{Monthly Notices of the Royal Astronomical Society}
  \bibinfo{volume}{296}, \bibinfo{pages}{44--55}.
\bibitem[{Gnedin et~al.(2008)Gnedin, Kravtsov and Chen}]{gnedin_escape_2008}
\bibinfo{author}{Gnedin, N.Y.}, \bibinfo{author}{Kravtsov, A.V.},
  \bibinfo{author}{Chen, H.}, \bibinfo{year}{2008}.
\newblock \bibinfo{journal}{Astrophysical Journal} \bibinfo{volume}{672},
  \bibinfo{pages}{765--775}.
\bibitem[{Gnedin et~al.(2009)Gnedin, Tassis and
  Kravtsov}]{gnedin_modeling_2009}
\bibinfo{author}{Gnedin, N.Y.}, \bibinfo{author}{Tassis, K.},
  \bibinfo{author}{Kravtsov, A.V.}, \bibinfo{year}{2009}.
\newblock \bibinfo{journal}{Astrophysical Journal} \bibinfo{volume}{697},
  \bibinfo{pages}{55--67}.
\bibitem[{Gnedin(2003)}]{gnedin_tidal_2003}
\bibinfo{author}{Gnedin, O.Y.}, \bibinfo{year}{2003}.
\newblock \bibinfo{journal}{Astrophysical Journal} \bibinfo{volume}{582},
  \bibinfo{pages}{141--161}.
\bibitem[{Gnedin et~al.(2004)Gnedin, Kravtsov, Klypin and
  Nagai}]{gnedin_response_2004}
\bibinfo{author}{Gnedin, O.Y.}, \bibinfo{author}{Kravtsov, A.V.},
  \bibinfo{author}{Klypin, A.A.}, \bibinfo{author}{Nagai, D.},
  \bibinfo{year}{2004}.
\newblock \bibinfo{journal}{Astrophysical Journal} \bibinfo{volume}{616},
  \bibinfo{pages}{16--26}.
\bibitem[{Gnedin and Ostriker(1999)}]{gnedin_self-consistent_1999}
\bibinfo{author}{Gnedin, O.Y.}, \bibinfo{author}{Ostriker, J.P.},
  \bibinfo{year}{1999}.
\newblock \bibinfo{journal}{Astrophysical Journal} \bibinfo{volume}{513},
  \bibinfo{pages}{626--637}.
\bibitem[{Goerdt et~al.(2007)Goerdt, Gnedin, Moore, Diemand and
  Stadel}]{goerdt_survival_2007}
\bibinfo{author}{Goerdt, T.}, \bibinfo{author}{Gnedin, O.Y.},
  \bibinfo{author}{Moore, B.}, \bibinfo{author}{Diemand, J.},
  \bibinfo{author}{Stadel, J.}, \bibinfo{year}{2007}.
\newblock \bibinfo{journal}{Monthly Notices of the Royal Astronomical Society}
  \bibinfo{volume}{375}, \bibinfo{pages}{191--198}.
\bibitem[{Goldreich and {Lynden-Bell}(1965)}]{goldreich_ii._1965}
\bibinfo{author}{Goldreich, P.}, \bibinfo{author}{{Lynden-Bell}, D.},
  \bibinfo{year}{1965}.
\newblock \bibinfo{journal}{Monthly Notices of the Royal Astronomical Society}
  \bibinfo{volume}{130}, \bibinfo{pages}{125}.
\bibitem[{Gon\c{c}alves and Fria\c{c}a(1999)}]{gonccalves_magnetic_1999}
\bibinfo{author}{Gon\c{c}alves, D.R.}, \bibinfo{author}{Fria\c{c}a, A.C.S.},
  \bibinfo{year}{1999}.
\newblock \bibinfo{journal}{Monthly Notices of the Royal Astronomical Society}
  \bibinfo{volume}{309}, \bibinfo{pages}{651--658}.
\bibitem[{{Gonzalez-Perez} et~al.(2008){Gonzalez-Perez}, Baugh, Lacey and
  Almeida}]{gonzalez-perez_massive_2008}
\bibinfo{author}{{Gonzalez-Perez}, V.}, \bibinfo{author}{Baugh, C.M.},
  \bibinfo{author}{Lacey, C.G.}, \bibinfo{author}{Almeida, C.},
  \bibinfo{year}{2008}.
\bibitem[{Gould and Rix(2000)}]{gould_binary_2000}
\bibinfo{author}{Gould, A.}, \bibinfo{author}{Rix, H.}, \bibinfo{year}{2000}.
\newblock \bibinfo{journal}{Astrophysical Journal} \bibinfo{volume}{532},
  \bibinfo{pages}{L29--L32}.
\bibitem[{Governato et~al.(1998)Governato, Baugh, Frenk, Cole, Lacey, Quinn and
  Stadel}]{governato_seeds_1998}
\bibinfo{author}{Governato, F.}, \bibinfo{author}{Baugh, C.M.},
  \bibinfo{author}{Frenk, C.S.}, \bibinfo{author}{Cole, S.},
  \bibinfo{author}{Lacey, C.G.}, \bibinfo{author}{Quinn, T.},
  \bibinfo{author}{Stadel, J.}, \bibinfo{year}{1998}.
\newblock \bibinfo{journal}{Nature} \bibinfo{volume}{392},
  \bibinfo{pages}{359--361}.
\bibitem[{Governato et~al.(1999)Governato, Gardner, Stadel, Quinn and
  Lake}]{governato_origin_1999}
\bibinfo{author}{Governato, F.}, \bibinfo{author}{Gardner, J.P.},
  \bibinfo{author}{Stadel, J.}, \bibinfo{author}{Quinn, T.},
  \bibinfo{author}{Lake, G.}, \bibinfo{year}{1999}.
\newblock \bibinfo{journal}{Astronomical Journal} \bibinfo{volume}{117},
  \bibinfo{pages}{1651--1656}.
\bibitem[{Governato et~al.(2007)Governato, Willman, Mayer, Brooks, Stinson,
  Valenzuela, Wadsley and Quinn}]{governato_forming_2007}
\bibinfo{author}{Governato, F.}, \bibinfo{author}{Willman, B.},
  \bibinfo{author}{Mayer, L.}, \bibinfo{author}{Brooks, A.},
  \bibinfo{author}{Stinson, G.}, \bibinfo{author}{Valenzuela, O.},
  \bibinfo{author}{Wadsley, J.}, \bibinfo{author}{Quinn, T.},
  \bibinfo{year}{2007}.
\newblock \bibinfo{journal}{Monthly Notices of the Royal Astronomical Society}
  \bibinfo{volume}{374}, \bibinfo{pages}{1479--1494}.
\bibitem[{Granato et~al.(2000)Granato, Lacey, Silva, Bressan, Baugh, Cole and
  Frenk}]{granato_infrared_2000}
\bibinfo{author}{Granato, G.L.}, \bibinfo{author}{Lacey, C.G.},
  \bibinfo{author}{Silva, L.}, \bibinfo{author}{Bressan, A.},
  \bibinfo{author}{Baugh, C.M.}, \bibinfo{author}{Cole, S.},
  \bibinfo{author}{Frenk, C.S.}, \bibinfo{year}{2000}.
\newblock \bibinfo{journal}{{ApJ}} \bibinfo{volume}{542},
  \bibinfo{pages}{710--730}.
\bibitem[{Greif et~al.(2008)Greif, Johnson, Klessen and
  Bromm}]{greif_first_2008}
\bibinfo{author}{Greif, T.H.}, \bibinfo{author}{Johnson, J.L.},
  \bibinfo{author}{Klessen, R.S.}, \bibinfo{author}{Bromm, V.},
  \bibinfo{year}{2008}.
\newblock \bibinfo{journal}{Monthly Notices of the Royal Astronomical Society}
  \bibinfo{volume}{387}, \bibinfo{pages}{1021--1036}.
\bibitem[{Guiderdoni et~al.(1998)Guiderdoni, Hivon, Bouchet and
  Maffei}]{guiderdoni_semi-analytic_1998}
\bibinfo{author}{Guiderdoni, B.}, \bibinfo{author}{Hivon, E.},
  \bibinfo{author}{Bouchet, F.R.}, \bibinfo{author}{Maffei, B.},
  \bibinfo{year}{1998}.
\newblock \bibinfo{journal}{{MNRAS}} \bibinfo{volume}{295},
  \bibinfo{pages}{877--898}.
\bibitem[{Gunn(1977)}]{gunn_massive_1977}
\bibinfo{author}{Gunn, J.E.}, \bibinfo{year}{1977}.
\newblock \bibinfo{journal}{The Astrophysical Journal} \bibinfo{volume}{218},
  \bibinfo{pages}{592--598}.
\bibitem[{Gunn and Gott(1972)}]{gunn_infall_1972}
\bibinfo{author}{Gunn, J.E.}, \bibinfo{author}{Gott, J.R.},
  \bibinfo{year}{1972}.
\newblock \bibinfo{journal}{Astrophysical Journal} \bibinfo{volume}{176},
  \bibinfo{pages}{1}.
\bibitem[{Guo and Oh(2008)}]{guo_feedback_2008}
\bibinfo{author}{Guo, F.}, \bibinfo{author}{Oh, S.P.}, \bibinfo{year}{2008}.
\newblock \bibinfo{journal}{Monthly Notices of the Royal Astronomical Society}
  \bibinfo{volume}{384}, \bibinfo{pages}{251--266}.
\bibitem[{Gustafsson et~al.(2006)Gustafsson, Fairbairn and
  {Sommer-Larsen}}]{gustafsson_baryonic_2006}
\bibinfo{author}{Gustafsson, M.}, \bibinfo{author}{Fairbairn, M.},
  \bibinfo{author}{{Sommer-Larsen}, J.}, \bibinfo{year}{2006}.
\newblock \bibinfo{journal}{Physical Review D} \bibinfo{volume}{74},
  \bibinfo{pages}{123522}.
\bibitem[{Haardt and Madau(1996)}]{haardt_radiative_1996}
\bibinfo{author}{Haardt, F.}, \bibinfo{author}{Madau, P.},
  \bibinfo{year}{1996}.
\newblock \bibinfo{journal}{Astrophysical Journal} \bibinfo{volume}{461},
  \bibinfo{pages}{20}.
\bibitem[{Hahn et~al.(2010)Hahn, Teyssier and Carollo}]{hahn_large-scale_2010}
\bibinfo{author}{Hahn, O.}, \bibinfo{author}{Teyssier, R.},
  \bibinfo{author}{Carollo, C.M.}, \bibinfo{year}{2010}.
\newblock \bibinfo{journal}{Monthly Notices of the Royal Astronomical Society}
  \bibinfo{volume}{405}, \bibinfo{pages}{274--290}.
\bibitem[{Hamilton et~al.(1991)Hamilton, Kumar, Lu and
  Matthews}]{hamilton_reconstructingprimordial_1991}
\bibinfo{author}{Hamilton, A.J.S.}, \bibinfo{author}{Kumar, P.},
  \bibinfo{author}{Lu, E.}, \bibinfo{author}{Matthews, A.},
  \bibinfo{year}{1991}.
\newblock \bibinfo{journal}{The Astrophysical Journal} \bibinfo{volume}{374},
  \bibinfo{pages}{L1--L4}.
\bibitem[{Harris et~al.(2007)Harris, {Lynas-Gray}, Miller and
  Tennyson}]{harris_non-grey_2007}
\bibinfo{author}{Harris, G.J.}, \bibinfo{author}{{Lynas-Gray}, A.E.},
  \bibinfo{author}{Miller, S.}, \bibinfo{author}{Tennyson, J.},
  \bibinfo{year}{2007}.
\newblock \bibinfo{journal}{Monthly Notices of the Royal Astronomical Society}
  \bibinfo{volume}{374}, \bibinfo{pages}{337--343}.
\bibitem[{Hatton et~al.(2003)Hatton, Devriendt, Ninin, Bouchet, Guiderdoni and
  Vibert}]{hatton_galics-_2003}
\bibinfo{author}{Hatton, S.}, \bibinfo{author}{Devriendt, J.E.G.},
  \bibinfo{author}{Ninin, S.}, \bibinfo{author}{Bouchet, F.R.},
  \bibinfo{author}{Guiderdoni, B.}, \bibinfo{author}{Vibert, D.},
  \bibinfo{year}{2003}.
\newblock \bibinfo{journal}{Monthly Notices of the Royal Astronomical Society}
  \bibinfo{volume}{343}, \bibinfo{pages}{75--106}.
\bibitem[{Heitmann et~al.(2009)Heitmann, Higdon, White, Habib, Williams,
  Lawrence and Wagner}]{heitmann_coyote_2009}
\bibinfo{author}{Heitmann, K.}, \bibinfo{author}{Higdon, D.},
  \bibinfo{author}{White, M.}, \bibinfo{author}{Habib, S.},
  \bibinfo{author}{Williams, B.J.}, \bibinfo{author}{Lawrence, E.},
  \bibinfo{author}{Wagner, C.}, \bibinfo{year}{2009}.
\newblock \bibinfo{journal}{The Astrophysical Journal} \bibinfo{volume}{705},
  \bibinfo{pages}{156--174}.
\bibitem[{Heitmann et~al.(2008)Heitmann, White, Wagner, Habib and
  Higdon}]{heitmann_coyote_2008}
\bibinfo{author}{Heitmann, K.}, \bibinfo{author}{White, M.},
  \bibinfo{author}{Wagner, C.}, \bibinfo{author}{Habib, S.},
  \bibinfo{author}{Higdon, D.}, \bibinfo{year}{2008}.
\bibitem[{Helly et~al.(2003a)Helly, Cole, Frenk, Baugh, Benson and
  Lacey}]{helly_galaxy_2003}
\bibinfo{author}{Helly, J.C.}, \bibinfo{author}{Cole, S.},
  \bibinfo{author}{Frenk, C.S.}, \bibinfo{author}{Baugh, C.M.},
  \bibinfo{author}{Benson, A.}, \bibinfo{author}{Lacey, C.},
  \bibinfo{year}{2003}a.
\newblock \bibinfo{journal}{Monthly Notices of the Royal Astronomical Society}
  \bibinfo{volume}{338}, \bibinfo{pages}{903--912}.
\bibitem[{Helly et~al.(2003b)Helly, Cole, Frenk, Baugh, Benson, Lacey and
  Pearce}]{helly_comparison_2003}
\bibinfo{author}{Helly, J.C.}, \bibinfo{author}{Cole, S.},
  \bibinfo{author}{Frenk, C.S.}, \bibinfo{author}{Baugh, C.M.},
  \bibinfo{author}{Benson, A.}, \bibinfo{author}{Lacey, C.},
  \bibinfo{author}{Pearce, F.R.}, \bibinfo{year}{2003}b.
\newblock \bibinfo{journal}{{MNRAS}} \bibinfo{volume}{338},
  \bibinfo{pages}{913--925}.
\bibitem[{Henriques et~al.(2009)Henriques, Thomas, Oliver and
  Roseboom}]{henriques_monte_2009}
\bibinfo{author}{Henriques, B.M.B.}, \bibinfo{author}{Thomas, P.A.},
  \bibinfo{author}{Oliver, S.}, \bibinfo{author}{Roseboom, I.},
  \bibinfo{year}{2009}.
\newblock \bibinfo{journal}{Monthly Notices of the Royal Astronomical Society}
  \bibinfo{volume}{396}, \bibinfo{pages}{535--547}.
\bibitem[{Hernquist et~al.(1993)Hernquist, Spergel and
  Heyl}]{hernquist_structure_1993}
\bibinfo{author}{Hernquist, L.}, \bibinfo{author}{Spergel, D.N.},
  \bibinfo{author}{Heyl, J.S.}, \bibinfo{year}{1993}.
\newblock \bibinfo{journal}{Astrophysical Journal} \bibinfo{volume}{416},
  \bibinfo{pages}{415}.
\bibitem[{Hollenbach and {McKee}(1979)}]{hollenbach_molecule_1979}
\bibinfo{author}{Hollenbach, D.}, \bibinfo{author}{{McKee}, C.F.},
  \bibinfo{year}{1979}.
\newblock \bibinfo{journal}{{ApJ} Supplement Series} \bibinfo{volume}{41},
  \bibinfo{pages}{555--592}.
\bibitem[{Hopkins et~al.(2010)Hopkins, Croton, Bundy, Khochfar, van~den Bosch,
  Somerville, Wetzel, Keres, Hernquist, Stewart, Younger, Genel and
  Ma}]{hopkins_mergers_2010}
\bibinfo{author}{Hopkins, P.F.}, \bibinfo{author}{Croton, D.},
  \bibinfo{author}{Bundy, K.}, \bibinfo{author}{Khochfar, S.},
  \bibinfo{author}{van~den Bosch, F.}, \bibinfo{author}{Somerville, R.S.},
  \bibinfo{author}{Wetzel, A.}, \bibinfo{author}{Keres, D.},
  \bibinfo{author}{Hernquist, L.}, \bibinfo{author}{Stewart, K.},
  \bibinfo{author}{Younger, J.D.}, \bibinfo{author}{Genel, S.},
  \bibinfo{author}{Ma, C.}, \bibinfo{year}{2010}.
\newblock
  \bibinfo{howpublished}{{http://adsabs.harvard.edu/abs/2010arXiv1004.2708H}}.
\bibitem[{Hoyle(1949)}]{hoyle_origin_1949}
\bibinfo{author}{Hoyle, F.}, \bibinfo{year}{1949}.
\newblock in: \bibinfo{booktitle}{Problems in Cosmical Aerodynamics},
  \bibinfo{publisher}{Central Air Documents Officem Ohio}.
\bibitem[{Hoyle and Lyttleton(1939)}]{hoyle_effect_1939}
\bibinfo{author}{Hoyle, F.}, \bibinfo{author}{Lyttleton, R.A.},
  \bibinfo{year}{1939}.
\newblock p. \bibinfo{pages}{405}.
\bibitem[{Hubble(1929)}]{hubble_spiral_1929}
\bibinfo{author}{Hubble, E.P.}, \bibinfo{year}{1929}.
\newblock \bibinfo{journal}{Astrophysical Journal} \bibinfo{volume}{69},
  \bibinfo{pages}{103--158}.
\bibitem[{Hubble(1936)}]{hubble_realm_1936}
\bibinfo{author}{Hubble, E.P.}, \bibinfo{year}{1936}.
\bibitem[{Jahnke and Maccio(2010)}]{jahnke_non-causal_2010}
\bibinfo{author}{Jahnke, K.}, \bibinfo{author}{Maccio, A.},
  \bibinfo{year}{2010}.
\newblock
  \bibinfo{howpublished}{{http://adsabs.harvard.edu/abs/2010arXiv1006.0482J}}.
\bibitem[{Jenkins and Binney(1990)}]{jenkins_spiral_1990}
\bibinfo{author}{Jenkins, A.}, \bibinfo{author}{Binney, J.},
  \bibinfo{year}{1990}.
\newblock \bibinfo{journal}{Monthly Notices of the Royal Astronomical Society}
  \bibinfo{volume}{245}, \bibinfo{pages}{305--317}.
\bibitem[{Jenkins et~al.(1998)Jenkins, Frenk, Pearce, Thomas, Colberg, White,
  Couchman, Peacock, Efstathiou and Nelson}]{jenkins_evolution_1998}
\bibinfo{author}{Jenkins, A.}, \bibinfo{author}{Frenk, C.S.},
  \bibinfo{author}{Pearce, F.R.}, \bibinfo{author}{Thomas, P.A.},
  \bibinfo{author}{Colberg, J.M.}, \bibinfo{author}{White, S.D.M.},
  \bibinfo{author}{Couchman, H.M.P.}, \bibinfo{author}{Peacock, J.A.},
  \bibinfo{author}{Efstathiou, G.}, \bibinfo{author}{Nelson, A.H.},
  \bibinfo{year}{1998}.
\newblock \bibinfo{journal}{Astrophysical Journal} \bibinfo{volume}{499},
  \bibinfo{pages}{20}.
\bibitem[{Jiang et~al.(2008)Jiang, Jing, Faltenbacher, Lin and
  Li}]{jiang_fitting_2008}
\bibinfo{author}{Jiang, C.Y.}, \bibinfo{author}{Jing, Y.P.},
  \bibinfo{author}{Faltenbacher, A.}, \bibinfo{author}{Lin, W.P.},
  \bibinfo{author}{Li, C.}, \bibinfo{year}{2008}.
\newblock \bibinfo{journal}{{ApJ}} \bibinfo{volume}{675},
  \bibinfo{pages}{1095--1105}.
\bibitem[{Jing and Suto(2002)}]{jing_triaxial_2002}
\bibinfo{author}{Jing, Y.P.}, \bibinfo{author}{Suto, Y.}, \bibinfo{year}{2002}.
\newblock \bibinfo{journal}{Astrophysical Journal} \bibinfo{volume}{574},
  \bibinfo{pages}{538--553}.
\bibitem[{Jog and Solomon(1984)}]{jog_two-fluid_1984}
\bibinfo{author}{Jog, C.J.}, \bibinfo{author}{Solomon, P.M.},
  \bibinfo{year}{1984}.
\newblock \bibinfo{journal}{Astrophysical Journal} \bibinfo{volume}{276},
  \bibinfo{pages}{114--126}.
\bibitem[{Johnston et~al.(2008)Johnston, Bullock, Sharma, Font, Robertson and
  Leitner}]{johnston_tracing_2008}
\bibinfo{author}{Johnston, K.V.}, \bibinfo{author}{Bullock, J.S.},
  \bibinfo{author}{Sharma, S.}, \bibinfo{author}{Font, A.},
  \bibinfo{author}{Robertson, B.E.}, \bibinfo{author}{Leitner, S.N.},
  \bibinfo{year}{2008}.
\newblock \bibinfo{journal}{Astrophysical Journal} \bibinfo{volume}{689},
  \bibinfo{pages}{936--957}.
\bibitem[{Jonsson(2006)}]{jonsson_sunrise:_2006}
\bibinfo{author}{Jonsson, P.}, \bibinfo{year}{2006}.
\newblock \bibinfo{journal}{Monthly Notices of the Royal Astronomical Society}
  \bibinfo{volume}{372}, \bibinfo{pages}{2--20}.
\bibitem[{Jubelgas et~al.(2004)Jubelgas, Springel and
  Dolag}]{jubelgas_thermal_2004}
\bibinfo{author}{Jubelgas, M.}, \bibinfo{author}{Springel, V.},
  \bibinfo{author}{Dolag, K.}, \bibinfo{year}{2004}.
\newblock \bibinfo{journal}{Monthly Notices of the Royal Astronomical Society}
  \bibinfo{volume}{351}, \bibinfo{pages}{423--435}.
\bibitem[{Kampakoglou and Benson(2007)}]{kampakoglou_tidal_2007}
\bibinfo{author}{Kampakoglou, M.}, \bibinfo{author}{Benson, A.J.},
  \bibinfo{year}{2007}.
\newblock \bibinfo{journal}{Monthly Notices of the Royal Astronomical Society}
  \bibinfo{volume}{374}, \bibinfo{pages}{775--786}.
\bibitem[{Kant(1755)}]{kant_allgemeine_1755}
\bibinfo{author}{Kant, I.}, \bibinfo{year}{1755}.
\newblock \bibinfo{publisher}{J. F. Peterson}, \bibinfo{address}{K\"onigsberg
  and Leipzig}.
\bibitem[{Katz(1992)}]{katz_dissipational_1992}
\bibinfo{author}{Katz, N.}, \bibinfo{year}{1992}.
\newblock \bibinfo{journal}{Astrophysical Journal} \bibinfo{volume}{391},
  \bibinfo{pages}{502--517}.
\bibitem[{Katz and White(1993)}]{katz_hierarchical_1993}
\bibinfo{author}{Katz, N.}, \bibinfo{author}{White, S.D.M.},
  \bibinfo{year}{1993}.
\newblock \bibinfo{journal}{Astrophysical Journal} \bibinfo{volume}{412},
  \bibinfo{pages}{455--478}.
\bibitem[{Kauffmann(1996)}]{kauffmann_age_1996}
\bibinfo{author}{Kauffmann, G.}, \bibinfo{year}{1996}.
\newblock \bibinfo{journal}{{MNRAS}} \bibinfo{volume}{281},
  \bibinfo{pages}{487--492}.
\bibitem[{Kauffmann and Charlot(1998)}]{kauffmann_chemical_1998}
\bibinfo{author}{Kauffmann, G.}, \bibinfo{author}{Charlot, S.},
  \bibinfo{year}{1998}.
\newblock \bibinfo{journal}{{MNRAS}} \bibinfo{volume}{294},
  \bibinfo{pages}{705}.
\bibitem[{Kauffmann et~al.(1999a)Kauffmann, Colberg, Diaferio and
  White}]{kauffmann_clustering_1999}
\bibinfo{author}{Kauffmann, G.}, \bibinfo{author}{Colberg, J.M.},
  \bibinfo{author}{Diaferio, A.}, \bibinfo{author}{White, S.D.M.},
  \bibinfo{year}{1999}a.
\newblock \bibinfo{journal}{{MNRAS}} \bibinfo{volume}{303},
  \bibinfo{pages}{188--206}.
\bibitem[{Kauffmann et~al.(1999b)Kauffmann, Colberg, Diaferio and
  White}]{kauffmann_clustering_1999-1}
\bibinfo{author}{Kauffmann, G.}, \bibinfo{author}{Colberg, J.M.},
  \bibinfo{author}{Diaferio, A.}, \bibinfo{author}{White, S.D.M.},
  \bibinfo{year}{1999}b.
\newblock \bibinfo{journal}{{MNRAS}} \bibinfo{volume}{307},
  \bibinfo{pages}{529--536}.
\bibitem[{Kauffmann et~al.(1994)Kauffmann, Guiderdoni and
  White}]{kauffmann_faint_1994}
\bibinfo{author}{Kauffmann, G.}, \bibinfo{author}{Guiderdoni, B.},
  \bibinfo{author}{White, S.D.M.}, \bibinfo{year}{1994}.
\newblock \bibinfo{journal}{{MNRAS}} \bibinfo{volume}{267},
  \bibinfo{pages}{981}.
\bibitem[{Kauffmann and Haehnelt(2000)}]{kauffmann_unified_2000}
\bibinfo{author}{Kauffmann, G.}, \bibinfo{author}{Haehnelt, M.},
  \bibinfo{year}{2000}.
\newblock \bibinfo{journal}{{MNRAS}} \bibinfo{volume}{311},
  \bibinfo{pages}{576--588}.
\bibitem[{Kauffmann and White(1993)}]{kauffmann_merging_1993}
\bibinfo{author}{Kauffmann, G.}, \bibinfo{author}{White, S.D.M.},
  \bibinfo{year}{1993}.
\newblock \bibinfo{journal}{Monthly Notices of the Royal Astronomical Society}
  \bibinfo{volume}{261}, \bibinfo{pages}{921--928}.
\bibitem[{Kauffmann et~al.(1993)Kauffmann, White and
  Guiderdoni}]{kauffmann_formation_1993}
\bibinfo{author}{Kauffmann, G.}, \bibinfo{author}{White, S.D.M.},
  \bibinfo{author}{Guiderdoni, B.}, \bibinfo{year}{1993}.
\newblock \bibinfo{journal}{{MNRAS}} \bibinfo{volume}{264},
  \bibinfo{pages}{201}.
\bibitem[{Kaufmann et~al.(2009)Kaufmann, Bullock, Maller, Fang and
  Wadsley}]{kaufmann_redistributing_2009}
\bibinfo{author}{Kaufmann, T.}, \bibinfo{author}{Bullock, J.S.},
  \bibinfo{author}{Maller, A.H.}, \bibinfo{author}{Fang, T.},
  \bibinfo{author}{Wadsley, J.}, \bibinfo{year}{2009}.
\newblock \bibinfo{journal}{Monthly Notices of the Royal Astronomical Society}
  \bibinfo{volume}{396}, \bibinfo{pages}{191--202}.
\bibitem[{Kay et~al.(2002)Kay, Pearce, Frenk and Jenkins}]{kay_including_2002}
\bibinfo{author}{Kay, S.T.}, \bibinfo{author}{Pearce, F.R.},
  \bibinfo{author}{Frenk, C.S.}, \bibinfo{author}{Jenkins, A.},
  \bibinfo{year}{2002}.
\newblock \bibinfo{journal}{Monthly Notices of the Royal Astronomical Society}
  \bibinfo{volume}{330}, \bibinfo{pages}{113--128}.
\bibitem[{Kazantzidis et~al.(2008)Kazantzidis, Bullock, Zentner, Kravtsov and
  Moustakas}]{kazantzidis_cold_2008}
\bibinfo{author}{Kazantzidis, S.}, \bibinfo{author}{Bullock, J.S.},
  \bibinfo{author}{Zentner, A.R.}, \bibinfo{author}{Kravtsov, A.V.},
  \bibinfo{author}{Moustakas, L.A.}, \bibinfo{year}{2008}.
\newblock \bibinfo{journal}{Astrophysical Journal} \bibinfo{volume}{688},
  \bibinfo{pages}{254--276}.
\bibitem[{Kazantzidis et~al.(2009)Kazantzidis, Zentner, Kravtsov, Bullock and
  Debattista}]{kazantzidis_cold_2009}
\bibinfo{author}{Kazantzidis, S.}, \bibinfo{author}{Zentner, A.R.},
  \bibinfo{author}{Kravtsov, A.V.}, \bibinfo{author}{Bullock, J.S.},
  \bibinfo{author}{Debattista, V.P.}, \bibinfo{year}{2009}.
\bibitem[{Kennicutt(1989)}]{kennicutt_star_1989}
\bibinfo{author}{Kennicutt, R.C.}, \bibinfo{year}{1989}.
\newblock \bibinfo{journal}{Astrophysical Journal} \bibinfo{volume}{344},
  \bibinfo{pages}{685--703}.
\bibitem[{Kennicutt(1998)}]{kennicutt_global_1998}
\bibinfo{author}{Kennicutt, R.C.}, \bibinfo{year}{1998}.
\newblock \bibinfo{journal}{Astrophysical Journal} \bibinfo{volume}{498},
  \bibinfo{pages}{541}.
\bibitem[{Kere\v{s} et~al.(2009)Kere\v{s}, Katz, Fardal, Dav\'e and
  Weinberg}]{kerevs_galaxies_2009}
\bibinfo{author}{Kere\v{s}, D.}, \bibinfo{author}{Katz, N.},
  \bibinfo{author}{Fardal, M.}, \bibinfo{author}{Dav\'e, R.},
  \bibinfo{author}{Weinberg, D.}, \bibinfo{year}{2009}.
\newblock \bibinfo{journal}{{MNRAS}} \bibinfo{volume}{395},
  \bibinfo{pages}{160}.
\bibitem[{Kere\v{s} et~al.(2005)Kere\v{s}, Katz, Weinberg and
  Dav\'e}]{kerevs_do_2005}
\bibinfo{author}{Kere\v{s}, D.}, \bibinfo{author}{Katz, N.},
  \bibinfo{author}{Weinberg, D.H.}, \bibinfo{author}{Dav\'e, R.},
  \bibinfo{year}{2005}.
\newblock \bibinfo{journal}{Monthly Notices of the Royal Astronomical Society}
  \bibinfo{volume}{363}, \bibinfo{pages}{2--28}.
\bibitem[{King(2003)}]{king_black_2003}
\bibinfo{author}{King, A.}, \bibinfo{year}{2003}.
\newblock \bibinfo{journal}{Astrophysical Journal} \bibinfo{volume}{596},
  \bibinfo{pages}{L27--L29}.
\bibitem[{King(2005)}]{king_agn-starburst_2005}
\bibinfo{author}{King, A.}, \bibinfo{year}{2005}.
\newblock \bibinfo{journal}{Astrophysical Journal} \bibinfo{volume}{635},
  \bibinfo{pages}{L121--L123}.
\bibitem[{King(2009)}]{king_heating_2009}
\bibinfo{author}{King, A.}, \bibinfo{year}{2009}.
\newblock \bibinfo{journal}{Astrophysical Journal} \bibinfo{volume}{695},
  \bibinfo{pages}{L107--L110}.
\bibitem[{Klypin et~al.(1999)Klypin, Gottlöber, Kravtsov and
  Khokhlov}]{klypin_galaxies_1999}
\bibinfo{author}{Klypin, A.}, \bibinfo{author}{Gottlöber, S.},
  \bibinfo{author}{Kravtsov, A.V.}, \bibinfo{author}{Khokhlov, A.M.},
  \bibinfo{year}{1999}.
\newblock \bibinfo{journal}{The Astrophysical Journal} \bibinfo{volume}{516},
  \bibinfo{pages}{530--551}.
\bibitem[{Klypin et~al.(1993)Klypin, Holtzman, Primack and
  Regos}]{klypin_structure_1993}
\bibinfo{author}{Klypin, A.}, \bibinfo{author}{Holtzman, J.},
  \bibinfo{author}{Primack, J.}, \bibinfo{author}{Regos, E.},
  \bibinfo{year}{1993}.
\newblock \bibinfo{journal}{The Astrophysical Journal} \bibinfo{volume}{416},
  \bibinfo{pages}{1}.
\bibitem[{Klypin and Shandarin(1983)}]{klypin_three-dimensional_1983}
\bibinfo{author}{Klypin, A.A.}, \bibinfo{author}{Shandarin, S.F.},
  \bibinfo{year}{1983}.
\newblock \bibinfo{journal}{Monthly Notices of the Royal Astronomical Society}
  \bibinfo{volume}{204}, \bibinfo{pages}{891--907}.
\bibitem[{Kobayashi et~al.(2007)Kobayashi, Springel and
  White}]{kobayashi_simulations_2007}
\bibinfo{author}{Kobayashi, C.}, \bibinfo{author}{Springel, V.},
  \bibinfo{author}{White, S.D.M.}, \bibinfo{year}{2007}.
\newblock \bibinfo{journal}{Monthly Notices of the Royal Astronomical Society}
  \bibinfo{volume}{376}, \bibinfo{pages}{1465--1479}.
\bibitem[{Komatsu et~al.(2009)Komatsu, Dunkley, Nolta, Bennett, Gold, Hinshaw,
  Jarosik, Larson, Limon, Page, Spergel, Halpern, Hill, Kogut, Meyer, Tucker,
  Weiland, Wollack and Wright}]{komatsu_five-year_2009}
\bibinfo{author}{Komatsu, E.}, \bibinfo{author}{Dunkley, J.},
  \bibinfo{author}{Nolta, M.R.}, \bibinfo{author}{Bennett, C.L.},
  \bibinfo{author}{Gold, B.}, \bibinfo{author}{Hinshaw, G.},
  \bibinfo{author}{Jarosik, N.}, \bibinfo{author}{Larson, D.},
  \bibinfo{author}{Limon, M.}, \bibinfo{author}{Page, L.},
  \bibinfo{author}{Spergel, D.N.}, \bibinfo{author}{Halpern, M.},
  \bibinfo{author}{Hill, R.S.}, \bibinfo{author}{Kogut, A.},
  \bibinfo{author}{Meyer, S.S.}, \bibinfo{author}{Tucker, G.S.},
  \bibinfo{author}{Weiland, J.L.}, \bibinfo{author}{Wollack, E.},
  \bibinfo{author}{Wright, E.L.}, \bibinfo{year}{2009}.
\newblock \bibinfo{journal}{Astrophysical Journal Supplement Series}
  \bibinfo{volume}{180}, \bibinfo{pages}{330--376}.
\bibitem[{Komatsu et~al.(2010)Komatsu, Smith, Dunkley, Bennett, Gold, Hinshaw,
  Jarosik, Larson, Nolta, Page, Spergel, Halpern, Hill, Kogut, Limon, Meyer,
  Odegard, Tucker, Weiland, Wollack and Wright}]{komatsu_seven-year_2010}
\bibinfo{author}{Komatsu, E.}, \bibinfo{author}{Smith, K.M.},
  \bibinfo{author}{Dunkley, J.}, \bibinfo{author}{Bennett, C.L.},
  \bibinfo{author}{Gold, B.}, \bibinfo{author}{Hinshaw, G.},
  \bibinfo{author}{Jarosik, N.}, \bibinfo{author}{Larson, D.},
  \bibinfo{author}{Nolta, M.R.}, \bibinfo{author}{Page, L.},
  \bibinfo{author}{Spergel, D.N.}, \bibinfo{author}{Halpern, M.},
  \bibinfo{author}{Hill, R.S.}, \bibinfo{author}{Kogut, A.},
  \bibinfo{author}{Limon, M.}, \bibinfo{author}{Meyer, S.S.},
  \bibinfo{author}{Odegard, N.}, \bibinfo{author}{Tucker, G.S.},
  \bibinfo{author}{Weiland, J.L.}, \bibinfo{author}{Wollack, E.},
  \bibinfo{author}{Wright, E.L.}, \bibinfo{year}{2010}.
\newblock
  \bibinfo{howpublished}{{http://adsabs.harvard.edu/abs/2010arXiv1001.4538K}}.
\bibitem[{Komissarov et~al.(2007)Komissarov, Barkov, Vlahakis and
  K\"onigl}]{komissarov_magnetic_2007}
\bibinfo{author}{Komissarov, S.S.}, \bibinfo{author}{Barkov, M.V.},
  \bibinfo{author}{Vlahakis, N.}, \bibinfo{author}{K\"onigl, A.},
  \bibinfo{year}{2007}.
\newblock \bibinfo{journal}{Monthly Notices of the Royal Astronomical Society}
  \bibinfo{volume}{380}, \bibinfo{pages}{51--70}.
\bibitem[{Kormendy et~al.(2009)Kormendy, Fisher, Cornell and
  Bender}]{kormendy_structure_2009}
\bibinfo{author}{Kormendy, J.}, \bibinfo{author}{Fisher, D.B.},
  \bibinfo{author}{Cornell, M.E.}, \bibinfo{author}{Bender, R.},
  \bibinfo{year}{2009}.
\newblock \bibinfo{journal}{The Astrophysical Journal Supplement Series}
  \bibinfo{volume}{182}, \bibinfo{pages}{216--309}.
\bibitem[{Kormendy and Freeman(2004)}]{kormendy_scaling_2004}
\bibinfo{author}{Kormendy, J.}, \bibinfo{author}{Freeman, K.C.},
  \bibinfo{year}{2004}.
\newblock p. \bibinfo{pages}{377}.
\bibitem[{Kormendy and Kennicutt(2004)}]{kormendy_secular_2004}
\bibinfo{author}{Kormendy, J.}, \bibinfo{author}{Kennicutt, R.C.},
  \bibinfo{year}{2004}.
\newblock \bibinfo{journal}{Annual Review of Astronomy and Astrophysics}
  \bibinfo{volume}{42}, \bibinfo{pages}{603--683}.
\bibitem[{Kowalski et~al.(2008)Kowalski, Rubin, Aldering, Agostinho, Amadon,
  Amanullah, Balland, Barbary, Blanc, Challis, Conley, Connolly, Covarrubias,
  Dawson, Deustua, Ellis, Fabbro, Fadeyev, Fan, Farris, Folatelli, Frye,
  Garavini, Gates, Germany, Goldhaber, Goldman, Goobar, Groom, Haissinski,
  Hardin, Hook, Kent, Kim, Knop, Lidman, Linder, Mendez, Meyers, Miller,
  Moniez, ao, Newberg, Nobili, Nugent, Pain, Perdereau, Perlmutter, Phillips,
  Prasad, Quimby, Regnault, Rich, Rubenstein, {Ruiz-Lapuente}, Santos,
  Schaefer, Schommer, Smith, Soderberg, Spadafora, Strolger, Strovink,
  Suntzeff, Suzuki, Thomas, Walton, Wang, {Wood-Vasey} and
  Yun}]{kowalski_improved_2008}
\bibinfo{author}{Kowalski, M.}, \bibinfo{author}{Rubin, D.},
  \bibinfo{author}{Aldering, G.}, \bibinfo{author}{Agostinho, R.J.},
  \bibinfo{author}{Amadon, A.}, \bibinfo{author}{Amanullah, R.},
  \bibinfo{author}{Balland, C.}, \bibinfo{author}{Barbary, K.},
  \bibinfo{author}{Blanc, G.}, \bibinfo{author}{Challis, P.J.},
  \bibinfo{author}{Conley, A.}, \bibinfo{author}{Connolly, N.V.},
  \bibinfo{author}{Covarrubias, R.}, \bibinfo{author}{Dawson, K.S.},
  \bibinfo{author}{Deustua, S.E.}, \bibinfo{author}{Ellis, R.},
  \bibinfo{author}{Fabbro, S.}, \bibinfo{author}{Fadeyev, V.},
  \bibinfo{author}{Fan, X.}, \bibinfo{author}{Farris, B.},
  \bibinfo{author}{Folatelli, G.}, \bibinfo{author}{Frye, B.L.},
  \bibinfo{author}{Garavini, G.}, \bibinfo{author}{Gates, E.L.},
  \bibinfo{author}{Germany, L.}, \bibinfo{author}{Goldhaber, G.},
  \bibinfo{author}{Goldman, B.}, \bibinfo{author}{Goobar, A.},
  \bibinfo{author}{Groom, D.E.}, \bibinfo{author}{Haissinski, J.},
  \bibinfo{author}{Hardin, D.}, \bibinfo{author}{Hook, I.},
  \bibinfo{author}{Kent, S.}, \bibinfo{author}{Kim, A.G.},
  \bibinfo{author}{Knop, R.A.}, \bibinfo{author}{Lidman, C.},
  \bibinfo{author}{Linder, E.V.}, \bibinfo{author}{Mendez, J.},
  \bibinfo{author}{Meyers, J.}, \bibinfo{author}{Miller, G.J.},
  \bibinfo{author}{Moniez, M.}, \bibinfo{author}{ao, A.M.M.},
  \bibinfo{author}{Newberg, H.}, \bibinfo{author}{Nobili, S.},
  \bibinfo{author}{Nugent, P.E.}, \bibinfo{author}{Pain, R.},
  \bibinfo{author}{Perdereau, O.}, \bibinfo{author}{Perlmutter, S.},
  \bibinfo{author}{Phillips, M.M.}, \bibinfo{author}{Prasad, V.},
  \bibinfo{author}{Quimby, R.}, \bibinfo{author}{Regnault, N.},
  \bibinfo{author}{Rich, J.}, \bibinfo{author}{Rubenstein, E.P.},
  \bibinfo{author}{{Ruiz-Lapuente}, P.}, \bibinfo{author}{Santos, F.D.},
  \bibinfo{author}{Schaefer, B.E.}, \bibinfo{author}{Schommer, R.A.},
  \bibinfo{author}{Smith, R.C.}, \bibinfo{author}{Soderberg, A.M.},
  \bibinfo{author}{Spadafora, A.L.}, \bibinfo{author}{Strolger, L.},
  \bibinfo{author}{Strovink, M.}, \bibinfo{author}{Suntzeff, N.B.},
  \bibinfo{author}{Suzuki, N.}, \bibinfo{author}{Thomas, R.C.},
  \bibinfo{author}{Walton, N.A.}, \bibinfo{author}{Wang, L.},
  \bibinfo{author}{{Wood-Vasey}, W.M.}, \bibinfo{author}{Yun, J.L.},
  \bibinfo{year}{2008}.
\newblock \bibinfo{journal}{Astrophysical Journal} \bibinfo{volume}{686},
  \bibinfo{pages}{749--778}.
\bibitem[{Kravtsov et~al.(2004)Kravtsov, Berlind, Wechsler, Klypin, Gottlöber,
  Allgood and Primack}]{kravtsov_dark_2004}
\bibinfo{author}{Kravtsov, A.V.}, \bibinfo{author}{Berlind, A.A.},
  \bibinfo{author}{Wechsler, R.H.}, \bibinfo{author}{Klypin, A.A.},
  \bibinfo{author}{Gottlöber, S.}, \bibinfo{author}{Allgood, B.},
  \bibinfo{author}{Primack, J.R.}, \bibinfo{year}{2004}.
\newblock \bibinfo{journal}{The Astrophysical Journal} \bibinfo{volume}{609},
  \bibinfo{pages}{35--49}.
\bibitem[{Kravtsov et~al.(2002)Kravtsov, Klypin and
  Hoffman}]{kravtsov_constrained_2002}
\bibinfo{author}{Kravtsov, A.V.}, \bibinfo{author}{Klypin, A.},
  \bibinfo{author}{Hoffman, Y.}, \bibinfo{year}{2002}.
\newblock \bibinfo{journal}{The Astrophysical Journal} \bibinfo{volume}{571},
  \bibinfo{pages}{563--575}.
\bibitem[{Kravtsov et~al.(1997)Kravtsov, Klypin and
  Khokhlov}]{kravtsov_adaptive_1997}
\bibinfo{author}{Kravtsov, A.V.}, \bibinfo{author}{Klypin, A.A.},
  \bibinfo{author}{Khokhlov, A.M.}, \bibinfo{year}{1997}.
\newblock \bibinfo{journal}{Astrophysical Journal Supplement Series}
  \bibinfo{volume}{111}, \bibinfo{pages}{73}.
\bibitem[{Kriek et~al.(2009)Kriek, van Dokkum, Franx, Illingworth and
  Magee}]{kriek_hubble_2009}
\bibinfo{author}{Kriek, M.}, \bibinfo{author}{van Dokkum, P.G.},
  \bibinfo{author}{Franx, M.}, \bibinfo{author}{Illingworth, G.D.},
  \bibinfo{author}{Magee, D.K.}, \bibinfo{year}{2009}.
\newblock \bibinfo{journal}{The Astrophysical Journal} \bibinfo{volume}{705},
  \bibinfo{pages}{L71--L75}.
\bibitem[{Kroupa(2001)}]{kroupa_variation_2001}
\bibinfo{author}{Kroupa, P.}, \bibinfo{year}{2001}.
\newblock \bibinfo{journal}{Monthly Notices of the Royal Astronomical Society}
  \bibinfo{volume}{322}, \bibinfo{pages}{231--246}.
\bibitem[{Krumholz et~al.(2009)Krumholz, {McKee} and
  Tumlinson}]{krumholz_star_2009}
\bibinfo{author}{Krumholz, M.R.}, \bibinfo{author}{{McKee}, C.F.},
  \bibinfo{author}{Tumlinson, J.}, \bibinfo{year}{2009}.
\bibitem[{Krumholz and Tan(2007)}]{krumholz_slow_2007}
\bibinfo{author}{Krumholz, M.R.}, \bibinfo{author}{Tan, J.C.},
  \bibinfo{year}{2007}.
\newblock \bibinfo{journal}{Astrophysical Journal} \bibinfo{volume}{654},
  \bibinfo{pages}{304--315}.
\bibitem[{Kuhlen et~al.(2008)Kuhlen, Diemand, Madau and Zemp}]{kuhlen_via_2008}
\bibinfo{author}{Kuhlen, M.}, \bibinfo{author}{Diemand, J.},
  \bibinfo{author}{Madau, P.}, \bibinfo{author}{Zemp, M.},
  \bibinfo{year}{2008}.
\newblock \bibinfo{journal}{Journal of Physics Conference Series}
  \bibinfo{volume}{125}, \bibinfo{pages}{2008}.
\bibitem[{Kuijken and Gilmore(1989)}]{kuijken_mass_1989}
\bibinfo{author}{Kuijken, K.}, \bibinfo{author}{Gilmore, G.},
  \bibinfo{year}{1989}.
\newblock \bibinfo{journal}{Monthly Notices of the Royal Astronomical Society}
  \bibinfo{volume}{239}, \bibinfo{pages}{571--603}.
\bibitem[{Lacey and Cole(1993)}]{lacey_merger_1993}
\bibinfo{author}{Lacey, C.}, \bibinfo{author}{Cole, S.}, \bibinfo{year}{1993}.
\newblock \bibinfo{journal}{Monthly Notices of the Royal Astronomical Society}
  \bibinfo{volume}{262}, \bibinfo{pages}{627--649}.
\bibitem[{Lacey and Silk(1991)}]{lacey_tidally_1991}
\bibinfo{author}{Lacey, C.}, \bibinfo{author}{Silk, J.}, \bibinfo{year}{1991}.
\newblock \bibinfo{journal}{Astrophysical Journal} \bibinfo{volume}{381},
  \bibinfo{pages}{14--32}.
\bibitem[{Lacey(1984)}]{lacey_influence_1984}
\bibinfo{author}{Lacey, C.G.}, \bibinfo{year}{1984}.
\newblock \bibinfo{journal}{Monthly Notices of the Royal Astronomical Society}
  \bibinfo{volume}{208}, \bibinfo{pages}{687--707}.
\bibitem[{Lacey et~al.(2010)Lacey, Baugh, Frenk, Benson and
  .}]{lacey_evolution_2010}
\bibinfo{author}{Lacey, C.G.}, \bibinfo{author}{Baugh, C.M.},
  \bibinfo{author}{Frenk, C.S.}, \bibinfo{author}{Benson, A.J.},
  \bibinfo{author}{.}, \bibinfo{year}{2010}.
\newblock
  \bibinfo{howpublished}{{http://adsabs.harvard.edu/abs/2010arXiv1004.3545L}}.
\bibitem[{Lacey et~al.(2008)Lacey, Baugh, Frenk, Silva, Granato and
  Bressan}]{lacey_galaxy_2008}
\bibinfo{author}{Lacey, C.G.}, \bibinfo{author}{Baugh, C.M.},
  \bibinfo{author}{Frenk, C.S.}, \bibinfo{author}{Silva, L.},
  \bibinfo{author}{Granato, G.L.}, \bibinfo{author}{Bressan, A.},
  \bibinfo{year}{2008}.
\newblock \bibinfo{journal}{{MNRAS}} \bibinfo{volume}{385},
  \bibinfo{pages}{1155--1178}.
\bibitem[{Lanzoni et~al.(2005)Lanzoni, Guiderdoni, Mamon, Devriendt and
  Hatton}]{lanzoni_galics-_2005}
\bibinfo{author}{Lanzoni, B.}, \bibinfo{author}{Guiderdoni, B.},
  \bibinfo{author}{Mamon, G.A.}, \bibinfo{author}{Devriendt, J.},
  \bibinfo{author}{Hatton, S.}, \bibinfo{year}{2005}.
\newblock \bibinfo{journal}{Monthly Notices of the Royal Astronomical Society}
  \bibinfo{volume}{361}, \bibinfo{pages}{369--384}.
\bibitem[{Larson(1974)}]{larson_effects_1974}
\bibinfo{author}{Larson, R.B.}, \bibinfo{year}{1974}.
\newblock \bibinfo{journal}{Monthly Notices of the Royal Astronomical Society}
  \bibinfo{volume}{169}, \bibinfo{pages}{229--246}.
\bibitem[{Laursen et~al.(2009)Laursen, Razoumov and
  {Sommer-Larsen}}]{laursen_ly_2009}
\bibinfo{author}{Laursen, P.}, \bibinfo{author}{Razoumov, A.O.},
  \bibinfo{author}{{Sommer-Larsen}, J.}, \bibinfo{year}{2009}.
\newblock \bibinfo{journal}{Astrophysical Journal} \bibinfo{volume}{696},
  \bibinfo{pages}{853--869}.
\bibitem[{Law et~al.(2009)Law, Steidel, Erb, Larkin, Pettini, Shapley and
  Wright}]{law_kiloparsec-scale_2009}
\bibinfo{author}{Law, D.R.}, \bibinfo{author}{Steidel, C.C.},
  \bibinfo{author}{Erb, D.K.}, \bibinfo{author}{Larkin, J.E.},
  \bibinfo{author}{Pettini, M.}, \bibinfo{author}{Shapley, A.E.},
  \bibinfo{author}{Wright, S.A.}, \bibinfo{year}{2009}.
\newblock \bibinfo{journal}{Astrophysical Journal} \bibinfo{volume}{697},
  \bibinfo{pages}{2057--2082}.
\bibitem[{Lawlor et~al.(2008)Lawlor, Young, Johnson and
  {MacDonald}}]{lawlor_single_2008}
\bibinfo{author}{Lawlor, T.M.}, \bibinfo{author}{Young, T.R.},
  \bibinfo{author}{Johnson, T.A.}, \bibinfo{author}{{MacDonald}, J.},
  \bibinfo{year}{2008}.
\newblock \bibinfo{journal}{Monthly Notices of the Royal Astronomical Society}
  \bibinfo{volume}{384}, \bibinfo{pages}{1533--1543}.
\bibitem[{Lee et~al.(2009)Lee, Worthey, Dotter, Chaboyer, Jevremovic, Baron,
  Briley, Ferguson, Coelho and Trager}]{lee_stellar_2009}
\bibinfo{author}{Lee, H.}, \bibinfo{author}{Worthey, G.},
  \bibinfo{author}{Dotter, A.}, \bibinfo{author}{Chaboyer, B.},
  \bibinfo{author}{Jevremovic, D.}, \bibinfo{author}{Baron, E.},
  \bibinfo{author}{Briley, M.M.}, \bibinfo{author}{Ferguson, J.W.},
  \bibinfo{author}{Coelho, P.}, \bibinfo{author}{Trager, S.C.},
  \bibinfo{year}{2009}.
\newblock \bibinfo{journal}{Astrophysical Journal} \bibinfo{volume}{694},
  \bibinfo{pages}{902--923}.
\bibitem[{Lemson and Kauffmann(1999)}]{lemson_environmental_1999}
\bibinfo{author}{Lemson, G.}, \bibinfo{author}{Kauffmann, G.},
  \bibinfo{year}{1999}.
\newblock \bibinfo{journal}{{MNRAS}} \bibinfo{volume}{302},
  \bibinfo{pages}{111--117}.
\bibitem[{Li et~al.(2008)Li, Li and Cen}]{li_cosmomhd:cosmological_2008}
\bibinfo{author}{Li, S.}, \bibinfo{author}{Li, H.}, \bibinfo{author}{Cen, R.},
  \bibinfo{year}{2008}.
\newblock \bibinfo{journal}{Astrophysical Journal Supplement Series}
  \bibinfo{volume}{174}, \bibinfo{pages}{1--12}.
\bibitem[{Li et~al.(2010)Li, Lucia and Helmi}]{li_nature_2010}
\bibinfo{author}{Li, Y.}, \bibinfo{author}{Lucia, G.D.},
  \bibinfo{author}{Helmi, A.}, \bibinfo{year}{2010}.
\newblock \bibinfo{journal}{Monthly Notices of the Royal Astronomical Society}
  \bibinfo{volume}{401}, \bibinfo{pages}{2036--2052}.
\bibitem[{Libeskind et~al.(2006)Libeskind, Cole, Frenk and
  Helly}]{libeskind_effect_2006}
\bibinfo{author}{Libeskind, N.I.}, \bibinfo{author}{Cole, S.},
  \bibinfo{author}{Frenk, C.S.}, \bibinfo{author}{Helly, J.C.},
  \bibinfo{year}{2006}.
\newblock \bibinfo{journal}{Monthly Notices of the Royal Astronomical Society}
  \bibinfo{volume}{368}, \bibinfo{pages}{1381--1391}.
\bibitem[{Lin and Shu(1964)}]{lin_spiral_1964}
\bibinfo{author}{Lin, C.C.}, \bibinfo{author}{Shu, F.H.}, \bibinfo{year}{1964}.
\newblock \bibinfo{journal}{The Astrophysical Journal} \bibinfo{volume}{140},
  \bibinfo{pages}{646}.
\bibitem[{Lin and Pringle(1987)}]{lin_formation_1987}
\bibinfo{author}{Lin, D.N.C.}, \bibinfo{author}{Pringle, J.E.},
  \bibinfo{year}{1987}.
\newblock \bibinfo{journal}{Astrophysical Journal} \bibinfo{volume}{320},
  \bibinfo{pages}{L87--L91}.
\bibitem[{Lintott et~al.(2008)Lintott, Schawinski, Slosar, Land, Bamford,
  Thomas, Raddick, Nichol, Szalay, Andreescu, Murray and
  Vandenberg}]{lintott_galaxy_2008}
\bibinfo{author}{Lintott, C.J.}, \bibinfo{author}{Schawinski, K.},
  \bibinfo{author}{Slosar, A.}, \bibinfo{author}{Land, K.},
  \bibinfo{author}{Bamford, S.}, \bibinfo{author}{Thomas, D.},
  \bibinfo{author}{Raddick, M.J.}, \bibinfo{author}{Nichol, R.C.},
  \bibinfo{author}{Szalay, A.}, \bibinfo{author}{Andreescu, D.},
  \bibinfo{author}{Murray, P.}, \bibinfo{author}{Vandenberg, J.},
  \bibinfo{year}{2008}.
\newblock \bibinfo{journal}{Monthly Notices of the Royal Astronomical Society}
  \bibinfo{volume}{389}, \bibinfo{pages}{1179--1189}.
\bibitem[{Lotz et~al.(2008)Lotz, Jonsson, Cox and Primack}]{lotz_galaxy_2008}
\bibinfo{author}{Lotz, J.M.}, \bibinfo{author}{Jonsson, P.},
  \bibinfo{author}{Cox, T.J.}, \bibinfo{author}{Primack, J.R.},
  \bibinfo{year}{2008}.
\newblock \bibinfo{journal}{Monthly Notices of the Royal Astronomical Society}
  \bibinfo{volume}{391}, \bibinfo{pages}{1137--1162}.
\bibitem[{Lotz et~al.(2004)Lotz, Primack and Madau}]{lotz_new_2004}
\bibinfo{author}{Lotz, J.M.}, \bibinfo{author}{Primack, J.},
  \bibinfo{author}{Madau, P.}, \bibinfo{year}{2004}.
\newblock \bibinfo{journal}{Astronomical Journal} \bibinfo{volume}{128},
  \bibinfo{pages}{163--182}.
\bibitem[{Lucia and Blaizot(2007)}]{lucia_hierarchical_2007}
\bibinfo{author}{Lucia, G.D.}, \bibinfo{author}{Blaizot, J.},
  \bibinfo{year}{2007}.
\newblock \bibinfo{journal}{Monthly Notices of the Royal Astronomical Society}
  \bibinfo{volume}{375}, \bibinfo{pages}{2--14}.
\bibitem[{Lucia et~al.(2004)Lucia, Kauffmann and White}]{lucia_chemical_2004}
\bibinfo{author}{Lucia, G.D.}, \bibinfo{author}{Kauffmann, G.},
  \bibinfo{author}{White, S.D.M.}, \bibinfo{year}{2004}.
\newblock \bibinfo{journal}{{MNRAS}} \bibinfo{volume}{349},
  \bibinfo{pages}{1101--1116}.
\bibitem[{Lucia et~al.(2006)Lucia, Springel, White, Croton and
  Kauffmann}]{lucia_formation_2006}
\bibinfo{author}{Lucia, G.D.}, \bibinfo{author}{Springel, V.},
  \bibinfo{author}{White, S.D.M.}, \bibinfo{author}{Croton, D.},
  \bibinfo{author}{Kauffmann, G.}, \bibinfo{year}{2006}.
\newblock \bibinfo{journal}{{MNRAS}} \bibinfo{volume}{366},
  \bibinfo{pages}{499--509}.
\bibitem[{{Lynden-Bell}(1967)}]{lynden-bell_statistical_1967}
\bibinfo{author}{{Lynden-Bell}, D.}, \bibinfo{year}{1967}.
\newblock \bibinfo{journal}{Monthly Notices of the Royal Astronomical Society}
  \bibinfo{volume}{136}, \bibinfo{pages}{101}.
\bibitem[{{Lynden-Bell} and Kalnajs(1972)}]{lynden-bell_generating_1972}
\bibinfo{author}{{Lynden-Bell}, D.}, \bibinfo{author}{Kalnajs, A.J.},
  \bibinfo{year}{1972}.
\newblock \bibinfo{journal}{Monthly Notices of the Royal Astronomical Society}
  \bibinfo{volume}{157}, \bibinfo{pages}{1}.
\bibitem[{Macci\'o et~al.(2009)Macci\'o, Kang, Fontanot, Somerville, Koposov
  and Monaco}]{maccio_origin_2009}
\bibinfo{author}{Macci\'o, A.V.}, \bibinfo{author}{Kang, X.},
  \bibinfo{author}{Fontanot, F.}, \bibinfo{author}{Somerville, R.S.},
  \bibinfo{author}{Koposov, S.E.}, \bibinfo{author}{Monaco, P.},
  \bibinfo{year}{2009}.
\bibitem[{Machida et~al.(2008)Machida, Matsumoto and ichiro
  Inutsuka}]{machida_magnetohydrodynamics_2008}
\bibinfo{author}{Machida, M.N.}, \bibinfo{author}{Matsumoto, T.},
  \bibinfo{author}{ichiro Inutsuka, S.}, \bibinfo{year}{2008}.
\newblock \bibinfo{journal}{Astrophysical Journal} \bibinfo{volume}{685},
  \bibinfo{pages}{690--704}.
\bibitem[{Madau(1995)}]{madau_radiative_1995}
\bibinfo{author}{Madau, P.}, \bibinfo{year}{1995}.
\newblock \bibinfo{journal}{Astrophysical Journal} \bibinfo{volume}{441},
  \bibinfo{pages}{18--27}.
\bibitem[{Madau et~al.(1996)Madau, Ferguson, Dickinson, Giavalisco, Steidel and
  Fruchter}]{madau_high-redshift_1996}
\bibinfo{author}{Madau, P.}, \bibinfo{author}{Ferguson, H.C.},
  \bibinfo{author}{Dickinson, M.E.}, \bibinfo{author}{Giavalisco, M.},
  \bibinfo{author}{Steidel, C.C.}, \bibinfo{author}{Fruchter, A.},
  \bibinfo{year}{1996}.
\newblock \bibinfo{journal}{Monthly Notices of the Royal Astronomical Society}
  \bibinfo{volume}{283}, \bibinfo{pages}{1388--1404}.
\bibitem[{Madau and Haardt(2009)}]{madau_he_2009}
\bibinfo{author}{Madau, P.}, \bibinfo{author}{Haardt, F.},
  \bibinfo{year}{2009}.
\newblock \bibinfo{journal}{Astrophysical Journal} \bibinfo{volume}{693},
  \bibinfo{pages}{L100--L103}.
\bibitem[{Magorrian et~al.(1998)Magorrian, Tremaine, Richstone, Bender, Bower,
  Dressler, Faber, Gebhardt, Green, Grillmair, Kormendy and
  Lauer}]{magorrian_demography_1998}
\bibinfo{author}{Magorrian, J.}, \bibinfo{author}{Tremaine, S.},
  \bibinfo{author}{Richstone, D.}, \bibinfo{author}{Bender, R.},
  \bibinfo{author}{Bower, G.}, \bibinfo{author}{Dressler, A.},
  \bibinfo{author}{Faber, S.M.}, \bibinfo{author}{Gebhardt, K.},
  \bibinfo{author}{Green, R.}, \bibinfo{author}{Grillmair, C.},
  \bibinfo{author}{Kormendy, J.}, \bibinfo{author}{Lauer, T.},
  \bibinfo{year}{1998}.
\newblock \bibinfo{journal}{Astronomical Journal} \bibinfo{volume}{115},
  \bibinfo{pages}{2285--2305}.
\bibitem[{Malbon et~al.(2007)Malbon, Baugh, Frenk and
  Lacey}]{malbon_black_2007}
\bibinfo{author}{Malbon, R.K.}, \bibinfo{author}{Baugh, C.M.},
  \bibinfo{author}{Frenk, C.S.}, \bibinfo{author}{Lacey, C.G.},
  \bibinfo{year}{2007}.
\newblock \bibinfo{journal}{{MNRAS}} \bibinfo{volume}{382},
  \bibinfo{pages}{1394--1414}.
\bibitem[{Maller and Bullock(2004)}]{maller_multiphase_2004}
\bibinfo{author}{Maller, A.H.}, \bibinfo{author}{Bullock, J.S.},
  \bibinfo{year}{2004}.
\newblock \bibinfo{journal}{Monthly Notices of the Royal Astronomical Society}
  \bibinfo{volume}{355}, \bibinfo{pages}{694--712}.
\bibitem[{Maller et~al.(2002)Maller, Dekel and
  Somerville}]{maller_modelling_2002}
\bibinfo{author}{Maller, A.H.}, \bibinfo{author}{Dekel, A.},
  \bibinfo{author}{Somerville, R.}, \bibinfo{year}{2002}.
\newblock \bibinfo{journal}{{MNRAS}} \bibinfo{volume}{329},
  \bibinfo{pages}{423--430}.
\bibitem[{Maller et~al.(2001)Maller, Prochaska, Somerville and
  Primack}]{maller_damped_2001}
\bibinfo{author}{Maller, A.H.}, \bibinfo{author}{Prochaska, J.X.},
  \bibinfo{author}{Somerville, R.S.}, \bibinfo{author}{Primack, J.R.},
  \bibinfo{year}{2001}.
\newblock \bibinfo{journal}{{MNRAS}} \bibinfo{volume}{326},
  \bibinfo{pages}{1475--1488}.
\bibitem[{Maller et~al.(2003)Maller, Prochaska, Somerville and
  Primack}]{maller_damped_2003}
\bibinfo{author}{Maller, A.H.}, \bibinfo{author}{Prochaska, J.X.},
  \bibinfo{author}{Somerville, R.S.}, \bibinfo{author}{Primack, J.R.},
  \bibinfo{year}{2003}.
\newblock \bibinfo{journal}{{MNRAS}} \bibinfo{volume}{343},
  \bibinfo{pages}{268--278}.
\bibitem[{Mandelbaum et~al.(2006)Mandelbaum, Seljak, Cool, Blanton, Hirata and
  Brinkmann}]{mandelbaum_density_2006}
\bibinfo{author}{Mandelbaum, R.}, \bibinfo{author}{Seljak, U.},
  \bibinfo{author}{Cool, R.J.}, \bibinfo{author}{Blanton, M.},
  \bibinfo{author}{Hirata, C.M.}, \bibinfo{author}{Brinkmann, J.},
  \bibinfo{year}{2006}.
\newblock \bibinfo{journal}{Monthly Notices of the Royal Astronomical Society}
  \bibinfo{volume}{372}, \bibinfo{pages}{758--776}.
\bibitem[{Mantz et~al.(2009)Mantz, Allen, Rapetti and
  Ebeling}]{mantz_observed_2009}
\bibinfo{author}{Mantz, A.}, \bibinfo{author}{Allen, S.W.},
  \bibinfo{author}{Rapetti, D.}, \bibinfo{author}{Ebeling, H.},
  \bibinfo{year}{2009}.
\newblock
  \bibinfo{howpublished}{{http://adsabs.harvard.edu/abs/2009arXiv0909.3098M}}.
\bibitem[{Mao et~al.(1998)Mao, Mo and White}]{mao_evolution_1998}
\bibinfo{author}{Mao, S.}, \bibinfo{author}{Mo, H.J.}, \bibinfo{author}{White,
  S.D.M.}, \bibinfo{year}{1998}.
\newblock \bibinfo{journal}{Monthly Notices of the Royal Astronomical Society}
  \bibinfo{volume}{297}, \bibinfo{pages}{L71--L75}.
\bibitem[{Maraston(2005)}]{maraston_evolutionary_2005}
\bibinfo{author}{Maraston, C.}, \bibinfo{year}{2005}.
\newblock \bibinfo{journal}{Monthly Notices of the Royal Astronomical Society}
  \bibinfo{volume}{362}, \bibinfo{pages}{799--825}.
\bibitem[{Marigo(2001)}]{marigo_chemical_2001}
\bibinfo{author}{Marigo, P.}, \bibinfo{year}{2001}.
\newblock \bibinfo{journal}{{A\&A}} \bibinfo{volume}{370},
  \bibinfo{pages}{194--217}.
\bibitem[{Markevitch et~al.(2004)Markevitch, Gonzalez, Clowe and
  Vikhlinin}]{markevitch_constraintsdark_2004}
\bibinfo{author}{Markevitch, M.}, \bibinfo{author}{Gonzalez, A.},
  \bibinfo{author}{Clowe, D.}, \bibinfo{author}{Vikhlinin, A.},
  \bibinfo{year}{2004}.
\newblock p. \bibinfo{pages}{263}.
\bibitem[{Marochnik and Suchkov(1996)}]{marochnik_milky_1996}
\bibinfo{author}{Marochnik, L.S.}, \bibinfo{author}{Suchkov, A.A.},
  \bibinfo{year}{1996}.
\bibitem[{Marri and White(2003)}]{marri_smoothed_2003}
\bibinfo{author}{Marri, S.}, \bibinfo{author}{White, S.D.M.},
  \bibinfo{year}{2003}.
\newblock \bibinfo{journal}{Monthly Notices of the Royal Astronomical Society}
  \bibinfo{volume}{345}, \bibinfo{pages}{561--574}.
\bibitem[{Martin and Kennicutt(2001)}]{martin_star_2001}
\bibinfo{author}{Martin, C.L.}, \bibinfo{author}{Kennicutt, R.C.},
  \bibinfo{year}{2001}.
\newblock \bibinfo{journal}{Astrophysical Journal} \bibinfo{volume}{555},
  \bibinfo{pages}{301--321}.
\bibitem[{{Mart\'inez-Serrano} et~al.(2008){Mart\'inez-Serrano}, Serna,
  {Dom\'inguez-Tenreiro} and Moll\'a}]{martinez-serrano_chemical_2008}
\bibinfo{author}{{Mart\'inez-Serrano}, F.J.}, \bibinfo{author}{Serna, A.},
  \bibinfo{author}{{Dom\'inguez-Tenreiro}, R.}, \bibinfo{author}{Moll\'a, M.},
  \bibinfo{year}{2008}.
\newblock \bibinfo{journal}{{MNRAS}} \bibinfo{volume}{388},
  \bibinfo{pages}{39--55}.
\bibitem[{Mastropietro et~al.(2005a)Mastropietro, Moore, Mayer, Debattista,
  Piffaretti and Stadel}]{mastropietro_morphological_2005}
\bibinfo{author}{Mastropietro, C.}, \bibinfo{author}{Moore, B.},
  \bibinfo{author}{Mayer, L.}, \bibinfo{author}{Debattista, V.P.},
  \bibinfo{author}{Piffaretti, R.}, \bibinfo{author}{Stadel, J.},
  \bibinfo{year}{2005}a.
\newblock \bibinfo{journal}{Monthly Notices of the Royal Astronomical Society}
  \bibinfo{volume}{364}, \bibinfo{pages}{607--619}.
\bibitem[{Mastropietro et~al.(2005b)Mastropietro, Moore, Mayer, Wadsley and
  Stadel}]{mastropietro_gravitational_2005}
\bibinfo{author}{Mastropietro, C.}, \bibinfo{author}{Moore, B.},
  \bibinfo{author}{Mayer, L.}, \bibinfo{author}{Wadsley, J.},
  \bibinfo{author}{Stadel, J.}, \bibinfo{year}{2005}b.
\newblock \bibinfo{journal}{Monthly Notices of the Royal Astronomical Society}
  \bibinfo{volume}{363}, \bibinfo{pages}{509--520}.
\bibitem[{Mateo(1998)}]{mateo_dwarf_1998}
\bibinfo{author}{Mateo, M.L.}, \bibinfo{year}{1998}.
\newblock \bibinfo{journal}{Annual Review of Astronomy and Astrophysics}
  \bibinfo{volume}{36}, \bibinfo{pages}{435--506}.
\bibitem[{Matteo et~al.(2008)Matteo, Colberg, Springel, Hernquist and
  Sijacki}]{matteo_direct_2008}
\bibinfo{author}{Matteo, T.D.}, \bibinfo{author}{Colberg, J.},
  \bibinfo{author}{Springel, V.}, \bibinfo{author}{Hernquist, L.},
  \bibinfo{author}{Sijacki, D.}, \bibinfo{year}{2008}.
\newblock \bibinfo{journal}{Astrophysical Journal} \bibinfo{volume}{676},
  \bibinfo{pages}{33--53}.
\bibitem[{Matteo et~al.(2003)Matteo, Croft, Springel and
  Hernquist}]{matteo_black_2003}
\bibinfo{author}{Matteo, T.D.}, \bibinfo{author}{Croft, R.A.C.},
  \bibinfo{author}{Springel, V.}, \bibinfo{author}{Hernquist, L.},
  \bibinfo{year}{2003}.
\newblock \bibinfo{journal}{Astrophysical Journal} \bibinfo{volume}{593},
  \bibinfo{pages}{56--68}.
\bibitem[{Matteo et~al.(2005)Matteo, Springel and
  Hernquist}]{matteo_energy_2005}
\bibinfo{author}{Matteo, T.D.}, \bibinfo{author}{Springel, V.},
  \bibinfo{author}{Hernquist, L.}, \bibinfo{year}{2005}.
\newblock \bibinfo{journal}{Nature} \bibinfo{volume}{433},
  \bibinfo{pages}{604--607}.
\bibitem[{Matteucci and Gibson(1995)}]{matteucci_chemical_1995}
\bibinfo{author}{Matteucci, F.}, \bibinfo{author}{Gibson, B.K.},
  \bibinfo{year}{1995}.
\newblock \bibinfo{journal}{Astronomy and Astrophysics} \bibinfo{volume}{304},
  \bibinfo{pages}{11}.
\bibitem[{Matteucci and Greggio(1986)}]{matteucci_relative_1986}
\bibinfo{author}{Matteucci, F.}, \bibinfo{author}{Greggio, L.},
  \bibinfo{year}{1986}.
\newblock \bibinfo{journal}{Astronomy and Astrophysics} \bibinfo{volume}{154},
  \bibinfo{pages}{279--287}.
\bibitem[{Mayer et~al.(2001a)Mayer, Governato, Colpi, Moore, Quinn, Wadsley,
  Stadel and Lake}]{mayer_tidal_2001}
\bibinfo{author}{Mayer, L.}, \bibinfo{author}{Governato, F.},
  \bibinfo{author}{Colpi, M.}, \bibinfo{author}{Moore, B.},
  \bibinfo{author}{Quinn, T.}, \bibinfo{author}{Wadsley, J.},
  \bibinfo{author}{Stadel, J.}, \bibinfo{author}{Lake, G.},
  \bibinfo{year}{2001}a.
\newblock \bibinfo{journal}{Astrophysical Journal} \bibinfo{volume}{547},
  \bibinfo{pages}{L123--L127}.
\bibitem[{Mayer et~al.(2001b)Mayer, Governato, Colpi, Moore, Quinn, Wadsley,
  Stadel and Lake}]{mayer_metamorphosis_2001}
\bibinfo{author}{Mayer, L.}, \bibinfo{author}{Governato, F.},
  \bibinfo{author}{Colpi, M.}, \bibinfo{author}{Moore, B.},
  \bibinfo{author}{Quinn, T.}, \bibinfo{author}{Wadsley, J.},
  \bibinfo{author}{Stadel, J.}, \bibinfo{author}{Lake, G.},
  \bibinfo{year}{2001}b.
\newblock \bibinfo{journal}{Astrophysical Journal} \bibinfo{volume}{559},
  \bibinfo{pages}{754--784}.
\bibitem[{{McCarthy} et~al.(2007){McCarthy}, Bower, Balogh, Voit, Pearce,
  Theuns, Babul, Lacey and Frenk}]{mccarthy_modelling_2007}
\bibinfo{author}{{McCarthy}, I.G.}, \bibinfo{author}{Bower, R.G.},
  \bibinfo{author}{Balogh, M.L.}, \bibinfo{author}{Voit, G.M.},
  \bibinfo{author}{Pearce, F.R.}, \bibinfo{author}{Theuns, T.},
  \bibinfo{author}{Babul, A.}, \bibinfo{author}{Lacey, C.G.},
  \bibinfo{author}{Frenk, C.S.}, \bibinfo{year}{2007}.
\newblock \bibinfo{journal}{Monthly Notices of the Royal Astronomical Society}
  \bibinfo{volume}{376}, \bibinfo{pages}{497--522}.
\bibitem[{{McCarthy} et~al.(2008){McCarthy}, Frenk, Font, Lacey, Bower,
  Mitchell, Balogh and Theuns}]{mccarthy_ram_2008}
\bibinfo{author}{{McCarthy}, I.G.}, \bibinfo{author}{Frenk, C.S.},
  \bibinfo{author}{Font, A.S.}, \bibinfo{author}{Lacey, C.G.},
  \bibinfo{author}{Bower, R.G.}, \bibinfo{author}{Mitchell, N.L.},
  \bibinfo{author}{Balogh, M.L.}, \bibinfo{author}{Theuns, T.},
  \bibinfo{year}{2008}.
\newblock \bibinfo{journal}{Monthly Notices of the Royal Astronomical Society}
  \bibinfo{volume}{383}, \bibinfo{pages}{593--605}.
\bibitem[{{McDowell}(1961)}]{mcdowell_formation_1961}
\bibinfo{author}{{McDowell}, M.R.C.}, \bibinfo{year}{1961}.
\newblock \bibinfo{journal}{The Observatory} \bibinfo{volume}{81},
  \bibinfo{pages}{240--243}.
\bibitem[{{McKee} and Ostriker(2007)}]{mckee_theory_2007}
\bibinfo{author}{{McKee}, C.F.}, \bibinfo{author}{Ostriker, E.C.},
  \bibinfo{year}{2007}.
\newblock \bibinfo{journal}{Annual Review of Astronomy and Astrophysics}
  \bibinfo{volume}{45}, \bibinfo{pages}{565--687}.
\bibitem[{{McKee} and Ostriker(1977)}]{mckee_theory_1977}
\bibinfo{author}{{McKee}, C.F.}, \bibinfo{author}{Ostriker, J.P.},
  \bibinfo{year}{1977}.
\newblock \bibinfo{journal}{Astrophysical Journal} \bibinfo{volume}{218},
  \bibinfo{pages}{148--169}.
\bibitem[{{McKernan} et~al.(2010){McKernan}, Maller and
  Ford}]{mckernan_new_2010}
\bibinfo{author}{{McKernan}, B.}, \bibinfo{author}{Maller, A.},
  \bibinfo{author}{Ford, K.E.S.}, \bibinfo{year}{2010}.
\newblock
  \bibinfo{howpublished}{{http://adsabs.harvard.edu/abs/2010arXiv1006.0169M}}.
\bibitem[{Meier(1999)}]{meier_magnetically_1999}
\bibinfo{author}{Meier, D.L.}, \bibinfo{year}{1999}.
\newblock \bibinfo{journal}{Astrophysical Journal} \bibinfo{volume}{522},
  \bibinfo{pages}{753--766}.
\bibitem[{Meier(2001)}]{meier_association_2001}
\bibinfo{author}{Meier, D.L.}, \bibinfo{year}{2001}.
\newblock \bibinfo{journal}{Astrophysical Journal} \bibinfo{volume}{548},
  \bibinfo{pages}{L9--L12}.
\bibitem[{Meiksin(2006)}]{meiksin_colour_2006}
\bibinfo{author}{Meiksin, A.}, \bibinfo{year}{2006}.
\newblock \bibinfo{journal}{Monthly Notices of the Royal Astronomical Society}
  \bibinfo{volume}{365}, \bibinfo{pages}{807--812}.
\bibitem[{Merritt et~al.(2004)Merritt, Milosavljevic, Favata, Hughes and
  Holz}]{merritt_consequences_2004}
\bibinfo{author}{Merritt, D.}, \bibinfo{author}{Milosavljevic, M.},
  \bibinfo{author}{Favata, M.}, \bibinfo{author}{Hughes, S.A.},
  \bibinfo{author}{Holz, D.E.}, \bibinfo{year}{2004}.
\newblock \bibinfo{journal}{Astrophysical Journal} \bibinfo{volume}{607},
  \bibinfo{pages}{L9--L12}.
\bibitem[{Merritt et~al.(2005)Merritt, Navarro, Ludlow and
  Jenkins}]{merritt_universal_2005}
\bibinfo{author}{Merritt, D.}, \bibinfo{author}{Navarro, J.F.},
  \bibinfo{author}{Ludlow, A.}, \bibinfo{author}{Jenkins, A.},
  \bibinfo{year}{2005}.
\newblock \bibinfo{journal}{{ApJ}} \bibinfo{volume}{624},
  \bibinfo{pages}{L85--L88}.
\bibitem[{Milosavljevic and Merritt(2001)}]{milosavljevic_formation_2001}
\bibinfo{author}{Milosavljevic, M.}, \bibinfo{author}{Merritt, D.},
  \bibinfo{year}{2001}.
\newblock \bibinfo{journal}{Astrophysical Journal} \bibinfo{volume}{563},
  \bibinfo{pages}{34--62}.
\bibitem[{Milosavljevic and Merritt(2003)}]{milosavljevic_long-term_2003}
\bibinfo{author}{Milosavljevic, M.}, \bibinfo{author}{Merritt, D.},
  \bibinfo{year}{2003}.
\newblock \bibinfo{journal}{Astrophysical Journal} \bibinfo{volume}{596},
  \bibinfo{pages}{860--878}.
\bibitem[{Minchev and Quillen(2006)}]{minchev_radial_2006}
\bibinfo{author}{Minchev, I.}, \bibinfo{author}{Quillen, A.C.},
  \bibinfo{year}{2006}.
\newblock \bibinfo{journal}{Monthly Notices of the Royal Astronomical Society}
  \bibinfo{volume}{368}, \bibinfo{pages}{623--636}.
\bibitem[{Miranda and Macci\`o(2007)}]{miranda_constraining_2007}
\bibinfo{author}{Miranda, M.}, \bibinfo{author}{Macci\`o, A.V.},
  \bibinfo{year}{2007}.
\newblock \bibinfo{journal}{Monthly Notices of the Royal Astronomical Society}
  \bibinfo{volume}{382}, \bibinfo{pages}{1225--1232}.
\bibitem[{Mo et~al.(1998)Mo, Mao and White}]{mo_formation_1998}
\bibinfo{author}{Mo, H.J.}, \bibinfo{author}{Mao, S.}, \bibinfo{author}{White,
  S.D.M.}, \bibinfo{year}{1998}.
\newblock \bibinfo{journal}{Monthly Notices of the Royal Astronomical Society}
  \bibinfo{volume}{295}, \bibinfo{pages}{319--336}.
\bibitem[{Monaco et~al.(2007)Monaco, Fontanot and
  Taffoni}]{monaco_morgana_2007}
\bibinfo{author}{Monaco, P.}, \bibinfo{author}{Fontanot, F.},
  \bibinfo{author}{Taffoni, G.}, \bibinfo{year}{2007}.
\newblock \bibinfo{journal}{Monthly Notices of the Royal Astronomical Society}
  \bibinfo{volume}{375}, \bibinfo{pages}{1189--1219}.
\bibitem[{Monaco et~al.(2000)Monaco, Salucci and Danese}]{monaco_joint_2000}
\bibinfo{author}{Monaco, P.}, \bibinfo{author}{Salucci, P.},
  \bibinfo{author}{Danese, L.}, \bibinfo{year}{2000}.
\newblock \bibinfo{journal}{Monthly Notices of the Royal Astronomical Society}
  \bibinfo{volume}{311}, \bibinfo{pages}{279--296}.
\bibitem[{Monaco et~al.(2002)Monaco, Theuns and
  Taffoni}]{monaco_pinocchio_2002}
\bibinfo{author}{Monaco, P.}, \bibinfo{author}{Theuns, T.},
  \bibinfo{author}{Taffoni, G.}, \bibinfo{year}{2002}.
\newblock \bibinfo{journal}{Monthly Notices of the Royal Astronomical Society}
  \bibinfo{volume}{331}, \bibinfo{pages}{587--608}.
\bibitem[{Moore and Davis(1994)}]{moore_origin_1994}
\bibinfo{author}{Moore, B.}, \bibinfo{author}{Davis, M.}, \bibinfo{year}{1994}.
\newblock \bibinfo{journal}{Monthly Notices of the Royal Astronomical Society}
  \bibinfo{volume}{270}, \bibinfo{pages}{209}.
\bibitem[{Moore et~al.(1999)Moore, Ghigna, Governato, Lake, Quinn, Stadel and
  Tozzi}]{moore_dark_1999}
\bibinfo{author}{Moore, B.}, \bibinfo{author}{Ghigna, S.},
  \bibinfo{author}{Governato, F.}, \bibinfo{author}{Lake, G.},
  \bibinfo{author}{Quinn, T.}, \bibinfo{author}{Stadel, J.},
  \bibinfo{author}{Tozzi, P.}, \bibinfo{year}{1999}.
\newblock \bibinfo{journal}{Astrophysical Journal} \bibinfo{volume}{524},
  \bibinfo{pages}{L19--L22}.
\bibitem[{Moore et~al.(1996a)Moore, Katz and Lake}]{moore_destruction_1996}
\bibinfo{author}{Moore, B.}, \bibinfo{author}{Katz, N.}, \bibinfo{author}{Lake,
  G.}, \bibinfo{year}{1996}a.
\newblock \bibinfo{journal}{Astrophysical Journal} \bibinfo{volume}{457},
  \bibinfo{pages}{455}.
\bibitem[{Moore et~al.(1996b)Moore, Katz, Lake, Dressler and
  Oemler}]{moore_galaxy_1996}
\bibinfo{author}{Moore, B.}, \bibinfo{author}{Katz, N.}, \bibinfo{author}{Lake,
  G.}, \bibinfo{author}{Dressler, A.}, \bibinfo{author}{Oemler, A.},
  \bibinfo{year}{1996}b.
\newblock \bibinfo{journal}{Nature} \bibinfo{volume}{379},
  \bibinfo{pages}{613--616}.
\bibitem[{Moore et~al.(1998)Moore, Lake and Katz}]{moore_morphological_1998}
\bibinfo{author}{Moore, B.}, \bibinfo{author}{Lake, G.}, \bibinfo{author}{Katz,
  N.}, \bibinfo{year}{1998}.
\newblock \bibinfo{journal}{Astrophysical Journal} \bibinfo{volume}{495},
  \bibinfo{pages}{139}.
\bibitem[{Munroe(2009)}]{munroe_correlation_2009}
\bibinfo{author}{Munroe, R.}, \bibinfo{year}{2009}.
\bibitem[{Murali and Weinberg(1997a)}]{murali_effect_1997}
\bibinfo{author}{Murali, C.}, \bibinfo{author}{Weinberg, M.D.},
  \bibinfo{year}{1997}a.
\newblock \bibinfo{journal}{Monthly Notices of the Royal Astronomical Society}
  \bibinfo{volume}{288}, \bibinfo{pages}{749--766}.
\bibitem[{Murali and Weinberg(1997b)}]{murali_evolution_1997}
\bibinfo{author}{Murali, C.}, \bibinfo{author}{Weinberg, M.D.},
  \bibinfo{year}{1997}b.
\newblock \bibinfo{journal}{Monthly Notices of the Royal Astronomical Society}
  \bibinfo{volume}{291}, \bibinfo{pages}{717}.
\bibitem[{Murali and Weinberg(1997c)}]{murali_globular_1997}
\bibinfo{author}{Murali, C.}, \bibinfo{author}{Weinberg, M.D.},
  \bibinfo{year}{1997}c.
\newblock \bibinfo{journal}{Monthly Notices of the Royal Astronomical Society}
  \bibinfo{volume}{288}, \bibinfo{pages}{767--776}.
\bibitem[{Murray et~al.(2005)Murray, Quataert and
  Thompson}]{murray_maximum_2005}
\bibinfo{author}{Murray, N.}, \bibinfo{author}{Quataert, E.},
  \bibinfo{author}{Thompson, T.A.}, \bibinfo{year}{2005}.
\newblock \bibinfo{journal}{Astrophysical Journal} \bibinfo{volume}{618},
  \bibinfo{pages}{569--585}.
\bibitem[{Nagashima et~al.(2005a)Nagashima, Lacey, Baugh, Frenk and
  Cole}]{nagashima_metal_2005}
\bibinfo{author}{Nagashima, M.}, \bibinfo{author}{Lacey, C.G.},
  \bibinfo{author}{Baugh, C.M.}, \bibinfo{author}{Frenk, C.S.},
  \bibinfo{author}{Cole, S.}, \bibinfo{year}{2005}a.
\newblock \bibinfo{journal}{{MNRAS}} \bibinfo{volume}{358},
  \bibinfo{pages}{1247--1266}.
\bibitem[{Nagashima et~al.(2005b)Nagashima, Lacey, Okamoto, Baugh, Frenk and
  Cole}]{nagashima_metal_2005-1}
\bibinfo{author}{Nagashima, M.}, \bibinfo{author}{Lacey, C.G.},
  \bibinfo{author}{Okamoto, T.}, \bibinfo{author}{Baugh, C.M.},
  \bibinfo{author}{Frenk, C.S.}, \bibinfo{author}{Cole, S.},
  \bibinfo{year}{2005}b.
\newblock \bibinfo{journal}{{MNRAS}} \bibinfo{volume}{363},
  \bibinfo{pages}{L31--L35}.
\bibitem[{Narayan and Yi(1994)}]{narayan_advection-dominated_1994}
\bibinfo{author}{Narayan, R.}, \bibinfo{author}{Yi, I.}, \bibinfo{year}{1994}.
\newblock \bibinfo{journal}{Astrophysical Journal} \bibinfo{volume}{428},
  \bibinfo{pages}{L13--L16}.
\bibitem[{Narayanan et~al.(2010)Narayanan, Hayward, Cox, Hernquist, Jonsson,
  Younger and Groves}]{narayanan_formation_2010}
\bibinfo{author}{Narayanan, D.}, \bibinfo{author}{Hayward, C.C.},
  \bibinfo{author}{Cox, T.J.}, \bibinfo{author}{Hernquist, L.},
  \bibinfo{author}{Jonsson, P.}, \bibinfo{author}{Younger, J.D.},
  \bibinfo{author}{Groves, B.}, \bibinfo{year}{2010}.
\newblock \bibinfo{journal}{Monthly Notices of the Royal Astronomical Society}
  \bibinfo{volume}{401}, \bibinfo{pages}{1613--1619}.
\bibitem[{Narlikar and Padmanabhan(2001)}]{narlikar_standard_2001}
\bibinfo{author}{Narlikar, J.V.}, \bibinfo{author}{Padmanabhan, T.},
  \bibinfo{year}{2001}.
\newblock \bibinfo{journal}{Annual Review of Astronomy and Astrophysics}
  \bibinfo{volume}{39}, \bibinfo{pages}{211--248}.
\bibitem[{Navarro and Benz(1991)}]{navarro_dynamics_1991}
\bibinfo{author}{Navarro, J.F.}, \bibinfo{author}{Benz, W.},
  \bibinfo{year}{1991}.
\newblock \bibinfo{journal}{Astrophysical Journal} \bibinfo{volume}{380},
  \bibinfo{pages}{320--329}.
\bibitem[{Navarro et~al.(1995)Navarro, Frenk and White}]{navarro_assembly_1995}
\bibinfo{author}{Navarro, J.F.}, \bibinfo{author}{Frenk, C.S.},
  \bibinfo{author}{White, S.D.M.}, \bibinfo{year}{1995}.
\newblock \bibinfo{journal}{Monthly Notices of the Royal Astronomical Society}
  \bibinfo{volume}{275}, \bibinfo{pages}{56--66}.
\bibitem[{Navarro et~al.(1996)Navarro, Frenk and
  White}]{navarro_structure_1996}
\bibinfo{author}{Navarro, J.F.}, \bibinfo{author}{Frenk, C.S.},
  \bibinfo{author}{White, S.D.M.}, \bibinfo{year}{1996}.
\newblock \bibinfo{journal}{Astrophysical Journal} \bibinfo{volume}{462},
  \bibinfo{pages}{563}.
\bibitem[{Navarro et~al.(1997)Navarro, Frenk and
  White}]{navarro_universal_1997}
\bibinfo{author}{Navarro, J.F.}, \bibinfo{author}{Frenk, C.S.},
  \bibinfo{author}{White, S.D.M.}, \bibinfo{year}{1997}.
\newblock \bibinfo{journal}{{ApJ}} \bibinfo{volume}{490}, \bibinfo{pages}{493}.
\bibitem[{Navarro et~al.(2004)Navarro, Hayashi, Power, Jenkins, Frenk, White,
  Springel, Stadel and Quinn}]{navarro_inner_2004}
\bibinfo{author}{Navarro, J.F.}, \bibinfo{author}{Hayashi, E.},
  \bibinfo{author}{Power, C.}, \bibinfo{author}{Jenkins, A.R.},
  \bibinfo{author}{Frenk, C.S.}, \bibinfo{author}{White, S.D.M.},
  \bibinfo{author}{Springel, V.}, \bibinfo{author}{Stadel, J.},
  \bibinfo{author}{Quinn, T.R.}, \bibinfo{year}{2004}.
\newblock \bibinfo{journal}{{MNRAS}} \bibinfo{volume}{349},
  \bibinfo{pages}{1039--1051}.
\bibitem[{Navarro and White(1994)}]{navarro_simulations_1994}
\bibinfo{author}{Navarro, J.F.}, \bibinfo{author}{White, S.D.M.},
  \bibinfo{year}{1994}.
\newblock \bibinfo{journal}{Monthly Notices of the Royal Astronomical Society}
  \bibinfo{volume}{267}, \bibinfo{pages}{401--412}.
\bibitem[{Neistein and Dekel(2008)}]{neistein_merger_2008}
\bibinfo{author}{Neistein, E.}, \bibinfo{author}{Dekel, A.},
  \bibinfo{year}{2008}.
\newblock \bibinfo{journal}{Monthly Notices of the Royal Astronomical Society}
  \bibinfo{volume}{388}, \bibinfo{pages}{1792--1802}.
\bibitem[{Nemmen et~al.(2007)Nemmen, Bower, Babul and
  {Storchi-Bergmann}}]{nemmen_models_2007}
\bibinfo{author}{Nemmen, R.S.}, \bibinfo{author}{Bower, R.G.},
  \bibinfo{author}{Babul, A.}, \bibinfo{author}{{Storchi-Bergmann}, T.},
  \bibinfo{year}{2007}.
\newblock \bibinfo{journal}{Monthly Notices of the Royal Astronomical Society}
  \bibinfo{volume}{377}, \bibinfo{pages}{1652--1662}.
\bibitem[{Neyman and Scott(1952)}]{neyman_theory_1952}
\bibinfo{author}{Neyman, J.}, \bibinfo{author}{Scott, E.L.},
  \bibinfo{year}{1952}.
\newblock \bibinfo{journal}{The Astrophysical Journal} \bibinfo{volume}{116},
  \bibinfo{pages}{144}.
\bibitem[{Norman et~al.(2009)Norman, Reynolds and
  So}]{norman_cosmological_2009}
\bibinfo{author}{Norman, M.L.}, \bibinfo{author}{Reynolds, D.R.},
  \bibinfo{author}{So, G.C.}, \bibinfo{year}{2009}.
\bibitem[{Nulsen(1982)}]{nulsen_transport_1982}
\bibinfo{author}{Nulsen, P.E.J.}, \bibinfo{year}{1982}.
\newblock \bibinfo{journal}{Monthly Notices of the Royal Astronomical Society}
  \bibinfo{volume}{198}, \bibinfo{pages}{1007--1016}.
\bibitem[{Ocvirk et~al.(2008)Ocvirk, Pichon and Teyssier}]{ocvirk_bimodal_2008}
\bibinfo{author}{Ocvirk, P.}, \bibinfo{author}{Pichon, C.},
  \bibinfo{author}{Teyssier, R.}, \bibinfo{year}{2008}.
\newblock \bibinfo{journal}{Monthly Notices of the Royal Astronomical Society}
  \bibinfo{volume}{390}, \bibinfo{pages}{1326--1338}.
\bibitem[{Ohkubo et~al.(2009)Ohkubo, Nomoto, Umeda, Yoshida and
  Tsuruta}]{ohkubo_evolution_2009}
\bibinfo{author}{Ohkubo, T.}, \bibinfo{author}{Nomoto, K.},
  \bibinfo{author}{Umeda, H.}, \bibinfo{author}{Yoshida, N.},
  \bibinfo{author}{Tsuruta, S.}, \bibinfo{year}{2009}.
\bibitem[{Okamoto et~al.(2008a)Okamoto, Gao and Theuns}]{okamoto_mass_2008}
\bibinfo{author}{Okamoto, T.}, \bibinfo{author}{Gao, L.},
  \bibinfo{author}{Theuns, T.}, \bibinfo{year}{2008}a.
\newblock \bibinfo{journal}{{MNRAS}} \bibinfo{volume}{390},
  \bibinfo{pages}{920--928}.
\bibitem[{Okamoto et~al.(2008b)Okamoto, Nemmen and Bower}]{okamoto_impact_2008}
\bibinfo{author}{Okamoto, T.}, \bibinfo{author}{Nemmen, R.S.},
  \bibinfo{author}{Bower, R.G.}, \bibinfo{year}{2008}b.
\newblock \bibinfo{journal}{Monthly Notices of the Royal Astronomical Society}
  \bibinfo{volume}{385}, \bibinfo{pages}{161--180}.
\bibitem[{Omma et~al.(2004)Omma, Binney, Bryan and Slyz}]{omma_heating_2004}
\bibinfo{author}{Omma, H.}, \bibinfo{author}{Binney, J.},
  \bibinfo{author}{Bryan, G.}, \bibinfo{author}{Slyz, A.},
  \bibinfo{year}{2004}.
\newblock \bibinfo{journal}{Monthly Notices of the Royal Astronomical Society}
  \bibinfo{volume}{348}, \bibinfo{pages}{1105--1119}.
\bibitem[{Oppenheimer and Dav\'e(2008)}]{oppenheimer_mass_2008}
\bibinfo{author}{Oppenheimer, B.D.}, \bibinfo{author}{Dav\'e, R.},
  \bibinfo{year}{2008}.
\newblock \bibinfo{journal}{Monthly Notices of the Royal Astronomical Society}
  \bibinfo{volume}{387}, \bibinfo{pages}{577--600}.
\bibitem[{{O'Shea} et~al.(2004){O'Shea}, Bryan, Bordner, Norman, Abel, Harkness
  and Kritsuk}]{oshea_introducing_2004}
\bibinfo{author}{{O'Shea}, B.W.}, \bibinfo{author}{Bryan, G.},
  \bibinfo{author}{Bordner, J.}, \bibinfo{author}{Norman, M.L.},
  \bibinfo{author}{Abel, T.}, \bibinfo{author}{Harkness, R.},
  \bibinfo{author}{Kritsuk, A.}, \bibinfo{year}{2004}.
\bibitem[{{O'Shea} and Norman(2007)}]{oshea_population_2007}
\bibinfo{author}{{O'Shea}, B.W.}, \bibinfo{author}{Norman, M.L.},
  \bibinfo{year}{2007}.
\newblock \bibinfo{journal}{Astrophysical Journal} \bibinfo{volume}{654},
  \bibinfo{pages}{66--92}.
\bibitem[{{O'Shea} and Norman(2008)}]{oshea_population_2008}
\bibinfo{author}{{O'Shea}, B.W.}, \bibinfo{author}{Norman, M.L.},
  \bibinfo{year}{2008}.
\newblock \bibinfo{journal}{Astrophysical Journal} \bibinfo{volume}{673},
  \bibinfo{pages}{14--33}.
\bibitem[{Ostriker and Peebles(1973)}]{ostriker_numerical_1973}
\bibinfo{author}{Ostriker, J.P.}, \bibinfo{author}{Peebles, P.J.E.},
  \bibinfo{year}{1973}.
\newblock \bibinfo{journal}{Astrophysical Journal} \bibinfo{volume}{186},
  \bibinfo{pages}{467--480}.
\bibitem[{Owen et~al.(2000)Owen, Eilek and Kassim}]{owen_m87_2000}
\bibinfo{author}{Owen, F.N.}, \bibinfo{author}{Eilek, J.A.},
  \bibinfo{author}{Kassim, N.E.}, \bibinfo{year}{2000}.
\newblock \bibinfo{journal}{The Astrophysical Journal} \bibinfo{volume}{543},
  \bibinfo{pages}{611--619}.
\bibitem[{Padilla and Baugh(2002)}]{padilla_cluster_2002}
\bibinfo{author}{Padilla, N.D.}, \bibinfo{author}{Baugh, C.M.},
  \bibinfo{year}{2002}.
\newblock \bibinfo{journal}{Monthly Notices of the Royal Astronomical Society}
  \bibinfo{volume}{329}, \bibinfo{pages}{431--444}.
\bibitem[{Parkinson et~al.(2008)Parkinson, Cole and
  Helly}]{parkinson_generating_2008}
\bibinfo{author}{Parkinson, H.}, \bibinfo{author}{Cole, S.},
  \bibinfo{author}{Helly, J.}, \bibinfo{year}{2008}.
\newblock \bibinfo{journal}{Monthly Notices of the Royal Astronomical Society}
  \bibinfo{volume}{383}, \bibinfo{pages}{557--564}.
\bibitem[{Parrish et~al.(2009)Parrish, Quataert and
  Sharma}]{parrish_anisotropic_2009}
\bibinfo{author}{Parrish, I.J.}, \bibinfo{author}{Quataert, E.},
  \bibinfo{author}{Sharma, P.}, \bibinfo{year}{2009}.
\newblock \bibinfo{journal}{The Astrophysical Journal} \bibinfo{volume}{703},
  \bibinfo{pages}{96--108}.
\bibitem[{Parry et~al.(2009)Parry, Eke and Frenk}]{parry_galaxy_2009}
\bibinfo{author}{Parry, O.H.}, \bibinfo{author}{Eke, V.R.},
  \bibinfo{author}{Frenk, C.S.}, \bibinfo{year}{2009}.
\newblock \bibinfo{journal}{Monthly Notices of the Royal Astronomical Society}
  \bibinfo{volume}{396}, \bibinfo{pages}{1972--1984}.
\bibitem[{Peacock and Dodds(1996)}]{peacock_non-linear_1996}
\bibinfo{author}{Peacock, J.A.}, \bibinfo{author}{Dodds, S.J.},
  \bibinfo{year}{1996}.
\newblock \bibinfo{journal}{Monthly Notices of the Royal Astronomical Society}
  \bibinfo{volume}{280}, \bibinfo{pages}{L19--L26}.
\bibitem[{Peacock and Smith(2000)}]{peacock_halo_2000}
\bibinfo{author}{Peacock, J.A.}, \bibinfo{author}{Smith, R.E.},
  \bibinfo{year}{2000}.
\newblock \bibinfo{journal}{Monthly Notices of the Royal Astronomical Society}
  \bibinfo{volume}{318}, \bibinfo{pages}{1144--1156}.
\bibitem[{Peebles(1968)}]{peebles_recombination_1968}
\bibinfo{author}{Peebles, P.J.E.}, \bibinfo{year}{1968}.
\newblock \bibinfo{journal}{{ApJ}} \bibinfo{volume}{153}, \bibinfo{pages}{1}.
\bibitem[{Percival et~al.(2007a)Percival, Nichol, Eisenstein, Frieman,
  Fukugita, Loveday, Pope, Schneider, Szalay, Tegmark, Vogeley, Weinberg,
  Zehavi, Bahcall, Brinkmann, Connolly and Meiksin}]{percival_shape_2007}
\bibinfo{author}{Percival, W.J.}, \bibinfo{author}{Nichol, R.C.},
  \bibinfo{author}{Eisenstein, D.J.}, \bibinfo{author}{Frieman, J.A.},
  \bibinfo{author}{Fukugita, M.}, \bibinfo{author}{Loveday, J.},
  \bibinfo{author}{Pope, A.C.}, \bibinfo{author}{Schneider, D.P.},
  \bibinfo{author}{Szalay, A.S.}, \bibinfo{author}{Tegmark, M.},
  \bibinfo{author}{Vogeley, M.S.}, \bibinfo{author}{Weinberg, D.H.},
  \bibinfo{author}{Zehavi, I.}, \bibinfo{author}{Bahcall, N.A.},
  \bibinfo{author}{Brinkmann, J.}, \bibinfo{author}{Connolly, A.J.},
  \bibinfo{author}{Meiksin, A.}, \bibinfo{year}{2007}a.
\newblock \bibinfo{journal}{Astrophysical Journal} \bibinfo{volume}{657},
  \bibinfo{pages}{645--663}.
\bibitem[{Percival et~al.(2007b)Percival, Nichol, Eisenstein, Weinberg,
  Fukugita, Pope, Schneider, Szalay, Vogeley, Zehavi, Bahcall, Brinkmann,
  Connolly, Loveday and Meiksin}]{percival_measuringmatter_2007}
\bibinfo{author}{Percival, W.J.}, \bibinfo{author}{Nichol, R.C.},
  \bibinfo{author}{Eisenstein, D.J.}, \bibinfo{author}{Weinberg, D.H.},
  \bibinfo{author}{Fukugita, M.}, \bibinfo{author}{Pope, A.C.},
  \bibinfo{author}{Schneider, D.P.}, \bibinfo{author}{Szalay, A.S.},
  \bibinfo{author}{Vogeley, M.S.}, \bibinfo{author}{Zehavi, I.},
  \bibinfo{author}{Bahcall, N.A.}, \bibinfo{author}{Brinkmann, J.},
  \bibinfo{author}{Connolly, A.J.}, \bibinfo{author}{Loveday, J.},
  \bibinfo{author}{Meiksin, A.}, \bibinfo{year}{2007}b.
\newblock \bibinfo{journal}{Astrophysical Journal} \bibinfo{volume}{657},
  \bibinfo{pages}{51--55}.
\bibitem[{Peters(1964)}]{peters_gravitational_1964}
\bibinfo{author}{Peters, P.C.}, \bibinfo{year}{1964}.
\newblock \bibinfo{journal}{Physical Review} \bibinfo{volume}{136},
  \bibinfo{pages}{1224--1232}.
\bibitem[{Peters et~al.(2010)Peters, Klessen, Low and
  Banerjee}]{peters_limiting_2010}
\bibinfo{author}{Peters, T.}, \bibinfo{author}{Klessen, R.S.},
  \bibinfo{author}{Low, M.M.}, \bibinfo{author}{Banerjee, R.},
  \bibinfo{year}{2010}.
\newblock
  \bibinfo{howpublished}{{http://adsabs.harvard.edu/abs/2010arXiv1005.3271P}}.
\bibitem[{Petkova and Springel(2009)}]{petkova_implementation_2009}
\bibinfo{author}{Petkova, M.}, \bibinfo{author}{Springel, V.},
  \bibinfo{year}{2009}.
\newblock \bibinfo{journal}{Monthly Notices of the Royal Astronomical Society}
  \bibinfo{volume}{396}, \bibinfo{pages}{1383--1403}.
\bibitem[{Pipino et~al.(2008)Pipino, Devriendt, Thomas, Silk and
  Kaviraj}]{pipino_galics_2008}
\bibinfo{author}{Pipino, A.}, \bibinfo{author}{Devriendt, J.E.G.},
  \bibinfo{author}{Thomas, D.}, \bibinfo{author}{Silk, J.},
  \bibinfo{author}{Kaviraj, S.}, \bibinfo{year}{2008}.
\bibitem[{Pipino and Matteucci(2004)}]{pipino_photochemical_2004}
\bibinfo{author}{Pipino, A.}, \bibinfo{author}{Matteucci, F.},
  \bibinfo{year}{2004}.
\newblock \bibinfo{journal}{Monthly Notices of the Royal Astronomical Society}
  \bibinfo{volume}{347}, \bibinfo{pages}{968--984}.
\bibitem[{Pipino and Matteucci(2006)}]{pipino_photochemical_2006}
\bibinfo{author}{Pipino, A.}, \bibinfo{author}{Matteucci, F.},
  \bibinfo{year}{2006}.
\newblock \bibinfo{journal}{Monthly Notices of the Royal Astronomical Society}
  \bibinfo{volume}{365}, \bibinfo{pages}{1114--1122}.
\bibitem[{Plewa and M\"uller(2001)}]{plewa_amra:adaptive_2001}
\bibinfo{author}{Plewa, T.}, \bibinfo{author}{M\"uller, E.},
  \bibinfo{year}{2001}.
\newblock \bibinfo{journal}{Computer Physics Communications}
  \bibinfo{volume}{138}, \bibinfo{pages}{101--127}.
\bibitem[{Pontzen et~al.(2008)Pontzen, Governato, Pettini, Booth, Stinson,
  Wadsley, Brooks, Quinn and Haehnelt}]{pontzen_damped_2008}
\bibinfo{author}{Pontzen, A.}, \bibinfo{author}{Governato, F.},
  \bibinfo{author}{Pettini, M.}, \bibinfo{author}{Booth, C.M.},
  \bibinfo{author}{Stinson, G.}, \bibinfo{author}{Wadsley, J.},
  \bibinfo{author}{Brooks, A.}, \bibinfo{author}{Quinn, T.},
  \bibinfo{author}{Haehnelt, M.}, \bibinfo{year}{2008}.
\newblock \bibinfo{journal}{Monthly Notices of the Royal Astronomical Society}
  \bibinfo{volume}{390}, \bibinfo{pages}{1349--1371}.
\bibitem[{Pope et~al.(2005)Pope, Pavlovski, Kaiser and
  Fangohr}]{pope_effects_2005}
\bibinfo{author}{Pope, E.C.D.}, \bibinfo{author}{Pavlovski, G.},
  \bibinfo{author}{Kaiser, C.R.}, \bibinfo{author}{Fangohr, H.},
  \bibinfo{year}{2005}.
\newblock \bibinfo{journal}{Monthly Notices of the Royal Astronomical Society}
  \bibinfo{volume}{364}, \bibinfo{pages}{13--28}.
\bibitem[{Portinari et~al.(1998)Portinari, Chiosi and
  Bressan}]{portinari_galactic_1998}
\bibinfo{author}{Portinari, L.}, \bibinfo{author}{Chiosi, C.},
  \bibinfo{author}{Bressan, A.}, \bibinfo{year}{1998}.
\newblock \bibinfo{journal}{{A\&A}} \bibinfo{volume}{334},
  \bibinfo{pages}{505--539}.
\bibitem[{Prada et~al.(2006)Prada, Klypin, Simonneau, {Betancort-Rijo}, Patiri,
  Gottl\"ober and {Sanchez-Conde}}]{prada_far_2006}
\bibinfo{author}{Prada, F.}, \bibinfo{author}{Klypin, A.A.},
  \bibinfo{author}{Simonneau, E.}, \bibinfo{author}{{Betancort-Rijo}, J.},
  \bibinfo{author}{Patiri, S.}, \bibinfo{author}{Gottl\"ober, S.},
  \bibinfo{author}{{Sanchez-Conde}, M.A.}, \bibinfo{year}{2006}.
\newblock \bibinfo{journal}{{ApJ}} \bibinfo{volume}{645},
  \bibinfo{pages}{1001--1011}.
\bibitem[{Press and Schechter(1974)}]{press_formation_1974}
\bibinfo{author}{Press, W.H.}, \bibinfo{author}{Schechter, P.},
  \bibinfo{year}{1974}.
\newblock \bibinfo{journal}{Astrophysical Journal} \bibinfo{volume}{187},
  \bibinfo{pages}{425--438}.
\bibitem[{Purcell et~al.(2009)Purcell, Kazantzidis and
  Bullock}]{purcell_destruction_2009}
\bibinfo{author}{Purcell, C.W.}, \bibinfo{author}{Kazantzidis, S.},
  \bibinfo{author}{Bullock, J.S.}, \bibinfo{year}{2009}.
\newblock \bibinfo{journal}{Astrophysical Journal} \bibinfo{volume}{694},
  \bibinfo{pages}{L98--L102}.
\bibitem[{Putman et~al.(2003)Putman, {Bland-Hawthorn}, Veilleux, Gibson,
  Freeman and Maloney}]{putman_h_2003}
\bibinfo{author}{Putman, M.E.}, \bibinfo{author}{{Bland-Hawthorn}, J.},
  \bibinfo{author}{Veilleux, S.}, \bibinfo{author}{Gibson, B.K.},
  \bibinfo{author}{Freeman, K.C.}, \bibinfo{author}{Maloney, P.R.},
  \bibinfo{year}{2003}.
\newblock \bibinfo{journal}{The Astrophysical Journal} \bibinfo{volume}{597},
  \bibinfo{pages}{948--956}.
\bibitem[{Quilis(2004)}]{quilis_new_2004}
\bibinfo{author}{Quilis, V.}, \bibinfo{year}{2004}.
\newblock \bibinfo{journal}{Monthly Notices of the Royal Astronomical Society}
  \bibinfo{volume}{352}, \bibinfo{pages}{1426--1438}.
\bibitem[{Quinlan(1996)}]{quinlan_dynamical_1996}
\bibinfo{author}{Quinlan, G.D.}, \bibinfo{year}{1996}.
\newblock \bibinfo{journal}{New Astronomy} \bibinfo{volume}{1},
  \bibinfo{pages}{35--56}.
\bibitem[{Quinlan and Hernquist(1997)}]{quinlan_dynamical_1997}
\bibinfo{author}{Quinlan, G.D.}, \bibinfo{author}{Hernquist, L.},
  \bibinfo{year}{1997}.
\newblock \bibinfo{journal}{New Astronomy} \bibinfo{volume}{2},
  \bibinfo{pages}{533--554}.
\bibitem[{Quinn et~al.(1993)Quinn, Hernquist and Fullagar}]{quinn_heating_1993}
\bibinfo{author}{Quinn, P.J.}, \bibinfo{author}{Hernquist, L.},
  \bibinfo{author}{Fullagar, D.P.}, \bibinfo{year}{1993}.
\newblock \bibinfo{journal}{Astrophysical Journal} \bibinfo{volume}{403},
  \bibinfo{pages}{74--93}.
\bibitem[{Randall et~al.(2008)Randall, Markevitch, Clowe, Gonzalez and
  Bradac}]{randall_constraintsself-interaction_2008}
\bibinfo{author}{Randall, S.W.}, \bibinfo{author}{Markevitch, M.},
  \bibinfo{author}{Clowe, D.}, \bibinfo{author}{Gonzalez, A.H.},
  \bibinfo{author}{Bradac, M.}, \bibinfo{year}{2008}.
\newblock \bibinfo{journal}{Astrophysical Journal} \bibinfo{volume}{679},
  \bibinfo{pages}{1173--1180}.
\bibitem[{Rasmussen et~al.(2009)Rasmussen, {Sommer-Larsen}, Pedersen, Toft,
  Benson, Bower and Grove}]{rasmussen_hot_2009}
\bibinfo{author}{Rasmussen, J.}, \bibinfo{author}{{Sommer-Larsen}, J.},
  \bibinfo{author}{Pedersen, K.}, \bibinfo{author}{Toft, S.},
  \bibinfo{author}{Benson, A.}, \bibinfo{author}{Bower, R.G.},
  \bibinfo{author}{Grove, L.F.}, \bibinfo{year}{2009}.
\newblock \bibinfo{journal}{Astrophysical Journal} \bibinfo{volume}{697},
  \bibinfo{pages}{79--93}.
\bibitem[{Ratnatunga et~al.(1999)Ratnatunga, Griffiths and
  Ostrander}]{ratnatunga_disk_1999}
\bibinfo{author}{Ratnatunga, K.U.}, \bibinfo{author}{Griffiths, R.E.},
  \bibinfo{author}{Ostrander, E.J.}, \bibinfo{year}{1999}.
\newblock \bibinfo{journal}{Astronomical Journal} \bibinfo{volume}{118},
  \bibinfo{pages}{86--107}.
\bibitem[{Razoumov and {Sommer-Larsen}(2006)}]{razoumov_escape_2006}
\bibinfo{author}{Razoumov, A.O.}, \bibinfo{author}{{Sommer-Larsen}, J.},
  \bibinfo{year}{2006}.
\newblock \bibinfo{journal}{Astrophysical Journal} \bibinfo{volume}{651},
  \bibinfo{pages}{L89--L92}.
\bibitem[{Read et~al.(2008)Read, Lake, Agertz and Debattista}]{read_thin_2008}
\bibinfo{author}{Read, J.I.}, \bibinfo{author}{Lake, G.},
  \bibinfo{author}{Agertz, O.}, \bibinfo{author}{Debattista, V.P.},
  \bibinfo{year}{2008}.
\newblock \bibinfo{journal}{Monthly Notices of the Royal Astronomical Society}
  \bibinfo{volume}{389}, \bibinfo{pages}{1041--1057}.
\bibitem[{Read et~al.(2006)Read, Wilkinson, Evans, Gilmore and
  Kleyna}]{read_tidal_2006}
\bibinfo{author}{Read, J.I.}, \bibinfo{author}{Wilkinson, M.I.},
  \bibinfo{author}{Evans, N.W.}, \bibinfo{author}{Gilmore, G.},
  \bibinfo{author}{Kleyna, J.T.}, \bibinfo{year}{2006}.
\newblock \bibinfo{journal}{Monthly Notices of the Royal Astronomical Society}
  \bibinfo{volume}{366}, \bibinfo{pages}{429--437}.
\bibitem[{Reed et~al.(2007)Reed, Bower, Frenk, Jenkins and
  Theuns}]{reed_halo_2007}
\bibinfo{author}{Reed, D.S.}, \bibinfo{author}{Bower, R.},
  \bibinfo{author}{Frenk, C.S.}, \bibinfo{author}{Jenkins, A.},
  \bibinfo{author}{Theuns, T.}, \bibinfo{year}{2007}.
\newblock \bibinfo{journal}{{MNRAS}} \bibinfo{volume}{374},
  \bibinfo{pages}{2--15}.
\bibitem[{Reed et~al.(2009)Reed, Bower, Frenk, Jenkins and
  Theuns}]{reed_clustering_2009}
\bibinfo{author}{Reed, D.S.}, \bibinfo{author}{Bower, R.},
  \bibinfo{author}{Frenk, C.S.}, \bibinfo{author}{Jenkins, A.},
  \bibinfo{author}{Theuns, T.}, \bibinfo{year}{2009}.
\newblock \bibinfo{journal}{Monthly Notices of the Royal Astronomical Society}
  \bibinfo{volume}{394}, \bibinfo{pages}{624--632}.
\bibitem[{Rees and Ostriker(1977)}]{rees_cooling_1977}
\bibinfo{author}{Rees, M.J.}, \bibinfo{author}{Ostriker, J.P.},
  \bibinfo{year}{1977}.
\newblock \bibinfo{journal}{Monthly Notices of the Royal Astronomical Society}
  \bibinfo{volume}{179}, \bibinfo{pages}{541--559}.
\bibitem[{Ricker et~al.(2000)Ricker, Dodelson and
  Lamb}]{ricker_cosmos:hybrid_2000}
\bibinfo{author}{Ricker, P.M.}, \bibinfo{author}{Dodelson, S.},
  \bibinfo{author}{Lamb, D.Q.}, \bibinfo{year}{2000}.
\newblock \bibinfo{journal}{Astrophysical Journal} \bibinfo{volume}{536},
  \bibinfo{pages}{122--143}.
\bibitem[{Ricotti et~al.(2001)Ricotti, Gnedin and
  Shull}]{ricotti_feedbackgalaxy_2001}
\bibinfo{author}{Ricotti, M.}, \bibinfo{author}{Gnedin, N.Y.},
  \bibinfo{author}{Shull, J.M.}, \bibinfo{year}{2001}.
\newblock \bibinfo{journal}{Astrophysical Journal} \bibinfo{volume}{560},
  \bibinfo{pages}{580--591}.
\bibitem[{Ricotti et~al.(2002)Ricotti, Gnedin and Shull}]{ricotti_fate_2002}
\bibinfo{author}{Ricotti, M.}, \bibinfo{author}{Gnedin, N.Y.},
  \bibinfo{author}{Shull, J.M.}, \bibinfo{year}{2002}.
\newblock \bibinfo{journal}{Astrophysical Journal} \bibinfo{volume}{575},
  \bibinfo{pages}{33--48}.
\bibitem[{Ricotti and Shull(2000)}]{ricotti_feedbackgalaxy_2000}
\bibinfo{author}{Ricotti, M.}, \bibinfo{author}{Shull, J.M.},
  \bibinfo{year}{2000}.
\newblock \bibinfo{journal}{Astrophysical Journal} \bibinfo{volume}{542},
  \bibinfo{pages}{548--558}.
\bibitem[{Rijkhorst et~al.(2006)Rijkhorst, Plewa, Dubey and
  Mellema}]{rijkhorst_hybrid_2006}
\bibinfo{author}{Rijkhorst, E.}, \bibinfo{author}{Plewa, T.},
  \bibinfo{author}{Dubey, A.}, \bibinfo{author}{Mellema, G.},
  \bibinfo{year}{2006}.
\newblock \bibinfo{journal}{Astronomy and Astrophysics} \bibinfo{volume}{452},
  \bibinfo{pages}{907--920}.
\bibitem[{Robertson et~al.(2006a)Robertson, Bullock, Cox, Matteo, Hernquist,
  Springel and Yoshida}]{robertson_merger-driven_2006}
\bibinfo{author}{Robertson, B.}, \bibinfo{author}{Bullock, J.S.},
  \bibinfo{author}{Cox, T.J.}, \bibinfo{author}{Matteo, T.D.},
  \bibinfo{author}{Hernquist, L.}, \bibinfo{author}{Springel, V.},
  \bibinfo{author}{Yoshida, N.}, \bibinfo{year}{2006}a.
\newblock \bibinfo{journal}{Astrophysical Journal} \bibinfo{volume}{645},
  \bibinfo{pages}{986--1000}.
\bibitem[{Robertson et~al.(2005)Robertson, Bullock, Font, Johnston and
  Hernquist}]{robertson__2005}
\bibinfo{author}{Robertson, B.}, \bibinfo{author}{Bullock, J.S.},
  \bibinfo{author}{Font, A.S.}, \bibinfo{author}{Johnston, K.V.},
  \bibinfo{author}{Hernquist, L.}, \bibinfo{year}{2005}.
\newblock \bibinfo{journal}{Astrophysical Journal} \bibinfo{volume}{632},
  \bibinfo{pages}{872--881}.
\bibitem[{Robertson et~al.(2006b)Robertson, Cox, Hernquist, Franx, Hopkins,
  Martini and Springel}]{robertson_fundamental_2006}
\bibinfo{author}{Robertson, B.}, \bibinfo{author}{Cox, T.J.},
  \bibinfo{author}{Hernquist, L.}, \bibinfo{author}{Franx, M.},
  \bibinfo{author}{Hopkins, P.F.}, \bibinfo{author}{Martini, P.},
  \bibinfo{author}{Springel, V.}, \bibinfo{year}{2006}b.
\newblock \bibinfo{journal}{Astrophysical Journal} \bibinfo{volume}{641},
  \bibinfo{pages}{21--40}.
\bibitem[{Robertson and Kravtsov(2008)}]{robertson_molecular_2008}
\bibinfo{author}{Robertson, B.E.}, \bibinfo{author}{Kravtsov, A.V.},
  \bibinfo{year}{2008}.
\newblock \bibinfo{journal}{Astrophysical Journal} \bibinfo{volume}{680},
  \bibinfo{pages}{1083--1111}.
\bibitem[{Robertson et~al.(2009)Robertson, Kravtsov, Tinker and
  Zentner}]{robertson_collapse_2009}
\bibinfo{author}{Robertson, B.E.}, \bibinfo{author}{Kravtsov, A.V.},
  \bibinfo{author}{Tinker, J.}, \bibinfo{author}{Zentner, A.R.},
  \bibinfo{year}{2009}.
\newblock \bibinfo{journal}{Astrophysical Journal} \bibinfo{volume}{696},
  \bibinfo{pages}{636--652}.
\bibitem[{Robinson et~al.(2004)Robinson, Dursi, Ricker, Rosner, Calder,
  Zingale, Truran, Linde, Caceres, Fryxell, Olson, Riley, Siegel and
  Vladimirova}]{robinson_morphology_2004}
\bibinfo{author}{Robinson, K.}, \bibinfo{author}{Dursi, L.J.},
  \bibinfo{author}{Ricker, P.M.}, \bibinfo{author}{Rosner, R.},
  \bibinfo{author}{Calder, A.C.}, \bibinfo{author}{Zingale, M.},
  \bibinfo{author}{Truran, J.W.}, \bibinfo{author}{Linde, T.},
  \bibinfo{author}{Caceres, A.}, \bibinfo{author}{Fryxell, B.},
  \bibinfo{author}{Olson, K.}, \bibinfo{author}{Riley, K.},
  \bibinfo{author}{Siegel, A.}, \bibinfo{author}{Vladimirova, N.},
  \bibinfo{year}{2004}.
\newblock \bibinfo{journal}{The Astrophysical Journal} \bibinfo{volume}{601},
  \bibinfo{pages}{621--643}.
\bibitem[{Romano et~al.(2005)Romano, Chiappini, Matteucci and
  Tosi}]{romano_quantifyinguncertainties_2005}
\bibinfo{author}{Romano, D.}, \bibinfo{author}{Chiappini, C.},
  \bibinfo{author}{Matteucci, F.}, \bibinfo{author}{Tosi, M.},
  \bibinfo{year}{2005}.
\newblock \bibinfo{journal}{Astronomy and Astrophysics} \bibinfo{volume}{430},
  \bibinfo{pages}{491--505}.
\bibitem[{Romeo(1992)}]{romeo_stability_1992}
\bibinfo{author}{Romeo, A.B.}, \bibinfo{year}{1992}.
\newblock \bibinfo{journal}{Monthly Notices of the Royal Astronomical Society}
  \bibinfo{volume}{256}, \bibinfo{pages}{307--320}.
\bibitem[{Romeo(1994)}]{romeo_faithful_1994}
\bibinfo{author}{Romeo, A.B.}, \bibinfo{year}{1994}.
\newblock \bibinfo{journal}{Astronomy and Astrophysics} \bibinfo{volume}{286},
  \bibinfo{pages}{799--806}.
\bibitem[{Roskar et~al.(2008a)Roskar, Debattista, Quinn, Stinson and
  Wadsley}]{roskar_ridingspiral_2008}
\bibinfo{author}{Roskar, R.}, \bibinfo{author}{Debattista, V.P.},
  \bibinfo{author}{Quinn, T.R.}, \bibinfo{author}{Stinson, G.S.},
  \bibinfo{author}{Wadsley, J.}, \bibinfo{year}{2008}a.
\newblock \bibinfo{journal}{Astrophysical Journal} \bibinfo{volume}{684},
  \bibinfo{pages}{L79--L82}.
\bibitem[{Roskar et~al.(2008b)Roskar, Debattista, Stinson, Quinn, Kaufmann and
  Wadsley}]{roskar_beyond_2008}
\bibinfo{author}{Roskar, R.}, \bibinfo{author}{Debattista, V.P.},
  \bibinfo{author}{Stinson, G.S.}, \bibinfo{author}{Quinn, T.R.},
  \bibinfo{author}{Kaufmann, T.}, \bibinfo{author}{Wadsley, J.},
  \bibinfo{year}{2008}b.
\newblock \bibinfo{journal}{Astrophysical Journal} \bibinfo{volume}{675},
  \bibinfo{pages}{L65--L68}.
\bibitem[{Rosswog(2009)}]{rosswog_astrophysical_2009}
\bibinfo{author}{Rosswog, S.}, \bibinfo{year}{2009}.
\bibitem[{Roychowdhury et~al.(2004)Roychowdhury, Ruszkowski, Nath and
  Begelman}]{roychowdhury_entropy_2004}
\bibinfo{author}{Roychowdhury, S.}, \bibinfo{author}{Ruszkowski, M.},
  \bibinfo{author}{Nath, B.B.}, \bibinfo{author}{Begelman, M.C.},
  \bibinfo{year}{2004}.
\newblock \bibinfo{journal}{Astrophysical Journal} \bibinfo{volume}{615},
  \bibinfo{pages}{681--688}.
\bibitem[{Rozo et~al.(2010)Rozo, Wechsler, Rykoff, Annis, Becker, Evrard,
  Frieman, Hansen, Hao, Johnston, Koester, {McKay}, Sheldon and
  Weinberg}]{rozo_cosmological_2010}
\bibinfo{author}{Rozo, E.}, \bibinfo{author}{Wechsler, R.H.},
  \bibinfo{author}{Rykoff, E.S.}, \bibinfo{author}{Annis, J.T.},
  \bibinfo{author}{Becker, M.R.}, \bibinfo{author}{Evrard, A.E.},
  \bibinfo{author}{Frieman, J.A.}, \bibinfo{author}{Hansen, S.M.},
  \bibinfo{author}{Hao, J.}, \bibinfo{author}{Johnston, D.E.},
  \bibinfo{author}{Koester, B.P.}, \bibinfo{author}{{McKay}, T.A.},
  \bibinfo{author}{Sheldon, E.S.}, \bibinfo{author}{Weinberg, D.H.},
  \bibinfo{year}{2010}.
\newblock \bibinfo{journal}{The Astrophysical Journal} \bibinfo{volume}{708},
  \bibinfo{pages}{645--660}.
\bibitem[{Ruszkowski et~al.(2004)Ruszkowski, Br\"uggen and
  Begelman}]{ruszkowski_cluster_2004}
\bibinfo{author}{Ruszkowski, M.}, \bibinfo{author}{Br\"uggen, M.},
  \bibinfo{author}{Begelman, M.C.}, \bibinfo{year}{2004}.
\newblock \bibinfo{journal}{Astrophysical Journal} \bibinfo{volume}{611},
  \bibinfo{pages}{158--163}.
\bibitem[{Ruszkowski et~al.(2008)Ruszkowski, En\ss{}lin, Br\"uggen, Begelman
  and Churazov}]{ruszkowski_cosmic_2008}
\bibinfo{author}{Ruszkowski, M.}, \bibinfo{author}{En\ss{}lin, T.A.},
  \bibinfo{author}{Br\"uggen, M.}, \bibinfo{author}{Begelman, M.C.},
  \bibinfo{author}{Churazov, E.}, \bibinfo{year}{2008}.
\newblock \bibinfo{journal}{Monthly Notices of the Royal Astronomical Society}
  \bibinfo{volume}{383}, \bibinfo{pages}{1359--1365}.
\bibitem[{Sales et~al.(2007)Sales, Navarro, Abadi and
  Steinmetz}]{sales_cosmic_2007}
\bibinfo{author}{Sales, L.V.}, \bibinfo{author}{Navarro, J.F.},
  \bibinfo{author}{Abadi, M.G.}, \bibinfo{author}{Steinmetz, M.},
  \bibinfo{year}{2007}.
\newblock \bibinfo{journal}{Monthly Notices of the Royal Astronomical Society}
  \bibinfo{volume}{379}, \bibinfo{pages}{1475--1483}.
\bibitem[{Salpeter(1955)}]{salpeter_luminosity_1955}
\bibinfo{author}{Salpeter, E.E.}, \bibinfo{year}{1955}.
\newblock \bibinfo{journal}{The Astrophysical Journal} \bibinfo{volume}{121},
  \bibinfo{pages}{161}.
\bibitem[{Sanchez et~al.(2009)Sanchez, Crocce, Cabre, Baugh and
  Gaztanaga}]{sanchez_cosmological_2009}
\bibinfo{author}{Sanchez, A.G.}, \bibinfo{author}{Crocce, M.},
  \bibinfo{author}{Cabre, A.}, \bibinfo{author}{Baugh, C.M.},
  \bibinfo{author}{Gaztanaga, E.}, \bibinfo{year}{2009}.
\bibitem[{Santoro and Shull(2006)}]{santoro_critical_2006}
\bibinfo{author}{Santoro, F.}, \bibinfo{author}{Shull, J.M.},
  \bibinfo{year}{2006}.
\newblock \bibinfo{journal}{Astrophysical Journal} \bibinfo{volume}{643},
  \bibinfo{pages}{26--37}.
\bibitem[{Scannapieco et~al.(2005)Scannapieco, Tissera, White and
  Springel}]{scannapieco_feedback_2005}
\bibinfo{author}{Scannapieco, C.}, \bibinfo{author}{Tissera, P.B.},
  \bibinfo{author}{White, S.D.M.}, \bibinfo{author}{Springel, V.},
  \bibinfo{year}{2005}.
\newblock \bibinfo{journal}{Monthly Notices of the Royal Astronomical Society}
  \bibinfo{volume}{364}, \bibinfo{pages}{552--564}.
\bibitem[{Scannapieco et~al.(2006a)Scannapieco, Tissera, White and
  Springel}]{scannapieco_feedback_2006}
\bibinfo{author}{Scannapieco, C.}, \bibinfo{author}{Tissera, P.B.},
  \bibinfo{author}{White, S.D.M.}, \bibinfo{author}{Springel, V.},
  \bibinfo{year}{2006}a.
\newblock \bibinfo{journal}{Monthly Notices of the Royal Astronomical Society}
  \bibinfo{volume}{371}, \bibinfo{pages}{1125--1139}.
\bibitem[{Scannapieco et~al.(2006b)Scannapieco, Kawata, Brook, Schneider,
  Ferrara and Gibson}]{scannapieco_spatial_2006}
\bibinfo{author}{Scannapieco, E.}, \bibinfo{author}{Kawata, D.},
  \bibinfo{author}{Brook, C.B.}, \bibinfo{author}{Schneider, R.},
  \bibinfo{author}{Ferrara, A.}, \bibinfo{author}{Gibson, B.K.},
  \bibinfo{year}{2006}b.
\newblock \bibinfo{journal}{Astrophysical Journal} \bibinfo{volume}{653},
  \bibinfo{pages}{285--299}.
\bibitem[{Scannapieco and Oh(2004)}]{scannapieco_quasar_2004}
\bibinfo{author}{Scannapieco, E.}, \bibinfo{author}{Oh, S.P.},
  \bibinfo{year}{2004}.
\newblock \bibinfo{journal}{Astrophysical Journal} \bibinfo{volume}{608},
  \bibinfo{pages}{62--79}.
\bibitem[{Schade et~al.(1996)Schade, Carlberg, Yee, {Lopez-Cruz} and
  Ellingson}]{schade_evolution_1996}
\bibinfo{author}{Schade, D.}, \bibinfo{author}{Carlberg, R.G.},
  \bibinfo{author}{Yee, H.K.C.}, \bibinfo{author}{{Lopez-Cruz}, O.},
  \bibinfo{author}{Ellingson, E.}, \bibinfo{year}{1996}.
\newblock \bibinfo{journal}{Astrophysical Journal} \bibinfo{volume}{464},
  \bibinfo{pages}{L63}.
\bibitem[{Schawinski(2009)}]{schawinski_role_2009}
\bibinfo{author}{Schawinski, K.}, \bibinfo{year}{2009}.
\bibitem[{Schawinski et~al.(2007)Schawinski, Thomas, Sarzi, Maraston, Kaviraj,
  Joo, Yi and Silk}]{schawinski_observational_2007}
\bibinfo{author}{Schawinski, K.}, \bibinfo{author}{Thomas, D.},
  \bibinfo{author}{Sarzi, M.}, \bibinfo{author}{Maraston, C.},
  \bibinfo{author}{Kaviraj, S.}, \bibinfo{author}{Joo, S.},
  \bibinfo{author}{Yi, S.K.}, \bibinfo{author}{Silk, J.}, \bibinfo{year}{2007}.
\newblock \bibinfo{journal}{Monthly Notices of the Royal Astronomical Society}
  \bibinfo{volume}{382}, \bibinfo{pages}{1415--1431}.
\bibitem[{Schaye(2004)}]{schaye_star_2004}
\bibinfo{author}{Schaye, J.}, \bibinfo{year}{2004}.
\newblock \bibinfo{journal}{Astrophysical Journal} \bibinfo{volume}{609},
  \bibinfo{pages}{667--682}.
\bibitem[{Schmidt(1959)}]{schmidt_rate_1959}
\bibinfo{author}{Schmidt, M.}, \bibinfo{year}{1959}.
\newblock \bibinfo{journal}{Astrophysical Journal} \bibinfo{volume}{129},
  \bibinfo{pages}{243}.
\bibitem[{Schreiber et~al.(2009)Schreiber, Genzel, Bouche, Cresci, Davies,
  Buschkamp, Shapiro, Tacconi, Hicks, Genel, Shapley, Erb, Steidel, Lutz,
  Eisenhauer, Gillessen, Sternberg, Renzini, Cimatti, Daddi, Kurk, Lilly, Kong,
  Lehnert, Nesvadba, Verma, {McCracken}, Arimoto, Mignoli and
  Onodera}]{schreiber_sins_2009}
\bibinfo{author}{Schreiber, N.M.F.}, \bibinfo{author}{Genzel, R.},
  \bibinfo{author}{Bouche, N.}, \bibinfo{author}{Cresci, G.},
  \bibinfo{author}{Davies, R.}, \bibinfo{author}{Buschkamp, P.},
  \bibinfo{author}{Shapiro, K.}, \bibinfo{author}{Tacconi, L.J.},
  \bibinfo{author}{Hicks, E.K.S.}, \bibinfo{author}{Genel, S.},
  \bibinfo{author}{Shapley, A.E.}, \bibinfo{author}{Erb, D.K.},
  \bibinfo{author}{Steidel, C.C.}, \bibinfo{author}{Lutz, D.},
  \bibinfo{author}{Eisenhauer, F.}, \bibinfo{author}{Gillessen, S.},
  \bibinfo{author}{Sternberg, A.}, \bibinfo{author}{Renzini, A.},
  \bibinfo{author}{Cimatti, A.}, \bibinfo{author}{Daddi, E.},
  \bibinfo{author}{Kurk, J.}, \bibinfo{author}{Lilly, S.},
  \bibinfo{author}{Kong, X.}, \bibinfo{author}{Lehnert, M.D.},
  \bibinfo{author}{Nesvadba, N.}, \bibinfo{author}{Verma, A.},
  \bibinfo{author}{{McCracken}, H.}, \bibinfo{author}{Arimoto, N.},
  \bibinfo{author}{Mignoli, M.}, \bibinfo{author}{Onodera, M.},
  \bibinfo{year}{2009}.
\bibitem[{Seljak et~al.(2005)Seljak, Makarov, {McDonald}, Anderson, Bahcall,
  Brinkmann, Burles, Cen, Doi, Gunn, Ivezic, Kent, Loveday, Lupton, Munn,
  Nichol, Ostriker, Schlegel, Schneider, Tegmark, Berk, Weinberg and
  York}]{seljak_cosmological_2005}
\bibinfo{author}{Seljak, U.}, \bibinfo{author}{Makarov, A.},
  \bibinfo{author}{{McDonald}, P.}, \bibinfo{author}{Anderson, S.F.},
  \bibinfo{author}{Bahcall, N.A.}, \bibinfo{author}{Brinkmann, J.},
  \bibinfo{author}{Burles, S.}, \bibinfo{author}{Cen, R.},
  \bibinfo{author}{Doi, M.}, \bibinfo{author}{Gunn, J.E.},
  \bibinfo{author}{Ivezic, Z.}, \bibinfo{author}{Kent, S.},
  \bibinfo{author}{Loveday, J.}, \bibinfo{author}{Lupton, R.H.},
  \bibinfo{author}{Munn, J.A.}, \bibinfo{author}{Nichol, R.C.},
  \bibinfo{author}{Ostriker, J.P.}, \bibinfo{author}{Schlegel, D.J.},
  \bibinfo{author}{Schneider, D.P.}, \bibinfo{author}{Tegmark, M.},
  \bibinfo{author}{Berk, D.E.}, \bibinfo{author}{Weinberg, D.H.},
  \bibinfo{author}{York, D.G.}, \bibinfo{year}{2005}.
\newblock \bibinfo{journal}{Physical Review D} \bibinfo{volume}{71},
  \bibinfo{pages}{103515}.
\bibitem[{Sellwood(2010)}]{sellwood_dynamics_2010}
\bibinfo{author}{Sellwood, J.A.}, \bibinfo{year}{2010}.
\newblock
  \bibinfo{howpublished}{{http://adsabs.harvard.edu/abs/2010arXiv1006.4855S}}.
\bibitem[{Sellwood and Binney(2002)}]{sellwood_radial_2002}
\bibinfo{author}{Sellwood, J.A.}, \bibinfo{author}{Binney, J.J.},
  \bibinfo{year}{2002}.
\newblock \bibinfo{journal}{Monthly Notices of the Royal Astronomical Society}
  \bibinfo{volume}{336}, \bibinfo{pages}{785--796}.
\bibitem[{Sellwood et~al.(1998)Sellwood, Nelson and
  Tremaine}]{sellwood_resonant_1998}
\bibinfo{author}{Sellwood, J.A.}, \bibinfo{author}{Nelson, R.W.},
  \bibinfo{author}{Tremaine, S.}, \bibinfo{year}{1998}.
\newblock \bibinfo{journal}{Astrophysical Journal} \bibinfo{volume}{506},
  \bibinfo{pages}{590--599}.
\bibitem[{Sersic(1968)}]{sersic_atlas_1968}
\bibinfo{author}{Sersic, J.L.}, \bibinfo{year}{1968}.
\bibitem[{Sesana et~al.(2007)Sesana, Haardt and
  Madau}]{sesana_interaction_2007}
\bibinfo{author}{Sesana, A.}, \bibinfo{author}{Haardt, F.},
  \bibinfo{author}{Madau, P.}, \bibinfo{year}{2007}.
\newblock \bibinfo{journal}{Astrophysical Journal} \bibinfo{volume}{660},
  \bibinfo{pages}{546--555}.
\bibitem[{Sesana et~al.(2008)Sesana, Haardt and
  Madau}]{sesana_interaction_2008}
\bibinfo{author}{Sesana, A.}, \bibinfo{author}{Haardt, F.},
  \bibinfo{author}{Madau, P.}, \bibinfo{year}{2008}.
\newblock \bibinfo{journal}{Astrophysical Journal} \bibinfo{volume}{686},
  \bibinfo{pages}{432--447}.
\bibitem[{Shakura and Sunyaev(1973)}]{shakura_black_1973}
\bibinfo{author}{Shakura, N.I.}, \bibinfo{author}{Sunyaev, R.A.},
  \bibinfo{year}{1973}.
\newblock \bibinfo{journal}{Astronomy and Astrophysics} \bibinfo{volume}{24},
  \bibinfo{pages}{337--355}.
\bibitem[{Shapiro(2005)}]{shapiro_spin_2005}
\bibinfo{author}{Shapiro, S.L.}, \bibinfo{year}{2005}.
\newblock \bibinfo{journal}{Astrophysical Journal} \bibinfo{volume}{620},
  \bibinfo{pages}{59--68}.
\bibitem[{Sharma and Steinmetz(2005)}]{sharma_angular_2005}
\bibinfo{author}{Sharma, S.}, \bibinfo{author}{Steinmetz, M.},
  \bibinfo{year}{2005}.
\newblock \bibinfo{journal}{{ApJ}} \bibinfo{volume}{628},
  \bibinfo{pages}{21--44}.
\bibitem[{Shaw and Mota(2008)}]{shaw_improved_2008}
\bibinfo{author}{Shaw, D.J.}, \bibinfo{author}{Mota, D.F.},
  \bibinfo{year}{2008}.
\newblock \bibinfo{journal}{Astrophysical Journal Supplement Series}
  \bibinfo{volume}{174}, \bibinfo{pages}{277--281}.
\bibitem[{Shen and Sellwood(2004)}]{shen_destruction_2004}
\bibinfo{author}{Shen, J.}, \bibinfo{author}{Sellwood, J.A.},
  \bibinfo{year}{2004}.
\newblock \bibinfo{journal}{Astrophysical Journal} \bibinfo{volume}{604},
  \bibinfo{pages}{614--631}.
\bibitem[{Sheth et~al.(2001)Sheth, Mo and Tormen}]{sheth_ellipsoidal_2001}
\bibinfo{author}{Sheth, R.K.}, \bibinfo{author}{Mo, H.J.},
  \bibinfo{author}{Tormen, G.}, \bibinfo{year}{2001}.
\newblock \bibinfo{journal}{Monthly Notices of the Royal Astronomical Society}
  \bibinfo{volume}{323}, \bibinfo{pages}{1--12}.
\bibitem[{Shlosman and Noguchi(1993)}]{shlosman_effects_1993}
\bibinfo{author}{Shlosman, I.}, \bibinfo{author}{Noguchi, M.},
  \bibinfo{year}{1993}.
\newblock \bibinfo{journal}{Astrophysical Journal} \bibinfo{volume}{414},
  \bibinfo{pages}{474--486}.
\bibitem[{Sijacki et~al.(2009)Sijacki, Springel and
  Haehnelt}]{sijacki_growingfirst_2009}
\bibinfo{author}{Sijacki, D.}, \bibinfo{author}{Springel, V.},
  \bibinfo{author}{Haehnelt, M.G.}, \bibinfo{year}{2009}.
\bibitem[{Sijacki et~al.(2007)Sijacki, Springel, Matteo and
  Hernquist}]{sijacki_unified_2007}
\bibinfo{author}{Sijacki, D.}, \bibinfo{author}{Springel, V.},
  \bibinfo{author}{Matteo, T.D.}, \bibinfo{author}{Hernquist, L.},
  \bibinfo{year}{2007}.
\newblock \bibinfo{journal}{Monthly Notices of the Royal Astronomical Society}
  \bibinfo{volume}{380}, \bibinfo{pages}{877--900}.
\bibitem[{Silk(2001)}]{silk_formation_2001}
\bibinfo{author}{Silk, J.}, \bibinfo{year}{2001}.
\newblock \bibinfo{journal}{Monthly Notices of the Royal Astronomical Society}
  \bibinfo{volume}{324}, \bibinfo{pages}{313--318}.
\bibitem[{Silk and Norman(1981)}]{silk_dissipational_1981}
\bibinfo{author}{Silk, J.}, \bibinfo{author}{Norman, C.}, \bibinfo{year}{1981}.
\newblock \bibinfo{journal}{Astrophysical Journal} \bibinfo{volume}{247},
  \bibinfo{pages}{59--76}.
\bibitem[{Silk and Rees(1998)}]{silk_quasars_1998}
\bibinfo{author}{Silk, J.}, \bibinfo{author}{Rees, M.J.}, \bibinfo{year}{1998}.
\newblock \bibinfo{journal}{Astronomy and Astrophysics} \bibinfo{volume}{331},
  \bibinfo{pages}{L1--L4}.
\bibitem[{Silva et~al.(1998)Silva, Granato, Bressan and
  Danese}]{silva_modelingeffects_1998}
\bibinfo{author}{Silva, L.}, \bibinfo{author}{Granato, G.L.},
  \bibinfo{author}{Bressan, A.}, \bibinfo{author}{Danese, L.},
  \bibinfo{year}{1998}.
\newblock \bibinfo{journal}{{ApJ}} \bibinfo{volume}{509},
  \bibinfo{pages}{103--117}.
\bibitem[{Simard et~al.(2002)Simard, Willmer, Vogt, Sarajedini, Phillips,
  Weiner, Koo, Im, Illingworth and Faber}]{simard_deep_2002}
\bibinfo{author}{Simard, L.}, \bibinfo{author}{Willmer, C.N.A.},
  \bibinfo{author}{Vogt, N.P.}, \bibinfo{author}{Sarajedini, V.L.},
  \bibinfo{author}{Phillips, A.C.}, \bibinfo{author}{Weiner, B.J.},
  \bibinfo{author}{Koo, D.C.}, \bibinfo{author}{Im, M.},
  \bibinfo{author}{Illingworth, G.D.}, \bibinfo{author}{Faber, S.M.},
  \bibinfo{year}{2002}.
\newblock \bibinfo{journal}{Astrophysical Journal Supplement Series}
  \bibinfo{volume}{142}, \bibinfo{pages}{1--33}.
\bibitem[{Slosar et~al.(2007)Slosar, {McDonald} and
  Seljak}]{slosar_cosmological_2007}
\bibinfo{author}{Slosar, A.}, \bibinfo{author}{{McDonald}, P.},
  \bibinfo{author}{Seljak, U.}, \bibinfo{year}{2007}.
\newblock \bibinfo{journal}{New Astronomy Review} \bibinfo{volume}{51},
  \bibinfo{pages}{327--331}.
\bibitem[{Smith et~al.(2009)Smith, Turk, Sigurdsson, {O'Shea} and
  Norman}]{smith_three_2009}
\bibinfo{author}{Smith, B.D.}, \bibinfo{author}{Turk, M.J.},
  \bibinfo{author}{Sigurdsson, S.}, \bibinfo{author}{{O'Shea}, B.W.},
  \bibinfo{author}{Norman, M.L.}, \bibinfo{year}{2009}.
\newblock \bibinfo{journal}{The Astrophysical Journal} \bibinfo{volume}{691},
  \bibinfo{pages}{441--451}.
\bibitem[{Smith et~al.(2003)Smith, Peacock, Jenkins, White, Frenk, Pearce,
  Thomas, Efstathiou and Couchman}]{smith_stable_2003}
\bibinfo{author}{Smith, R.E.}, \bibinfo{author}{Peacock, J.A.},
  \bibinfo{author}{Jenkins, A.}, \bibinfo{author}{White, S.D.M.},
  \bibinfo{author}{Frenk, C.S.}, \bibinfo{author}{Pearce, F.R.},
  \bibinfo{author}{Thomas, P.A.}, \bibinfo{author}{Efstathiou, G.},
  \bibinfo{author}{Couchman, H.M.P.}, \bibinfo{year}{2003}.
\newblock \bibinfo{journal}{Monthly Notices of the Royal Astronomical Society}
  \bibinfo{volume}{341}, \bibinfo{pages}{1311--1332}.
\bibitem[{Smoluchowksi(1916)}]{smoluchowksi__1916}
\bibinfo{author}{Smoluchowksi, M.}, \bibinfo{year}{1916}.
\newblock \bibinfo{journal}{Phys. Zeit.} \bibinfo{volume}{17},
  \bibinfo{pages}{557}.
\bibitem[{Somerville(2002)}]{somerville_can_2002}
\bibinfo{author}{Somerville, R.S.}, \bibinfo{year}{2002}.
\newblock \bibinfo{journal}{{ApJ}} \bibinfo{volume}{572},
  \bibinfo{pages}{L23--L26}.
\bibitem[{Somerville et~al.(2008a)Somerville, Barden, Rix, Bell, Beckwith,
  Borch, Caldwell, H\"au\ss{}ler, Heymans, Jahnke, Jogee, {McIntosh},
  Meisenheimer, Peng, S\'anchez, Wisotzki and
  Wolf}]{somerville_explanation_2008}
\bibinfo{author}{Somerville, R.S.}, \bibinfo{author}{Barden, M.},
  \bibinfo{author}{Rix, H.}, \bibinfo{author}{Bell, E.F.},
  \bibinfo{author}{Beckwith, S.V.W.}, \bibinfo{author}{Borch, A.},
  \bibinfo{author}{Caldwell, J.A.R.}, \bibinfo{author}{H\"au\ss{}ler, B.},
  \bibinfo{author}{Heymans, C.}, \bibinfo{author}{Jahnke, K.},
  \bibinfo{author}{Jogee, S.}, \bibinfo{author}{{McIntosh}, D.H.},
  \bibinfo{author}{Meisenheimer, K.}, \bibinfo{author}{Peng, C.Y.},
  \bibinfo{author}{S\'anchez, S.F.}, \bibinfo{author}{Wisotzki, L.},
  \bibinfo{author}{Wolf, C.}, \bibinfo{year}{2008}a.
\newblock \bibinfo{journal}{{ApJ}} \bibinfo{volume}{672},
  \bibinfo{pages}{776--786}.
\bibitem[{Somerville et~al.(2003)Somerville, Bullock and
  Livio}]{somerville_epoch_2003}
\bibinfo{author}{Somerville, R.S.}, \bibinfo{author}{Bullock, J.S.},
  \bibinfo{author}{Livio, M.}, \bibinfo{year}{2003}.
\newblock \bibinfo{journal}{Astrophysical Journal} \bibinfo{volume}{593},
  \bibinfo{pages}{616--621}.
\bibitem[{Somerville et~al.(2008b)Somerville, Hopkins, Cox, Robertson and
  Hernquist}]{somerville_semi-analytic_2008}
\bibinfo{author}{Somerville, R.S.}, \bibinfo{author}{Hopkins, P.F.},
  \bibinfo{author}{Cox, T.J.}, \bibinfo{author}{Robertson, B.E.},
  \bibinfo{author}{Hernquist, L.}, \bibinfo{year}{2008}b.
\newblock \bibinfo{journal}{Monthly Notices of the Royal Astronomical Society}
  \bibinfo{volume}{391}, \bibinfo{pages}{481--506}.
\bibitem[{Somerville and Livio(2003)}]{somerville_star_2003}
\bibinfo{author}{Somerville, R.S.}, \bibinfo{author}{Livio, M.},
  \bibinfo{year}{2003}.
\newblock \bibinfo{journal}{{ApJ}} \bibinfo{volume}{593},
  \bibinfo{pages}{611--615}.
\bibitem[{Somerville and Primack(1999)}]{somerville_semi-analytic_1999}
\bibinfo{author}{Somerville, R.S.}, \bibinfo{author}{Primack, J.R.},
  \bibinfo{year}{1999}.
\newblock \bibinfo{journal}{Monthly Notices of the Royal Astronomical Society}
  \bibinfo{volume}{310}, \bibinfo{pages}{1087--1110}.
\bibitem[{Spitzer(1942)}]{spitzer_dynamics_1942}
\bibinfo{author}{Spitzer, L.}, \bibinfo{year}{1942}.
\newblock \bibinfo{journal}{Astrophysical Journal} \bibinfo{volume}{95},
  \bibinfo{pages}{329}.
\bibitem[{Spitzer(1962)}]{spitzer_physics_1962}
\bibinfo{author}{Spitzer, L.}, \bibinfo{year}{1962}.
\bibitem[{Spitzer and Schwarzschild(1953)}]{spitzer_possible_1953}
\bibinfo{author}{Spitzer, L.}, \bibinfo{author}{Schwarzschild, M.},
  \bibinfo{year}{1953}.
\newblock \bibinfo{journal}{Astrophysical Journal} \bibinfo{volume}{118},
  \bibinfo{pages}{106}.
\bibitem[{Springel(2005)}]{springel_cosmological_2005}
\bibinfo{author}{Springel, V.}, \bibinfo{year}{2005}.
\newblock \bibinfo{journal}{Monthly Notices of the Royal Astronomical Society}
  \bibinfo{volume}{364}, \bibinfo{pages}{1105--1134}.
\bibitem[{Springel and Hernquist(2005)}]{springel_formation_2005}
\bibinfo{author}{Springel, V.}, \bibinfo{author}{Hernquist, L.},
  \bibinfo{year}{2005}.
\newblock \bibinfo{journal}{Astrophysical Journal} \bibinfo{volume}{622},
  \bibinfo{pages}{L9--L12}.
\bibitem[{Springel et~al.(2005a)Springel, Matteo and
  Hernquist}]{springel_modelling_2005}
\bibinfo{author}{Springel, V.}, \bibinfo{author}{Matteo, T.D.},
  \bibinfo{author}{Hernquist, L.}, \bibinfo{year}{2005}a.
\newblock \bibinfo{journal}{Monthly Notices of the Royal Astronomical Society}
  \bibinfo{volume}{361}, \bibinfo{pages}{776--794}.
\bibitem[{Springel et~al.(2008)Springel, Wang, Vogelsberger, Ludlow, Jenkins,
  Helmi, Navarro, Frenk and White}]{springel_aquarius_2008}
\bibinfo{author}{Springel, V.}, \bibinfo{author}{Wang, J.},
  \bibinfo{author}{Vogelsberger, M.}, \bibinfo{author}{Ludlow, A.},
  \bibinfo{author}{Jenkins, A.}, \bibinfo{author}{Helmi, A.},
  \bibinfo{author}{Navarro, J.F.}, \bibinfo{author}{Frenk, C.S.},
  \bibinfo{author}{White, S.D.M.}, \bibinfo{year}{2008}.
\newblock \bibinfo{journal}{Monthly Notices of the Royal Astronomical Society}
  \bibinfo{volume}{391}, \bibinfo{pages}{1685--1711}.
\bibitem[{Springel et~al.(2005b)Springel, White, Jenkins, Frenk, Yoshida, Gao,
  Navarro, Thacker, Croton, Helly, Peacock, Cole, Thomas, Couchman, Evrard,
  Colberg and Pearce}]{springel_simulations_2005}
\bibinfo{author}{Springel, V.}, \bibinfo{author}{White, S.D.M.},
  \bibinfo{author}{Jenkins, A.}, \bibinfo{author}{Frenk, C.S.},
  \bibinfo{author}{Yoshida, N.}, \bibinfo{author}{Gao, L.},
  \bibinfo{author}{Navarro, J.}, \bibinfo{author}{Thacker, R.},
  \bibinfo{author}{Croton, D.}, \bibinfo{author}{Helly, J.},
  \bibinfo{author}{Peacock, J.A.}, \bibinfo{author}{Cole, S.},
  \bibinfo{author}{Thomas, P.}, \bibinfo{author}{Couchman, H.},
  \bibinfo{author}{Evrard, A.}, \bibinfo{author}{Colberg, J.},
  \bibinfo{author}{Pearce, F.}, \bibinfo{year}{2005}b.
\newblock \bibinfo{journal}{Nature} \bibinfo{volume}{435},
  \bibinfo{pages}{629--636}.
\bibitem[{Springel et~al.(2001)Springel, White, Tormen and
  Kauffmann}]{springel_populatingcluster_2001}
\bibinfo{author}{Springel, V.}, \bibinfo{author}{White, S.D.M.},
  \bibinfo{author}{Tormen, G.}, \bibinfo{author}{Kauffmann, G.},
  \bibinfo{year}{2001}.
\newblock \bibinfo{journal}{{MNRAS}} \bibinfo{volume}{328},
  \bibinfo{pages}{726--750}.
\bibitem[{Stacy and Bromm(2007)}]{stacy_impact_2007}
\bibinfo{author}{Stacy, A.}, \bibinfo{author}{Bromm, V.}, \bibinfo{year}{2007}.
\newblock \bibinfo{journal}{Monthly Notices of the Royal Astronomical Society}
  \bibinfo{volume}{382}, \bibinfo{pages}{229--238}.
\bibitem[{Steidel et~al.(2010)Steidel, Erb, Shapley, Pettini, Reddy,
  Bogosavljević, Rudie and Rakic}]{steidel_structure_2010}
\bibinfo{author}{Steidel, C.C.}, \bibinfo{author}{Erb, D.K.},
  \bibinfo{author}{Shapley, A.E.}, \bibinfo{author}{Pettini, M.},
  \bibinfo{author}{Reddy, N.A.}, \bibinfo{author}{Bogosavljević, M.},
  \bibinfo{author}{Rudie, G.C.}, \bibinfo{author}{Rakic, O.},
  \bibinfo{year}{2010}.
\newblock
  \bibinfo{howpublished}{{http://adsabs.harvard.edu/abs/2010arXiv1003.0679S}}.
\bibitem[{Steinmetz and Navarro(2002)}]{steinmetz_hierarchical_2002}
\bibinfo{author}{Steinmetz, M.}, \bibinfo{author}{Navarro, J.F.},
  \bibinfo{year}{2002}.
\newblock \bibinfo{journal}{New Astronomy} \bibinfo{volume}{7},
  \bibinfo{pages}{155--160}.
\bibitem[{Stewart et~al.(2009)Stewart, Bullock, Barton and
  Wechsler}]{stewart_galaxy_2009}
\bibinfo{author}{Stewart, K.R.}, \bibinfo{author}{Bullock, J.S.},
  \bibinfo{author}{Barton, E.J.}, \bibinfo{author}{Wechsler, R.H.},
  \bibinfo{year}{2009}.
\newblock \bibinfo{journal}{The Astrophysical Journal} \bibinfo{volume}{702},
  \bibinfo{pages}{1005--1015}.
\bibitem[{Stinson(2007)}]{stinson_supernova_2007}
\bibinfo{author}{Stinson, G.}, \bibinfo{year}{2007}.
\bibitem[{Stinson et~al.(2006)Stinson, Seth, Katz, Wadsley, Governato and
  Quinn}]{stinson_star_2006}
\bibinfo{author}{Stinson, G.}, \bibinfo{author}{Seth, A.},
  \bibinfo{author}{Katz, N.}, \bibinfo{author}{Wadsley, J.},
  \bibinfo{author}{Governato, F.}, \bibinfo{author}{Quinn, T.},
  \bibinfo{year}{2006}.
\newblock \bibinfo{journal}{Monthly Notices of the Royal Astronomical Society}
  \bibinfo{volume}{373}, \bibinfo{pages}{1074--1090}.
\bibitem[{Strigari et~al.(2008)Strigari, Bullock, Kaplinghat, Simon, Geha,
  Willman and Walker}]{strigari_common_2008}
\bibinfo{author}{Strigari, L.E.}, \bibinfo{author}{Bullock, J.S.},
  \bibinfo{author}{Kaplinghat, M.}, \bibinfo{author}{Simon, J.D.},
  \bibinfo{author}{Geha, M.}, \bibinfo{author}{Willman, B.},
  \bibinfo{author}{Walker, M.G.}, \bibinfo{year}{2008}.
\newblock \bibinfo{journal}{Nature} \bibinfo{volume}{454},
  \bibinfo{pages}{1096--1097}.
\bibitem[{Stringer et~al.(2010)Stringer, Brooks, Benson and
  Governato}]{stringer_analytic_2010}
\bibinfo{author}{Stringer, M.J.}, \bibinfo{author}{Brooks, A.M.},
  \bibinfo{author}{Benson, A.J.}, \bibinfo{author}{Governato, F.},
  \bibinfo{year}{2010}.
\newblock
  \bibinfo{howpublished}{{http://adsabs.harvard.edu/abs/2010arXiv1001.0594S}}.
\bibitem[{Summers et~al.(1995)Summers, Davis and Evrard}]{summers_galaxy_1995}
\bibinfo{author}{Summers, F.J.}, \bibinfo{author}{Davis, M.},
  \bibinfo{author}{Evrard, A.E.}, \bibinfo{year}{1995}.
\newblock \bibinfo{journal}{Astrophysical Journal} \bibinfo{volume}{454},
  \bibinfo{pages}{1}.
\bibitem[{Swinbank et~al.(2008)Swinbank, Lacey, Smail, Baugh, Frenk, Blain,
  Chapman, Coppin, Ivison, Gonzalez and Hainline}]{swinbank_properties_2008}
\bibinfo{author}{Swinbank, A.M.}, \bibinfo{author}{Lacey, C.G.},
  \bibinfo{author}{Smail, I.}, \bibinfo{author}{Baugh, C.M.},
  \bibinfo{author}{Frenk, C.S.}, \bibinfo{author}{Blain, A.W.},
  \bibinfo{author}{Chapman, S.C.}, \bibinfo{author}{Coppin, K.E.K.},
  \bibinfo{author}{Ivison, R.J.}, \bibinfo{author}{Gonzalez, J.E.},
  \bibinfo{author}{Hainline, L.J.}, \bibinfo{year}{2008}.
\newblock \bibinfo{journal}{Monthly Notices of the Royal Astronomical Society}
  \bibinfo{volume}{391}, \bibinfo{pages}{420--434}.
\bibitem[{Tasker and Bryan(2006)}]{tasker_simulating_2006}
\bibinfo{author}{Tasker, E.J.}, \bibinfo{author}{Bryan, G.L.},
  \bibinfo{year}{2006}.
\newblock \bibinfo{journal}{Astrophysical Journal} \bibinfo{volume}{641},
  \bibinfo{pages}{878--890}.
\bibitem[{Tasker and Bryan(2008)}]{tasker_effect_2008}
\bibinfo{author}{Tasker, E.J.}, \bibinfo{author}{Bryan, G.L.},
  \bibinfo{year}{2008}.
\newblock \bibinfo{journal}{The Astrophysical Journal} \bibinfo{volume}{673},
  \bibinfo{pages}{810--831}.
\bibitem[{Tassis et~al.(2008)Tassis, Kravtsov and Gnedin}]{tassis_scaling_2008}
\bibinfo{author}{Tassis, K.}, \bibinfo{author}{Kravtsov, A.V.},
  \bibinfo{author}{Gnedin, N.Y.}, \bibinfo{year}{2008}.
\newblock \bibinfo{journal}{Astrophysical Journal} \bibinfo{volume}{672},
  \bibinfo{pages}{888--903}.
\bibitem[{Taylor and Babul(2004)}]{taylor_evolution_2004}
\bibinfo{author}{Taylor, J.E.}, \bibinfo{author}{Babul, A.},
  \bibinfo{year}{2004}.
\newblock \bibinfo{journal}{Monthly Notices of the Royal Astronomical Society}
  \bibinfo{volume}{348}, \bibinfo{pages}{811--830}.
\bibitem[{Tegmark et~al.(2004)Tegmark, Blanton, Strauss, Hoyle, Schlegel,
  Scoccimarro, Vogeley, Weinberg, Zehavi, Berlind, Budavari, Connolly,
  Eisenstein, Finkbeiner, Frieman, Gunn, Hamilton, Hui, Jain, Johnston, Kent,
  Lin, Nakajima, Nichol, Ostriker, Pope, Scranton, Seljak, Sheth, Stebbins,
  Szalay, Szapudi, Verde, Xu, Annis, Bahcall, Brinkmann, Burles, Castander,
  Csabai, Loveday, Doi, Fukugita, Gott, Hennessy, Hogg, Ivezic, Knapp, Lamb,
  Lee, Lupton, {McKay}, Kunszt, Munn, {O'Connell}, Peoples, Pier, Richmond,
  Rockosi, Schneider, Stoughton, Tucker, Berk, Yanny and
  York}]{tegmark_three-dimensional_2004}
\bibinfo{author}{Tegmark, M.}, \bibinfo{author}{Blanton, M.R.},
  \bibinfo{author}{Strauss, M.A.}, \bibinfo{author}{Hoyle, F.},
  \bibinfo{author}{Schlegel, D.}, \bibinfo{author}{Scoccimarro, R.},
  \bibinfo{author}{Vogeley, M.S.}, \bibinfo{author}{Weinberg, D.H.},
  \bibinfo{author}{Zehavi, I.}, \bibinfo{author}{Berlind, A.},
  \bibinfo{author}{Budavari, T.}, \bibinfo{author}{Connolly, A.},
  \bibinfo{author}{Eisenstein, D.J.}, \bibinfo{author}{Finkbeiner, D.},
  \bibinfo{author}{Frieman, J.A.}, \bibinfo{author}{Gunn, J.E.},
  \bibinfo{author}{Hamilton, A.J.S.}, \bibinfo{author}{Hui, L.},
  \bibinfo{author}{Jain, B.}, \bibinfo{author}{Johnston, D.},
  \bibinfo{author}{Kent, S.}, \bibinfo{author}{Lin, H.},
  \bibinfo{author}{Nakajima, R.}, \bibinfo{author}{Nichol, R.C.},
  \bibinfo{author}{Ostriker, J.P.}, \bibinfo{author}{Pope, A.},
  \bibinfo{author}{Scranton, R.}, \bibinfo{author}{Seljak, U.},
  \bibinfo{author}{Sheth, R.K.}, \bibinfo{author}{Stebbins, A.},
  \bibinfo{author}{Szalay, A.S.}, \bibinfo{author}{Szapudi, I.},
  \bibinfo{author}{Verde, L.}, \bibinfo{author}{Xu, Y.},
  \bibinfo{author}{Annis, J.}, \bibinfo{author}{Bahcall, N.A.},
  \bibinfo{author}{Brinkmann, J.}, \bibinfo{author}{Burles, S.},
  \bibinfo{author}{Castander, F.J.}, \bibinfo{author}{Csabai, I.},
  \bibinfo{author}{Loveday, J.}, \bibinfo{author}{Doi, M.},
  \bibinfo{author}{Fukugita, M.}, \bibinfo{author}{Gott, J.R.},
  \bibinfo{author}{Hennessy, G.}, \bibinfo{author}{Hogg, D.W.},
  \bibinfo{author}{Ivezic, Z.}, \bibinfo{author}{Knapp, G.R.},
  \bibinfo{author}{Lamb, D.Q.}, \bibinfo{author}{Lee, B.C.},
  \bibinfo{author}{Lupton, R.H.}, \bibinfo{author}{{McKay}, T.A.},
  \bibinfo{author}{Kunszt, P.}, \bibinfo{author}{Munn, J.A.},
  \bibinfo{author}{{O'Connell}, L.}, \bibinfo{author}{Peoples, J.},
  \bibinfo{author}{Pier, J.R.}, \bibinfo{author}{Richmond, M.},
  \bibinfo{author}{Rockosi, C.}, \bibinfo{author}{Schneider, D.P.},
  \bibinfo{author}{Stoughton, C.}, \bibinfo{author}{Tucker, D.L.},
  \bibinfo{author}{Berk, D.E.V.}, \bibinfo{author}{Yanny, B.},
  \bibinfo{author}{York, D.G.}, \bibinfo{year}{2004}.
\newblock \bibinfo{journal}{Astrophysical Journal} \bibinfo{volume}{606},
  \bibinfo{pages}{702--740}.
\bibitem[{Tegmark et~al.(2006)Tegmark, Eisenstein, Strauss, Weinberg, Blanton,
  Frieman, Fukugita, Gunn, Hamilton, Knapp, Nichol, Ostriker, Padmanabhan,
  Percival, Schlegel, Schneider, Scoccimarro, Seljak, Seo, Swanson, Szalay,
  Vogeley, Yoo, Zehavi, Abazajian, Anderson, Annis, Bahcall, Bassett, Berlind,
  Brinkmann, Budavari, Castander, Connolly, Csabai, Doi, Finkbeiner, Gillespie,
  Glazebrook, Hennessy, Hogg, Ivezic, Jain, Johnston, Kent, Lamb, Lee, Lin,
  Loveday, Lupton, Munn, Pan, Park, Peoples, Pier, Pope, Richmond, Rockosi,
  Scranton, Sheth, Stebbins, Stoughton, Szapudi, Tucker, Berk, Yanny and
  York}]{tegmark_cosmological_2006}
\bibinfo{author}{Tegmark, M.}, \bibinfo{author}{Eisenstein, D.J.},
  \bibinfo{author}{Strauss, M.A.}, \bibinfo{author}{Weinberg, D.H.},
  \bibinfo{author}{Blanton, M.R.}, \bibinfo{author}{Frieman, J.A.},
  \bibinfo{author}{Fukugita, M.}, \bibinfo{author}{Gunn, J.E.},
  \bibinfo{author}{Hamilton, A.J.S.}, \bibinfo{author}{Knapp, G.R.},
  \bibinfo{author}{Nichol, R.C.}, \bibinfo{author}{Ostriker, J.P.},
  \bibinfo{author}{Padmanabhan, N.}, \bibinfo{author}{Percival, W.J.},
  \bibinfo{author}{Schlegel, D.J.}, \bibinfo{author}{Schneider, D.P.},
  \bibinfo{author}{Scoccimarro, R.}, \bibinfo{author}{Seljak, U.},
  \bibinfo{author}{Seo, H.}, \bibinfo{author}{Swanson, M.},
  \bibinfo{author}{Szalay, A.S.}, \bibinfo{author}{Vogeley, M.S.},
  \bibinfo{author}{Yoo, J.}, \bibinfo{author}{Zehavi, I.},
  \bibinfo{author}{Abazajian, K.}, \bibinfo{author}{Anderson, S.F.},
  \bibinfo{author}{Annis, J.}, \bibinfo{author}{Bahcall, N.A.},
  \bibinfo{author}{Bassett, B.}, \bibinfo{author}{Berlind, A.},
  \bibinfo{author}{Brinkmann, J.}, \bibinfo{author}{Budavari, T.},
  \bibinfo{author}{Castander, F.}, \bibinfo{author}{Connolly, A.},
  \bibinfo{author}{Csabai, I.}, \bibinfo{author}{Doi, M.},
  \bibinfo{author}{Finkbeiner, D.P.}, \bibinfo{author}{Gillespie, B.},
  \bibinfo{author}{Glazebrook, K.}, \bibinfo{author}{Hennessy, G.S.},
  \bibinfo{author}{Hogg, D.W.}, \bibinfo{author}{Ivezic, Z.},
  \bibinfo{author}{Jain, B.}, \bibinfo{author}{Johnston, D.},
  \bibinfo{author}{Kent, S.}, \bibinfo{author}{Lamb, D.Q.},
  \bibinfo{author}{Lee, B.C.}, \bibinfo{author}{Lin, H.},
  \bibinfo{author}{Loveday, J.}, \bibinfo{author}{Lupton, R.H.},
  \bibinfo{author}{Munn, J.A.}, \bibinfo{author}{Pan, K.},
  \bibinfo{author}{Park, C.}, \bibinfo{author}{Peoples, J.},
  \bibinfo{author}{Pier, J.R.}, \bibinfo{author}{Pope, A.},
  \bibinfo{author}{Richmond, M.}, \bibinfo{author}{Rockosi, C.},
  \bibinfo{author}{Scranton, R.}, \bibinfo{author}{Sheth, R.K.},
  \bibinfo{author}{Stebbins, A.}, \bibinfo{author}{Stoughton, C.},
  \bibinfo{author}{Szapudi, I.}, \bibinfo{author}{Tucker, D.L.},
  \bibinfo{author}{Berk, D.E.V.}, \bibinfo{author}{Yanny, B.},
  \bibinfo{author}{York, D.G.}, \bibinfo{year}{2006}.
\newblock \bibinfo{journal}{Physical Review D} \bibinfo{volume}{74},
  \bibinfo{pages}{123507}.
\bibitem[{Tegmark et~al.(1997)Tegmark, Silk, Rees, Blanchard, Abel and
  Palla}]{tegmark_small_1997}
\bibinfo{author}{Tegmark, M.}, \bibinfo{author}{Silk, J.},
  \bibinfo{author}{Rees, M.J.}, \bibinfo{author}{Blanchard, A.},
  \bibinfo{author}{Abel, T.}, \bibinfo{author}{Palla, F.},
  \bibinfo{year}{1997}.
\newblock \bibinfo{journal}{{ApJ}} \bibinfo{volume}{474}, \bibinfo{pages}{1}.
\bibitem[{Thacker and Couchman(2000)}]{thacker_implementing_2000}
\bibinfo{author}{Thacker, R.J.}, \bibinfo{author}{Couchman, H.M.P.},
  \bibinfo{year}{2000}.
\newblock \bibinfo{journal}{Astrophysical Journal} \bibinfo{volume}{545},
  \bibinfo{pages}{728--752}.
\bibitem[{Thacker and Couchman(2001)}]{thacker_star_2001}
\bibinfo{author}{Thacker, R.J.}, \bibinfo{author}{Couchman, H.M.P.},
  \bibinfo{year}{2001}.
\newblock \bibinfo{journal}{Astrophysical Journal} \bibinfo{volume}{555},
  \bibinfo{pages}{L17--L20}.
\bibitem[{Thorne(1974)}]{thorne_disk-accretion_1974}
\bibinfo{author}{Thorne, K.S.}, \bibinfo{year}{1974}.
\newblock \bibinfo{journal}{Astrophysical Journal} \bibinfo{volume}{191},
  \bibinfo{pages}{507--520}.
\bibitem[{Tichy and Marronetti(2007)}]{tichy_binary_2007}
\bibinfo{author}{Tichy, W.}, \bibinfo{author}{Marronetti, P.},
  \bibinfo{year}{2007}.
\newblock \bibinfo{journal}{Physical Review D {(Particles,} Fields,
  Gravitation, and Cosmology)} \bibinfo{volume}{76},
  \bibinfo{pages}{061502--5}.
\bibitem[{Timmes et~al.(1995)Timmes, Woosley and Weaver}]{timmes_galactic_1995}
\bibinfo{author}{Timmes, F.X.}, \bibinfo{author}{Woosley, S.E.},
  \bibinfo{author}{Weaver, T.A.}, \bibinfo{year}{1995}.
\newblock \bibinfo{journal}{Astrophysical Journal Supplement Series}
  \bibinfo{volume}{98}, \bibinfo{pages}{617--658}.
\bibitem[{Tinker et~al.(2008)Tinker, Kravtsov, Klypin, Abazajian, Warren,
  Yepes, Gottl\"ober and Holz}]{tinker_towardhalo_2008}
\bibinfo{author}{Tinker, J.}, \bibinfo{author}{Kravtsov, A.V.},
  \bibinfo{author}{Klypin, A.}, \bibinfo{author}{Abazajian, K.},
  \bibinfo{author}{Warren, M.}, \bibinfo{author}{Yepes, G.},
  \bibinfo{author}{Gottl\"ober, S.}, \bibinfo{author}{Holz, D.E.},
  \bibinfo{year}{2008}.
\newblock \bibinfo{journal}{The Astrophysical Journal} \bibinfo{volume}{688},
  \bibinfo{pages}{709--728}.
\bibitem[{Tinker et~al.(2010)Tinker, Wechsler and
  Zheng}]{tinker_interpretingclustering_2010}
\bibinfo{author}{Tinker, J.L.}, \bibinfo{author}{Wechsler, R.H.},
  \bibinfo{author}{Zheng, Z.}, \bibinfo{year}{2010}.
\newblock \bibinfo{journal}{The Astrophysical Journal} \bibinfo{volume}{709},
  \bibinfo{pages}{67--76}.
\bibitem[{Tollerud et~al.(2008)Tollerud, Bullock, Strigari and
  Willman}]{tollerud_hundreds_2008}
\bibinfo{author}{Tollerud, E.J.}, \bibinfo{author}{Bullock, J.S.},
  \bibinfo{author}{Strigari, L.E.}, \bibinfo{author}{Willman, B.},
  \bibinfo{year}{2008}.
\newblock \bibinfo{journal}{Astrophysical Journal} \bibinfo{volume}{688},
  \bibinfo{pages}{277--289}.
\bibitem[{Tolstoy et~al.(2009)Tolstoy, Hill and Tosi}]{tolstoy_star_2009}
\bibinfo{author}{Tolstoy, E.}, \bibinfo{author}{Hill, V.},
  \bibinfo{author}{Tosi, M.}, \bibinfo{year}{2009}.
\bibitem[{Toomre(1964)}]{toomre_gravitational_1964}
\bibinfo{author}{Toomre, A.}, \bibinfo{year}{1964}.
\newblock \bibinfo{journal}{Astrophysical Journal} \bibinfo{volume}{139},
  \bibinfo{pages}{1217--1238}.
\bibitem[{Tormen et~al.(1998)Tormen, Diaferio and Syer}]{tormen_survival_1998}
\bibinfo{author}{Tormen, G.}, \bibinfo{author}{Diaferio, A.},
  \bibinfo{author}{Syer, D.}, \bibinfo{year}{1998}.
\newblock \bibinfo{journal}{Monthly Notices of the Royal Astronomical Society}
  \bibinfo{volume}{299}, \bibinfo{pages}{728--742}.
\bibitem[{Toth and Ostriker(1992)}]{toth_galactic_1992}
\bibinfo{author}{Toth, G.}, \bibinfo{author}{Ostriker, J.P.},
  \bibinfo{year}{1992}.
\newblock \bibinfo{journal}{Astrophysical Journal} \bibinfo{volume}{389},
  \bibinfo{pages}{5--26}.
\bibitem[{Tozzi and Norman(2001)}]{tozzi_evolution_2001}
\bibinfo{author}{Tozzi, P.}, \bibinfo{author}{Norman, C.},
  \bibinfo{year}{2001}.
\newblock \bibinfo{journal}{Astrophysical Journal} \bibinfo{volume}{546},
  \bibinfo{pages}{63--84}.
\bibitem[{Tremaine et~al.(2002)Tremaine, Gebhardt, Bender, Bower, Dressler,
  Faber, Filippenko, Green, Grillmair, Ho, Kormendy, Lauer, Magorrian, Pinkney
  and Richstone}]{tremaine_slope_2002}
\bibinfo{author}{Tremaine, S.}, \bibinfo{author}{Gebhardt, K.},
  \bibinfo{author}{Bender, R.}, \bibinfo{author}{Bower, G.},
  \bibinfo{author}{Dressler, A.}, \bibinfo{author}{Faber, S.M.},
  \bibinfo{author}{Filippenko, A.V.}, \bibinfo{author}{Green, R.},
  \bibinfo{author}{Grillmair, C.}, \bibinfo{author}{Ho, L.C.},
  \bibinfo{author}{Kormendy, J.}, \bibinfo{author}{Lauer, T.R.},
  \bibinfo{author}{Magorrian, J.}, \bibinfo{author}{Pinkney, J.},
  \bibinfo{author}{Richstone, D.}, \bibinfo{year}{2002}.
\newblock \bibinfo{journal}{Astrophysical Journal} \bibinfo{volume}{574},
  \bibinfo{pages}{740--753}.
\bibitem[{Tremaine et~al.(1986)Tremaine, Henon and
  {Lynden-Bell}}]{tremaine_h-functions_1986}
\bibinfo{author}{Tremaine, S.}, \bibinfo{author}{Henon, M.},
  \bibinfo{author}{{Lynden-Bell}, D.}, \bibinfo{year}{1986}.
\newblock \bibinfo{journal}{Monthly Notices of the Royal Astronomical Society}
  \bibinfo{volume}{219}, \bibinfo{pages}{285--297}.
\bibitem[{Trenti and Stiavelli(2009)}]{trenti_formation_2009}
\bibinfo{author}{Trenti, M.}, \bibinfo{author}{Stiavelli, M.},
  \bibinfo{year}{2009}.
\newblock \bibinfo{journal}{Astrophysical Journal} \bibinfo{volume}{694},
  \bibinfo{pages}{879--892}.
\bibitem[{Tumlinson(2006)}]{tumlinson_chemical_2006}
\bibinfo{author}{Tumlinson, J.}, \bibinfo{year}{2006}.
\newblock \bibinfo{journal}{Astrophysical Journal} \bibinfo{volume}{641},
  \bibinfo{pages}{1--20}.
\bibitem[{Turk(2009)}]{turk_formation_2009}
\bibinfo{author}{Turk, M.J.}, \bibinfo{year}{2009}.
\newblock
  \bibinfo{howpublished}{{http://adsabs.harvard.edu/abs/2009PhDT.........5T}}.
\bibitem[{de~Vaucouleurs(1959)}]{vaucouleurs_classification_1959}
\bibinfo{author}{de~Vaucouleurs, G.}, \bibinfo{year}{1959}.
\newblock \bibinfo{journal}{Handbuch der Physik} \bibinfo{volume}{53},
  \bibinfo{pages}{275}.
\bibitem[{de~Vaucouleurs et~al.(1991)de~Vaucouleurs, de~Vaucouleurs, Corwin,
  Buta, Paturel and Fouque}]{vaucouleurs_third_1991}
\bibinfo{author}{de~Vaucouleurs, G.}, \bibinfo{author}{de~Vaucouleurs, A.},
  \bibinfo{author}{Corwin, H.G.}, \bibinfo{author}{Buta, R.J.},
  \bibinfo{author}{Paturel, G.}, \bibinfo{author}{Fouque, P.},
  \bibinfo{year}{1991}.
\bibitem[{Vecchia et~al.(2004)Vecchia, Bower, Theuns, Balogh, Mazzotta and
  Frenk}]{dalla_vecchia_quenching_2004}
\bibinfo{author}{Vecchia, C.D.}, \bibinfo{author}{Bower, R.G.},
  \bibinfo{author}{Theuns, T.}, \bibinfo{author}{Balogh, M.L.},
  \bibinfo{author}{Mazzotta, P.}, \bibinfo{author}{Frenk, C.S.},
  \bibinfo{year}{2004}.
\newblock \bibinfo{journal}{Monthly Notices of the Royal Astronomical Society}
  \bibinfo{volume}{355}, \bibinfo{pages}{995--1004}.
\bibitem[{Vecchia and Schaye(2008)}]{vecchia_simulating_2008}
\bibinfo{author}{Vecchia, C.D.}, \bibinfo{author}{Schaye, J.},
  \bibinfo{year}{2008}.
\newblock \bibinfo{journal}{Monthly Notices of the Royal Astronomical Society}
  \bibinfo{volume}{387}, \bibinfo{pages}{1431--1444}.
\bibitem[{Veilleux et~al.(2005)Veilleux, Cecil and
  {Bland-Hawthorn}}]{veilleux_galactic_2005}
\bibinfo{author}{Veilleux, S.}, \bibinfo{author}{Cecil, G.},
  \bibinfo{author}{{Bland-Hawthorn}, J.}, \bibinfo{year}{2005}.
\newblock \bibinfo{journal}{Annual Review of Astronomy and Astrophysics}
  \bibinfo{volume}{43}, \bibinfo{pages}{769--826}.
\bibitem[{Velazquez and White(1999)}]{velazquez_sinking_1999}
\bibinfo{author}{Velazquez, H.}, \bibinfo{author}{White, S.D.M.},
  \bibinfo{year}{1999}.
\newblock \bibinfo{journal}{Monthly Notices of the Royal Astronomical Society}
  \bibinfo{volume}{304}, \bibinfo{pages}{254--270}.
\bibitem[{Vernaleo and Reynolds(2006)}]{vernaleo_agn_2006}
\bibinfo{author}{Vernaleo, J.C.}, \bibinfo{author}{Reynolds, C.S.},
  \bibinfo{year}{2006}.
\newblock \bibinfo{journal}{Astrophysical Journal} \bibinfo{volume}{645},
  \bibinfo{pages}{83--94}.
\bibitem[{Viel et~al.(2008)Viel, Becker, Bolton, Haehnelt, Rauch and
  Sargent}]{viel_cold_2008}
\bibinfo{author}{Viel, M.}, \bibinfo{author}{Becker, G.D.},
  \bibinfo{author}{Bolton, J.S.}, \bibinfo{author}{Haehnelt, M.G.},
  \bibinfo{author}{Rauch, M.}, \bibinfo{author}{Sargent, W.L.W.},
  \bibinfo{year}{2008}.
\newblock \bibinfo{journal}{Physical Review Letters} \bibinfo{volume}{100},
  \bibinfo{pages}{41304}.
\bibitem[{Vikhlinin et~al.(2009)Vikhlinin, Kravtsov, Burenin, Ebeling, Forman,
  Hornstrup, Jones, Murray, Nagai, Quintana and
  Voevodkin}]{vikhlinin_chandra_2009}
\bibinfo{author}{Vikhlinin, A.}, \bibinfo{author}{Kravtsov, A.V.},
  \bibinfo{author}{Burenin, R.A.}, \bibinfo{author}{Ebeling, H.},
  \bibinfo{author}{Forman, W.R.}, \bibinfo{author}{Hornstrup, A.},
  \bibinfo{author}{Jones, C.}, \bibinfo{author}{Murray, S.S.},
  \bibinfo{author}{Nagai, D.}, \bibinfo{author}{Quintana, H.},
  \bibinfo{author}{Voevodkin, A.}, \bibinfo{year}{2009}.
\newblock \bibinfo{journal}{The Astrophysical Journal} \bibinfo{volume}{692},
  \bibinfo{pages}{1060--1074}.
\bibitem[{Villalobos and Helmi(2008)}]{villalobos_simulations_2008}
\bibinfo{author}{Villalobos, {\'Alvaro}.}, \bibinfo{author}{Helmi, A.},
  \bibinfo{year}{2008}.
\newblock \bibinfo{journal}{Monthly Notices of the Royal Astronomical Society}
  \bibinfo{volume}{391}, \bibinfo{pages}{1806--1827}.
\bibitem[{Villiers et~al.(2005)Villiers, Hawley, Krolik and
  Hirose}]{villiers_magnetically_2005}
\bibinfo{author}{Villiers, J.D.}, \bibinfo{author}{Hawley, J.F.},
  \bibinfo{author}{Krolik, J.H.}, \bibinfo{author}{Hirose, S.},
  \bibinfo{year}{2005}.
\newblock \bibinfo{journal}{Astrophysical Journal} \bibinfo{volume}{620},
  \bibinfo{pages}{878--888}.
\bibitem[{Vitvitska et~al.(2002)Vitvitska, Klypin, Kravtsov, Wechsler, Primack
  and Bullock}]{vitvitska_origin_2002}
\bibinfo{author}{Vitvitska, M.}, \bibinfo{author}{Klypin, A.A.},
  \bibinfo{author}{Kravtsov, A.V.}, \bibinfo{author}{Wechsler, R.H.},
  \bibinfo{author}{Primack, J.R.}, \bibinfo{author}{Bullock, J.S.},
  \bibinfo{year}{2002}.
\newblock \bibinfo{journal}{Astrophysical Journal} \bibinfo{volume}{581},
  \bibinfo{pages}{799--809}.
\bibitem[{Voit et~al.(2003)Voit, Balogh, Bower, Lacey and
  Bryan}]{voit_origin_2003}
\bibinfo{author}{Voit, G.M.}, \bibinfo{author}{Balogh, M.L.},
  \bibinfo{author}{Bower, R.G.}, \bibinfo{author}{Lacey, C.G.},
  \bibinfo{author}{Bryan, G.L.}, \bibinfo{year}{2003}.
\newblock \bibinfo{journal}{Astrophysical Journal} \bibinfo{volume}{593},
  \bibinfo{pages}{272--290}.
\bibitem[{Volonteri et~al.(2003)Volonteri, Haardt and
  Madau}]{volonteri_assembly_2003}
\bibinfo{author}{Volonteri, M.}, \bibinfo{author}{Haardt, F.},
  \bibinfo{author}{Madau, P.}, \bibinfo{year}{2003}.
\newblock \bibinfo{journal}{Astrophysical Journal} \bibinfo{volume}{582},
  \bibinfo{pages}{559--573}.
\bibitem[{Volonteri et~al.(2008)Volonteri, Lodato and
  Natarajan}]{volonteri_evolution_2008}
\bibinfo{author}{Volonteri, M.}, \bibinfo{author}{Lodato, G.},
  \bibinfo{author}{Natarajan, P.}, \bibinfo{year}{2008}.
\newblock \bibinfo{journal}{Monthly Notices of the Royal Astronomical Society}
  \bibinfo{volume}{383}, \bibinfo{pages}{1079--1088}.
\bibitem[{Volonteri et~al.(2005)Volonteri, Madau, Quataert and
  Rees}]{volonteri_distribution_2005}
\bibinfo{author}{Volonteri, M.}, \bibinfo{author}{Madau, P.},
  \bibinfo{author}{Quataert, E.}, \bibinfo{author}{Rees, M.J.},
  \bibinfo{year}{2005}.
\newblock \bibinfo{journal}{Astrophysical Journal} \bibinfo{volume}{620},
  \bibinfo{pages}{69--77}.
\bibitem[{Volonteri and Natarajan(2009)}]{volonteri_journey_2009}
\bibinfo{author}{Volonteri, M.}, \bibinfo{author}{Natarajan, P.},
  \bibinfo{year}{2009}.
\bibitem[{Wadepuhl and Springel(2010)}]{wadepuhl_satellite_2010}
\bibinfo{author}{Wadepuhl, M.}, \bibinfo{author}{Springel, V.},
  \bibinfo{year}{2010}.
\newblock
  \bibinfo{howpublished}{{http://adsabs.harvard.edu/abs/2010arXiv1004.3217W}}.
\bibitem[{Wadsley et~al.(2004)Wadsley, Stadel and
  Quinn}]{wadsley_gasoline:flexible_2004}
\bibinfo{author}{Wadsley, J.W.}, \bibinfo{author}{Stadel, J.},
  \bibinfo{author}{Quinn, T.}, \bibinfo{year}{2004}.
\newblock \bibinfo{journal}{New Astronomy} \bibinfo{volume}{9},
  \bibinfo{pages}{137--158}.
\bibitem[{Wang and Silk(1994)}]{wang_gravitational_1994}
\bibinfo{author}{Wang, B.}, \bibinfo{author}{Silk, J.}, \bibinfo{year}{1994}.
\newblock \bibinfo{journal}{Astrophysical Journal} \bibinfo{volume}{427},
  \bibinfo{pages}{759--769}.
\bibitem[{Wang and Abel(2009)}]{wang_magnetohydrodynamic_2009}
\bibinfo{author}{Wang, P.}, \bibinfo{author}{Abel, T.}, \bibinfo{year}{2009}.
\newblock \bibinfo{journal}{The Astrophysical Journal} \bibinfo{volume}{696},
  \bibinfo{pages}{96--109}.
\bibitem[{Wang et~al.(2005)Wang, Yao, Tripp, Fang, Cui, Nicastro, Mathur,
  Williams, Song and Croft}]{wang_warm-hot_2005}
\bibinfo{author}{Wang, Q.D.}, \bibinfo{author}{Yao, Y.},
  \bibinfo{author}{Tripp, T.M.}, \bibinfo{author}{Fang, T.},
  \bibinfo{author}{Cui, W.}, \bibinfo{author}{Nicastro, F.},
  \bibinfo{author}{Mathur, S.}, \bibinfo{author}{Williams, R.J.},
  \bibinfo{author}{Song, L.}, \bibinfo{author}{Croft, R.},
  \bibinfo{year}{2005}.
\newblock \bibinfo{journal}{Astrophysical Journal} \bibinfo{volume}{635},
  \bibinfo{pages}{386--395}.
\bibitem[{Warren et~al.(1992)Warren, Quinn, Salmon and
  Zurek}]{warren_dark_1992}
\bibinfo{author}{Warren, M.S.}, \bibinfo{author}{Quinn, P.J.},
  \bibinfo{author}{Salmon, J.K.}, \bibinfo{author}{Zurek, W.H.},
  \bibinfo{year}{1992}.
\newblock \bibinfo{journal}{{ApJ}} \bibinfo{volume}{399},
  \bibinfo{pages}{405--425}.
\bibitem[{Watanabe et~al.(1985)Watanabe, Kodaira and
  Okamura}]{watanabe_digital_1985}
\bibinfo{author}{Watanabe, M.}, \bibinfo{author}{Kodaira, K.},
  \bibinfo{author}{Okamura, S.}, \bibinfo{year}{1985}.
\newblock \bibinfo{journal}{Astrophysical Journal} \bibinfo{volume}{292},
  \bibinfo{pages}{72--78}.
\bibitem[{Wechsler et~al.(2002)Wechsler, Bullock, Primack, Kravtsov and
  Dekel}]{wechsler_concentrations_2002}
\bibinfo{author}{Wechsler, R.H.}, \bibinfo{author}{Bullock, J.S.},
  \bibinfo{author}{Primack, J.R.}, \bibinfo{author}{Kravtsov, A.V.},
  \bibinfo{author}{Dekel, A.}, \bibinfo{year}{2002}.
\newblock \bibinfo{journal}{The Astrophysical Journal} \bibinfo{volume}{568},
  \bibinfo{pages}{52--70}.
\bibitem[{Wechsler et~al.(2001)Wechsler, Somerville, Bullock, Kolatt, Primack,
  Blumenthal and Dekel}]{wechsler_galaxy_2001}
\bibinfo{author}{Wechsler, R.H.}, \bibinfo{author}{Somerville, R.S.},
  \bibinfo{author}{Bullock, J.S.}, \bibinfo{author}{Kolatt, T.S.},
  \bibinfo{author}{Primack, J.R.}, \bibinfo{author}{Blumenthal, G.R.},
  \bibinfo{author}{Dekel, A.}, \bibinfo{year}{2001}.
\newblock \bibinfo{journal}{{ApJ}} \bibinfo{volume}{554},
  \bibinfo{pages}{85--103}.
\bibitem[{Wen and Zhao(2004)}]{wen_thick_2004}
\bibinfo{author}{Wen, W.}, \bibinfo{author}{Zhao, J.}, \bibinfo{year}{2004}.
\newblock \bibinfo{journal}{Progress in Astronomy} \bibinfo{volume}{22},
  \bibinfo{pages}{235--244}.
\bibitem[{Wetzel et~al.(2008)Wetzel, Cohn and White}]{wetzel_clustering_2008}
\bibinfo{author}{Wetzel, A.R.}, \bibinfo{author}{Cohn, J.D.},
  \bibinfo{author}{White, M.}, \bibinfo{year}{2008}.
\newblock \bibinfo{journal}{0810.3650} .
\bibitem[{White(2004)}]{white_feedback_2004}
\bibinfo{author}{White, S.}, \bibinfo{year}{2004}.
\newblock p.~\bibinfo{pages}{30}.
\bibitem[{White et~al.(1987a)White, Davis, Efstathiou and
  Frenk}]{white_galaxy_1987}
\bibinfo{author}{White, S.D.M.}, \bibinfo{author}{Davis, M.},
  \bibinfo{author}{Efstathiou, G.}, \bibinfo{author}{Frenk, C.S.},
  \bibinfo{year}{1987}a.
\newblock \bibinfo{journal}{Nature} \bibinfo{volume}{330},
  \bibinfo{pages}{451--453}.
\bibitem[{White et~al.(1984)White, Davis and Frenk}]{white_size_1984}
\bibinfo{author}{White, S.D.M.}, \bibinfo{author}{Davis, M.},
  \bibinfo{author}{Frenk, C.S.}, \bibinfo{year}{1984}.
\newblock \bibinfo{journal}{Monthly Notices of the Royal Astronomical Society}
  \bibinfo{volume}{209}, \bibinfo{pages}{27P--31P}.
\bibitem[{White and Frenk(1991)}]{white_galaxy_1991}
\bibinfo{author}{White, S.D.M.}, \bibinfo{author}{Frenk, C.S.},
  \bibinfo{year}{1991}.
\newblock \bibinfo{journal}{Astrophysical Journal} \bibinfo{volume}{379},
  \bibinfo{pages}{52--79}.
\bibitem[{White et~al.(1987b)White, Frenk, Davis and
  Efstathiou}]{white_clusters_1987}
\bibinfo{author}{White, S.D.M.}, \bibinfo{author}{Frenk, C.S.},
  \bibinfo{author}{Davis, M.}, \bibinfo{author}{Efstathiou, G.},
  \bibinfo{year}{1987}b.
\newblock \bibinfo{journal}{Astrophysical Journal} \bibinfo{volume}{313},
  \bibinfo{pages}{505--516}.
\bibitem[{White and Rees(1978)}]{white_core_1978}
\bibinfo{author}{White, S.D.M.}, \bibinfo{author}{Rees, M.J.},
  \bibinfo{year}{1978}.
\newblock \bibinfo{journal}{{MNRAS}} \bibinfo{volume}{183},
  \bibinfo{pages}{341--358}.
\bibitem[{Williams et~al.(2006)Williams, Mathur and
  Nicastro}]{williams_chandra_2006}
\bibinfo{author}{Williams, R.J.}, \bibinfo{author}{Mathur, S.},
  \bibinfo{author}{Nicastro, F.}, \bibinfo{year}{2006}.
\newblock \bibinfo{journal}{Astrophysical Journal} \bibinfo{volume}{645},
  \bibinfo{pages}{179--185}.
\bibitem[{Willman et~al.(2005)Willman, Dalcanton, {Martinez-Delgado}, West,
  Blanton, Hogg, Barentine, Brewington, Harvanek, Kleinman, Krzesinski, Long,
  Neilsen, Nitta and Snedden}]{willman_new_2005}
\bibinfo{author}{Willman, B.}, \bibinfo{author}{Dalcanton, J.J.},
  \bibinfo{author}{{Martinez-Delgado}, D.}, \bibinfo{author}{West, A.A.},
  \bibinfo{author}{Blanton, M.R.}, \bibinfo{author}{Hogg, D.W.},
  \bibinfo{author}{Barentine, J.C.}, \bibinfo{author}{Brewington, H.J.},
  \bibinfo{author}{Harvanek, M.}, \bibinfo{author}{Kleinman, S.J.},
  \bibinfo{author}{Krzesinski, J.}, \bibinfo{author}{Long, D.},
  \bibinfo{author}{Neilsen, E.H.}, \bibinfo{author}{Nitta, A.},
  \bibinfo{author}{Snedden, S.A.}, \bibinfo{year}{2005}.
\newblock \bibinfo{journal}{Astrophysical Journal} \bibinfo{volume}{626},
  \bibinfo{pages}{L85--L88}.
\bibitem[{Wise and Abel(2007a)}]{wise_suppression_2007}
\bibinfo{author}{Wise, J.H.}, \bibinfo{author}{Abel, T.},
  \bibinfo{year}{2007}a.
\newblock \bibinfo{journal}{Astrophysical Journal} \bibinfo{volume}{671},
  \bibinfo{pages}{1559--1567}.
\bibitem[{Wise and Abel(2007b)}]{wise_resolvingformation_2007}
\bibinfo{author}{Wise, J.H.}, \bibinfo{author}{Abel, T.},
  \bibinfo{year}{2007}b.
\newblock \bibinfo{journal}{Astrophysical Journal} \bibinfo{volume}{665},
  \bibinfo{pages}{899--910}.
\bibitem[{Wise and Abel(2008a)}]{wise_very_2008}
\bibinfo{author}{Wise, J.H.}, \bibinfo{author}{Abel, T.},
  \bibinfo{year}{2008}a.
\newblock \bibinfo{journal}{Astrophysical Journal} \bibinfo{volume}{684},
  \bibinfo{pages}{1--17}.
\bibitem[{Wise and Abel(2008b)}]{wise_resolvingformation_2008}
\bibinfo{author}{Wise, J.H.}, \bibinfo{author}{Abel, T.},
  \bibinfo{year}{2008}b.
\newblock \bibinfo{journal}{Astrophysical Journal} \bibinfo{volume}{685},
  \bibinfo{pages}{40--56}.
\bibitem[{Wise and Cen(2009)}]{wise_ionizing_2009}
\bibinfo{author}{Wise, J.H.}, \bibinfo{author}{Cen, R.}, \bibinfo{year}{2009}.
\newblock \bibinfo{journal}{Astrophysical Journal} \bibinfo{volume}{693},
  \bibinfo{pages}{984--999}.
\bibitem[{Wise et~al.(2008)Wise, Turk and
  Abel}]{wise_resolvingformation_2008-1}
\bibinfo{author}{Wise, J.H.}, \bibinfo{author}{Turk, M.J.},
  \bibinfo{author}{Abel, T.}, \bibinfo{year}{2008}.
\newblock \bibinfo{journal}{Astrophysical Journal} \bibinfo{volume}{682},
  \bibinfo{pages}{745--757}.
\bibitem[{Wright(1750)}]{wright_original_1750}
\bibinfo{author}{Wright, T.}, \bibinfo{year}{1750}.
\newblock \bibinfo{address}{London}.
\bibitem[{Wuyts et~al.(2009)Wuyts, Franx, Cox, {{F\"orster} Schreiber},
  Hayward, Hernquist, Hopkins, Labb\'e, Marchesini, Robertson, Toft and van
  Dokkum}]{wuyts_color_2009}
\bibinfo{author}{Wuyts, S.}, \bibinfo{author}{Franx, M.}, \bibinfo{author}{Cox,
  T.J.}, \bibinfo{author}{{{F\"orster} Schreiber}, N.M.},
  \bibinfo{author}{Hayward, C.C.}, \bibinfo{author}{Hernquist, L.},
  \bibinfo{author}{Hopkins, P.F.}, \bibinfo{author}{Labb\'e, I.},
  \bibinfo{author}{Marchesini, D.}, \bibinfo{author}{Robertson, B.E.},
  \bibinfo{author}{Toft, S.}, \bibinfo{author}{van Dokkum, P.G.},
  \bibinfo{year}{2009}.
\newblock \bibinfo{journal}{The Astrophysical Journal} \bibinfo{volume}{700},
  \bibinfo{pages}{799--819}.
\bibitem[{Wyithe and Loeb(2003)}]{wyithe_self-regulated_2003}
\bibinfo{author}{Wyithe, J.S.B.}, \bibinfo{author}{Loeb, A.},
  \bibinfo{year}{2003}.
\newblock \bibinfo{journal}{Astrophysical Journal} \bibinfo{volume}{595},
  \bibinfo{pages}{614--623}.
\bibitem[{Yajima et~al.(2009)Yajima, Umemura, Mori and
  Nakamoto}]{yajima_escape_2009}
\bibinfo{author}{Yajima, H.}, \bibinfo{author}{Umemura, M.},
  \bibinfo{author}{Mori, M.}, \bibinfo{author}{Nakamoto, T.},
  \bibinfo{year}{2009}.
\newblock \bibinfo{journal}{Monthly Notices of the Royal Astronomical Society}
  \bibinfo{volume}{398}, \bibinfo{pages}{715--721}.
\bibitem[{Yang et~al.(2003)Yang, Mo and van~den Bosch}]{yang_constraining_2003}
\bibinfo{author}{Yang, X.}, \bibinfo{author}{Mo, H.J.},
  \bibinfo{author}{van~den Bosch, F.C.}, \bibinfo{year}{2003}.
\newblock \bibinfo{journal}{Monthly Notices of the Royal Astronomical Society}
  \bibinfo{volume}{339}, \bibinfo{pages}{1057--1080}.
\bibitem[{Yoshida et~al.(2003)Yoshida, Abel, Hernquist and
  Sugiyama}]{yoshida_simulations_2003}
\bibinfo{author}{Yoshida, N.}, \bibinfo{author}{Abel, T.},
  \bibinfo{author}{Hernquist, L.}, \bibinfo{author}{Sugiyama, N.},
  \bibinfo{year}{2003}.
\newblock \bibinfo{journal}{Astrophysical Journal} \bibinfo{volume}{592},
  \bibinfo{pages}{645--663}.
\bibitem[{Yoshida et~al.(2006)Yoshida, Omukai, Hernquist and
  Abel}]{yoshida_formation_2006}
\bibinfo{author}{Yoshida, N.}, \bibinfo{author}{Omukai, K.},
  \bibinfo{author}{Hernquist, L.}, \bibinfo{author}{Abel, T.},
  \bibinfo{year}{2006}.
\newblock \bibinfo{journal}{Astrophysical Journal} \bibinfo{volume}{652},
  \bibinfo{pages}{6--25}.
\bibitem[{Yoshida et~al.(2002)Yoshida, Stoehr, Springel and
  White}]{yoshida_gas_2002}
\bibinfo{author}{Yoshida, N.}, \bibinfo{author}{Stoehr, F.},
  \bibinfo{author}{Springel, V.}, \bibinfo{author}{White, S.D.M.},
  \bibinfo{year}{2002}.
\newblock \bibinfo{journal}{Monthly Notices of the Royal Astronomical Society}
  \bibinfo{volume}{335}, \bibinfo{pages}{762--772}.
\bibitem[{Younger and Bryan(2007)}]{younger_cosmological_2007}
\bibinfo{author}{Younger, J.D.}, \bibinfo{author}{Bryan, G.L.},
  \bibinfo{year}{2007}.
\newblock \bibinfo{journal}{Astrophysical Journal} \bibinfo{volume}{666},
  \bibinfo{pages}{647--657}.
\bibitem[{Younger et~al.(2009)Younger, Hayward, Narayanan, Cox, Hernquist and
  Jonsson}]{younger_merger-driven_2009}
\bibinfo{author}{Younger, J.D.}, \bibinfo{author}{Hayward, C.C.},
  \bibinfo{author}{Narayanan, D.}, \bibinfo{author}{Cox, T.J.},
  \bibinfo{author}{Hernquist, L.}, \bibinfo{author}{Jonsson, P.},
  \bibinfo{year}{2009}.
\newblock \bibinfo{journal}{Monthly Notices of the Royal Astronomical Society}
  \bibinfo{volume}{396}, \bibinfo{pages}{L66--L70}.
\bibitem[{Yu(2002)}]{yu_evolution_2002}
\bibinfo{author}{Yu, Q.}, \bibinfo{year}{2002}.
\newblock \bibinfo{journal}{Monthly Notices of the Royal Astronomical Society}
  \bibinfo{volume}{331}, \bibinfo{pages}{935--958}.
\bibitem[{Y\"uksel et~al.(2008)Y\"uksel, Beacom and
  Watson}]{yuksel_strong_2008}
\bibinfo{author}{Y\"uksel, H.}, \bibinfo{author}{Beacom, J.F.},
  \bibinfo{author}{Watson, C.R.}, \bibinfo{year}{2008}.
\newblock \bibinfo{journal}{Physical Review Letters} \bibinfo{volume}{101},
  \bibinfo{pages}{121301}.
\bibitem[{Zavala et~al.(2008)Zavala, Okamoto and Frenk}]{zavala_bulges_2008}
\bibinfo{author}{Zavala, J.}, \bibinfo{author}{Okamoto, T.},
  \bibinfo{author}{Frenk, C.S.}, \bibinfo{year}{2008}.
\newblock \bibinfo{journal}{Monthly Notices of the Royal Astronomical Society}
  \bibinfo{volume}{387}, \bibinfo{pages}{364--370}.
\bibitem[{Zhang et~al.(2008)Zhang, Fakhouri and Ma}]{zhang_to_2008}
\bibinfo{author}{Zhang, J.}, \bibinfo{author}{Fakhouri, O.},
  \bibinfo{author}{Ma, C.}, \bibinfo{year}{2008}.
\newblock \bibinfo{journal}{Monthly Notices of the Royal Astronomical Society}
  \bibinfo{volume}{389}, \bibinfo{pages}{1521--1538}.
\bibitem[{Zhao et~al.(2009)Zhao, Jing, Mo and B\"orner}]{zhao_accurate_2009}
\bibinfo{author}{Zhao, D.H.}, \bibinfo{author}{Jing, Y.P.},
  \bibinfo{author}{Mo, H.J.}, \bibinfo{author}{B\"orner, G.},
  \bibinfo{year}{2009}.
\newblock \bibinfo{journal}{Astrophysical Journal} \bibinfo{volume}{707},
  \bibinfo{pages}{354--369}.
\bibitem[{Zhao et~al.(2007)Zhao, Hooper, Angus, Taylor and
  Silk}]{zhao_tidal_2007}
\bibinfo{author}{Zhao, H.}, \bibinfo{author}{Hooper, D.},
  \bibinfo{author}{Angus, G.W.}, \bibinfo{author}{Taylor, J.E.},
  \bibinfo{author}{Silk, J.}, \bibinfo{year}{2007}.
\newblock \bibinfo{journal}{Astrophysical Journal} \bibinfo{volume}{654},
  \bibinfo{pages}{697--701}.
\bibitem[{Zheng et~al.(2007)Zheng, Coil and Zehavi}]{zheng_galaxy_2007}
\bibinfo{author}{Zheng, Z.}, \bibinfo{author}{Coil, A.L.},
  \bibinfo{author}{Zehavi, I.}, \bibinfo{year}{2007}.
\newblock \bibinfo{journal}{The Astrophysical Journal} \bibinfo{volume}{667},
  \bibinfo{pages}{760--779}.

\end{thebibliography}

\end{document}